\documentclass[useAMS,usenatbib]{mn2e}

\usepackage[dvips]{graphicx}

\title[Warm Molecular Gas in Abell 2597 and Sersic 159-03]{The Distribution and Condition of the Warm Molecular Gas in Abell 2597 and Sersic 159-03}

\author[J. B. R. Oonk et al.]{J. B. R. Oonk$^{1}$, W. Jaffe$^{1}$, M. N. Bremer$^{2}$, R. J. van Weeren$^{1}$\\
$^1$Leiden Observatory, Leiden University, P.B. 9513, Leiden, 2300 RA, The Netherlands\\
$^2$Department of Physics, H.H. Wills Physics Laboratory, Bristol University, Tyndall Avenue, Bristol BS8 ITL, United Kingdom}
\begin{document}

\date{Accepted 2010 February 15}

\pagerange{\pageref{firstpage}--\pageref{lastpage}} \pubyear{2010}

\maketitle

\label{firstpage}

\begin{abstract}
We have used the SINFONI integral field spectrograph to map the near-infrared K-band emission lines of molecular and ionised hydrogen in the central regions of two cool core galaxy clusters, Abell 2597 and Sersic 159-03. Gas is detected out to 20~$\mathrm{kpc}$ from the nuclei of the brightest cluster galaxies and found to be distributed in clumps and filaments around it. The ionised and molecular gas phases trace each other closely in extent and dynamical state. Both gas phases show signs of interaction with the active nucleus. Within the nuclear regions the kinetic luminosity of this gas is found to be somewhat smaller than the current radio luminosity. Outside the nuclear region the gas has a low velocity dispersion and shows smooth velocity gradients. There is no strong correlation between the intensity of the molecular and ionised gas emission and either the radio or X-ray emission. The molecular gas in Abell 2597 and Sersic 159-03 is well described by a gas in local thermal equilibrium (LTE) with a single excitation temperature $\textit{T}_{\mathrm{exc}}$~$\sim$~2300~$\mathrm{K}$. The emission line ratios do not vary strongly as function of position, with the exception of the nuclear regions where the ionised to molecular gas ratio is found decrease. These constant line ratios imply a single source of heating and excitation for both gas phases.
\end{abstract}


\section{Introduction}
Cool cores are regions at the centre of rich clusters where the hot thermal X-ray emitting gas ($\textit{T}$~$\sim$ 10$^{8}$~$\mathrm{K}$) is dense enough to cool radiatively within a Hubble time \citep[see][for reviews]{P06,F94}. Cooling rates of the order of 100~$\mathrm{M_{\odot}}$~$\mathrm{yr}^{-1}$ and up to 1000~$\mathrm{M_{\odot}}$~$\mathrm{yr}^{-1}$ have been estimated for this hot X-ray gas \citep[e.g.,][]{P98}. However, recent \textit{Chandra} and \textit{XMM-Newton} X-ray spectra show that little or no X-ray emitting gas ($<$10\%) cools below one third of the virial temperature \citep[e.g.,][]{K01,P06}. The solution most often invoked in the literature is that some form of reheating balances the radiative cooling of the X-ray gas.

Substantial cooler gas and dust components exist in the cores of these galaxy clusters \citep*{E01,I01,S03,O08}. Extended $10^{4}$~$\mathrm{K}$ emission-line nebulae are found surrounding Brightest Cluster Galaxies (BCG) \citep[][herafter J05]{H89,C99,J05}. These nebulae are observed to extend at least up to 50~$\mathrm{kpc}$ from the BCG nuclei (J05). This component at $\textit{T}$~$\sim$ $10^{4}$~$\mathrm{K}$ emits far more energy than can be explained by the simple cooling of the intracluster gas through this temperature i.e., additional heating is needed \citep{F81,H89}.

More recently, work in the infrared and mm-wavelengths has shown that there are large quantities of molecular gas at the centre of these clusters with temperatures between 15 and 2500~$\mathrm{K}$ extending at least up to 20~$\mathrm{kpc}$ from the BCG nuclei \citep*[e.g., J05;][]{J97,J01,F98,E01,E02,W02,S03,H05,J07,W09}. The molecular gas has a cooling time of order years \citep{L83,B87,M96}. Without some form of heating one would expect this gas to collapse rapidly and form stars. Although there is strong evidence that starformation does take place at the centres of cool core galaxy clusters it is not yet observed to match the extended gas nebulae \citep[][Oonk et al. in prep.]{M92,O08}.

The heating and cooling of the molecular and ionised gas phases are important elements in the energetics of the cool core region. An energy source comparable to that needed to stop the hot X-ray gas from cooling is necessary to heat these colder phases (J05). The primary source of ionisation and heating of the H$_{\mathrm{2}}$ and HII must be local to the gas \citep[J05;][]{J88}, consistent with a stellar photoionising source. However, stars are unable to explain the high temperature of the ionised gas \citep[][hereafter VD97]{V97}. The molecular hydrogen lines are much stronger relative to the ionised hydrogen lines than in other types of extragalactic sources, such as AGN or starburst galaxies \citep[e.g., J05;][]{D03,D05}.

The ratio of H$_{\mathrm{2}}$ to HII emission lines \citep[J05;][]{H05}, as well as detailed analysis of the mid-infrared and optical line ratios \citep[VD97;][]{J07} indicate that to explain both heating and ionisation balance, photons harder than those available from O-stars are needed. However, often very high ionisation lines are missing (e.g., [Ne~V] -- typical of AGN spectra. If present, the source of these photons is elusive. Alternatively shock heating has been considered, however the characteristic [O~III] 4363 Angstrom line is missing (VD97). High energy particles have been invoked to explain this problem \citep{F09}. It is of great importance to pinpoint the nature of the heating mechanism and include it in models of cooling flows in galaxy clusters as well as models of galaxy growth and evolution.

Cool core clusters are the low redshift cluster scale analogs of high redshift galaxy scale cooling flows. To understand the formation of massive galaxies at high redshift and the feeding and feedback mechanisms in AGNs it is important to understand the heating of the cool gas in BCGs.

All gas phases observed in the intracluster medium require reheating to avoid catastrophic cooling. A variety of heat sources to counteract this cooling gas have been proposed over the years: AGN feedback \citep*[e.g.,][]{C01,B03,D04,M01,Bi04}, low velocity shock waves \citep{F06}, conduction (VD97), hot stars (VD97) and energetic particles \citep{F09}. None of these heat sources have so far been able to match the detailed characteristics of cool core galaxy clusters. Whatever the heat mechanism, the cooling region extends over hundreds of $\mathrm{kpc}$ across the cluster core, and heating is unlikely to balance the cooling exactly over such a large region. Some residual cooling will occur and presumably corresponds to the emission line nebulae and star clusters surrounding brightest cluster galaxies (BCG) at the centres of cool core galaxy clusters \citep[J05;][]{H05,R06}.
\\
\\
The BCGs at centres of cooling clusters fall within a region of the BPT diagram \citep*[][VD97]{B81,C99} that is occupied by LINERs and AGN. In our previous samples \citep[J05;][]{J97,J01} we have focused on the LINER-like BCGs and we continue to do so here. These clusters were originally selected based on their high cooling rates, strong H$\alpha$, H$_{\mathrm{2}}$ emission and low ionisation radiation. LINER-like BCGs were chosen because we wish to minimise the role that their AGN have on the global radiation field. In the work presented here we focus on two LINER-like BCGs from our previous samples, PGC 071390 in Abell 2597 (hereafter A2597) and ESO 0291-G009 in Sersic 159-03 (hereafter S159). Optical ([O~III]/H$\beta$ and [O~I]/H$\alpha$, \citet[][VD97;]{BTh}) and mid-infrared spectra ([Ne~III]/[Ne~II] and [Ne~V]/[Ne~II], Jaffe \& Bremer in prep.) of these BCGs indicate that their ionising spectrum is very soft i.e. they are extreme LINERs \citep[VD97;][]{BTh}.

However, these BCGs do contain radio-loud AGN. Their 1.4~$\mathrm{GHz}$ radio specific luminosity is, 29.3$\times$10$^{31}$ and 1.6$\times$10$^{31}$~$\mathrm{erg}$~$\mathrm{s}^{-1}$~$\mathrm{Hz}^{-1}$ for A2597 and S159 \citep{Bi08} respectively, which are typical for BCGs in cool core galaxy clusters \citep{Q08}. In this work we will concern ourselves with the extended molecular and ionised gas surrounding the BCGs in A2597 and S159.

A2597 and to a lesser extent S159 have been the subject of numerous investigations and have been observed at many wavelengths from radio to X-rays \citep[e.g., J05; VD97;][]{O94,O04,C05}. In both clusters ionised and molecular gas was observed to at least 50~$\mathrm{kpc}$ and 20~$\mathrm{kpc}$ from their BCG nuclei respectively \citep[J05;][]{H89}. Previous investigations of these objects made use of narrowband imaging and longslit spectra. Using long slit observations we were only able to sample parts of the extended line emission and with narrowband imaging no information on the dynamics of the gas is obtained. Furthermore the emission sampled with longslits in previous observations is often strongly dominated by strong emission from the BCG nucleus. As we will show below, the emission away from the nuclei has very different characteristics.

There are a number of kinematic problems concerning the cooler gas phases in cool core clusters. In nearby clusters ionised and molecular gas is found in thin, long lived, multi-$\mathrm{kpc}$ scale, filamentary structures surrounding the BCG \citep[e.g.,][]{F08,H05,Cr05}. These structures show smooth velocity gradients, but no rotation beyond the central few $\mathrm{kpc}$ \citep*{W06,H07}. The molecular clouds in these structures are dense and without kinematic support should fall to the galaxy centre. However, they show no signs of infalling. Velocities barely exceed 100~$\mathrm{km}$~$\mathrm{s}^{-1}$ \citep*[J05;][and this work]{W09}, whereas infall velocities should exceed $\sim$600~$\mathrm{km}$~$\mathrm{s}^{-1}$. Magnetic support has been invoked by \citet{F08} but there is no observational evidence yet for the required ordered magnetic fields in clusters. There has also been speculation whether or not all the molecular and ionised gas is locked up in these dense filaments or if a more diffuse underlying component exists. 

\subsection{This project}
Here we present the first deep K-band integral field (IFU) spectroscopic observations of the cluster cores in A2597 and S159, taken with the Spectrograph for INtegral Field Observations in the Near-Infrared (SINFONI) on the Very Large Telescope (VLT). With these observations we sample the molecular and ionised gas phases over a major fraction of each cluster's BCG. Our observations are able to provide information on the distribution, kinematics and temperature of this gas. Using these measurements we can compare in detail the kinematic and thermal structure of the gas with the X-ray and radio structures, which represent respectively the primary source of mass in the environment and the primary source of local energy input. Similar data on three other cool core clusters has recently been presented by \citet{W09}, but we are the first to make a detailed comparison of such data with radio and X-ray observations of cool core clusters

In Section 2 we describe the observations and the data reduction. We discuss the morphology and kinematics of the molecular and ionised gas in A2597 in Sections 3 and 4 and similarly for S159 in Sections 5 and 6. In Section 7 we will discuss the excitation of the molecular gas and in Section 8 we compare the observed gas structure to high resolution X-ray and Radio maps. We summarise our results in Section 9 and present our conclusions in Section 10.

Throughout this paper we will assume the following cosmology; $\textit{H}_{\mathrm{0}}$=72~$\mathrm{km}$~$\mathrm{s}^{-1}$~$\mathrm{Mpc}^{-1}$, $\Omega_{\mathrm{m}}$=0.3 and $\Omega_{\mathrm{\Lambda}}$=0.7. For Abell 2597 at z=0.0821 (VD97) this gives a luminosity distance 363~$\mathrm{Mpc}$ and angular size scale 1.5~$\mathrm{kpc}$~$\mathrm{arcsec}^{-1}$. For Sersic159-03 at z=0.0564 \citep{M87} this gives a luminosity distance 245~$\mathrm{Mpc}$ and angular size scale 1.1~$\mathrm{kpc}$~$\mathrm{arcsec}^{-1}$.

\section{Observations and reduction}

\subsection{Near Infrared Data}
The near infrared (NIR) observations were performed in the K-band with the integral field spectrograph SINFONI \citep{E03,Bo04} on the Very Large Telescope (VLT). SINFONI is an image slicer, slicing the image into strips before dispersing the light using 32 slitlets. The instrument has a spectral resolving power of $\textit{R}\approx$~4000 in the K-band. Opting for a 8$\arcsec\times$8$\arcsec$ field of view (FOV) the spatial pixels each cover an area of 0.125$\arcsec\times$0.250$\arcsec$. Each spectral pixel covers 2.45$\times$10$^{-4}$~$\mu \mathrm{m}$ in wavelength, oversampling the resolution by a factor two (i.e. Nyquist sampling). The total on-source exposure time for each source is listed in Table \ref{tabl_obs_summ}.

The observations were done in a 'ABBA' pattern (A=source, B=sky) and each set was followed by a pointing observation to keep track of pointing drifts. All observations were done such that the FOV was oriented with north pointing up. The spatial extent of each slitlet is then oriented east-west. Equal amounts of time were spend on the sky and on the source. Each science observation has an exposure time of 600 seconds. Each pointing observation has an exposure time of 60 seconds. The observations were carried out in July and August of 2005 in photometric sky conditions with a typical seeing of about 0.9 arcsec in K-band.

Four fields were observed for A2597 and three fields for S159. These fields were selected to lie within areas known to have extended H$\alpha$ emission (J05). The observed spectral and spatial resolution, as measured from telluric lines and standard star observations, is summarised in Table \ref{tabl_smooth_resolution}.

\subsubsection{Initial Reduction and Slit Definition}
The reduction of the data was done using a combination of the ESO SINFONI pipeline recipes (\textit{SINFONI pipeline version 1.7.1} and \textit{CPL version 3.6.1}), ECLIPSE \citep{D01} and a set of dedicated IDL procedures. From the SINFONI pipeline we obtain a masterdark frame, masterflat frame, hot pixel map and a slit curvature model. Wavelength calibration, hot pixel removal, slit edge detection, distortion correction, sky removal and illumination correction as given by the pipeline were found to be inadequate for our purposes and therefore an additional set of reduction tasks was written in IDL.

The reduction was carried out as follows. Source and sky frames are corrected for dark current and flat fielded using the masterflat and masterdark frames from the SINFONI pipeline. Having estimated the slit edges (by eye) the different slits are defined and cut out of the science frames. Each of these slits is then treated independently in the subsequent reduction steps. CCD artefacts are removed from the data. Hot pixels and those affected by cosmic rays are interpolated over.

We correct for the spatial curvature of the slit optics as imaged on the detector by applying the curvature model obtained by the SINFONI pipeline using the ECLIPSE task \textit{warping} \citep{D01}. Slit columns which do not contain information across the full wavelength range are removed. This also removes the overlap between different slits as imaged on the CCD.

The spectra are wavelength calibrated using a set of 19 identified telluric OH lines in the wavelength range 1.95-2.30~$\mu \mathrm{m}$ \citep{R00}. The output wavelength scale is set to 2.45$\times$10$^{-4}$~$\mu \mathrm{m}$ per pixel thereby Nyquist sampling the data.

\subsubsection{Sky Removal}
The K-band night sky is variable on short time scales. We have rather long exposure times, as compared to the variations in the sky background. This means that there is a complicated relationship between the sky contribution in our source frame and the sky observed in our corresponding sky frame. This is readily observed by subtracting two sky frames taken directly after each other, and leads to systematic residuals of up to 10\% in the peak heights of telluric lines. A scaling between the source and its corresponding sky frame thus needs to be performed in order to decrease these residuals. The standard sky scaling performed by the ESO SINFONI pipeline reduces these residuals to about 5\% and the special SINFONI pipeline sky correction utility reduces the residuals further to about 4\%. In both cases it was noted that the sky removal suffered from poor wavelength matching between the sky and source frames due to flexure of the instrument.

In our approach we remove the sky emission after detailed wavelength calibration using the telluric lines. We a apply a simple scaling function to the sky frame before subtracting it from the source frame. This scaling function consists of a constant and a small, linear, wavelength dependent factor. The constant is determined from the relative heights, above the continuum, of the telluric lines and assumed to hold at 2.1~$\mu \mathrm{m}$. The small, linear, wavelength dependent factor is the slope of a linear fit to the ratio of the object spectrum and its corresponding sky spectrum.

The full wavelength dependent behaviour of the sky emission between an object and a sky frame is often more complex than the simple linear function used. Here we are only interested in line emission and as such a small residual gradient in the continuum emission does not affect the analysis performed below. Our method leads to residuals that are $\le$2\% in the peak heights of telluric lines. This is a significant improvement over the other methods mentioned above. In the final analysis of the line emission we checked our results carefully for telluric line residuals and removed wavelength regions affected by these from our analysis.

\subsubsection{Illumination Correction}
After correcting for any residual distortion we collapse the sky and the sky subtracted source spectra into cubes with pixel size (0.125$\arcsec$,0.125$\arcsec$,0.000245~$\mu \mathrm{m}$). It is known that SINFONI, after all reduction steps described above, still has a varying illumination across its FOV and that this illumination is a function of wavelength (J. Reunanen priv. comm.). This is mostly due to a difference in the illumination of the various slitlets and most easily observed in the sky cubes. Defining a reference slitlet in the sky cubes we determine the variation in illumination across the FOV for each wavelength. We then correct for this variation in the sky subtracted source cubes. The correction is typically less than 10\% and particularly affects wavelengths below 2.1~$\mu \mathrm{m}$.

\subsubsection{Flux Calibration}
Flux calibration is carried out using one or, if available, multiple standard star observations per night. The standard star observations are reduced in the same way as outlined for the object observations above. All standard stars observed were either O or B stars, and brighter than 8th magnitude in the K-band. The atmospheric transmission function was determined by dividing the spatially summed standard star spectrum with a black body spectrum of the appropriate temperature. The absolute flux scale is set by using 2MASS K-band magnitudes, these are accurate to 0.05 magnitude in K-band.

\subsubsection{Extracting the line emission}
\begin{figure*}
    \includegraphics[width=0.40\textwidth, angle=90]{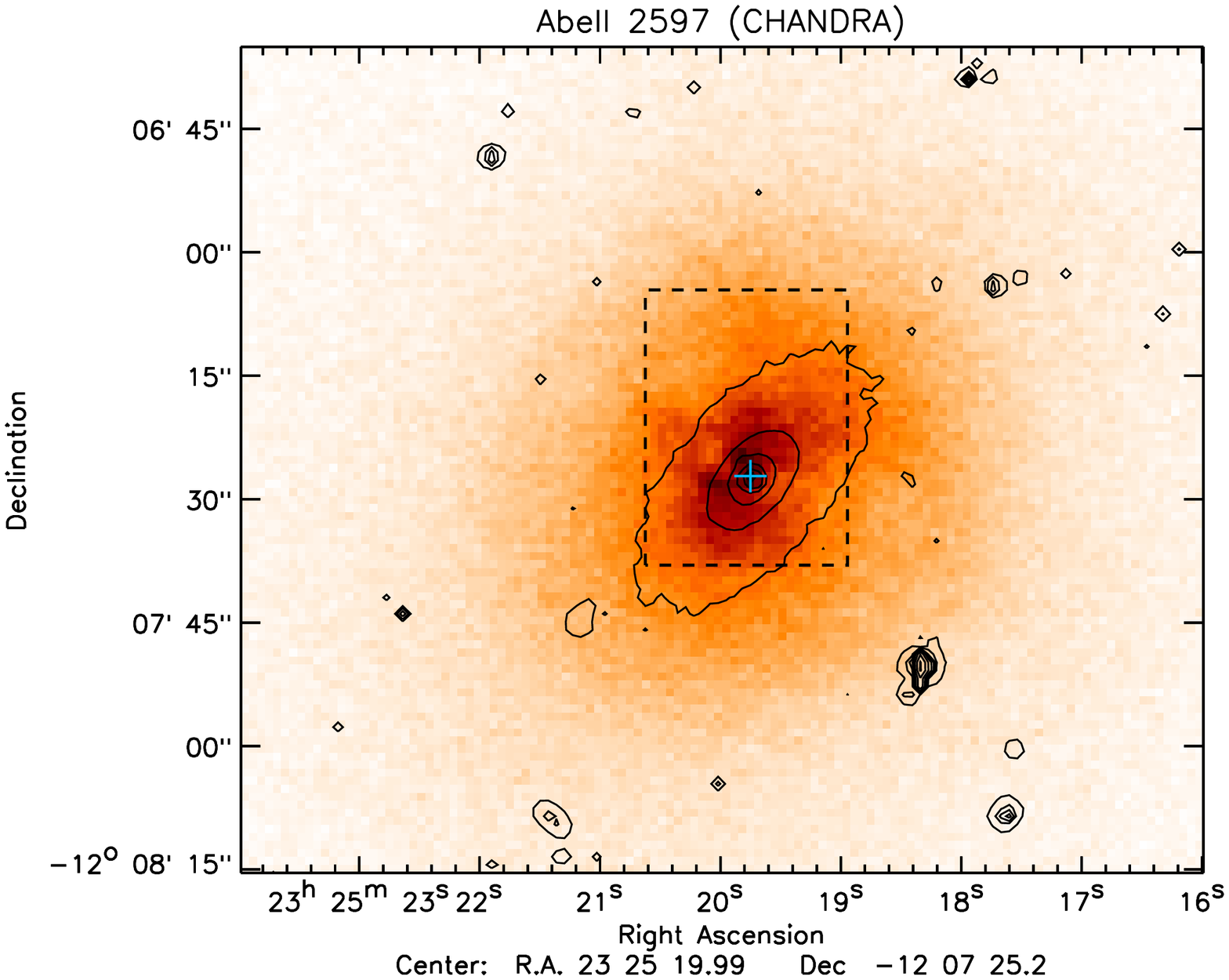}
    \includegraphics[width=0.40\textwidth, angle=90]{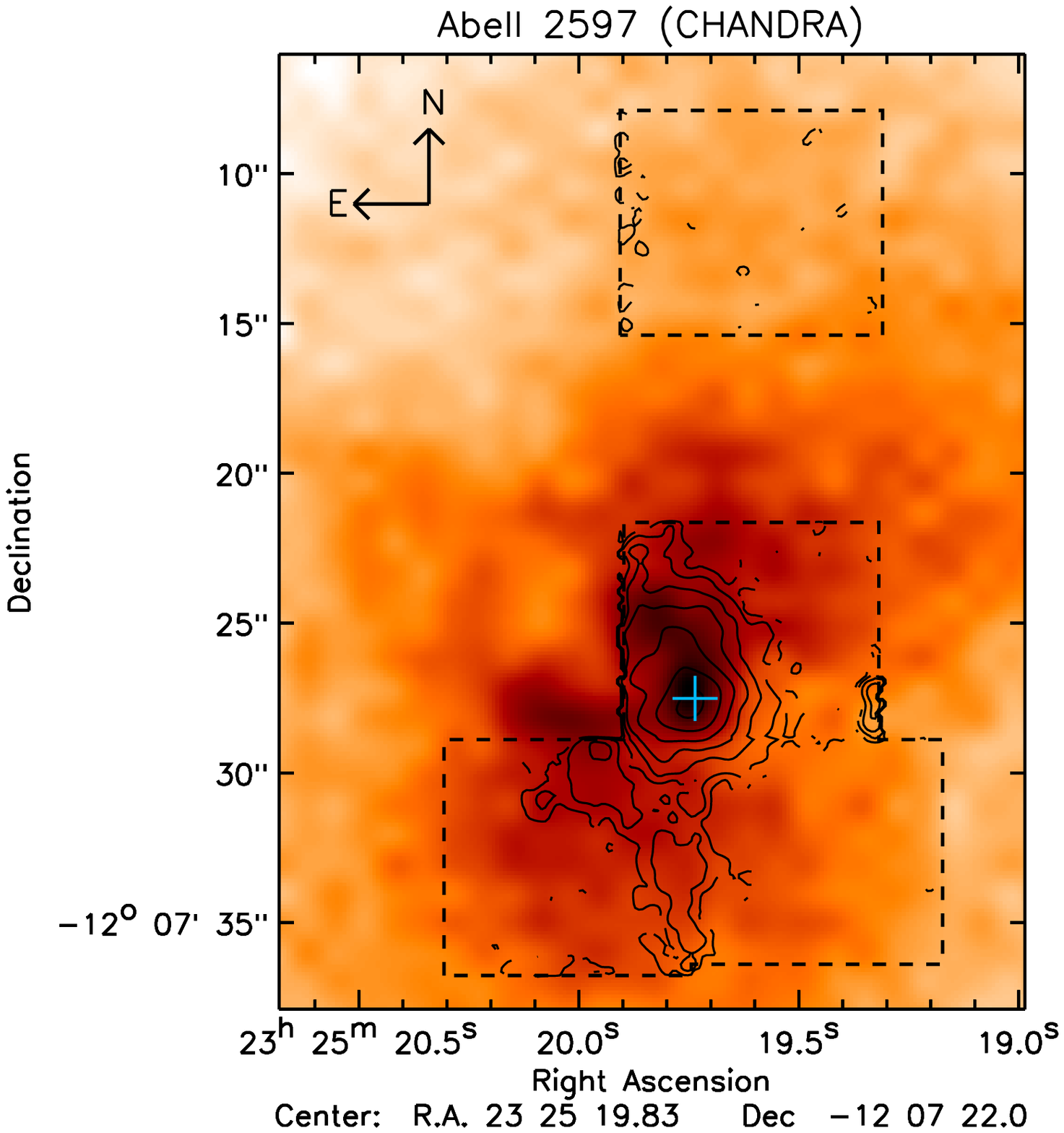}
  \vspace{0.5cm}
  \caption{ABELL 2597 Observed fields. (\textit{left}) extent of the X-ray emission in the cluster core as observed by \textit{CHANDRA} (exposure time 153.7 $\mathrm{ks}$). VLT FORS \textit{R}-band contours are overlayed in black (solid line). (\textit{right}) zoom in of the region marked by the black dashed box in the left image. The fields observed by SINFONI are indicated with a dashed line in the right image. H$_{\mathrm{2}}$ 1-0 S(3) contours are overlayed in black (solid line), drawn starting at 1.5 $\times$ 10$^{-17}$ $\mathrm{erg}$ $\mathrm{s}^{-1}$ $\mathrm{cm}^{-2}$ $\mathrm{arcsec}^{-2}$ in steps of 2$^{\mathrm{n}}$ with n=0,1,2,... The blue plus sign marks the position of the stellar nucleus.}\label{fig_area_a2597_ss44}
\end{figure*}

\begin{figure*}
    \includegraphics[width=0.40\textwidth, angle=90]{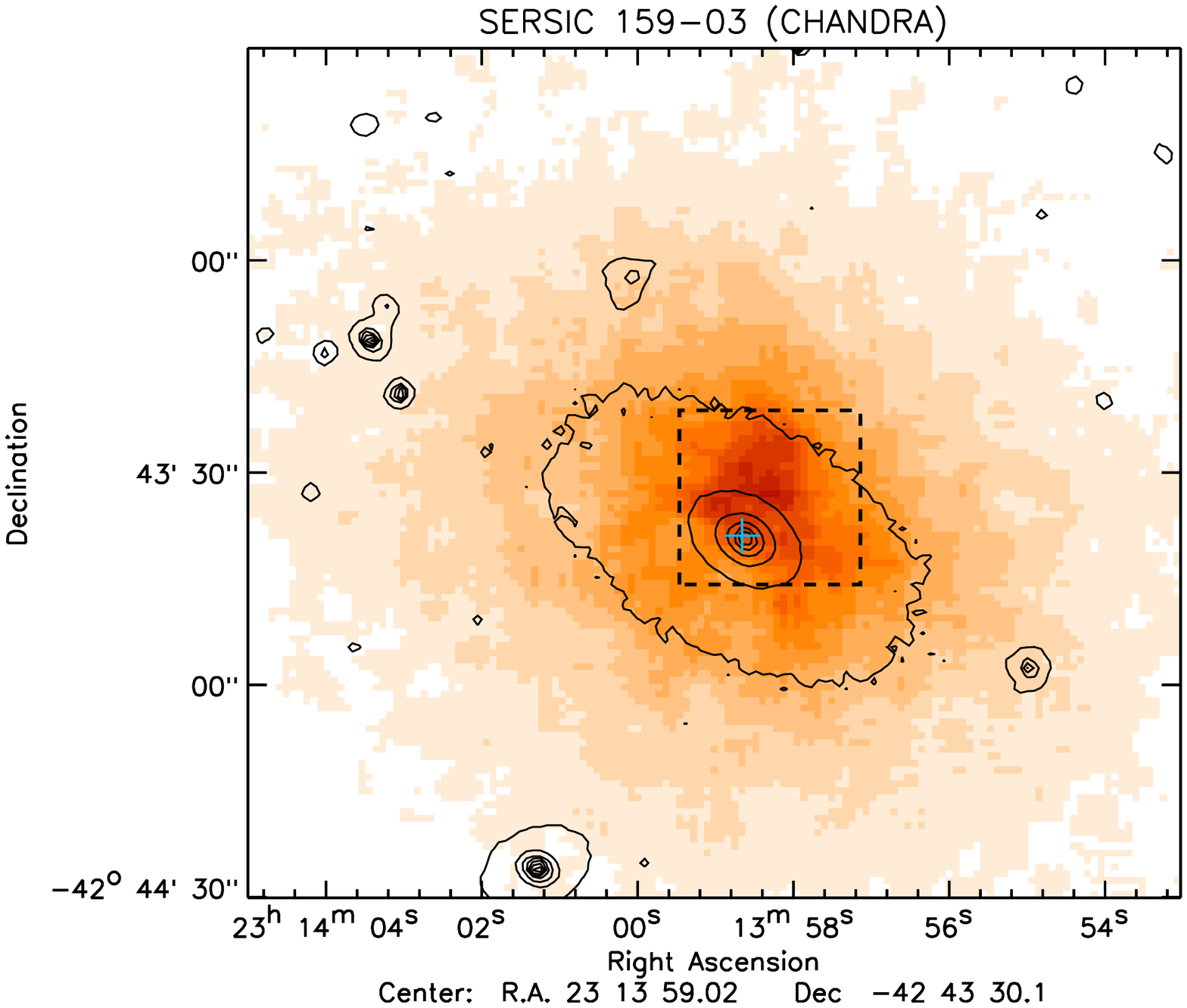}
    \includegraphics[width=0.40\textwidth, angle=90]{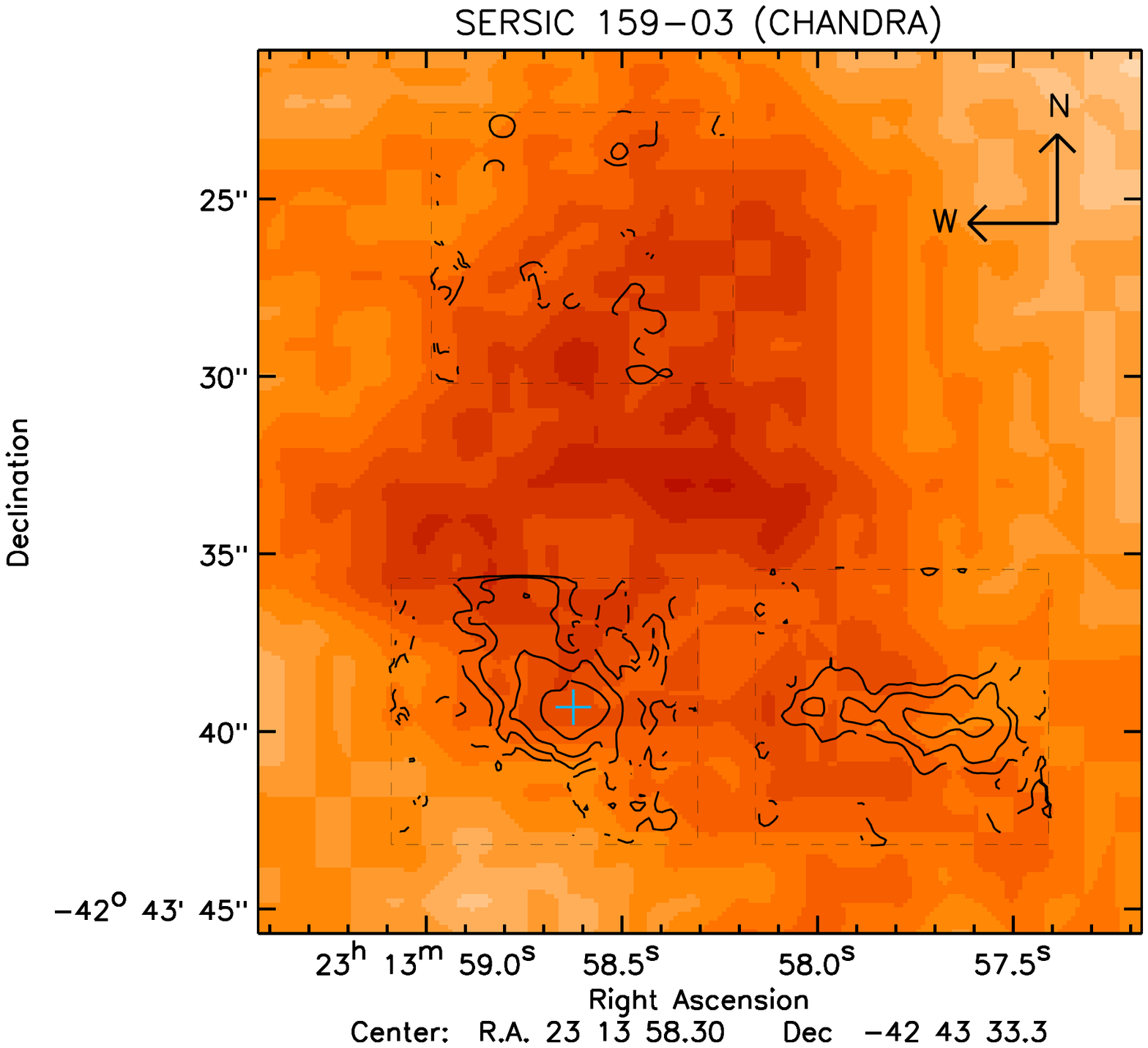}
  \vspace{0.5cm}
  \caption{SERSIC 159-03 Observed fields . (\textit{left}) the extent of the X-ray emission in the cluster core as observed by \textit{CHANDRA} (exposure time 10.1 ks), has been smoothed slightly. VLT FORS \textit{R}-band contours are overlayed in black (solid line). (\textit{right}) zoom in of the region marked by the black dashed box in the left image. The fields observed by SINFONI are indicated with a dashed line in the right image. H$_{\mathrm{2}}$ 1-0 S(3) contours are overlayed in black (solid line), drawn starting at 1.5~$\times$~10$^{-17}$~$\mathrm{erg}$~$\mathrm{s}^{-1}$~$\mathrm{cm}^{-2}$~$\mathrm{arcsec}^{-2}$ in steps of 2$^{\mathrm{n}}$ with n=0,1,2,... The blue plus sign marks the position of the stellar nucleus.}\label{fig_area_sersic_ss44}
\end{figure*}

Following flux calibration the source cubes are combined. The northern and southern edges of the exposures for the different fields overlap well. The eastern and western edges overlap less well leading to a higher noise here. The most northern and southern slitlet have very low signal to noise and were removed from the data. Any remaining telluric emission is removed by defining off-source regions. These are marked by the dotted lines in Figs. \ref{fig_area_a2597_ss44} and \ref{fig_area_sersic_ss44}.

Continuum emission is removed by fitting the continuum in the immediate neighbourhood of the science line. The continuum subtracted line feature is fit by a single Gaussian function, using the $\textit{mpfitpeak}$ routine \citep{M09} within IDL. It is observed that a single Gaussian always provides a good description of the observed line profile. The line flux, centre and width are determined from this Gaussian fit. For selected regions line profiles and Gaussian fits to them are shown in Appendix \ref{app_tb}. The errors quoted for the fitted line properties are based on Monte-Carlo simulations.

\subsubsection{Constructing the line maps}
In order to visualise the surface brightness of the line emission we performed a Gaussian smoothing of four pixels full width at half maximum (FWHM) in both the spatial (4 pixels=0.5$\arcsec$) and the spectral plane (4 pixels=9.80$\times$10$^{-4}$~$\mu \mathrm{m}$). To visualise the kinematics of the line emitting gas a two pixel FWHM Gaussian spatial smoothing and no spectral smoothing was found to be adequate for A2597, whereas for S159 a two pixel FWHM Gaussian smoothing in both the spatial and spectral planes was performed. The degradation of the spatial and spectral resolution as a function of the smoothing kernel used is given in Table \ref{tabl_smooth_resolution}. The corresponding noise is given in Tables \ref{tabl_smooth_sensitivity_a2597} and \ref{tabl_smooth_sensitivity_sersic}.

Surface brightness maps for all lines that could be mapped on a pixel to pixel basis are shown in Appendix \ref{app_lmp_a2597} and \ref{app_lmp_sersic}. For A2597 the northern field is not shown as the signal to noise here is inadequate to show the emission on the same spatial resolution as the central and southern fields. Velocity and velocity dispersion are shown only for the strongest detected ionised and molecular gas line. We note that the velocity structure observed in all detected emission lines agrees with that shown for these lines.

\subsection{X-ray Data}
Cool core clusters were first discovered by analysing their X-ray emission. These observations lead to the still unresolved cooling flow problem for the hot X-ray gas \citep[e.g.][]{P06}. In this paper we are concerned with the cooler HII and H$_{\mathrm{2}}$ gas phases and will not focus on the cooling flow problem related to the hot X-ray gas. However, in Section 8 we will investigate whether there is a spatial correlation between the X-ray emitting gas and cooler gas phases. In order to do so we have obtained all available X-ray data from the \textit{CHANDRA} archive. The A2597 image, Fig. \ref{fig_area_a2597_ss44}, is a co-add of three separate observations having a combined exposure time of 153.7~$\mathrm{ks}$ (project codes 7329; 6934; 922). The S159 image, Fig. \ref{fig_area_sersic_ss44}, consists of only one shallow 10.1~$\mathrm{ks}$ observation (project code 1668).

\textit{CHANDRA} data for A2597 and S159 has previously been published by \citet{M01} and J05. A very notable difference in the X-ray emission for the two clusters is that the peak emission in A2597 is well aligned with the nucleus of the BCG, whereas in S159 the peak emission is about 8$\arcsec$ north of the BCG nucleus.

\subsection{Radio Data}
Out of the many heat sources proposed, AGN feedback has received the most attention in recent years. The observed anti-correlation between X-ray and radio emission, also referred to as \textit{X-ray cavities} and \textit{Radio bubbles}, has led to models in which the AGN outflows interact strongly with its surrounding medium \citep[e.g.][]{Su07}. The kinetic energy carried by these outflows has been calculated based on these X-ray cavities and recent results show that the mechanical power of the jet that created the X-ray cavities can be orders of magnitude larger than its radio inferred radiative power \citep{Bi04,Bi08,D06}. In Section 8 we will compare our SINFONI results with high resolution radio maps to investigate how the current AGN outflows interact with the cooler gas in the cores of A2597 and S159.

A2597: A VLA 8.4 $\mathrm{GHz}$ map of A2597 was obtained from C. Sarazin \citep{S95}. This map has a beam size of 0.26$\arcsec\times$0.21$\arcsec$. The one sigma noise is 50~$\mu\mathrm{Jy}$~$\mathrm{beam}^{-1}$. It was noted that there is a significant offset of $\sim$0.1 seconds in Right Ascension as compared to the 2MASS position of the BCG. Two more 8.4~$\mathrm{GHz}$ maps have been published \citep{Bi08,D00}. These have a much better agreement with the 2MASS position. We thus conclude that this offset is an error and shift the 8.4~$\mathrm{GHz}$ map accordingly. Detailed radio maps at lower frequencies have been published by \citet{C05} and show that there is more low level radio emission present than apparent from the 8.4~$\mathrm{GHz}$ map. However, the 8.4~$\mathrm{GHz}$ map does give a good indication of the current AGN outflows. A much higher resolution radio map at 1.3 and 5.0~$\mathrm{GHz}$ using very long baseline array (VLBA) interferometry has been published by \citet{T99}. These observations show that the current jet has a position angle (PA) of 70 degrees.

S159: Archival VLA 8.4 GHz observations of S159 (project code: AB1190) were reduced with the NRAO Astronomical Image Processing System (AIPS). The B-configuration observations were taken in single channel continuum mode with two IFs centred at 8435 and 8485 MHz. The total on source time was 103 min. The data was flux calibrated using the primary calibrator 0137+331. We used the Perley \& Taylor 1999 extension to the \citet{B77} scale to set the absolute flux scale. Amplitude and phase variations were tracked using the secondary calibrator 2314-449 and applied these to the data. The data was imaged using robust weighting set to 0.5, giving a beam size of 3.26$\arcsec\times$0.67$\arcsec$. The one sigma map noise is 25~$\mu\mathrm{Jy}$~$\mathrm{beam}^{-1}$. Radio maps of S159 at 1.4, 5.0 and 8.4~$\mathrm{GHz}$ were previously published by \citet{Bi08}.

\section{Abell 2597 -- Gas Distribution}
Four 8$\arcsec\times$8$\arcsec$ fields were observed on and surrounding the BCG PGC071390 in A2597, see Fig. \ref{fig_area_a2597_ss44}. The integration time for each exposure is 600 seconds. The central field, which includes the nucleus of PGC071390, contains 13 exposures. The south-eastern (SE) and south-western (SW) field contain 8 and 15 exposures respectively. The northern field contains 13 exposures. The overlap region between the central and southern fields is sufficient for the line emission to be mapped without problems. However, the SE and SW fields do not completely overlap everywhere. In various locations along the overlap area there are small gaps between the fields of one to two pixels (1 spatial pixel=0.125$\arcsec$). We interpolated these before mapping the emission. Despite this, due to the increased noise at the east, west edges of each field, this overlap region (about 10 pixels in width) between the southern fields has a rather poor signal to noise. The northern field is offset by about 6$\arcsec$ from the central field.

A four pixel spatial and spectral smoothing was applied to the data prior to fitting the lines. A single Gaussian function provides a good fit to the observed line profile. Surface brightness maps for all other lines that could be mapped on a pixel to pixel basis are shown in Appendix \ref{app_lmp_a2597}. The northern field is not shown in these images because the signal to noise here is inadequate to show the emission on the same spatial resolution as the central and southern fields.

\subsection{Molecular gas}
The integrated line fluxes for all lines detected within the observed fields for A2597 are given in Table \ref{tabl_line_summ_a2597_ss44}. All H$_{\mathrm{2}}$~1-0 and H$_{\mathrm{2}}$~2-1 S-transitions redshifted into the K-band (1.95-2.40~$\mathrm{\mu m}$) are detected. A flux value for the H$_{\mathrm{2}}$ 2-1 S(4) line has been omitted due to uncertain continuum subtraction. The H$_{\mathrm{2}}$~2-1~S(5) and the Br~$\delta$ line are too closein wavelength to be separated by our observations. None of the H$_{\mathrm{2}}$~3-2 S-transitions were detected. As an example of the fidelity of the data we show the full K-band spectrum of the nuclear region and the south eastern filament in Fig. \ref{fig_fs_a2597_ss44}.

The H$_{\mathrm{2}}$ surface brightness maps all show the same structure. As an example of the molecular gas distribution we show the surface brightness map for the H$_{\mathrm{2}}$~1-0~S(3) line in Fig. \ref{fig_sign_maps_a2597_ss44}. This map clearly shows that the peak of the molecular gas emission coincides with the stellar nucleus of PGC071390. Two extended gas structures away from the nucleus are observed. One extends north from the nucleus and hence we will refer to this structure as the northern filament. The second structure extends from the north-east to the south-west in the SE field, just south from the nucleus and hence we will refer to this structure as the southern filament.

We observe that the surface brightness of the molecular gas varies rather smoothly within the nuclear region. However, from higher spatial resolution HST imaging by \citet[][hereafter D00]{D00} we know, that the molecular and ionised line emission in the central 4$\arcsec\times$4$\arcsec$ is concentrated in narrow clumpy, filamentary structures. Here we do not have the resolution to resolve these structures. We do note some enhanced emission features, embedded within the central field, extending to the north and east away from the nucleus which are roughly coincidental with some of structures observed by D00.

The northern filament extends at least up to the northern edge of the central field, i.e.,  6$\arcsec$ (9~$\mathrm{kpc}$) from the nucleus. This is well beyond the region in which molecular emission was detected by D00. Using deep K-band longslit spectra J05 have previously observed that the H$_{\mathrm{2}}$ emission extends at least up to 20 $\mathrm{kpc}$ towards the north from the nucleus. We will see below that molecular gas can still be found in the northern field observed by us, i.e., at a distance of about 22 $\mathrm{kpc}$ from the nucleus, thus confirming the J05 results.

The southern filament is clearly detected in the emission of the stronger lines. This southern filament has not been observed in D00, but J05 also find molecular gas south of the nucleus (see their figures 8 and 11). The extent of the northern and southern filaments observed here is bounded by the edges of the observed fields, and it is likely that these continue beyond the regions mapped by us.

\subsection{Ionised gas}
The Pa~$\alpha$ line is redshifted into the K-band for both galaxy clusters studied here. The line is redshifted into a region of poor atmospheric transmission, but it is the strongest ionised gas line by far in our spectra and unambiguously detected in both clusters. In A2597 the Pa~$\alpha$ emission globally follows the H$_{\mathrm{2}}$ emission closely, Fig. \ref{fig_sign_maps_a2597_ss44}. Within the nuclear region enhanced emission is again observed towards the north and east. These features are roughly coincidental with the emission line filaments observed in D00. Beyond the nuclear region the emission extends along the northern and southern filaments, peaking in the same locations as the H$_{\mathrm{2}}$ emission.

We also detect the Br~$\gamma$, Br~$\delta$ and Fe~II~(1.8100~$\mu \mathrm{m}$ rest wavelength) lines. The Br~$\delta$ line is blended with H2 2-1 S(5) and these cannot be disentangled directly by our observations. In the central field we find that Br~$\gamma$/Pa~$\alpha$ = 0.082. This ratio agrees with the dust-free Case B recombination ratio of these lines for $\textit{n}_{\mathrm{e}}$~=~10$^{2}$~$\mathrm{cm}^{-3}$ and $\textit{T}$~=~10$^{4}$~$\mathrm{K}$ \citep{O06}. The Case B scenario then implies that Br~$\gamma$/Br~$\delta$ = 1.5, and we use this ratio to disentangle the Br~$\delta$, H$_{\mathrm{2}}$ 2-1 S(5) blend. In the nuclesar region we find that Br~$\delta$ and H$_{\mathrm{2}}$~1-0~S(5) are of similar strength.

A small dust lane has been observed in the nuclear region of A2597 \citep[D00;][]{K99}. We have investigated whether differential extinction in the K-band may affect our emission line ratios. From the above value for the Br~$\gamma$/Pa~$\alpha$ ratio we find that differential extinction is unimportant in the K-band. This is confirmed by deep optical spectroscopy of A2597 by VD97 and \citet{BTh}. They find a V-band extinction A$_{\mathrm{V}}$~$\sim$~1 across the nebulosity. Assuming standard galactic dust (R$_{\mathrm{V}}$=3.1) an A$_{\mathrm{V}}$~$\sim$~1 translates in to A$_{\mathrm{K}}$~$\sim$~0.1. This amount of extinction is negligible.

The Fe~II~(1.81~$\mu \mathrm{m}$) line is redshifted to the short wavelength edge of our observed spectrum. It is unambiguously detected in the nuclear region. The decrease in the Fe II emission outwards from the nucleus, in both intensity and dispersion, appears to be much faster than for either the HII lines or the H$_{\mathrm{2}}$ lines. The HII emission drops by a factor of 3 and the H$_{\mathrm{2}}$ flux drops by a factor 4 from the nuclear region to a region just north of the nucleus. The Fe II emission drops by a factor of 10 for the same regions. If the Fe II emission has a different origin than hydrogen lines, for example if it is preferably coming from the AGN and the associated jet instead of the gaseous filaments, this may explain the difference. Our observations do not have the spatial resolution to investigate this in detail.

We have searched our spectra for the presence of even higher ionisation lines, such as the Si~VI~(1.9634~$\mu \mathrm{m}$) line, which one would expect from typical hard AGN spectra. None of these higher ionisation lines were detected. This once more confirms the LINER nature of PGC071390. It may also indicate that the active nucleus is not the main source of ionisation of the gas observed in the core of A2597. Alternatively it would have to have an atypically soft ionising spectrum.

\section{Abell 2597 -- Gas Kinematics}\label{sect_a2597_kinm}
A single Gaussian function gives a good description of the observed line profiles, see Appendix \ref{app_tb}. From the Gaussian fits of these line profiles we obtain information about the kinematical structure of the molecular and ionised gas in A2597. The velocity, with respect to the systemic velocity of PGC071390, and the velocity dispersion of the gas have been derived for all emission lines. These all show the same global kinematical structure. The velocity and velocity dispersion maps shown below differ from the surface brightness maps in that only a two pixel spatial smoothing has been applied, as supposed to a four pixel spatial and spectral smoothing. This is done to preserve as much of the velocity structure as possible and provides us with a velocity resolution of 38~$\mathrm{km}$~$\mathrm{s}^{-1}$.

\subsection{Molecular gas}
The molecular gas in A2597 shows a wealth of small scale kinematical structure. Velocity and velocity dispersion maps were made for all H$_{\mathrm{2}}$ lines. All show the same kinematical structure on all scales observed. As an example of this structure we have displayed the velocity and velocity dispersion maps for the H$_{\mathrm{2}}$~1-0~S(3) line in Figs. \ref{fig_kinm_velc_maps_a2597_ss20} and \ref{fig_kinm_disp_maps_a2597_ss20}. The nuclear region contains a blueshifted and a redshifted gas component at about $\pm$80~$\mathrm{km}$~$\mathrm{s}^{-1}$. This is seen more clearly if we place a pseudo slit with a width of 1$\arcsec$ and a PA of 105.5 degree, centred 1~$\mathrm{kpc}$ south of the stellar nucleus. The corresponding position-velocity diagram is shown in Fig. \ref{fig_a2597_pv}. The velocity structure observed in Figs. \ref{fig_kinm_velc_maps_a2597_ss20} and \ref{fig_a2597_pv} is reminiscent of gas rotating around the nucleus and does not appear to be related to an expanding shell or AGN outflows.

The average velocity of gas in the nuclear region is approximately zero with respect to the systemic velocity of PGC071390 (z=0.0821, VD97). This shows that the gas here is situated at or near the stellar nucleus. The reason for placing the pseudo slit slightly south of the nucleus is because east of the nucleus there is a small strongly redshifted feature at +150~$\mathrm{km}$~$\mathrm{s}^{-1}$. Whether this feature is part of the global gas flow or a single event is unclear. It shows up prominently in all velocity maps. Projected on to the sky, the feature appears to be coincident with the north-eastern radio jet of PKS2322-12 the radio source in PGC071390, see Fig. \ref{fig_kinm_velc_maps_a2597_ss20}. The filamentary structures extending towards the north and the south from the nucleus show smooth velocity gradients and these will be discussed in more detail below.

The velocity dispersion of the molecular gas also shows interesting structure. Globally the dispersion of the gas decreases with distance from the nucleus. It drops from an average of about 220~$\mathrm{km}$~$\mathrm{s}^{-1}$ in the nuclear region to about 100~$\mathrm{km}$~$\mathrm{s}^{-1}$ a few $\mathrm{kpc}$ north and the south of the nucleus. The velocity dispersion is very high in two narrow structures extending towards the east and south of the nucleus. The two-dimensional data allows us to determine the area which is disturbed to be an elongated structure of about 2~$\mathrm{kpc}$ by 5~$\mathrm{kpc}$ oriented at a PA of about 45 degrees.

Projected on to the sky these high dispersion structures appear to run along the inner, South East edge of the curved radio lobes of PKS2322-12, see Fig. \ref{fig_kinm_disp_maps_a2597_ss20}.  If we interpret the lobe morphology as a Wide Angle Tail, caused by the relative motion of the AGN and the external medium, then the dispersion map illustrates for the first time the turbulent wake expected from this motion. Alternatively, the region of maximum dispersion, at PA$\sim$70 degrees from the nucleus, may represent the interaction of the current, VLBI radio jet with the surrounding medium, as has been seen in Centaurus A \citep{N07}. In this picture we must assume that the counter-jet has been deflected near the nucleus to the South, causing the high dispersion region and radio lobe in this direction.  There is, however, no evidence for a major kinematic disturbance at the point of deflection.

The highest velocity dispersion is found for the small, strongly red-shifted feature east of the nucleus. This high velocity and dispersion for this feature can be explained if this is gas that is entrained within the AGN outflow. The feature aligns well with the current, projected jet axis \citep[PA=70 degrees,][]{T99}.  

\subsection{Ionised gas}
Velocity and the velocity dispersion maps for the ionised gas in A2597, as traced by the Pa~$\alpha$ line, are shown in Figs. \ref{fig_kinm_velc_maps_a2597_ss20} and \ref{fig_kinm_disp_maps_a2597_ss20}. We observe two key features when we compare the Pa~$\alpha$ and H$_{\mathrm{2}}$ derived kinematics. Firstly, globally we find that the Pa~$\alpha$ derived kinematics follows the H$_{\mathrm{2}}$ derived kinematics tightly. Secondly, the velocity dispersion of the Pa~$\alpha$ emitting gas appears on average to be slightly higher than the H$_{\mathrm{2}}$ emitting gas, especially in the nuclear region.

It may be possible that on scales below the resolution of our observations the ionised gas has a different distribution than the H$_{\mathrm{2}}$ emitting gas. This may be especially true in the nuclear region where the active nucleus appears to be strongly interacting with the gas. The position-velocity diagram shown in Fig. \ref{fig_a2597_pv} also shows that the ionised gas, as traced by the Pa~$\alpha$ and Fe~II lines, reaches slightly higher velocities in the nuclear region. D00 show that within the nuclear region the H$_{\mathrm{2}}$ and HII gas has a very complicated and disturbed morphology and it is difficult to say how well these two trace each other on small scales here.

The kinematics of the molecular and ionised gas for A2597 derived here agrees well with previous long slit investigations by J05 and \citet{H89}. \citet[][hereafter O94]{O94} detected HI in absorption against the radio continuum source PKS2322-12 in A2597. The absorption observation represents a line of sight of a few arcsec along the central radio source. They find a spatially resolved broad HI component with $\sigma\sim$174~$\mathrm{km}$~$\mathrm{s}^{-1}$ and a narrow unresolved HI component with $\sigma\sim$93~$\mathrm{km}$~$\mathrm{s}^{-1}$ at the position of the nucleus. The width of the broad component is somewhat smaller than the width observed in HII and H$_{\mathrm{2}}$. O94 find that the widths are consistent if one takes into account that the HI absorption measurements only sample the gas in front of the radio source, whereas the HII and $H_{\mathrm{2}}$ measurements sample all of the gas along the line of sight.

As in our data O94 find a narow and a broad component, but the relative strength of narrow component is much stronger in their observation. We do not see the narrow component on the nucleus. The dominance of the narrow component in the HI observations is probably caused by the 1/$\textit{T}_{\mathrm{s}}$ dependence of the HI absorption, as pointed out by O94. $\textit{T}_{\mathrm{s}}$ being the spin temperature of the HI gas. In the HI absorption spectra the cold gas at large radii in front of the nucleus is probably over-represented relative to the HII and H2 emission spectra. We conclude, as do O94, that there is no evidence for a kinematically distinct HI component.

\subsection{Filaments}\label{a2575_film_reg}
In Fig. \ref{fig_film_a2597_ss20} we show the surface brightness, velocity and velocity dispersion along the two filamentary structures we identified in our observations of A2597. The regions used for this investigation are marked by the green and red squares in Fig. \ref{fig_h2_area_ss44}. The black points in Fig. \ref{fig_film_a2597_ss20} correspond to green squares and the red points to the red squares. Following the northern filament from slightly south of the nucleus towards the northern edge of the central field we find that the Pa~$\alpha$/H$_{\mathrm{2}}$~1-0~S(3) is approximately equal to 0.75 in the nuclear region and rapidly increases to unity outwards. The northern filament shows a smooth velocity gradient from south to north across the nucleus, as the velocity decreases from +50~$\mathrm{km}$~$\mathrm{s}^{-1}$ to -50~$\mathrm{km}$~$\mathrm{s}^{-1}$~2$\arcsec$ north of the nucleus. At this point the velocity gradient reverses and the velocity increases again to +50~$\mathrm{km}$~$\mathrm{s}^{-1}$ towards the northern edge of the central field.

Velocity gradients and even reversals for this filament may be explained in terms of bending and stretching of the filament, perhaps due to a combination of its proper motion and gravitational forces. However, it is more likely that we observe multiple filaments, each with its own characteristic motion, along the line of sight. Our data shows that the eastern part of the northern filament is predominantly blue shifted whereas the western part is red shifted. Higher spatial resolution images taken with HST by \citet{O04} and Oonk et al. (in prep.) show evidence that the northern filament observed here consists of at least two filamentary structures. We thus favour the latter explanation for the observed velocity structure of the northern filament

This interpretation also agrees with what is observed for more nearby galaxy clusters such as Perseus and Centaurus, where a multitude of long, thin filaments are observed along the line of sight \citep{F08,Cr05,H05}. The narrow spatial and velocity range observed here for the filaments however still suggest that any substructure in it will likely have a common origin. If the gas observed in the northern filament is connected to the gas detected in the northern field its velocity continues to increase to about +150 $\mathrm{km}$~$\mathrm{s}^{-1}$, as also shown by J05. From the J05 observations it appears that the gas in the central field is joined smoothly with that in the northern field, in terms of both surface brightness and dynamics.

The velocity dispersion along the northern filament decreases smoothly from 220~$\mathrm{km}$~$\mathrm{s}^{-1}$ to 100~$\mathrm{km}$~$\mathrm{s}^{-1}$, from the nucleus to the edges of the central field. This decrease is fastest near the nucleus and slows down beyond 3~$\mathrm{kpc}$ north of the nucleus. This point may mark a change in the influence of the AGN upon the dynamical state of the gas.

The southern filament has a much lower surface brightness and is hence detected at a lower signal to noise. Variations along this filament are thus more difficult to detect. Following this filament from the north-east (NE) to the south-west (SW) we find that the surface brightness is highest at its NE edge whereafter it decreases slightly and becomes approximately constant. The velocity decreases from +50~$\mathrm{km}$~$\mathrm{s}^{-1}$ to about -40~$\mathrm{km}$~$\mathrm{s}^{-1}$. The velocity dispersion remains constant at about 100~$\mathrm{km}$~$\mathrm{s}^{-1}$ along the filament. We will discuss the stability of the observed filaments in more detail below.

\section{Sersic 159-03 -- Gas Distribution}
Three 8$\arcsec\times$8$\arcsec$ fields were observed on and surrounding the BCG ESO291-G009 in S159, Fig. \ref{fig_area_sersic_ss44}. The integration time for each exposure is 600s. The south-eastern (SE) and south-western (SW) fields contain 8 and 9 exposures respectively. The northern field contains 8 exposures. The SE field contains the nucleus of ESO 291-G009. There is no overlap between the three fields observed. A four pixel spatial and spectral smoothing was applied to the data prior to fitting the lines. A single Gaussian function provides a good fit to the observed line profiles. Surface brightness maps for all detected emission lines that could be mapped on pixel to pixel basis are shown in Appendix \ref{app_lmp_sersic}. 

\subsection{Molecular gas}
Only two out of five H$_{\mathrm{2}}$~1-0 S-transition lines redshifted into the K-band are unambiguously detected for S159. These are the H$_{\mathrm{2}}$~1-0~S(1) and H$_{\mathrm{2}}$~1-0~S(3) lines. Their resulting integrated line fluxes are given in Table \ref{tabl_line_summ_sersic_ss44}. The presence of the H$_{\mathrm{2}}$~1-0~S(2) line can also be confirmed, see also Fig. \ref{fig_fs_sersic_ss44}. This line cannot be fitted reliably due to a strong telluric line residual in the red wing of the line profile and as such a flux value has been omitted. Tentative evidence is found for the presence of the H$_{\mathrm{2}}$~1-0~S(0) and H$_{\mathrm{2}}$~1-0~S(4) lines. However, both lines lie in spectral regions of poor atmospheric transmission and are close to strong telluric features, preventing us from claiming detections. We have searched for the H$_{\mathrm{2}}$~2-1~S-transitions, but these and higher H$_{\mathrm{2}}$ transitions remain undetected in our spectra. Except for the above mentioned two H$_{\mathrm{2}}$ lines only the Pa $\alpha$ line is detected. Full K-band spectra of the nuclear region and the western filament are presented in Fig. \ref{fig_fs_sersic_ss44}.

The two H$_{\mathrm{2}}$ surface brightness maps show the same structure and as an example of this we show the map for the H$_{\mathrm{2}}$~1-0~S(3) line in Fig. \ref{fig_sign_maps_sersic_ss44}. This map clearly shows that the peak of the H$_{\mathrm{2}}$ emission is well aligned with the stellar nucleus of ESO0291-G009. A filament of molecular emission extends north-east from the nucleus up to the edge of the SE field. We will refer to this structure as the northern filament. Clumpy emission extends towards the west and the south of the nucleus, up to the edges of the SE field.

The SW field shows a strong filament of gas having an east-west elongation, originally discovered by \citet{C92}. We will refer to this structure as the western filament. The northern field shows two features (i) low signal to noise clumpy emission in the southern and central part of the field, and (ii) a stronger, somewhat larger emission feature at the northern edge of the field. Both features are treated in more detail below. Whether the clumpy emission observed in the northern field is part of the northern filament observed in the SE field is not clear from our observations.

The spatial extent of the gas for both the western and northern filament is bounded by the edges of our observed fields and it is likely that these filaments continue beyond the regions mapped here, as seems to be the case from narrowband H$\alpha$+[NII] imaging by J05 and \citet{H00}. 

\subsection{Ionised gas}
Pa~$\alpha$ is the only ionised gas line detected in our K-band spectra for S159. It is redshifted into a region of poor atmospheric transmission. Strong Pa~$\alpha$ emission is found along the northern and western filaments, Fig. \ref{fig_sign_maps_sersic_ss44}. We detect Pa~$\alpha$ in all places where H$_{\mathrm{2}}$ emission is detected. A noticeable difference concerns the nuclear region. Almost no Pa~$\alpha$ emission appears to be associated with the stellar nucleus. As we will see below, some Pa~$\alpha$ emission is found here, but there is a strong decrease of it relative to molecular hydrogen emission. The Pa~$\alpha$/H$_{\mathrm{2}}$~1-0~S(3) ratio is observed to drop from about 0.7 in the filaments to about 0.2 in the nuclear region. We note that the nuclear region has a rather high spectral noise, due to the strong stellar continuum, which affects our ability to detect emission lines here. The detection of the Pa~$\alpha$ line is furthermore complicated by it being in a region of poor atmospheric transmission.

Pa~$\alpha$ is also present within the northern field and, like the H$_{\mathrm{2}}$ gas, appears to be clumpy. The strong emission feature observed in H$_{\mathrm{2}}$ towards the northern edge of this field is also confirmed by Pa~$\alpha$ emission.

In S159 the Pa~$\alpha$ line is the only ionised gas line detected. An estimate of differential extinction within the K-band can thus not be made directly from our observations. \citet{BTh} finds finds little variation from A$_{\mathrm{V}}$~$\sim$~1 across the nebulosity. Applying the same arguments to S159 as for A2597, we conclude that differential extinction in the K-band is negligible for this cluster.

\section{Sersic 159-03 -- Gas Kinematics}
A single Gaussian function gives a good description of the observed line profiles, see Appendix \ref{app_tb}. The velocity, with respect to the systemic velocity of ESO 0291-G009, and the velocity dispersion of the gas have been derived for all emission lines. These all show the same global kinematical structure. The velocity and velocity dispersion maps shown below differ from the surface brightness maps in that only a two pixel spatial and spectral smoothing has been applied. This is done to preserve as much of the velocity structure as possible and provides us with a velocity resolution of 51~$\mathrm{km}$~$\mathrm{s}^{-1}$.

\subsection{Molecular gas}
The kinematical structure observed in the H$_{\mathrm{2}}$~1-0~S(1) and H$_{\mathrm{2}}$~1-0~S(3) maps is the same. As an example of this structure we show the velocity and velocity dispersion of the H$_{\mathrm{2}}$ gas as traced by the H$_{\mathrm{2}}$~1-0~S(3) line in Figs. \ref{fig_kinm_velc_maps_sersic_ss22} and \ref{fig_kinm_disp_maps_sersic_ss22}. The nuclear region contains a blue- and redshifted gas component at about $\pm$120~$\mathrm{km}$~$\mathrm{s}^{-1}$. This velocity structure is reminiscent of gas rotating around the nucleus. However, the two gas filaments coming in from the north-east and west towards the nucleus may also explain the observed velocity structure. The velocity along the filaments will be treated in more detail below. The gas extending towards the west and south from the nucleus appears to be predominantly blueshifted. The average velocity of the gas in the nuclear region is equal to the systemic velocity of ESO 291-G009 \citep[$\textit{z}$=0.0564,][]{M87} showing that the gas is situated at or near the stellar nucleus.

Globally the dispersion of the gas is low and decreases with distance from the nucleus. The dispersion of the gas in the nuclear region is about 230~$\mathrm{km}$~$\mathrm{s}^{-1}$, but drops rapidly to about 90~$\mathrm{km}$~$\mathrm{s}^{-1}$ along the two filaments. This is similar to what is observed in A2597. In projection the high dispersion structure around the nucleus appears coincidental with the radio jets of ESO0291-G009, see Fig. \ref{fig_kinm_disp_maps_sersic_ss22}. The increase in the velocity dispersion here again indicates that the radio jets are stirring up the gas.

From Figs. \ref{fig_kinm_velc_maps_sersic_ss22} and \ref{fig_kinm_disp_maps_sersic_ss22} it is difficult to draw conclusions on the average velocity and velocity dispersion of the clumpy low signal to noise emission in the northern field. From the previous investigation by J05 it appears that the gas observed here is connected to the strong filament extending north from the nucleus. Below, we will see that the clumpy emission has velocities varying between -20 and -60~$\mathrm{km}$~$\mathrm{s}^{-1}$ and a velocity dispersion less than 100~$\mathrm{km}$~$\mathrm{s}^{-1}$. If this emission is connected to the filament extending north this implies that the line emission continues to decrease with distance to nucleus as was also shown by J05.

\subsection{Ionised gas}
Velocity and velocity dispersion maps for the ionised gas in S159 as traced by the Pa~$\alpha$ line are shown in Figs. \ref{fig_kinm_velc_maps_sersic_ss22} and \ref{fig_kinm_disp_maps_sersic_ss22}. As for A2597 we find that the Pa~$\alpha$ derived kinematics follows that of the H$_{\mathrm{2}}$ lines closely. The only exception in S159 being the high dispersion nuclear region observed in H$_{\mathrm{2}}$ emission, which appears to be missed by the Pa~$\alpha$ emission. The high noise in the nuclear region of S159 combined with the poor atmospheric transmission in the wavelength range of the Pa~$\alpha$ line makes the fit of this line here rather difficult.

\subsection{Filaments}\label{sersic_film_reg}
In Fig. \ref{fig_film_sersic_ss22} we show the surface brightness, velocity and velocity dispersion along the two filaments observed in S159. The regions used for this investigation are marked by the green squares in Fig. \ref{fig_h2_area_ss44}. Following the northern filament from the north-eastern edge in the SE field southwards toward the nucleus and subsequently to the eastern edge of the SE field, we find that the Pa~$\alpha$/H$_{\mathrm{2}}$~1-0~S(3) decreases strongly in the nuclear region and is approximately constant outside of it. The filament has a smooth velocity gradient. North-east of the nucleus the velocity decreases very slightly with distance from about +130~$\mathrm{km}$~$\mathrm{s}^{-1}$ to about +90~$\mathrm{km}$~$\mathrm{s}^{-1}$. Across the nucleus the velocity changes from about +100~$\mathrm{km}$~$\mathrm{s}^{-1}$ to -100~$\mathrm{km}$~$\mathrm{s}^{-1}$. Whether this velocity reversal is due to rotation or due to the filamentary structure of the gas can not be clearly distinguished.

The velocity dispersion of the gas in the northern filament is low everywhere, except within the nuclear region. All detected lines show an increase in the velocity dispersion near the nucleus, but the increase in the H$_{\mathrm{2}}$ lines is much higher than for the Pa~$\alpha$ line. The decrease in the dispersion east of the nucleus is difficult to measure due to low signal to noise here.

The surface brightness of the western filament has two prominent peaks about 11$\arcsec$ from the nucleus. It has a very smooth velocity gradient along the filament, from about -100~$\mathrm{km}$~$\mathrm{s}^{-1}$ to about +20~$\mathrm{km}$~$\mathrm{s}^{-1}$. This velocity structure agrees with the possibility that it is connected to the nuclear region. The dispersion of the gas in this filament is low everywhere. All three lines detected show the same flux and velocity structure along the western filament. 

\section{Physical Conditions in the Warm Molecular Gas}
Molecular hydrogen emission can be excited by various processes, (1) thermal excitation produced by kinetic energy injection into the gas due to for example shock heating \citep{Ho89}, (2) fluorescence by soft-UV photons, i.e. Photo dissociation regions (PDR) \citep*{B87,S89}, (3) high energy X-ray photons, i.e. X-ray dissociation regions (XDR) \citep{M96} and (4) high energy particles \citep{L83,F09}. If the density of the molecular gas is above the critical density the gas is in local thermal equilibrium (LTE). It is then not possible to distinguish between the various H$_{\mathrm{2}}$ excitation mechanisms and one would observe a thermal H$_{\mathrm{2}}$ spectrum where the excitation temperature is equal to the kinetic temperature of this gas. For the photon and particle processes mentioned above the gas can be thermalised, via heating through secondary (suprathermal) electrons.

\begin{figure*}
    \includegraphics[width=0.45\textwidth, angle=90]{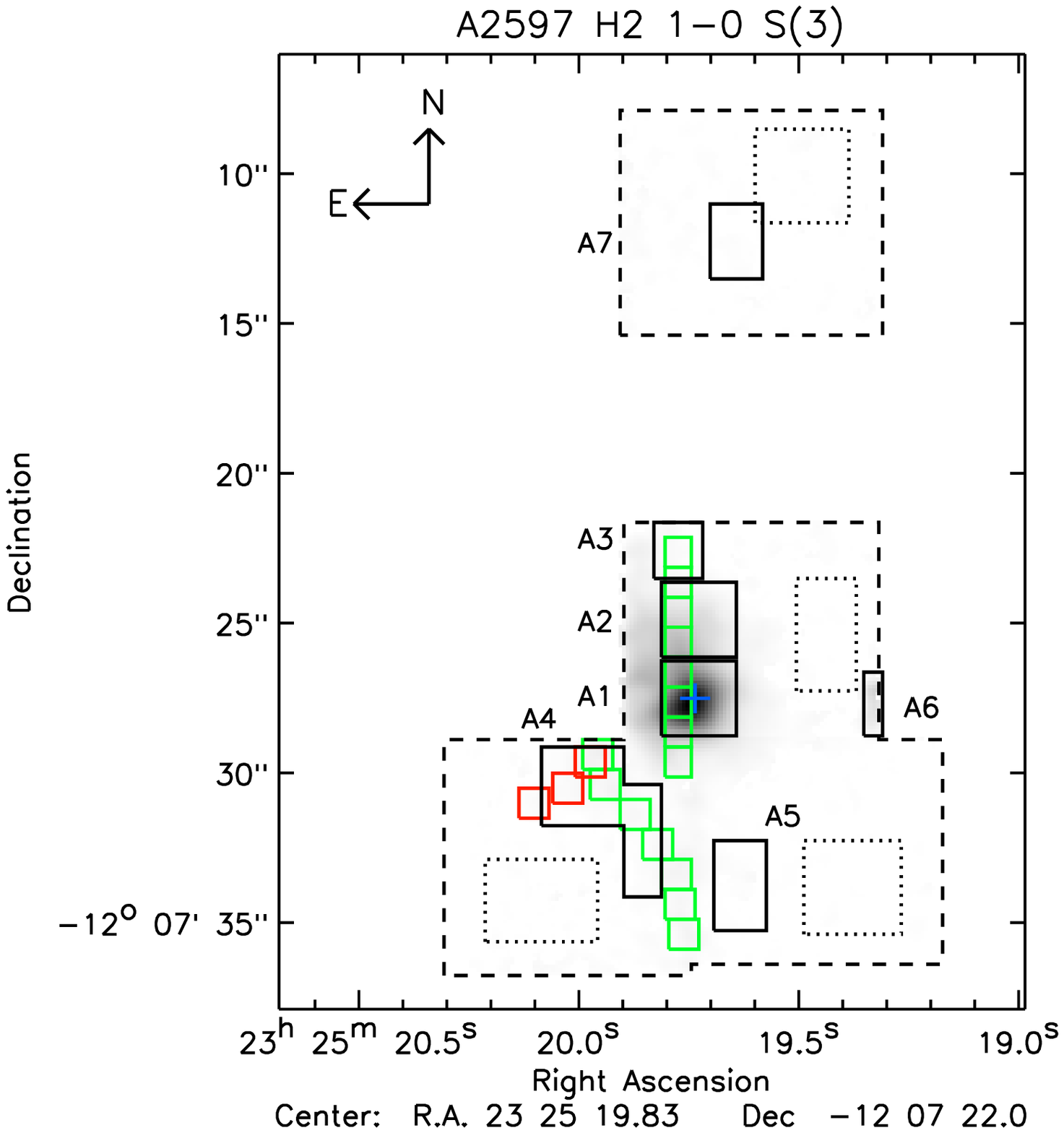}
    \includegraphics[width=0.45\textwidth, angle=90]{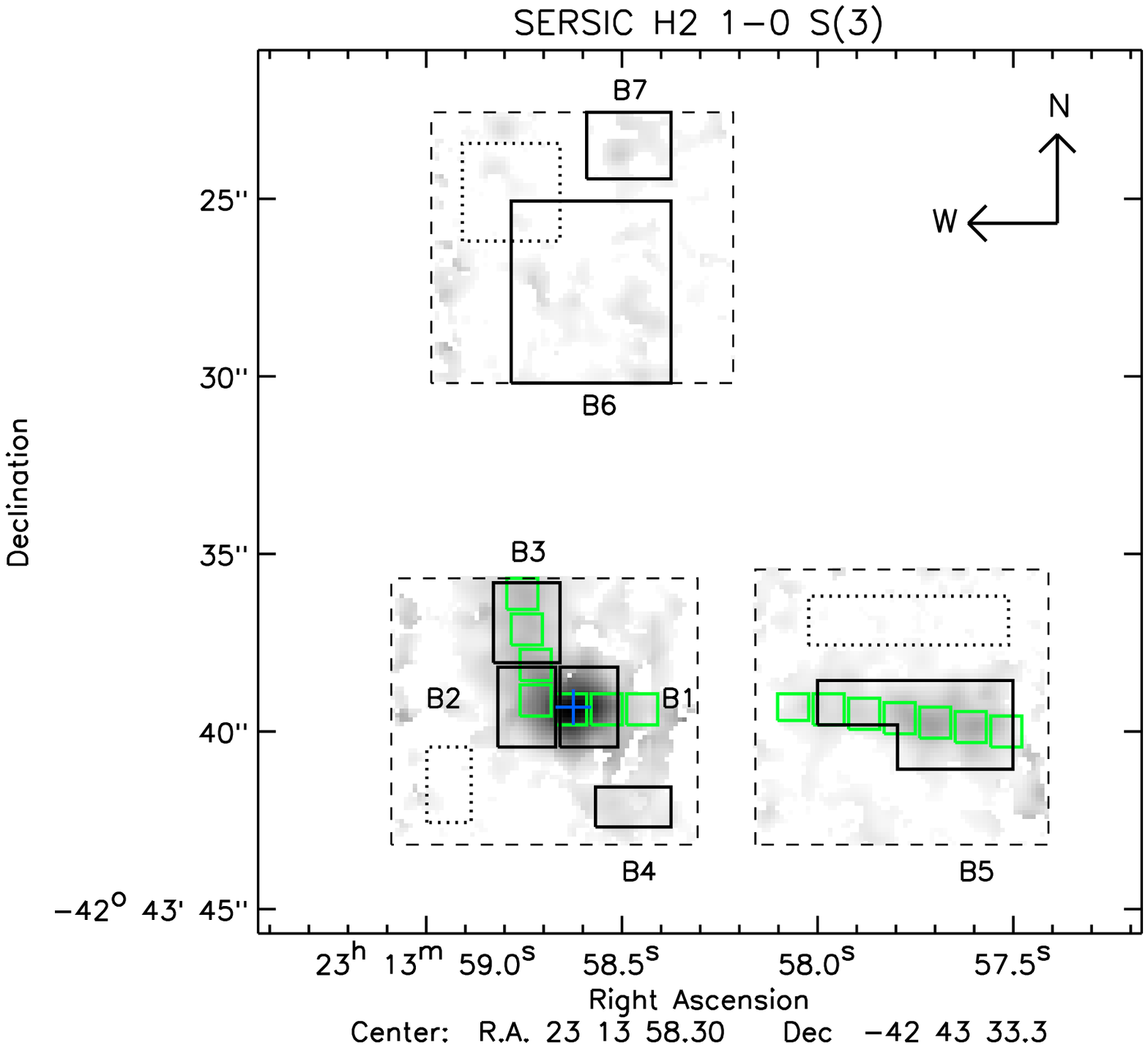}
  \vspace{0.5cm}
  \caption{Selected regions in ABELL 2597 and SERSIC 159-03. The regions are shown overlayed on SINFONI H$_{\mathrm{2}}$~1-0~S(3) surface brightness maps. The left image shows A2597 and the right image S159. The fields observed by SINFONI are indicated with a dashed black line. The dotted black lines show the regions used for obtaining the off-source spatial median in these fields. The solid black lines show the selected regions A1-A7 for A2597 and B1-B7 for S159, see Sections \ref{a2575_sel_reg} and \ref{sersic_sel_reg}. The small green and red squares show the regions investigated along the filaments, see Sections \ref{a2575_film_reg} and \ref{sersic_film_reg}. The blue plus sign marks the position of the stellar nucleus.}\label{fig_h2_area_ss44}
\end{figure*}

\begin{figure*}
    \includegraphics[width=0.30\textwidth, angle=90]{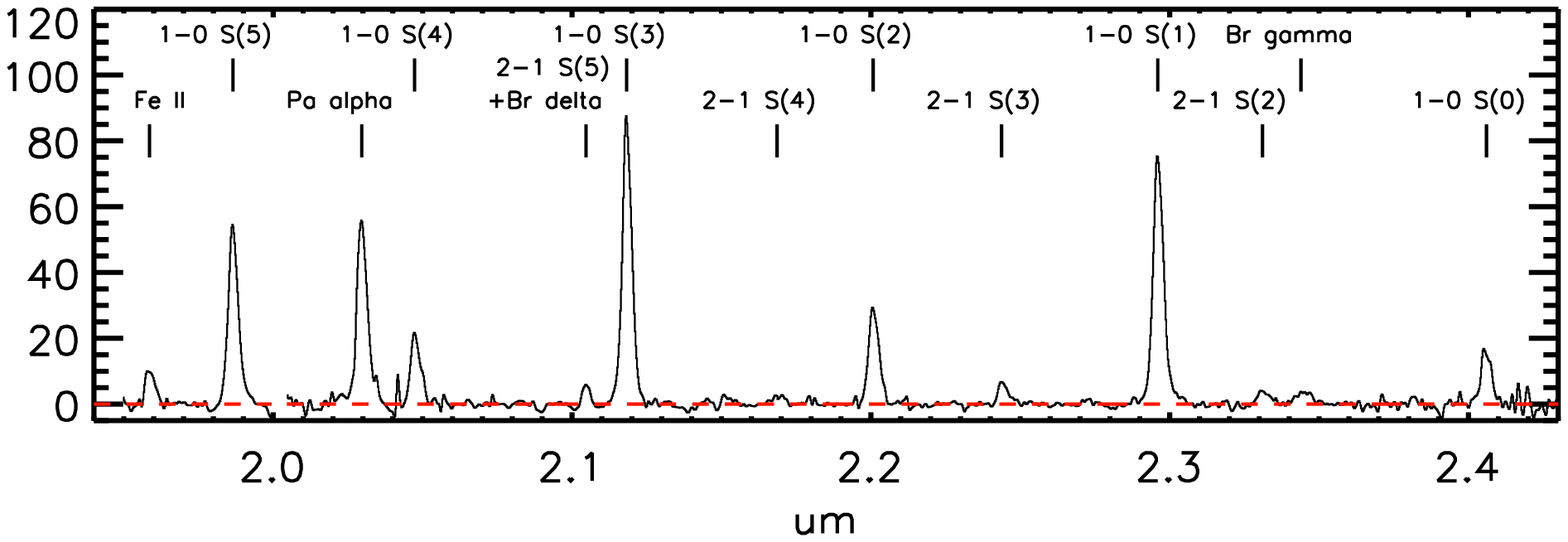}
    \includegraphics[width=0.30\textwidth, angle=90]{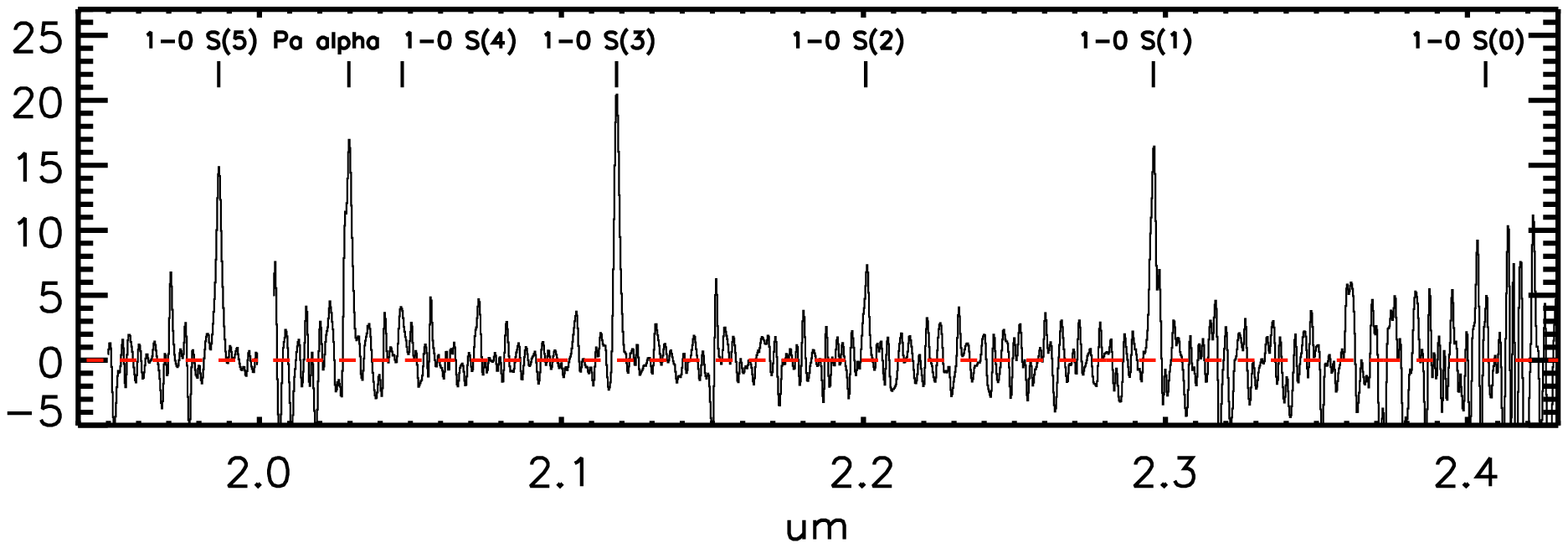}
  \vspace{0.5cm}
  \caption{ABELL 2597 full K-band spectra. As an example of the quality of our SINFONI spectra we show full K-band spectra for two regions. The top image shows the spectrum for region A1 and the bottom image shows the spectrum for region A4, see Fig. \ref{fig_h2_area_ss44} and Section \ref{a2575_sel_reg}. These spectra were obtained after smoothing the data by four pixels in both the spatial and spectral planes. The horizontal axis is given in units of $\mu$m and the vertical axis is given in units of 10$^{-14}$~$\mathrm{erg}$~$\mathrm{cm}^{-2}$~$\mathrm{s}^{-1}$~$\mu \mathrm{m}^{-1}$.}\label{fig_fs_a2597_ss44}
\end{figure*}

\begin{figure*}
    \includegraphics[width=0.30\textwidth, angle=90]{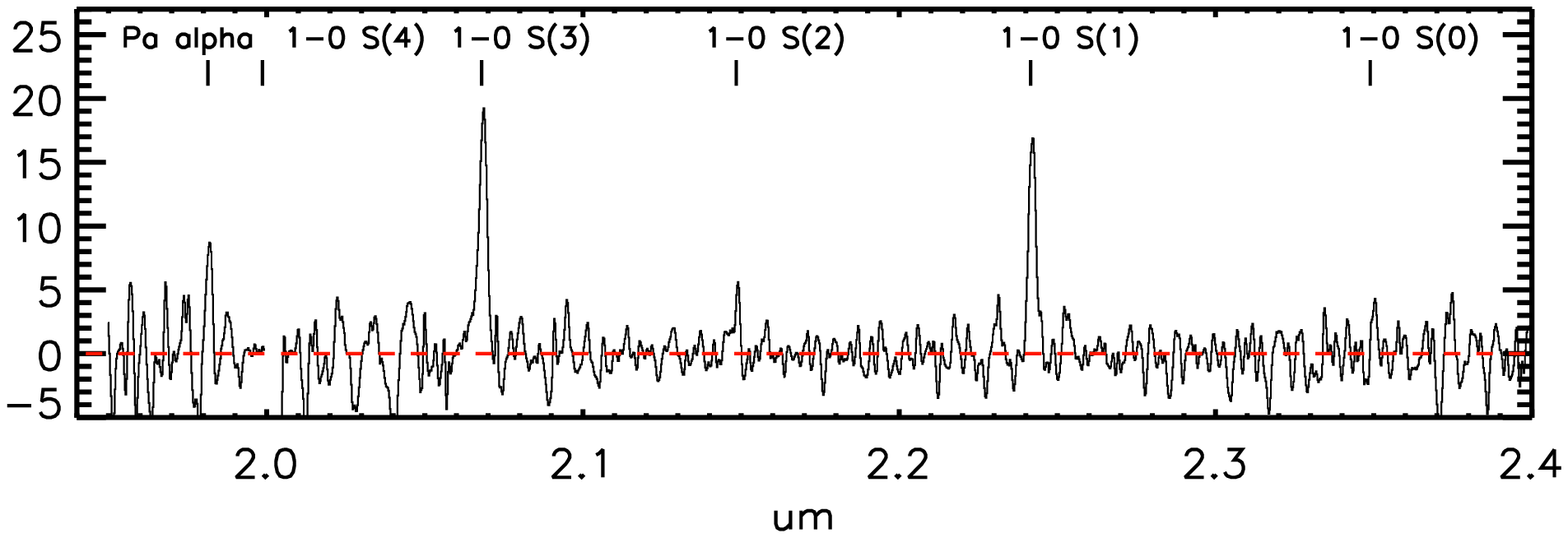}
    \includegraphics[width=0.30\textwidth, angle=90]{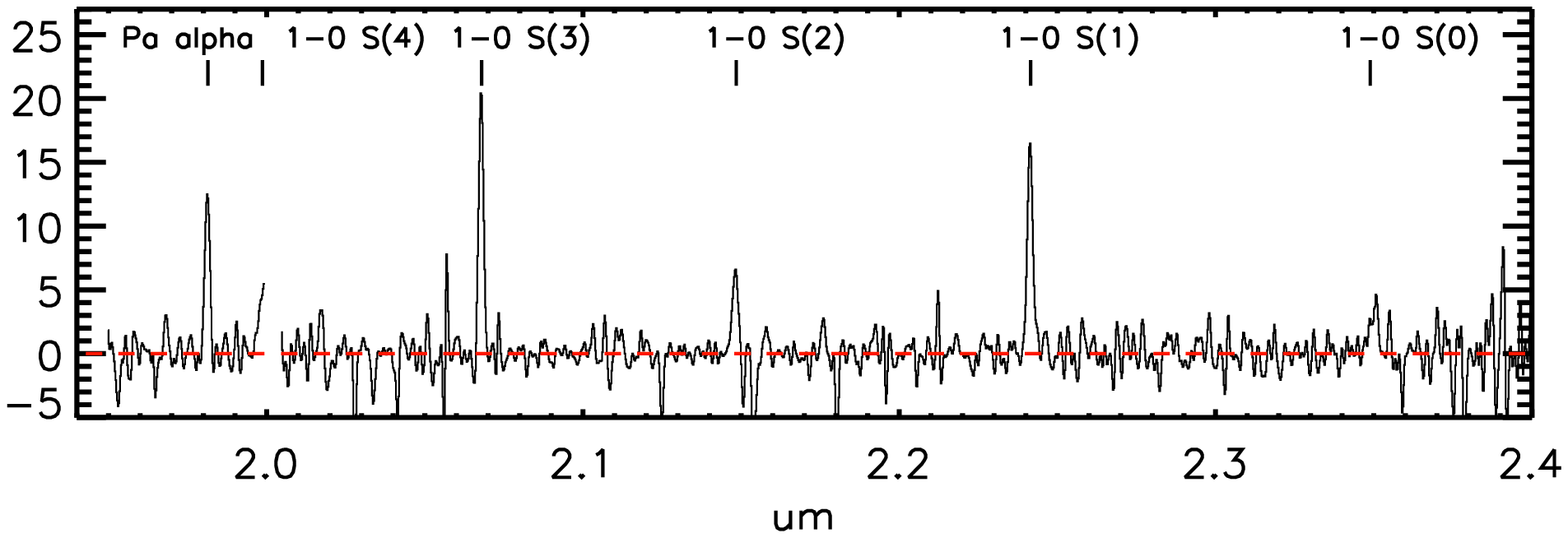}
  \vspace{0.5cm}
  \caption{SERSIC159-03 full K-band spectra. As an example of the quality of our SINFONI spectra we show full K-band spectra for two regions. The top image shows the spectrum for the combined regions B1 and B2. The bottom image shows the spectrum for region B5, see Fig. \ref{fig_h2_area_ss44} and Section \ref{sersic_sel_reg}. These spectra were obtained after smoothing the data by four pixels in both the spatial and spectral planes. The horizontal axis is given in units of $\mu$m and the vertical axis is given in units of 10$^{-14}$~$\mathrm{erg}$~$\mathrm{cm}^{-2}$~$\mathrm{s}^{-1}$~$\mu \mathrm{m}^{-1}$.}\label{fig_fs_sersic_ss44}
\end{figure*}

\begin{figure*}
    \includegraphics[width=0.60\textwidth, angle=90]{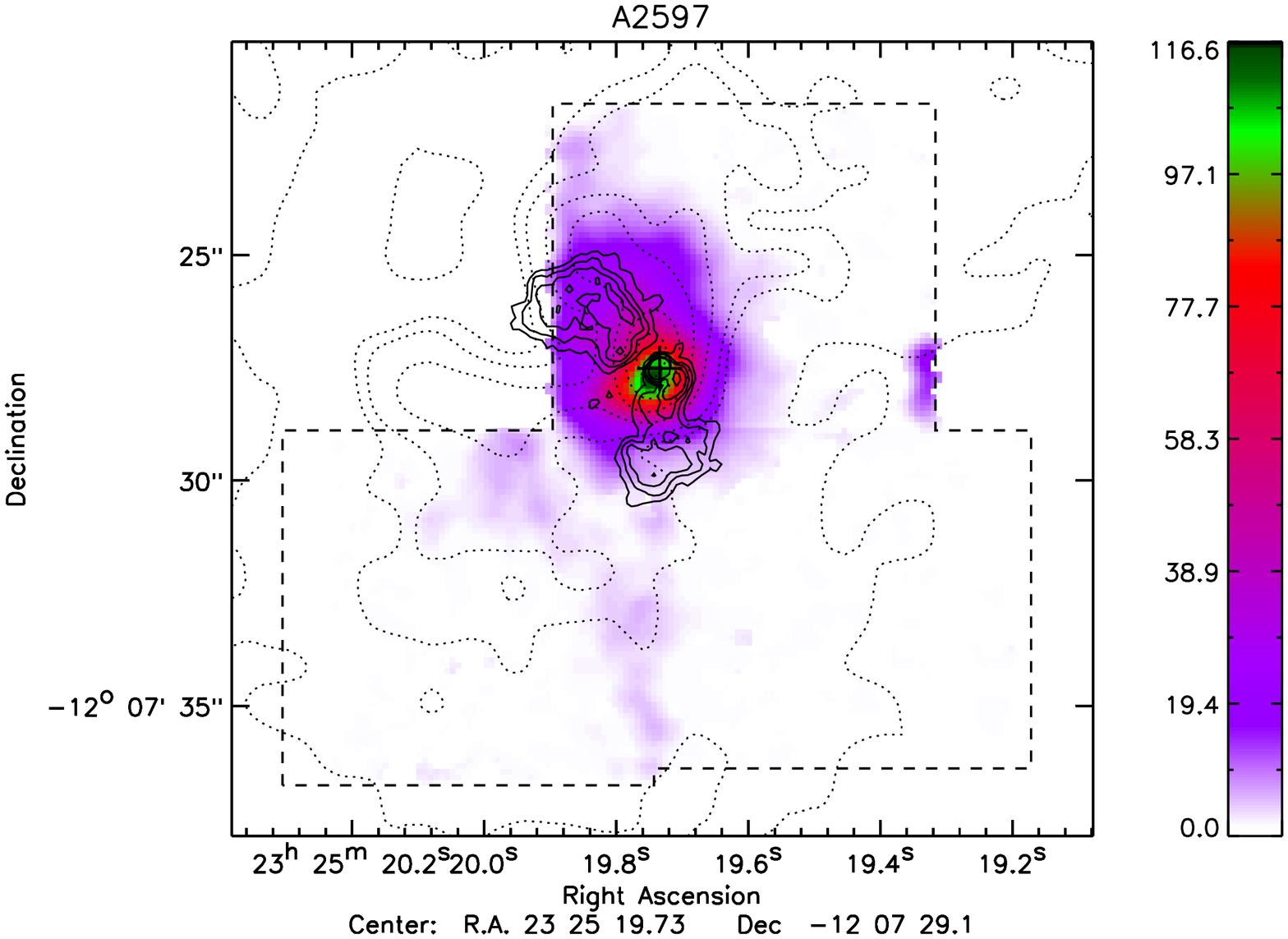}
    \includegraphics[width=0.60\textwidth, angle=90]{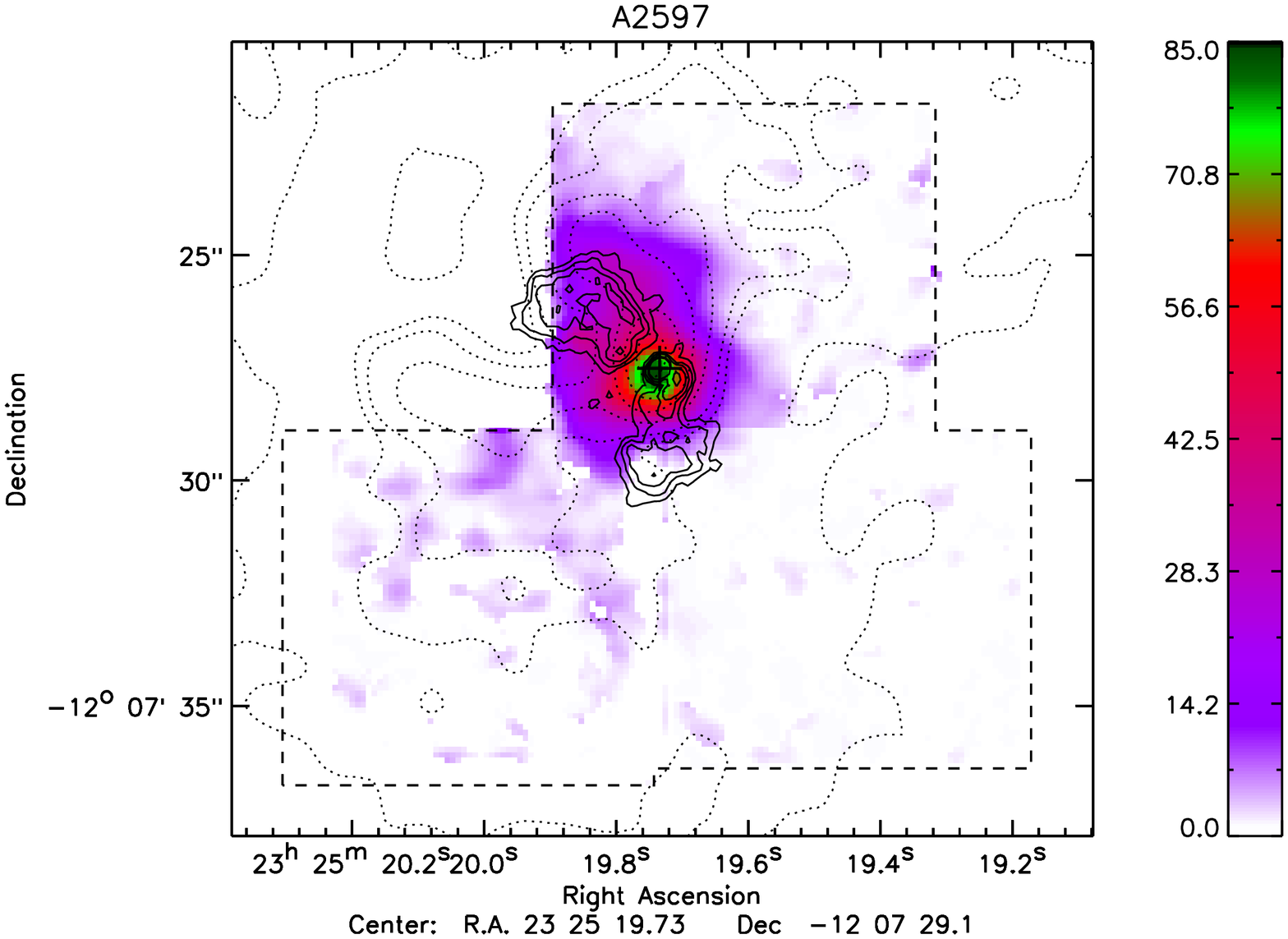}
  \vspace{0.5cm}
  \caption{ABELL 2597 Surface brightness. We show surface brightness maps for the H$_{\mathrm{2}}$~1-0~S(3) (top) and Pa~$\alpha$ (bottom) lines in units of 10$^{-17}$~$\mathrm{erg}$~$\mathrm{s}^{-1}$~$\mathrm{cm}^{-2}$~$\mathrm{arcsec}^{-2}$. The maps are obtained by fitting the spectrum for each spatial pixel in the data cube by a single Gaussian. The data was smoothed by four pixels in both the spatial and spectral planes. The stellar nucleus is indicated by the black cross. VLA 8.4~$\mathrm{GHz}$ Radio Continuum contours \citep[solid black line,][]{S95} and \textit{CHANDRA} X-ray contours (dotted black line) are overlayed. The Radio contours start at 4$\sigma$ (1$\sigma$~=~50~$\mu \mathrm{Jy}$). Consecutive Radio and X-ray contours double in value. Surface brightness maps for all detected lines are shown in Appendix \ref{app_lmp_a2597}. The northern field is not shown.}\label{fig_sign_maps_a2597_ss44}
\end{figure*}

\begin{figure*}
    \includegraphics[width=0.60\textwidth, angle=90]{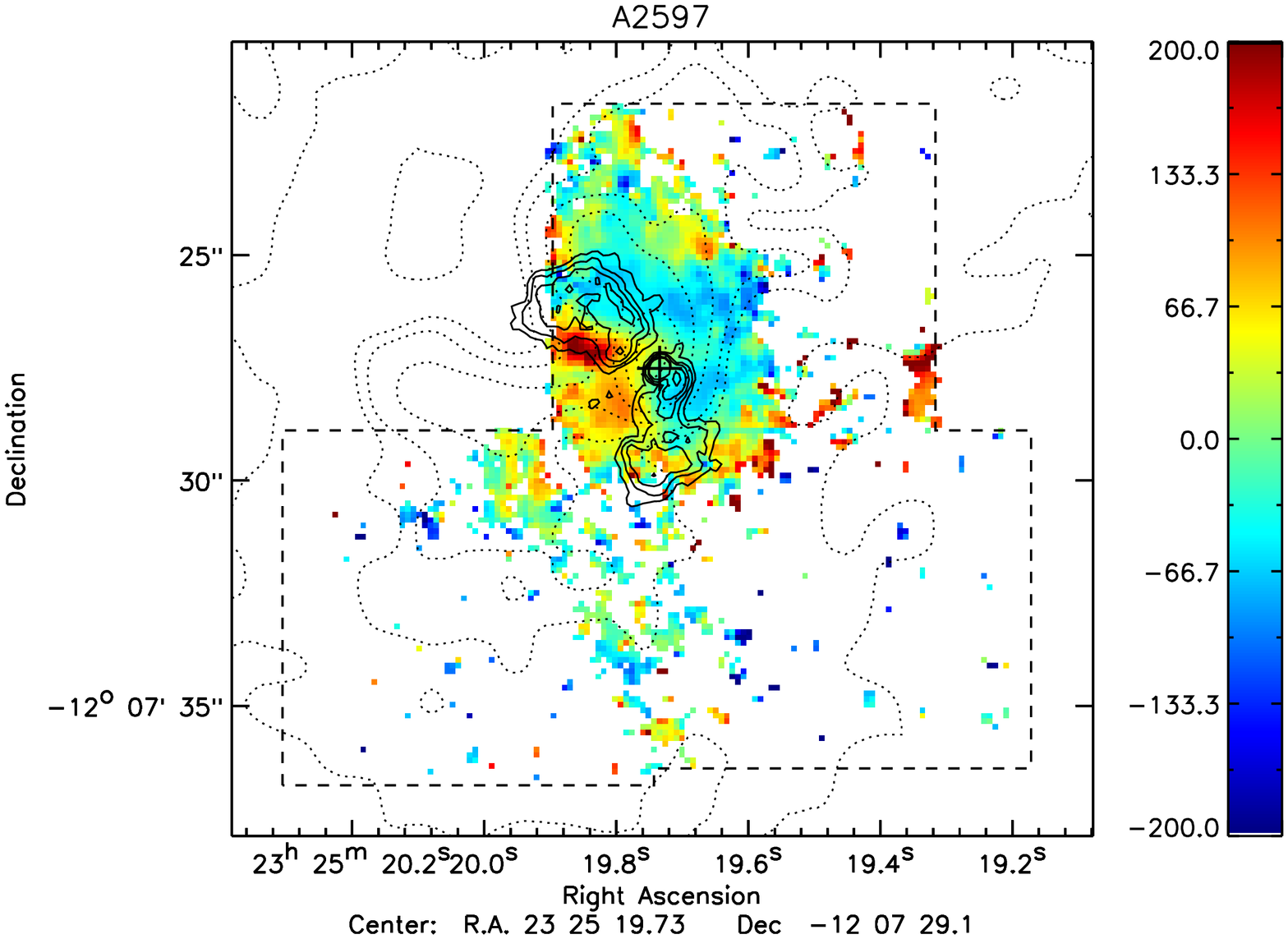}
    \includegraphics[width=0.60\textwidth, angle=90]{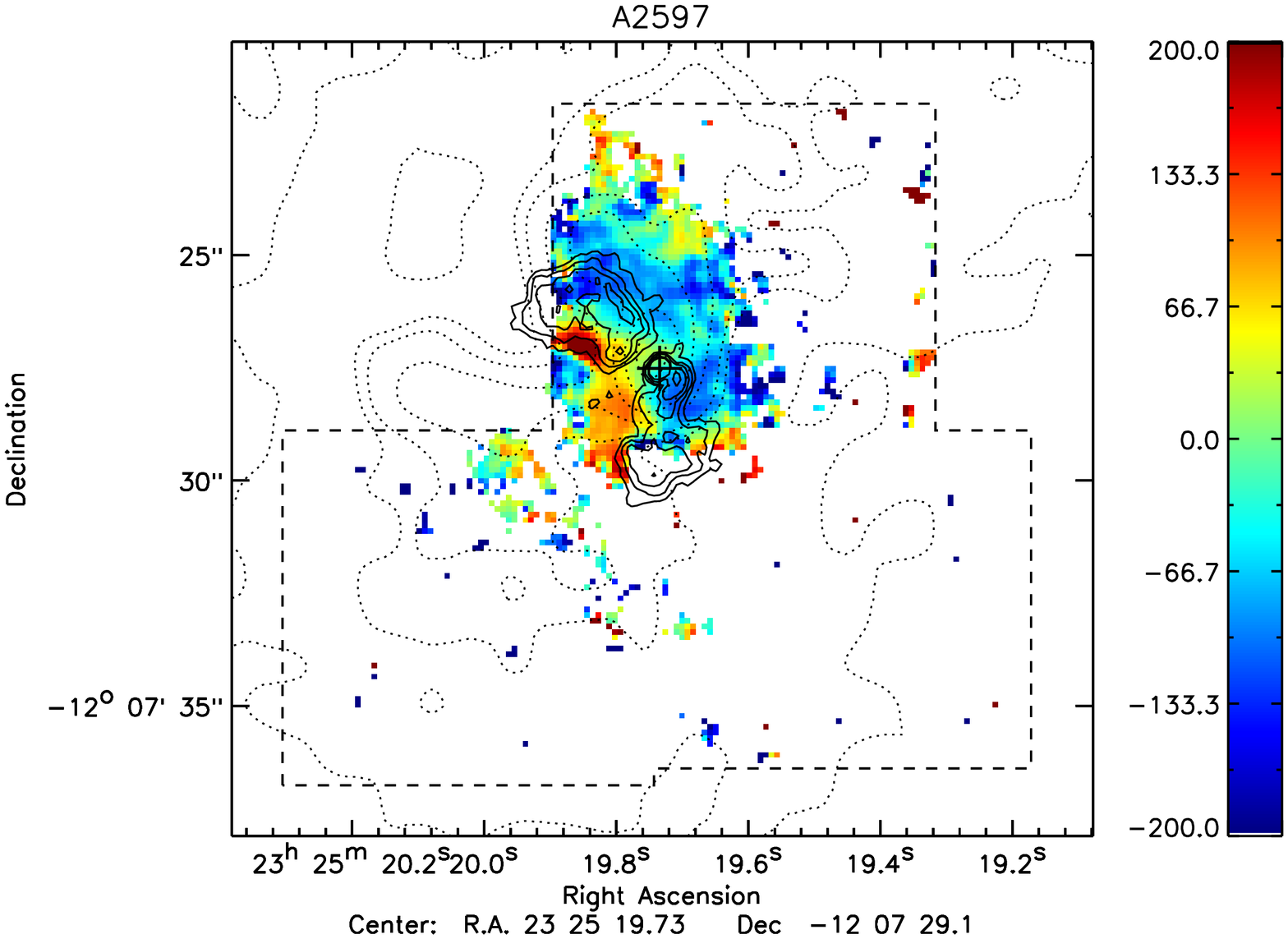}
  \vspace{0.5cm}
  \caption{ABELL 2597 Velocity. We show velocity maps for the H$_{\mathrm{2}}$~1-0~S(3) (top) and Pa~$\alpha$ (bottom) lines in units of $\mathrm{km}$~$\mathrm{s}^{-1}$ with respect to the systemic velocity of the BCG. The velocity of the gas is derived by fitting the spectrum for each spatial pixel in the data cube by a single Gaussian. The data was smoothed by two pixels in the spatial plane. The stellar nucleus is indicated by the black cross. VLA 8.4~$\mathrm{GHz}$ Radio Continuum contours \citep[solid black line,][]{S95} and \textit{CHANDRA} X-ray contours (dotted black line) are overlayed. The Radio contours start at 4$\sigma$ (1$\sigma$~=~50~$\mu$Jy). Consecutive Radio and X-ray contours double in value. Velocity maps were made for all detected emission lines. These all show the same structure and are thus not all shown here. The northern field is not shown.}\label{fig_kinm_velc_maps_a2597_ss20}
\end{figure*}

\begin{figure*}
    \includegraphics[width=0.60\textwidth, angle=90]{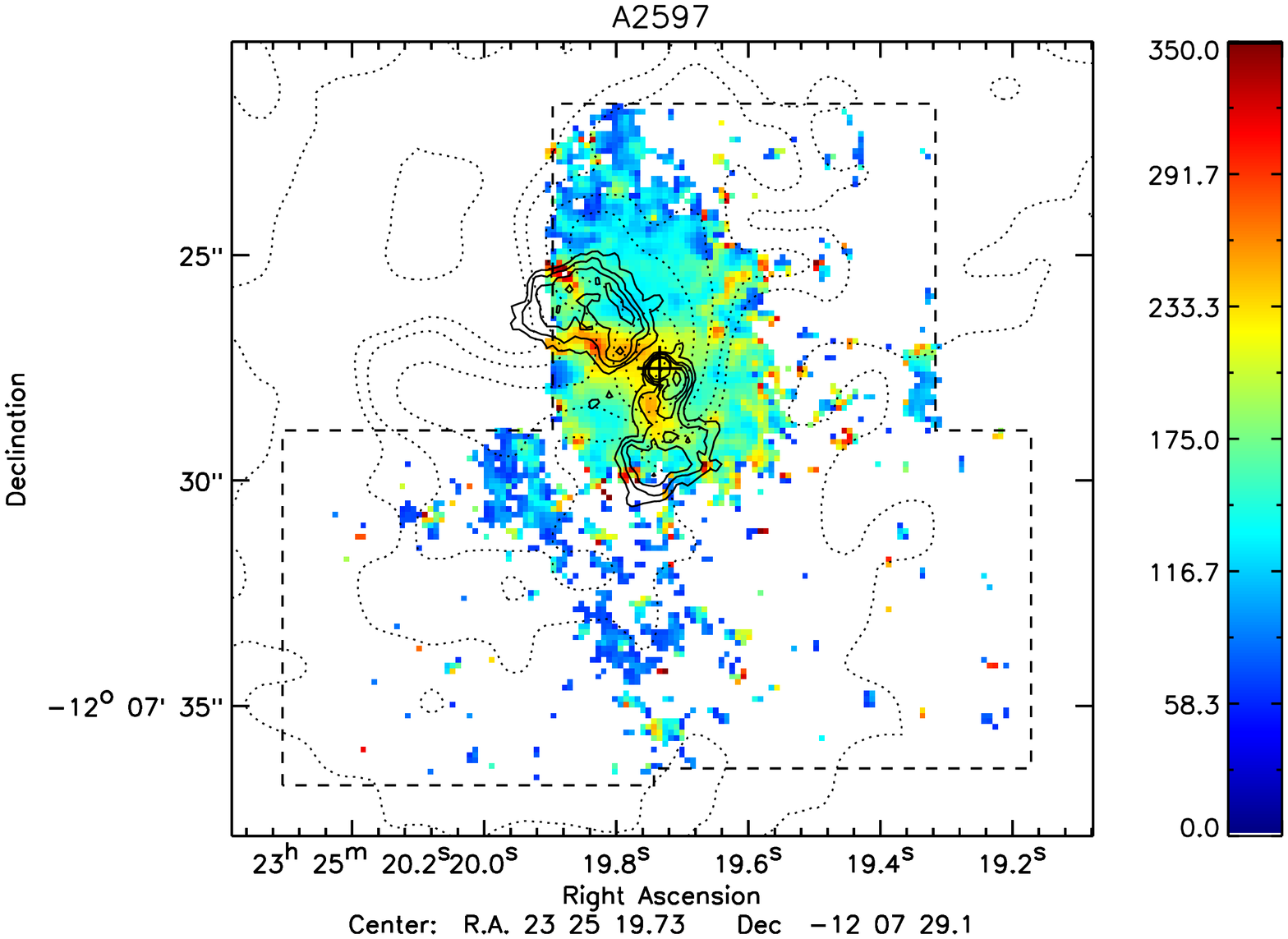}
    \includegraphics[width=0.60\textwidth, angle=90]{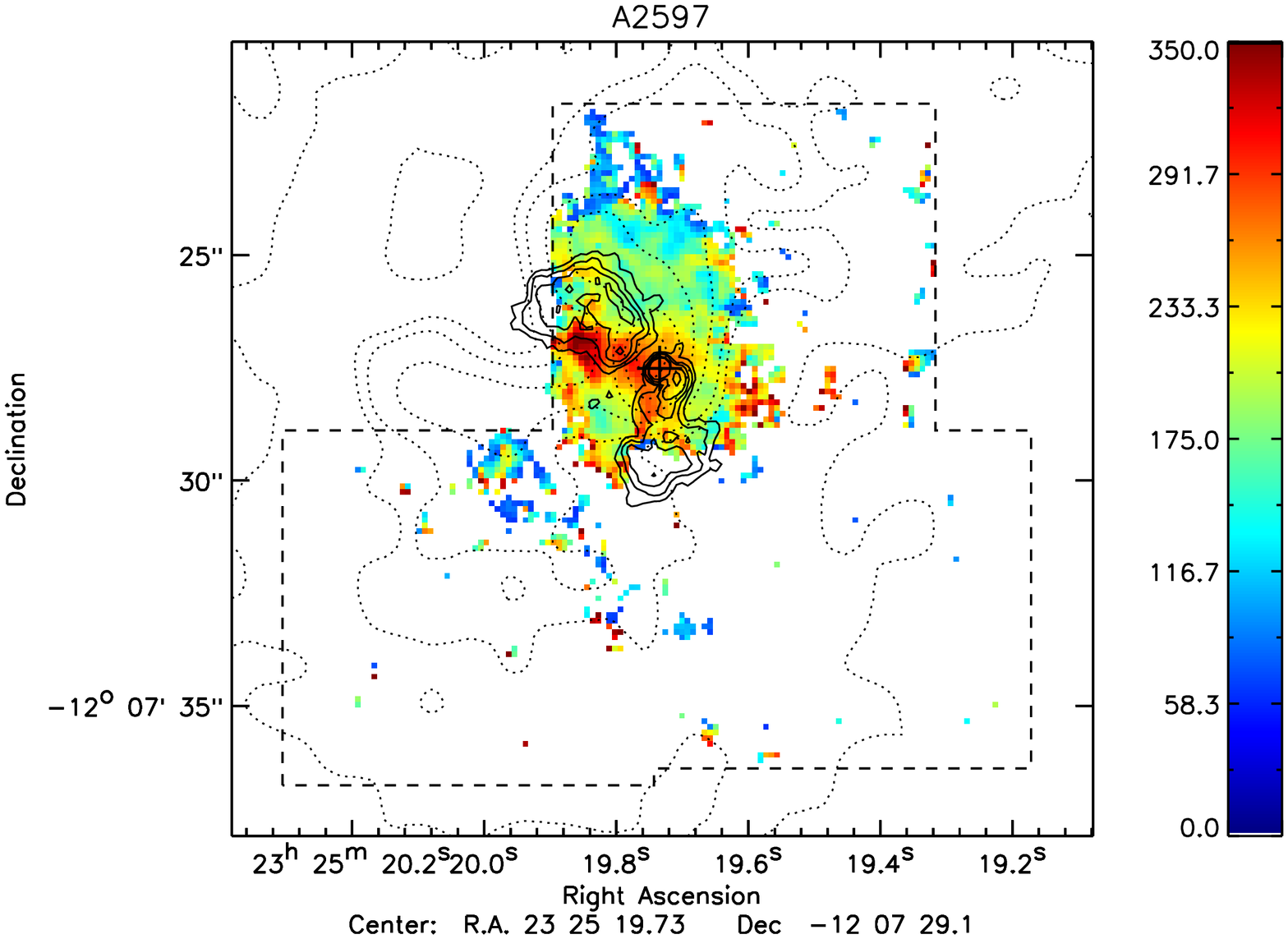}
  \vspace{0.5cm}
  \caption{ABELL 2597 Velocity dispersion. We show velocity dispersion maps for the H$_{\mathrm{2}}$~1-0~S(3) (top) and the Pa~$\alpha$ (bottom) lines in units of $\mathrm{km}$~$\mathrm{s}^{-1}$. The velocity dispersion $\sigma_{v}$  of the gas is derived by fitting the spectrum for each spatial pixel in the data cube by a single Gaussian. The data cube was smoothed by two pixels in the spatial plane. The stellar nucleus is indicated by the black cross. VLA 8.4~$\mathrm{GHz}$ Radio Continuum contours \citep[solid black line,][]{S95} and \textit{CHANDRA} X-ray contours (dotted black line) are overlayed on the dispersion maps. The Radio contours start at 4$\sigma$ (1$\sigma$~=~50~$\mu \mathrm{Jy}$). Consecutive Radio and X-ray contours double in value. Gas dispersion maps were made for all detected emission lines. All show the same structure and are thus not all shown here. The northern field is not shown.}\label{fig_kinm_disp_maps_a2597_ss20}
\end{figure*}

\begin{figure*}
    \includegraphics[width=0.60\textwidth, angle=90]{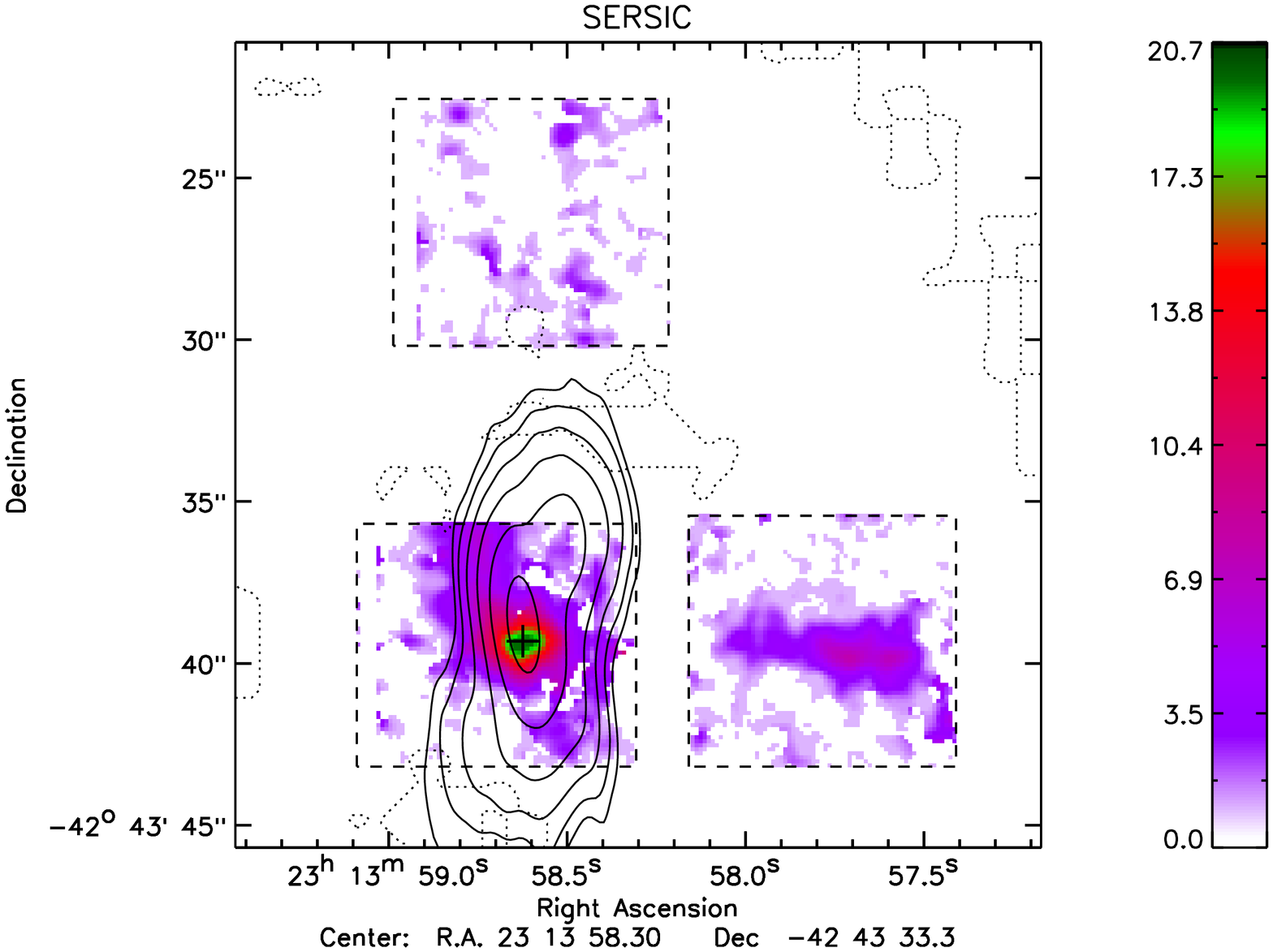}
    \includegraphics[width=0.60\textwidth, angle=90]{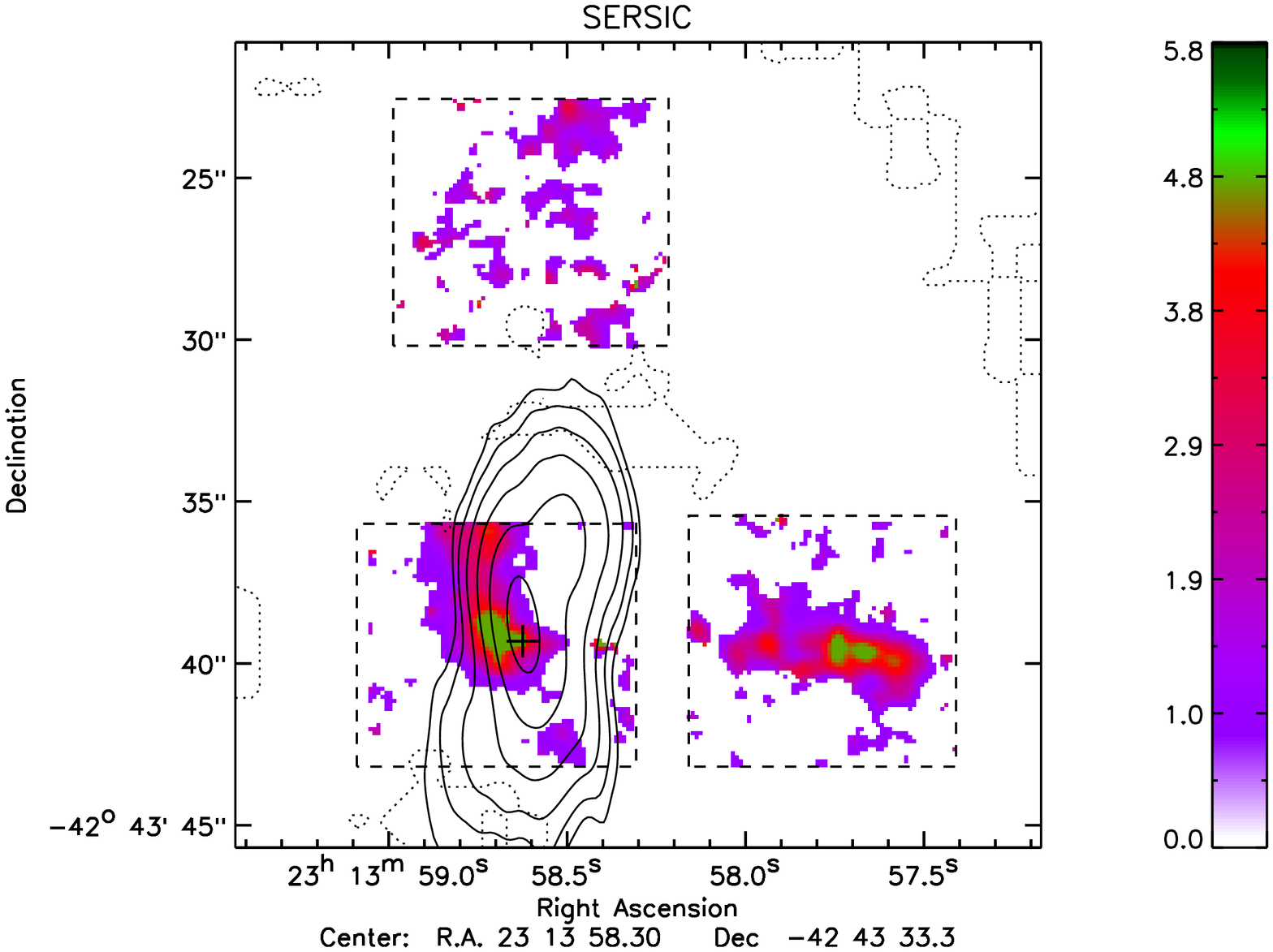}
  \vspace{0.5cm}
  \caption{SERSIC 159-03 Surface brightness. We show surface brightness maps for the H$_{\mathrm{2}}$~1-0~S(3) (top) and Pa~$\alpha$ (bottom) lines in units of 10$^{-17}$~$\mathrm{erg}$~$\mathrm{s}^{-1}$~$\mathrm{cm}^{-2}$~$\mathrm{arcsec}^{-2}$. The maps are obtained by fitting the spectrum for each spatial pixel in the data cube by a single Gaussian. The data cube was smoothed by four pixels in both the spatial and spectral planes.  The stellar nucleus is indicated by the black cross. VLA 8.4~$\mathrm{GHz}$ Radio Continuum contours (solid black line) and \textit{CHANDRA} X-ray contours (dotted black line) are overlayed. The Radio contours start at 4$\sigma$ (1$\sigma$~=~25~$\mu \mathrm{Jy}$). Consecutive Radio and X-ray contours double in value. Surface brightness maps for all detected lines are shown in Appendix \ref{app_lmp_sersic}.}\label{fig_sign_maps_sersic_ss44}
\end{figure*}

\begin{figure*}
    \includegraphics[width=0.60\textwidth, angle=90]{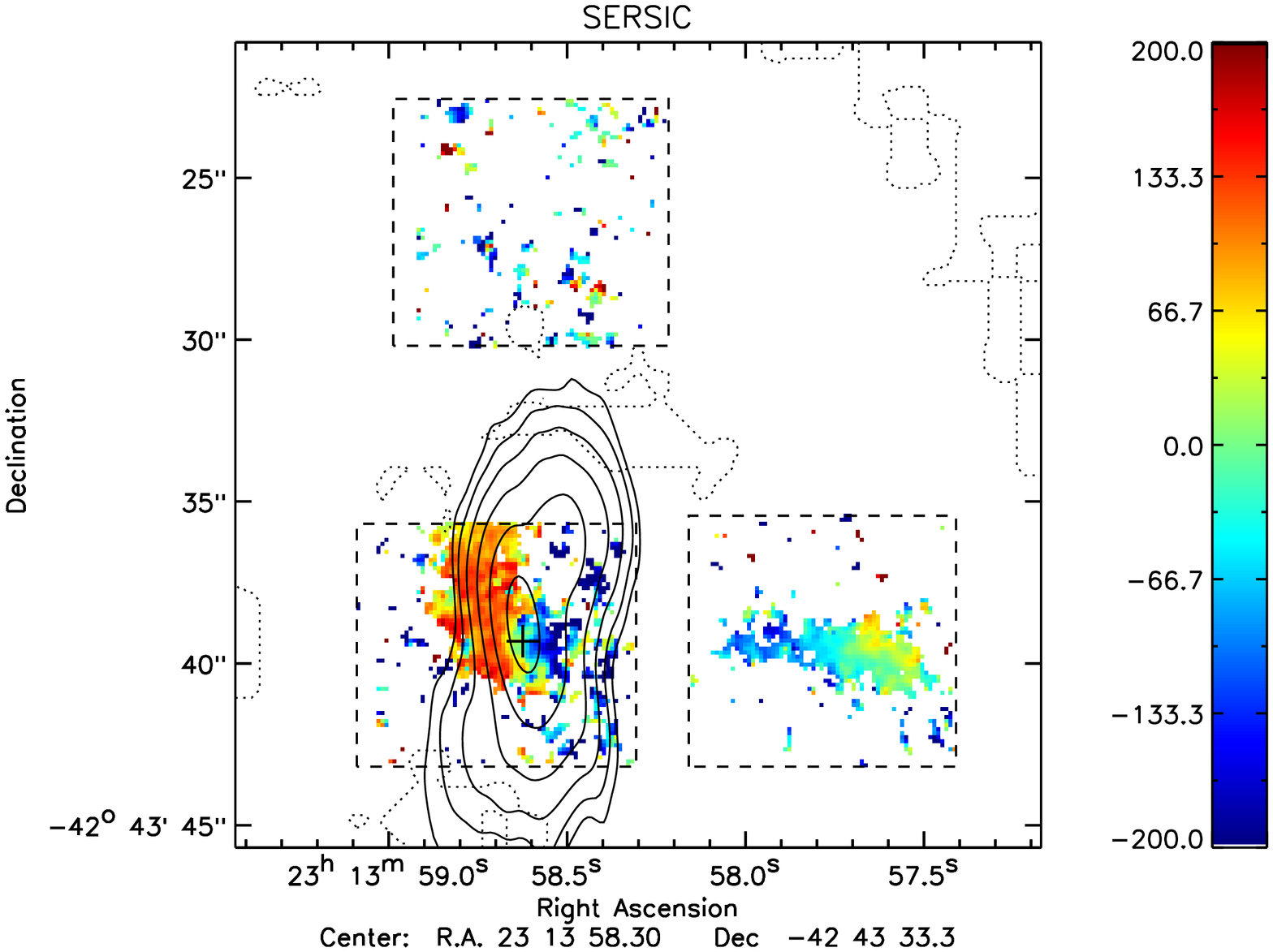}
    \includegraphics[width=0.60\textwidth, angle=90]{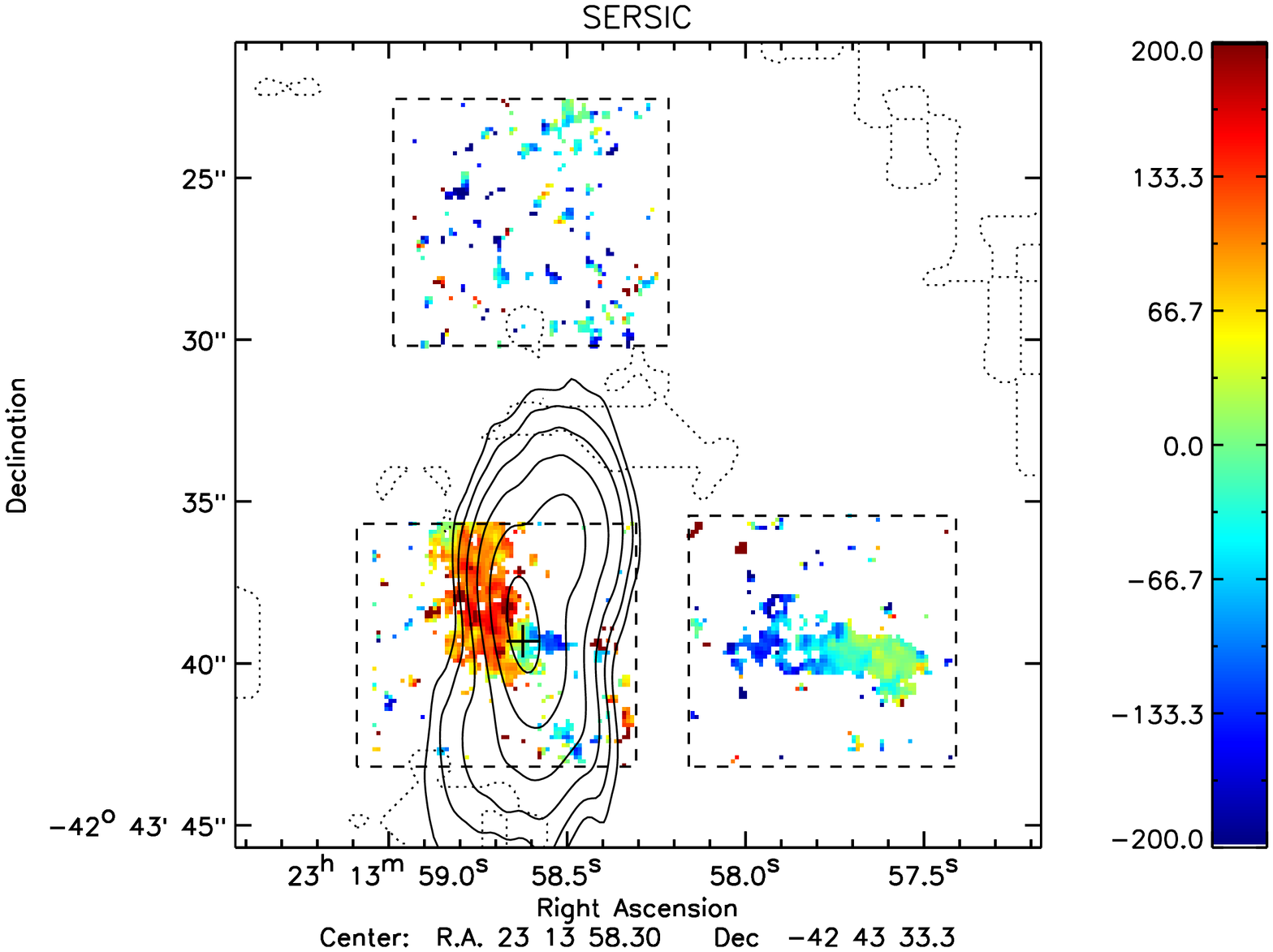}
    \vspace{0.5cm}
  \caption{SERSIC 159-03 Velocity. We show velocity maps for the H$_{\mathrm{2}}$~1-0~S(3) (top) and Pa~$\alpha$ (bottom) lines in units of $\mathrm{km}$~$\mathrm{s}^{-1}$ with respect to the systemic velocity of the BCG. The velocity of the gas is derived by fitting the spectrum for each spatial pixel in the data cube by a single Gaussian. The data was smoothed by two pixels in the spatial plane. The stellar nucleus is indicated by the black cross. VLA 8.4~$\mathrm{GHz}$ Radio Continuum contours (solid black line) and \textit{CHANDRA} X-ray contours (dotted black line) are overlayed. The Radio contours start at 4$\sigma$ (1$\sigma$~=~25~$\mu$Jy). Consecutive Radio and X-ray contours double in value. Velocity maps were made for all detected emission lines. These all show the same structure and are thus not all shown here.}\label{fig_kinm_velc_maps_sersic_ss22}
\end{figure*}

\begin{figure*}
    \includegraphics[width=0.60\textwidth, angle=90]{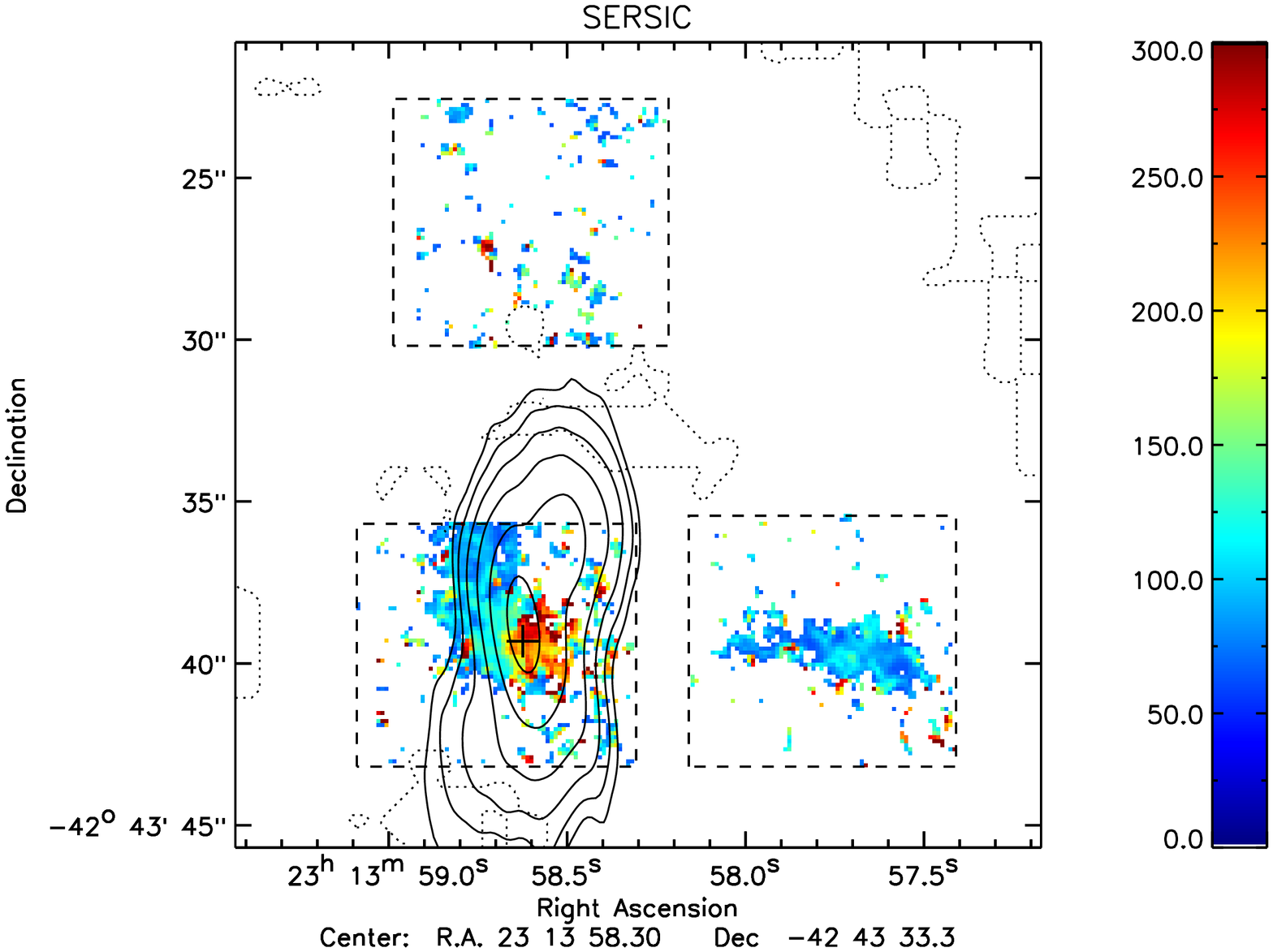}
    \includegraphics[width=0.60\textwidth, angle=90]{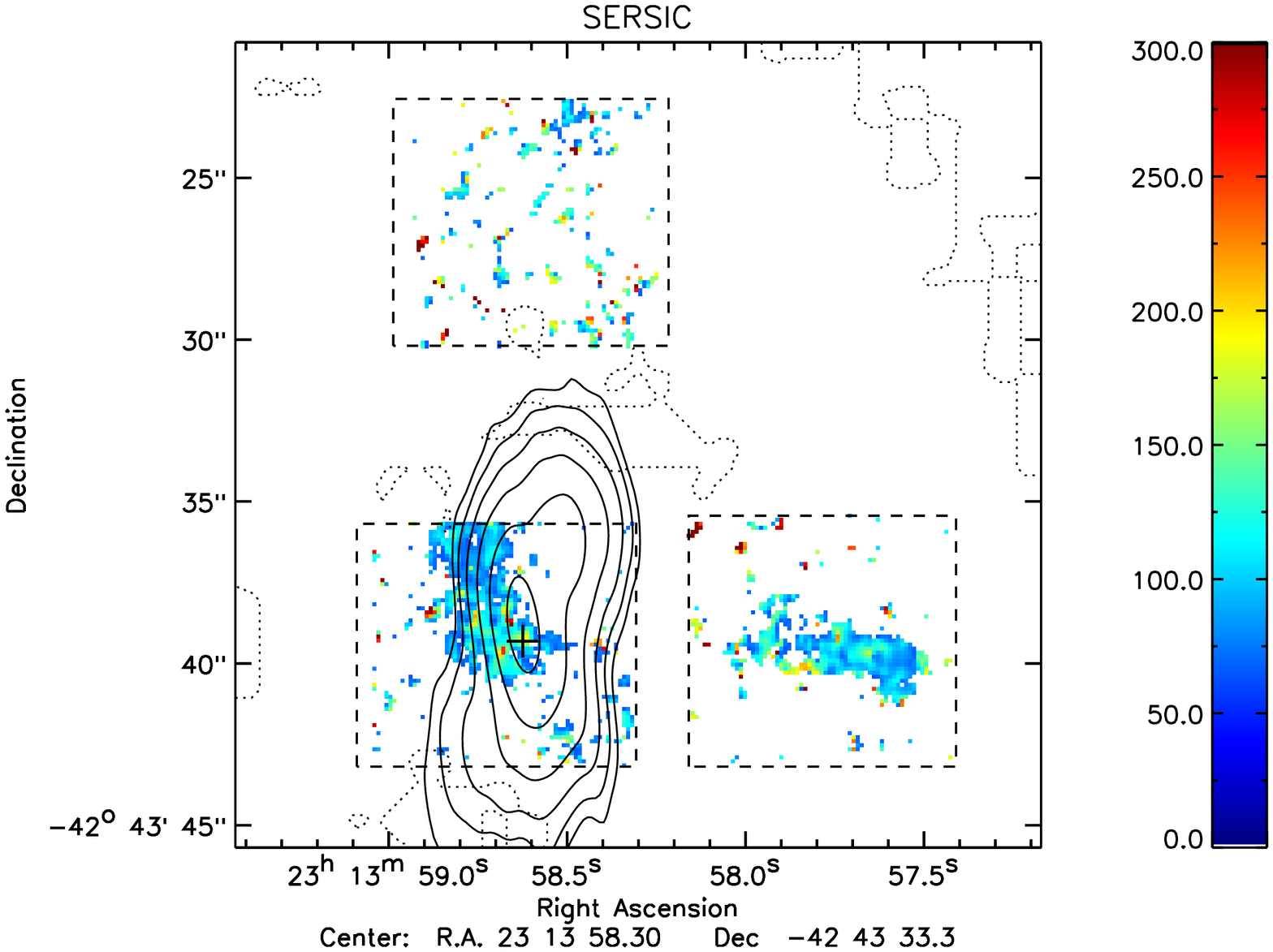}
  \vspace{0.5cm}
  \caption{SERSIC 159-03 Velocity dispersion. We show velocity dispersion maps for the H$_{\mathrm{2}}$~1-0~S(3) (top) and the Pa~$\alpha$ (bottom) lines in units of $\mathrm{km}$~$\mathrm{s}^{-1}$. The velocity dispersion $\sigma_{v}$ of the gas is derived by fitting the spectrum for each spatial pixel in the data cube by a single Gaussian. The data cube was smoothed by two pixels in the spatial plane.  The stellar nucleus is indicated by the black cross. VLA 8.4~$\mathrm{GHz}$ Radio Continuum contours (solid black line) and \textit{CHANDRA} X-ray contours (dotted black line) are overlayed on the dispersion maps. The Radio contours start at 4$\sigma$ (1$\sigma$~=~25~$\mu \mathrm{Jy}$). Consecutive Radio and X-ray contours double in value. Gas dispersion maps were made for all detected emission lines. All show the same structure and are thus not all shown here.}\label{fig_kinm_disp_maps_sersic_ss22}
\end{figure*}

In order to investigate the H$_{\mathrm{2}}$ excitation mechanism we have constructed H$_{\mathrm{2}}$ excitation diagrams for a number of regions in A2597. Seven regions, A1-A7, were selected based on having distinct physical properties in terms of either their emissivity, spatial location or kinematical structure, see Fig. \ref{fig_h2_area_ss44}. Similarly seven regions, B1-B7, were selected in S159, see Fig. \ref{fig_h2_area_ss44}. Excitation diagrams were not made for S159 since we only have reliable detections for two H$_{\mathrm{2}}$ lines.

All line detections for regions A1-A7 and B1-B7 are presented in Appendix \ref{app_tb}. The line profiles are well described by a single Gaussian function. For the lines detected in these regions their integrated fluxes are given in Tables \ref{tabl_line_area_a2597_ss20} and \ref{tabl_line_area_sersic_ss22}. Kinematical information for these regions are presented for the H$_{\mathrm{2}}$~1-0~S(3) and Pa~$\alpha$ lines only and these are given in Tables \ref{tabl_kinm_area_a2597_ss20} and \ref{tabl_kinm_area_sersic_ss22}. Gas temperatures and masses for the selected regions are given in Tables \ref{tabl_phys_area_a2597_ss20} and \ref{tabl_phys_area_sersic_ss22}. We will first shortly describe the selected regions below.

\subsection{A2597: Selected regions}\label{a2575_sel_reg}
In the central field we have selected four regions, A1, A2, A3 and A6. Region A1 corresponds to the nuclear region, A2 samples the region where the northern filament connects to the nucleus. A3, just north of A2, samples that part of the northern filament that appears to not be influenced (directly) by the nucleus anymore. Lastly, in the central field a clump of strong emission is noted on the western edge of the field, which we selected as region A6. There is tentative evidence for a narrow ridge connecting A6 to the nucleus, but we cannot confirm this with the present observations. The emission in region A6 itself is also uncertain, due to the increased noise at the edge of the field. It is only observed significantly in two H$_{\mathrm{2}}$ lines and therefore we have not constructed an excitation diagram for this region.

In the south-eastern field we have selected one region, A4, which captures most of the emission in the southern filament. This region was selected such to avoid the noisy overlap region between the SE and SW fields. The strongest emission lines showed weak evidence in their surface brightness maps that the south-eastern filament may extend across the overlap region into the SW field. Region A5 was selected to test this. Significant line emission is found in this region. The kinematical structure of the gas observed in A5 is similar to that measured in A4 and thus a connection between the two regions seems plausible.

The northern field also contains significant emission for the strongest emission lines. A systematic search for line emission in the northern field was performed using various binnings. Region A7 was selected to show that emission does exist at a significant level in the northern field. J05 previously showed that molecular and ionised gas existed out to 20~$\mathrm{kpc}$ north from the nucleus using long slit spectra. The H$_{\mathrm{2}}$~1-0~S(1), 1-0~S(3) and Pa~$\alpha$ lines are all detected at the 3.0 sigma level, and the H$_{\mathrm{2}}$~1-0~S(5) line is observed at the 2 sigma level. All four detected lines show the same velocity structure.

The velocity dispersion of the gas in A7 is more difficult to constrain. Of the four lines the H$_{\mathrm{2}}$~1-0~S(3) line is observed at the highest significance and has the most reliable fit. This line has a dispersion of about 60~$\mathrm{km}$~$\mathrm{s}^{-1}$ using a spatial smoothing of two pixels. The other lines have higher fitted velocity dispersions ranging from 70 to 110~$\mathrm{km}$~$\mathrm{s}^{-1}$, however within their large errors (40-60~$\mathrm{km}$~$\mathrm{s}^{-1}$) they agree with the H$_{\mathrm{2}}$~1-0~S(3) result.

The positive velocity and low velocity dispersion of the gas makes it plausible that the gas in A7 is connected with the northern filament. This interpretation agrees with the J05 results. We conclude that molecular and ionised gas is present at least up to region A7, i.e. 22.5~$\mathrm{kpc}$ from the nucleus, in good agreement with the J05 results. There is tentative evidence from regions investigated north and south of A7 within the northern field that line emission is present there as well.

\subsection{S159: Selected regions}\label{sersic_sel_reg}
In the SE field we have selected four regions, B1, B2, B3 and B4. Region B1 corresponds to the nuclear region. B2 samples the tail of the northern filament as it connects to the nucleus. Region B3, just north of B2, is selected to sample the filament where it is no longer (directly) influenced by the nucleus. Lastly, B4 is selected to contain the clump of emission just to the south of the nucleus. Whether this clump is part of a filament or not our observations can not confirm.

As discussed above, the nuclear region B1 is significantly noisier than surrounding regions. The spectral noise is higher here by a factor two to three. The H$_{\mathrm{2}}$~1-0~S(1) and 1-0~S(3) lines are still easily detected here. The H$_{\mathrm{2}}$~1-0~S(3) line is considerably stronger than the H$_{\mathrm{2}}$~1-0~S(1) line in this region as compared to any of the other regions. The Pa~$\alpha$ line is detected at a much lower significance and at a much lower flux level. In the SW field we have selected one region, B5, which captures most of the western filament.

We have selected two regions in the the northern field, B6 and B7, to investigate the low level clumpy emission here. Region B7 was selected to capture the strong clump of emission at the northern edge of this field. Region B6 was selected to investigate the remaining emission. The summed spectra clearly show that line emission is present in the northern field. We can thus conclude that molecular and ionised emission is present at least up to 18.0~$\mathrm{kpc}$ from the nucleus. Whether the gas in the northern field is directly connected to that in the SE field cannot be confirmed by our observations, although it seems plausible from the J05 results. The velocity and low velocity dispersion of the gas are such that this gas can be connected smoothly to that in the northern filament.

\subsection{Thermal excitation of the molecular gas}
\begin{figure*}
    \includegraphics[width=0.35\textwidth, angle=90]{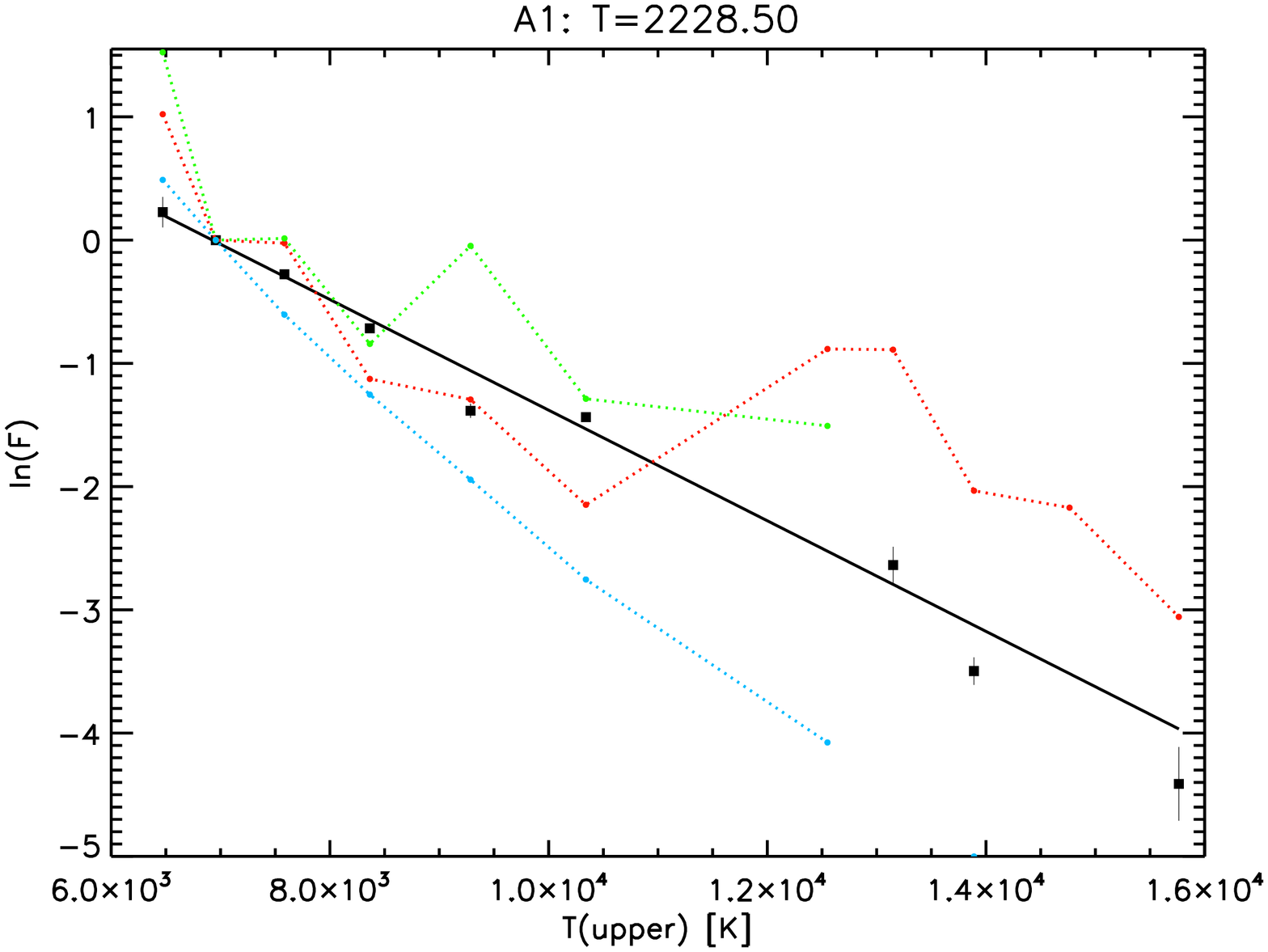}
    \includegraphics[width=0.35\textwidth, angle=90]{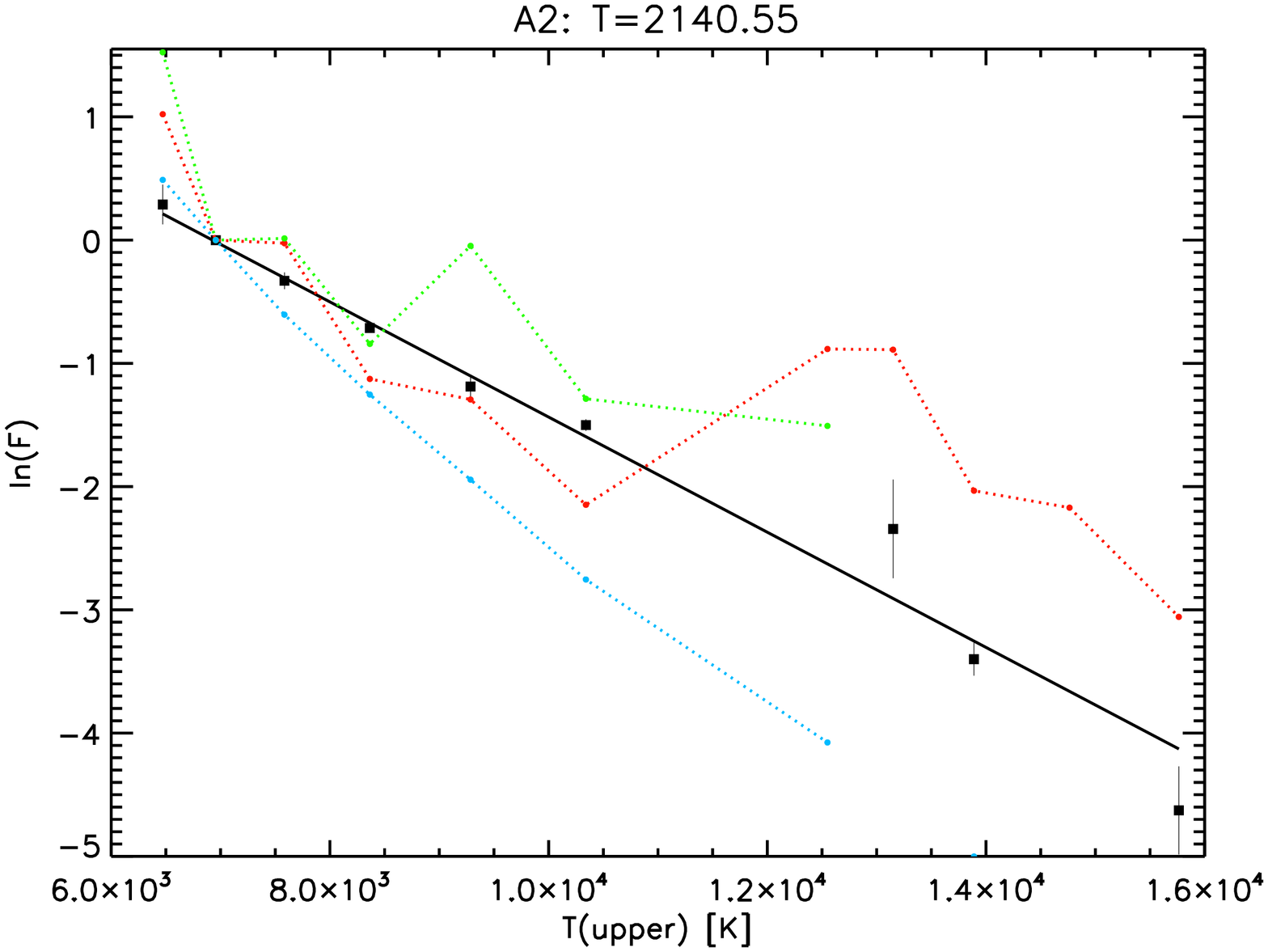}
    \includegraphics[width=0.35\textwidth, angle=90]{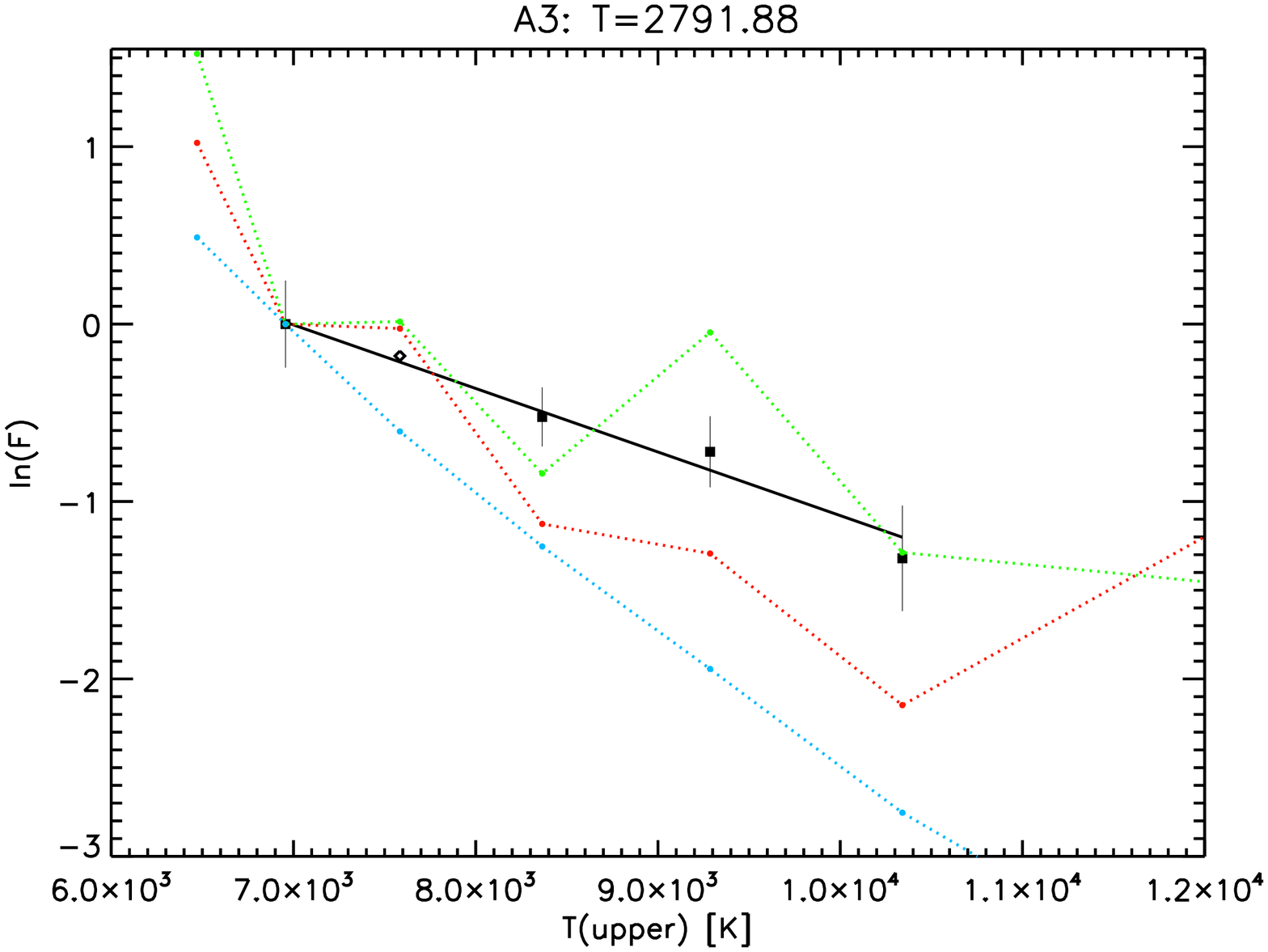}
    \includegraphics[width=0.35\textwidth, angle=90]{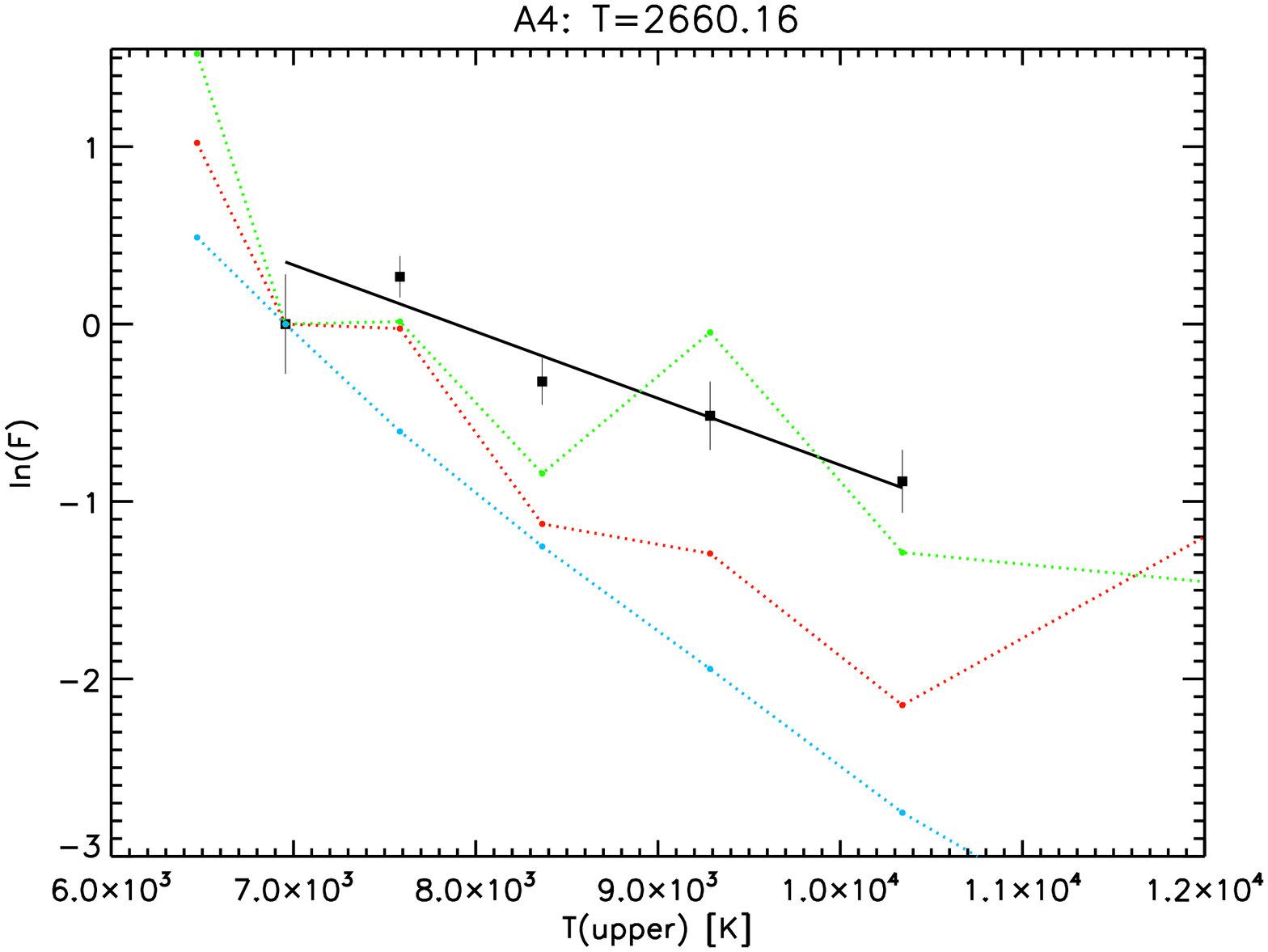}
    \includegraphics[width=0.35\textwidth, angle=90]{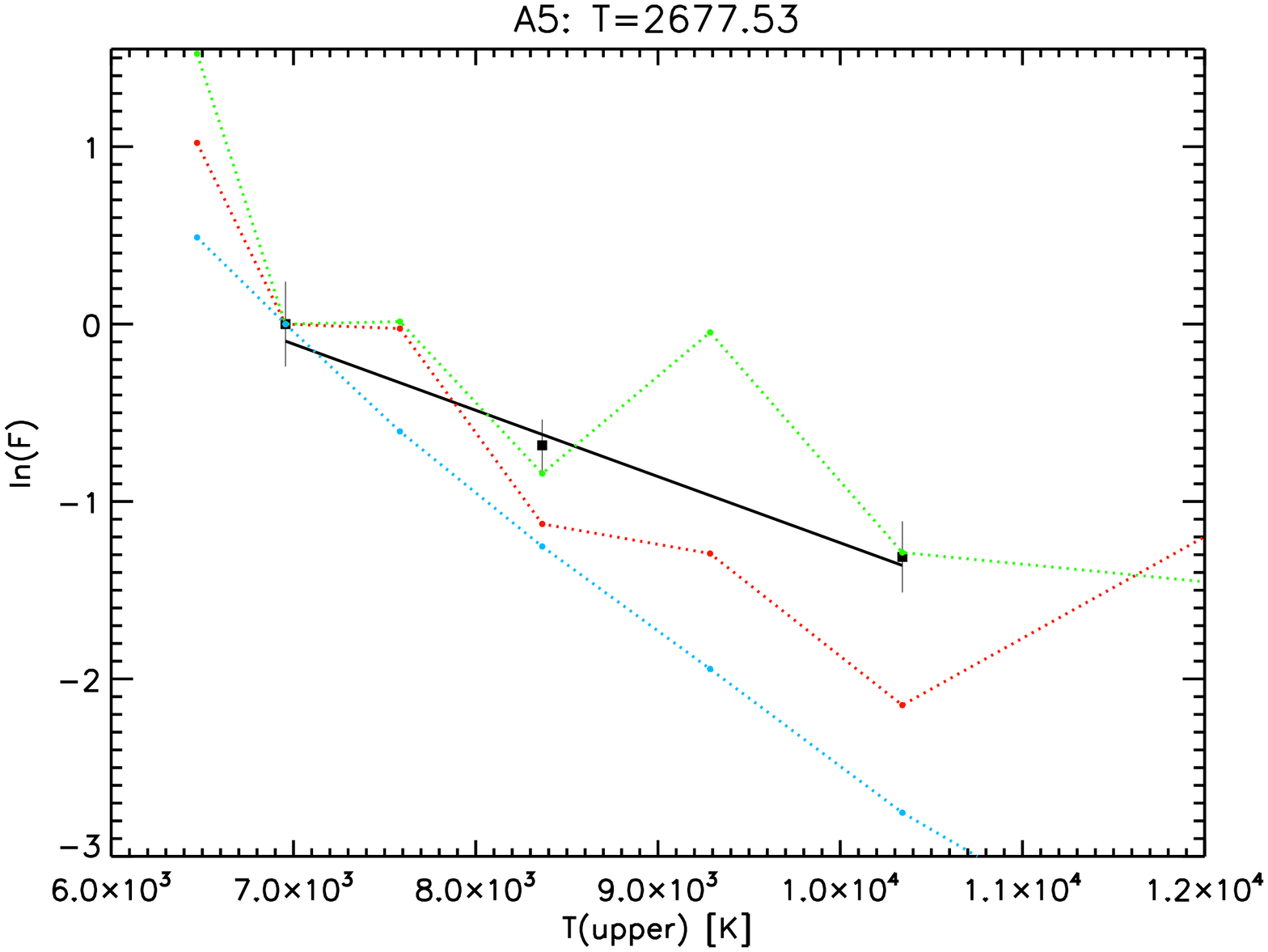}
    \includegraphics[width=0.35\textwidth, angle=90]{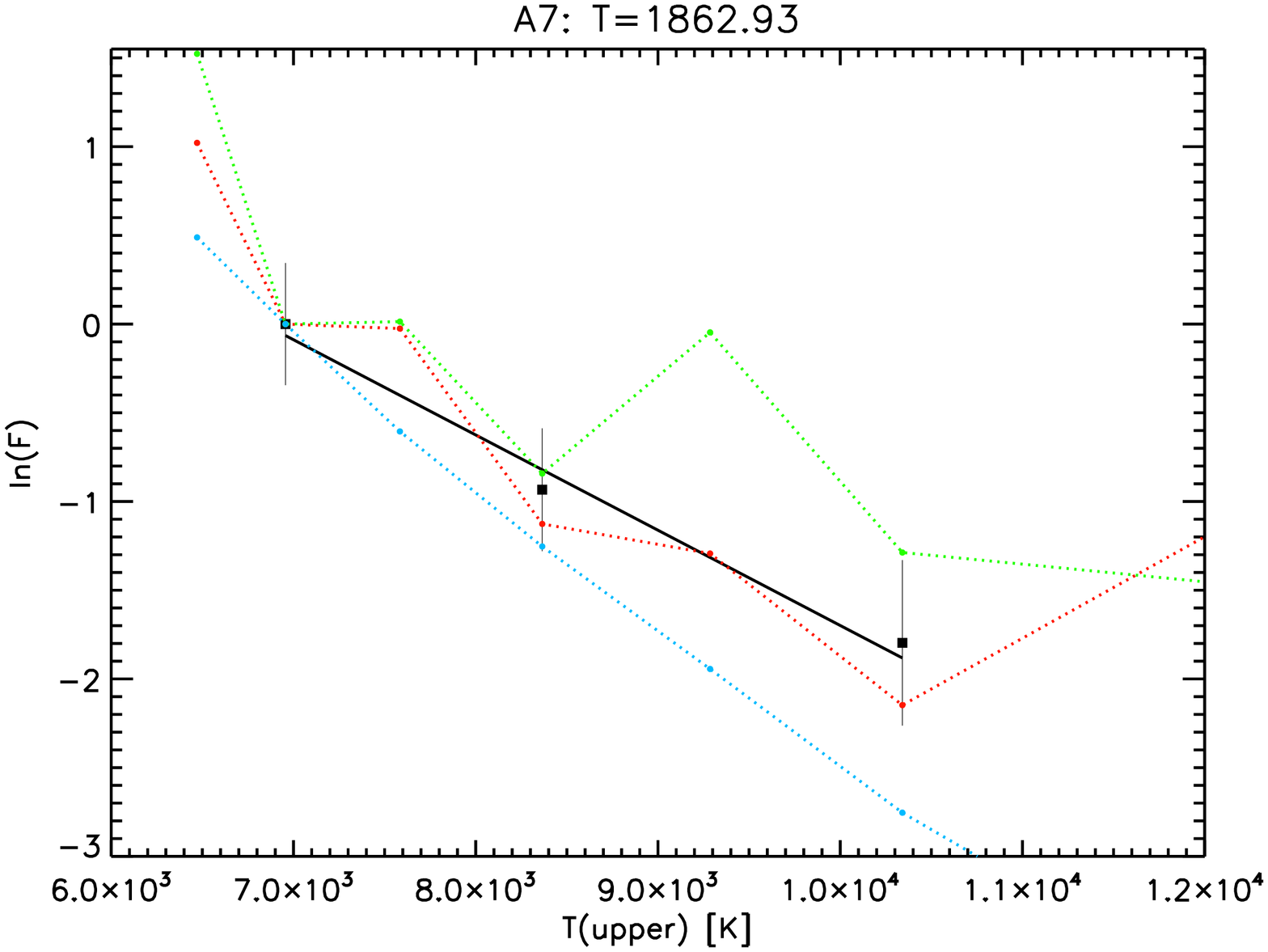}
  \vspace{0.5cm}
  \caption{ABELL 2597 Molecular excitation. Excitation diagrams for the H$_{\mathrm{2}}$ lines in regions A1, A2, A3, A4, A5 and A7. The figures show the natural logarithm of the normalised line flux $\textit{ln(F)}$ versus the upper state temperature $\textit{T}_{\mathrm{u}}$. The best-fitting LTE model is given by the solid black line line. For qualitative comparison reasons we plot three other H$_{\mathrm{2}}$ excitation models. The dotted red line shows a low-density UV fluorescence model by Black and van Dishoeck 1987 (their model 14; $\textit{n}$~=~3$\times$10$^{3}$~$\mathrm{cm}^{-3}$, $\textit{T}$~=~100~$\mathrm{K}$, $\textit{I}_{\mathrm{UV}}$~=~1$\times$10$^{3}$ relative to $\textit{I}_{\mathrm{tot}}$). The dotted blue line shows a high-density UV fluorescence model by Sternberg \& Dalgarno 1989 (their model 2D; $\textit{n}$~=~1$\times$10$^{6}$~$\mathrm{cm}^{-3}$, $\textit{T}\sim$1000~$\mathrm{K}$, $\textit{I}_{\mathrm{UV}}$~=~1$\times$10$^{2}$ relative to $\textit{I}_{\mathrm{tot}}$). The green line shows the Ferland et al. 2009 cosmic ray model for the Perseus cluster. For region A3 the uncertain H$_{\mathrm{2}}$~1-0~S(2) flux value is shown, but it is not used for the fit.}\label{fig_lte_mod_a2597_ss20}
\end{figure*}

\begin{figure*}
    \includegraphics[width=0.35\textwidth, angle=90]{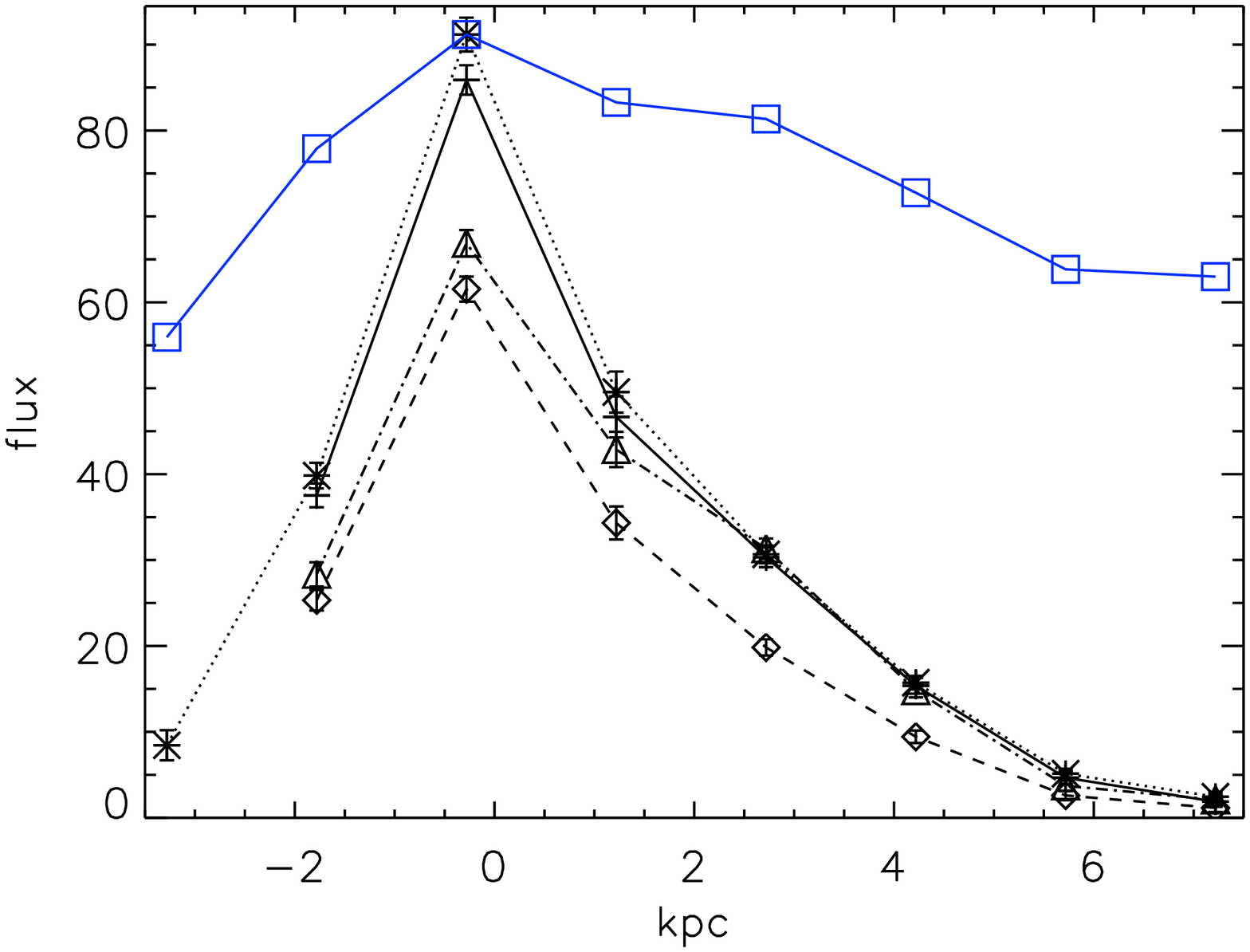}
    \includegraphics[width=0.35\textwidth, angle=90]{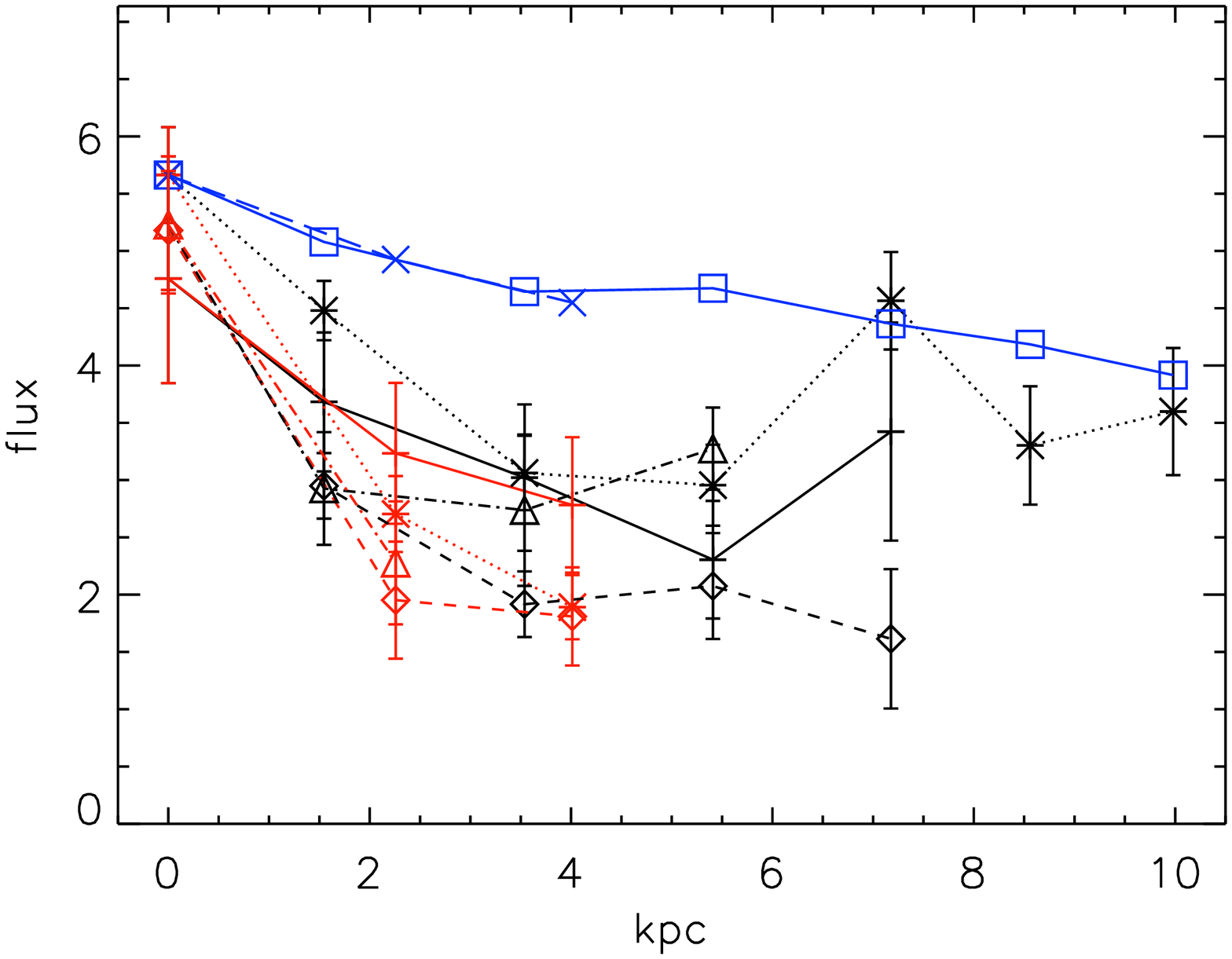}
    \includegraphics[width=0.35\textwidth, angle=90]{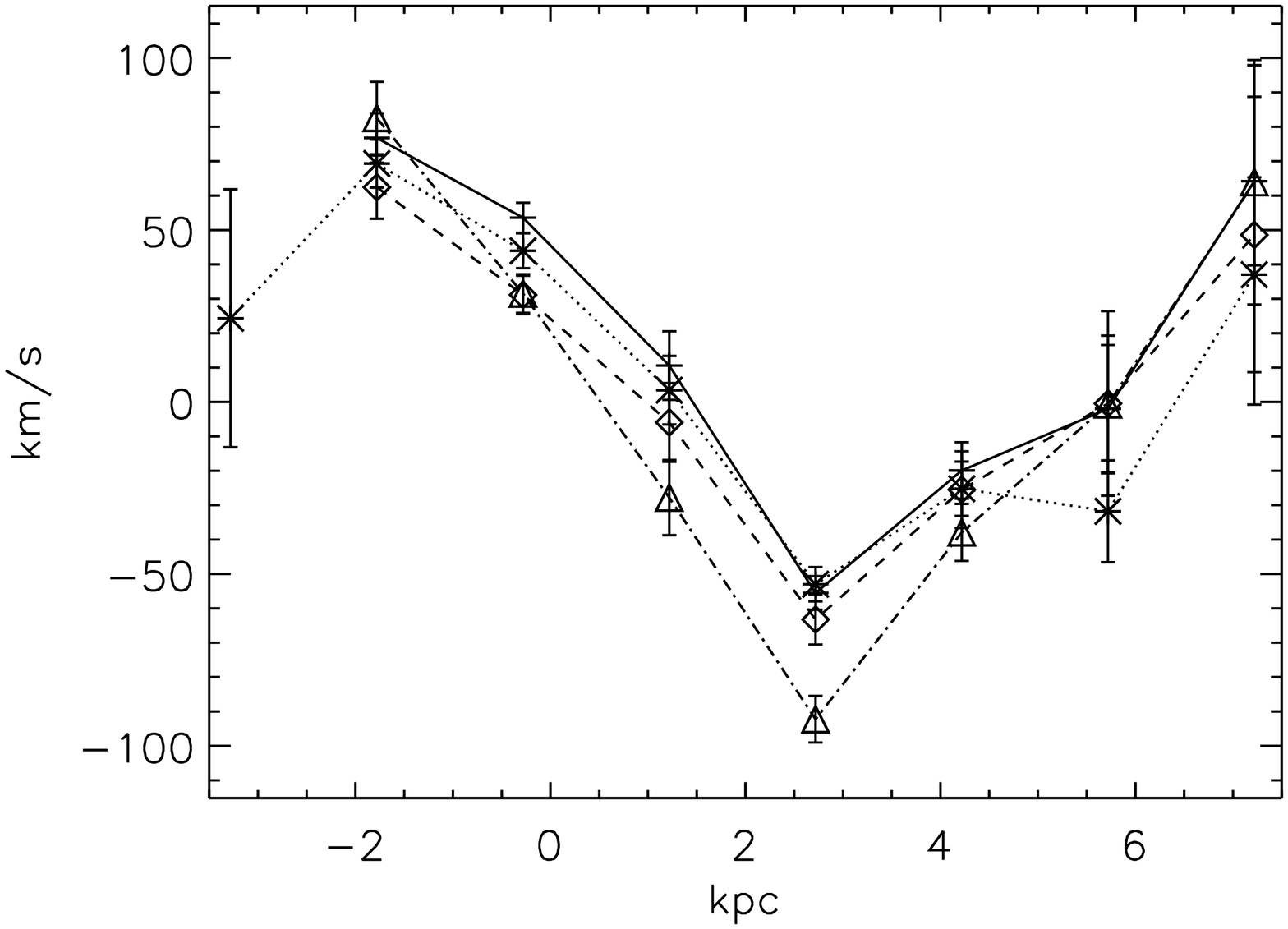}
    \includegraphics[width=0.35\textwidth, angle=90]{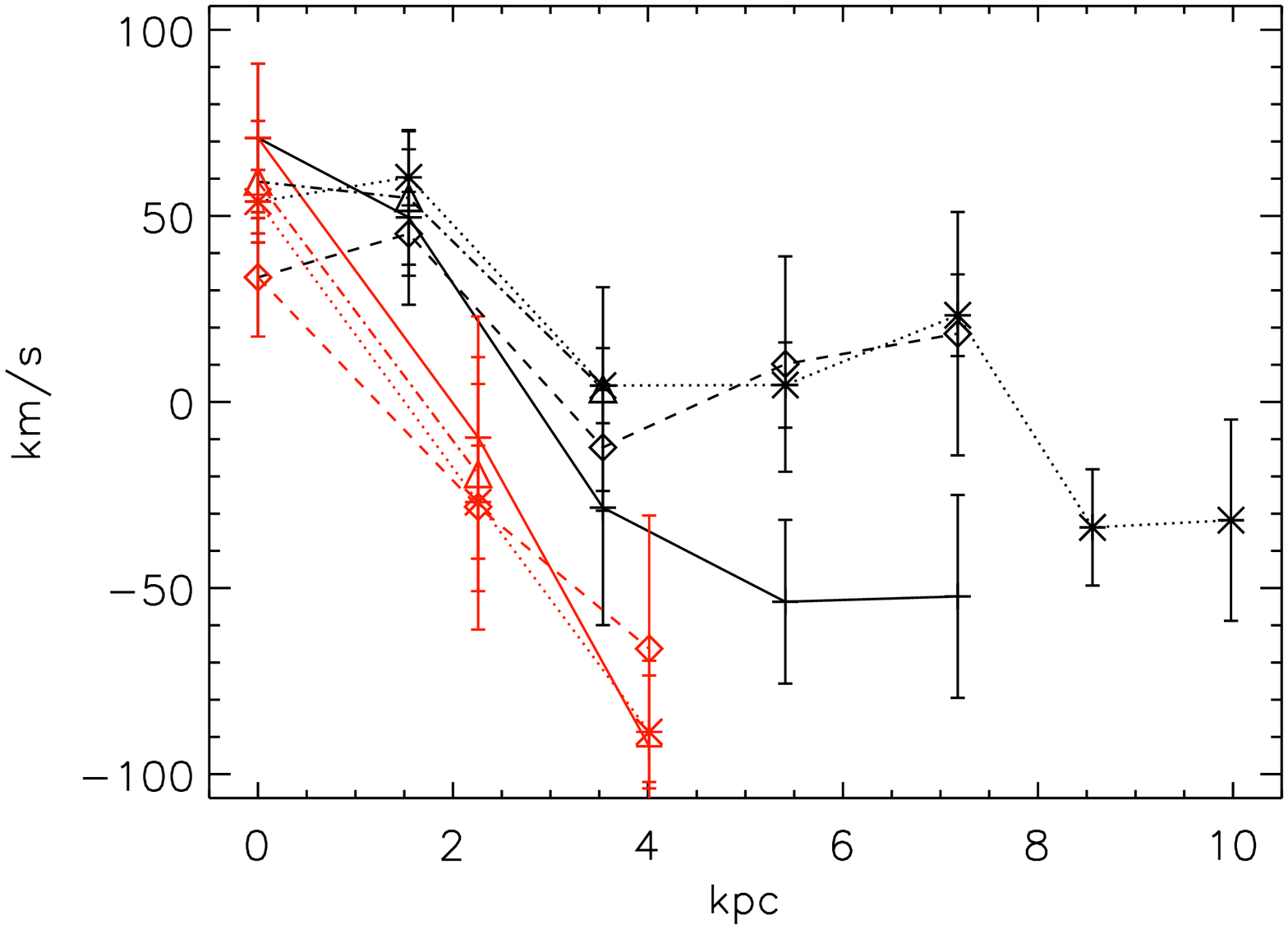}
    \includegraphics[width=0.35\textwidth, angle=90]{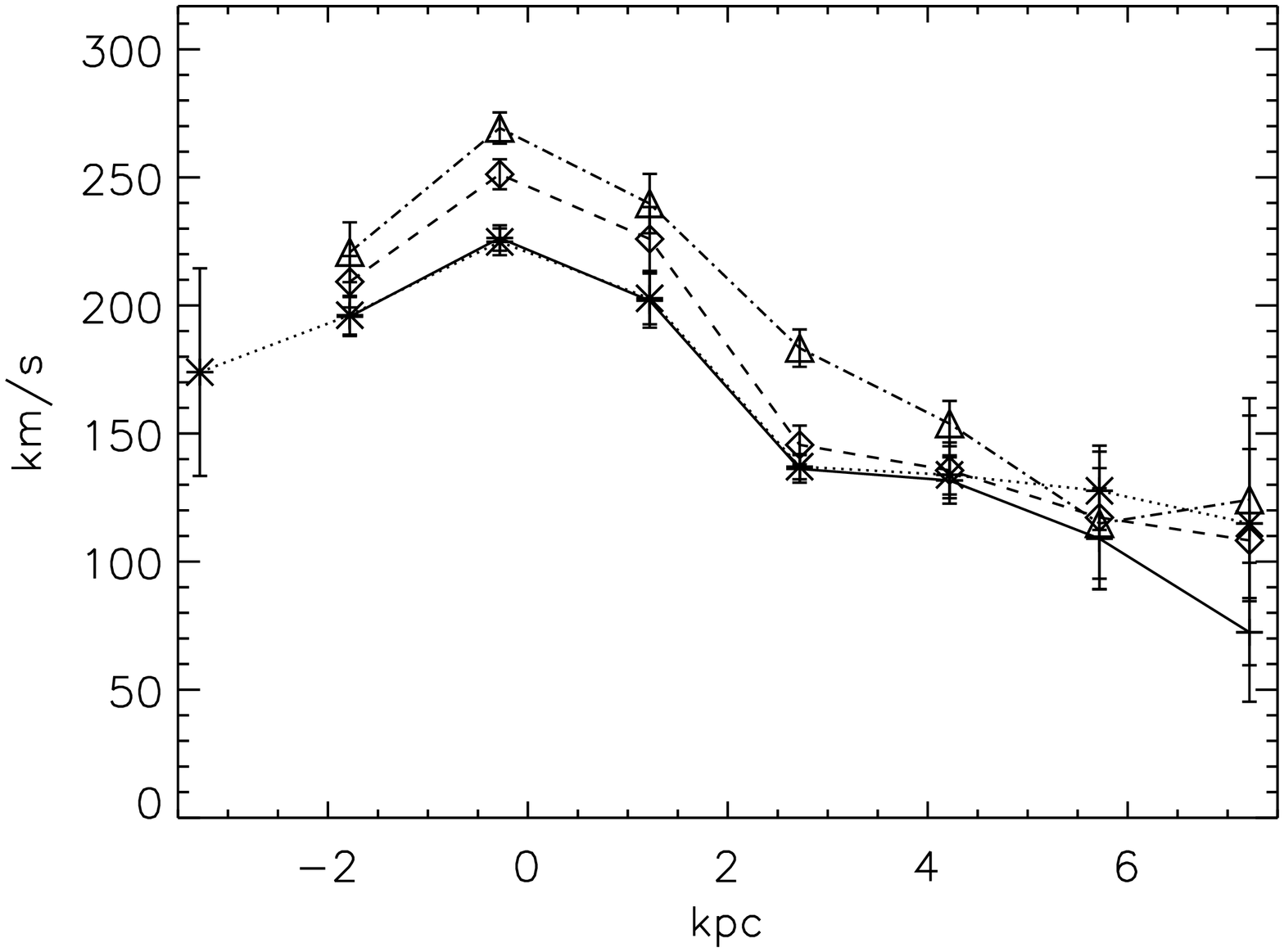}
    \includegraphics[width=0.35\textwidth, angle=90]{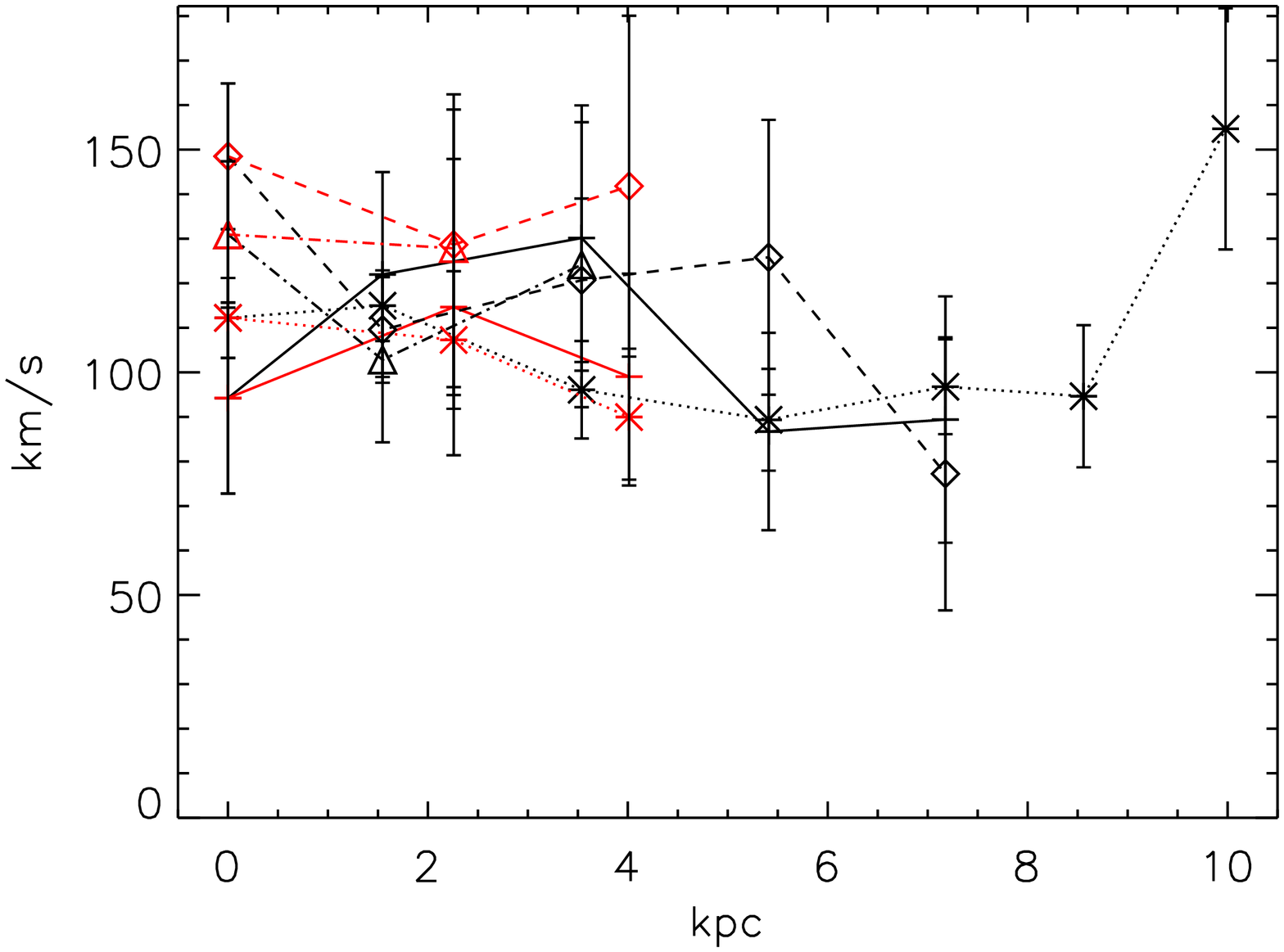}
  \vspace{0.5cm}
  \caption{ABELL 2597 Filaments. We show from top to bottom the flux, velocity and velocity dispersion along the filaments in A2597. The left images show the northern filament and the right images the southern filament. Each point shown represents a 1$\arcsec\times$1$\arcsec$ area along a filament. These areas are marked by the green and red squares in Fig. \ref{fig_h2_area_ss44}. The black points correspond to green squares and the red points to the red squares.  The distance of the points along the northern filament are given in $\mathrm{kpc}$ from the nucleus, here north is positive. The distance of the points along the southern filament are given in $\mathrm{kpc}$ from the north-eastern tip of this filament. The black points trace the southern filament from its north-eastern tip towards the south-west. The red points trace the southern filament from its north-eastern tip towards the south-east. The plus, asterisk, diamond and triangle symbols indicate values obtained for the H$_{\mathrm{2}}$~1-0~S(1), 1-0~S(3), 1-0~S(5) and the Pa~$\alpha$ lines respectively. The blue lines and symbols indicate values obtained for the X-ray emission as observed by \textit{CHANDRA}. These X-ray points have been normalised with respect to the maximum value of the H$_{\mathrm{2}}$~1-0~S(3) line.}\label{fig_film_a2597_ss20}
\end{figure*}

\begin{figure*}
    \includegraphics[width=0.35\textwidth, angle=90]{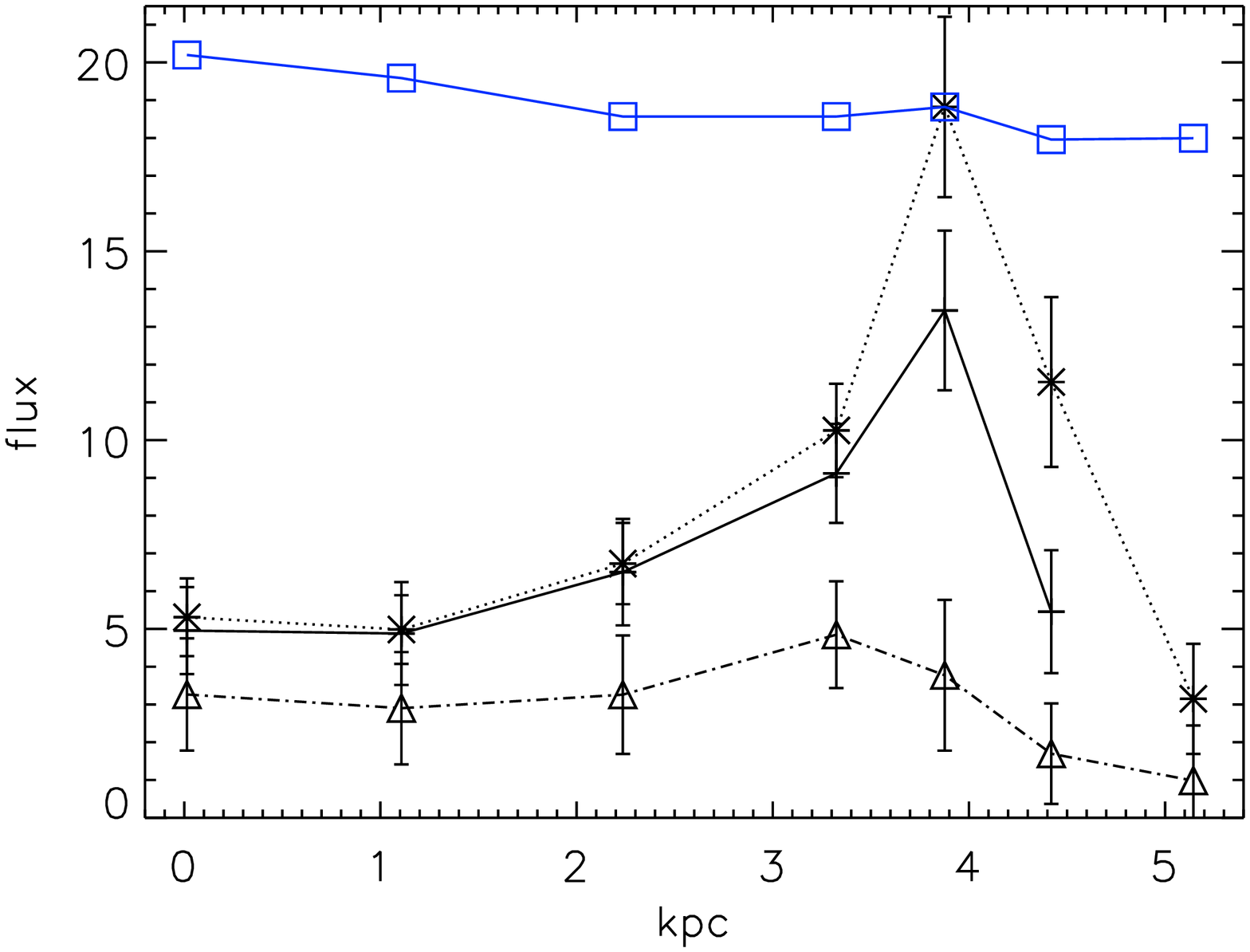}
    \includegraphics[width=0.35\textwidth, angle=90]{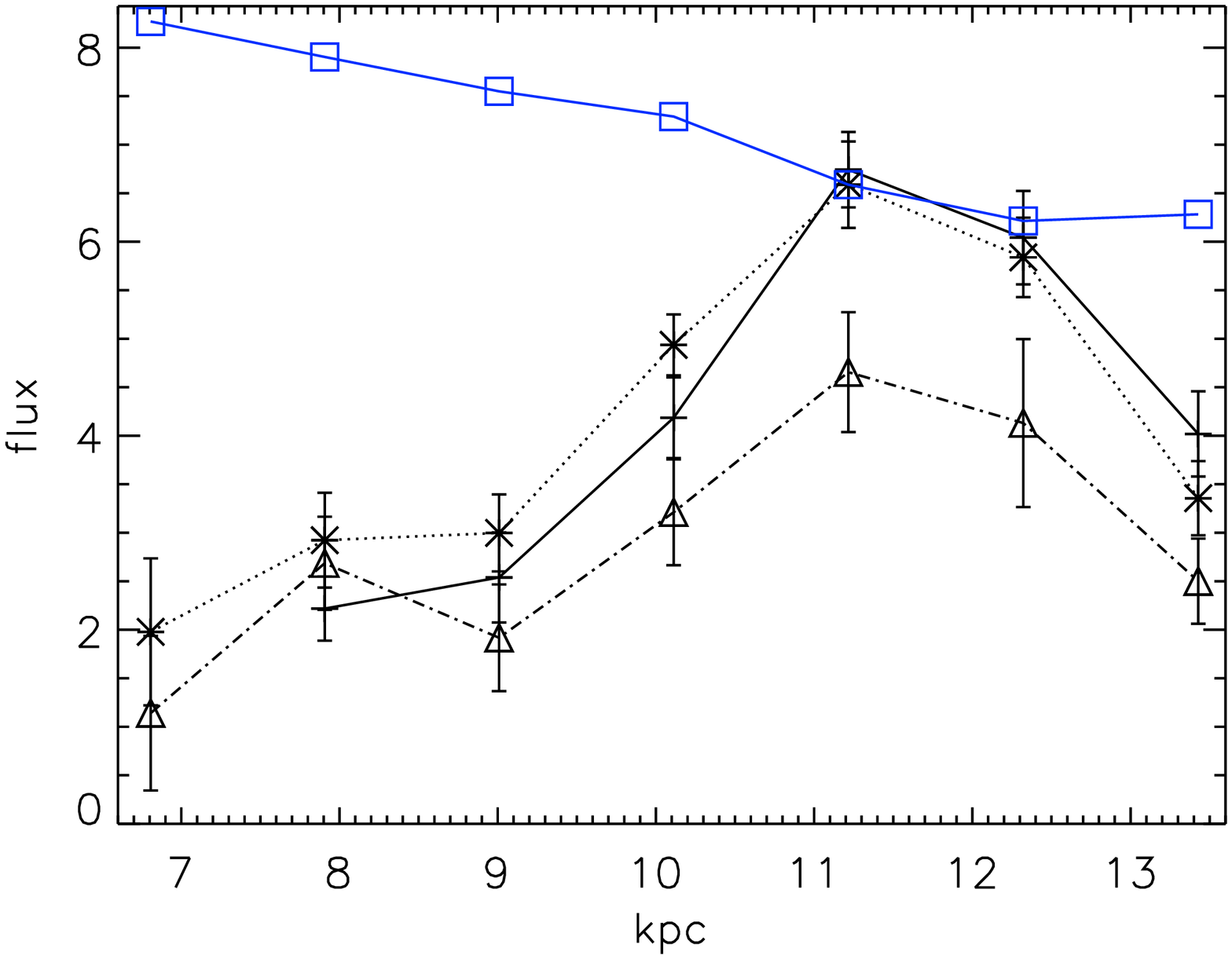}
    \includegraphics[width=0.35\textwidth, angle=90]{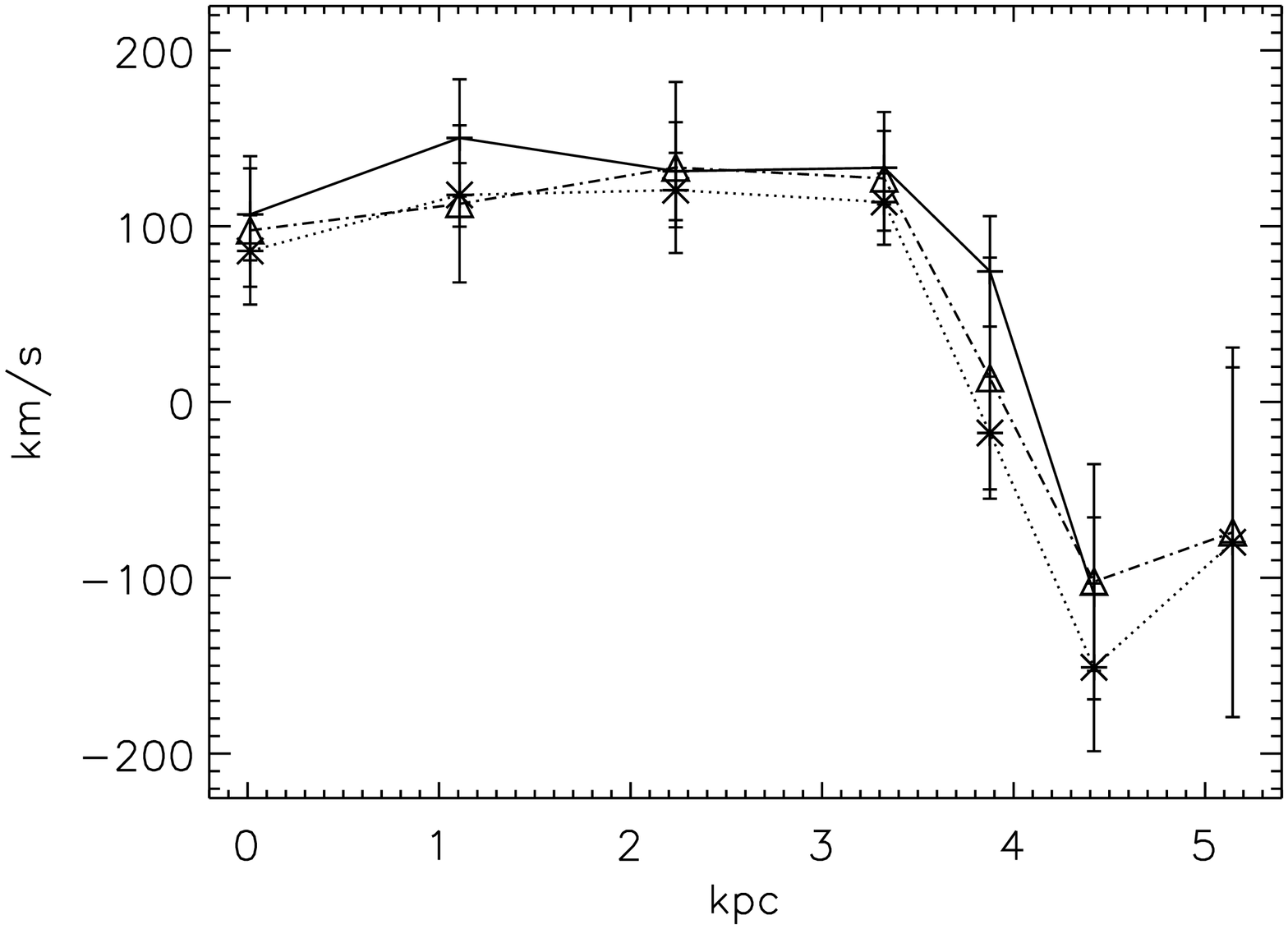}
    \includegraphics[width=0.35\textwidth, angle=90]{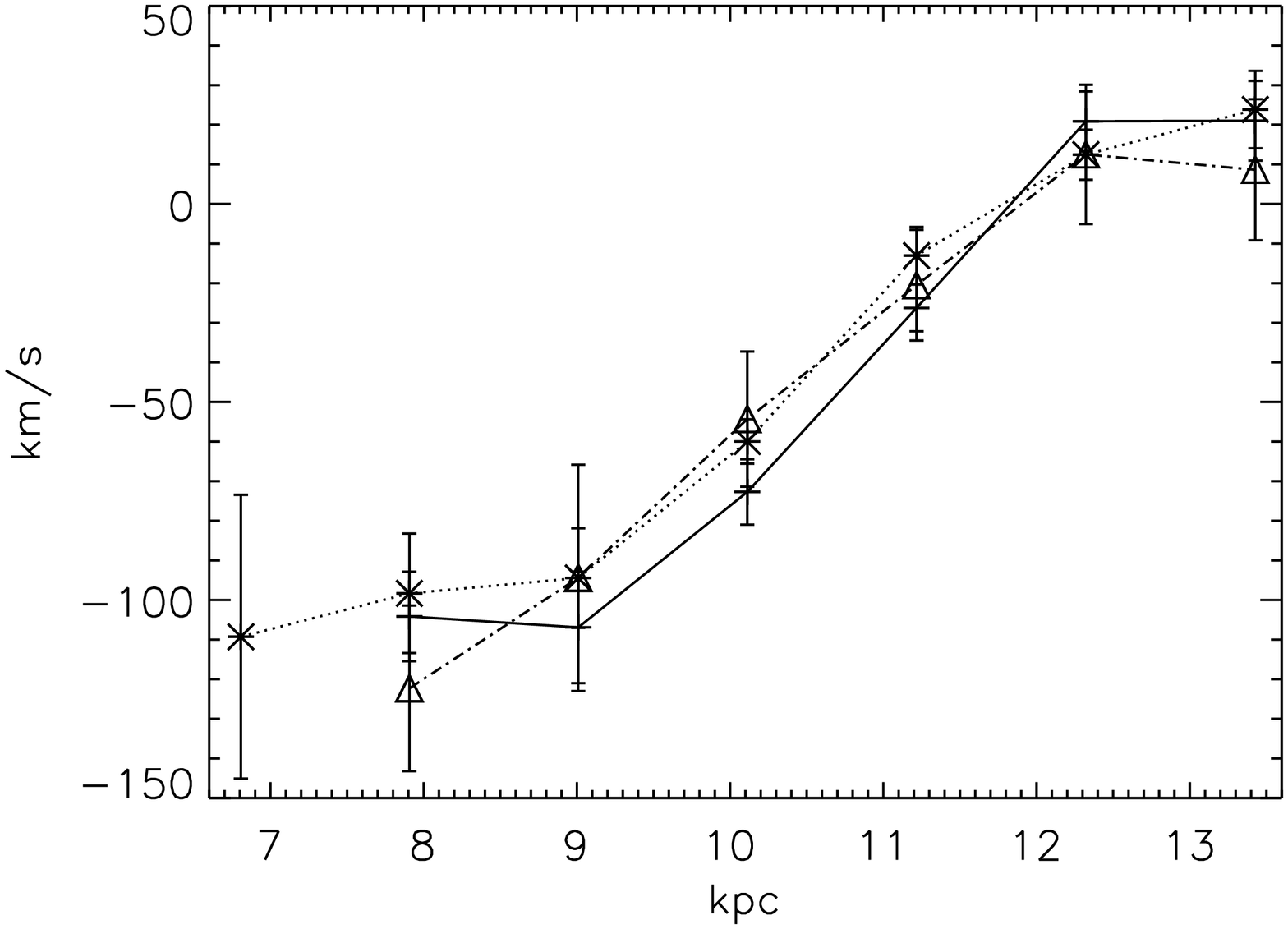}
    \includegraphics[width=0.35\textwidth, angle=90]{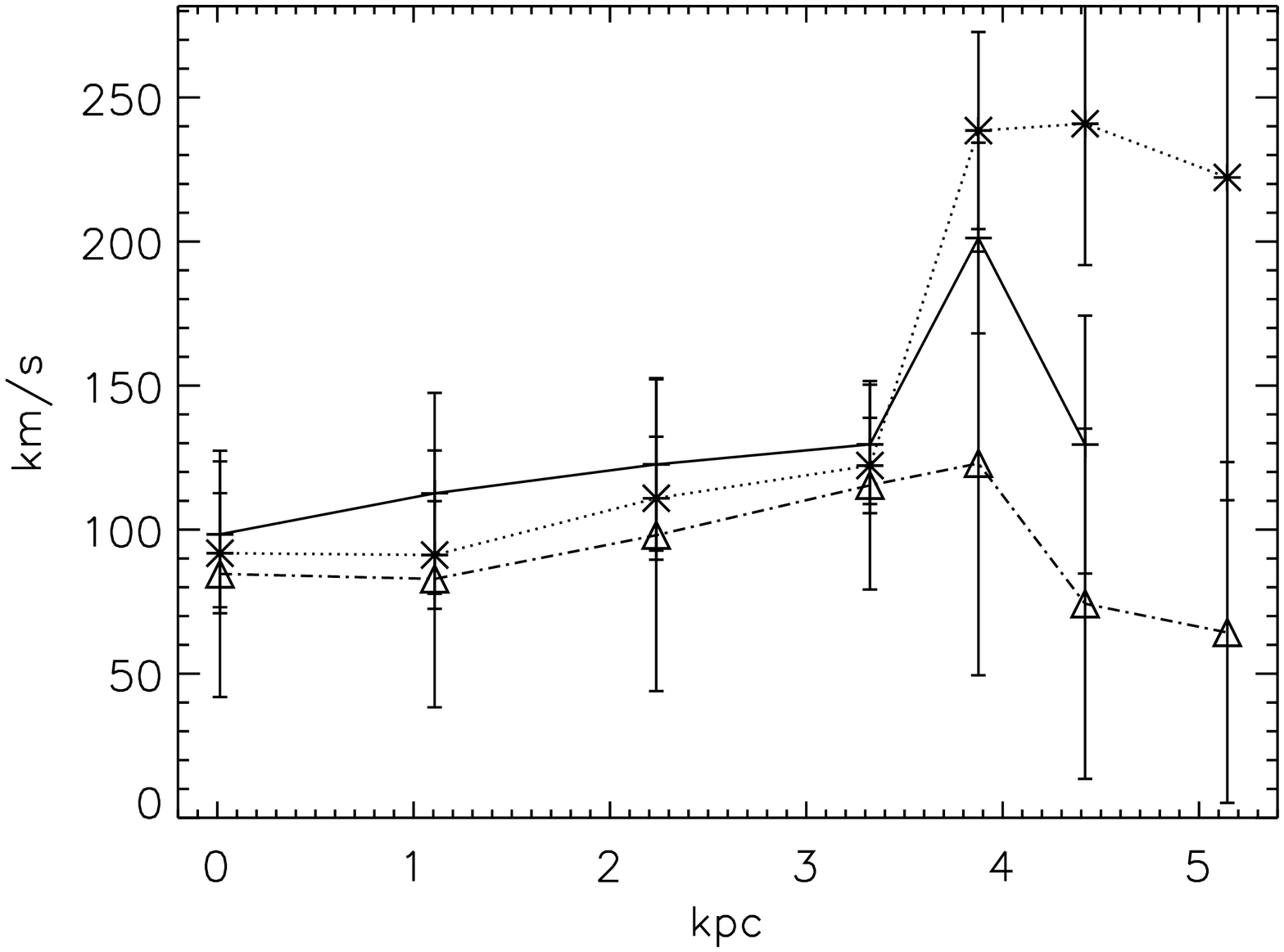}
    \includegraphics[width=0.35\textwidth, angle=90]{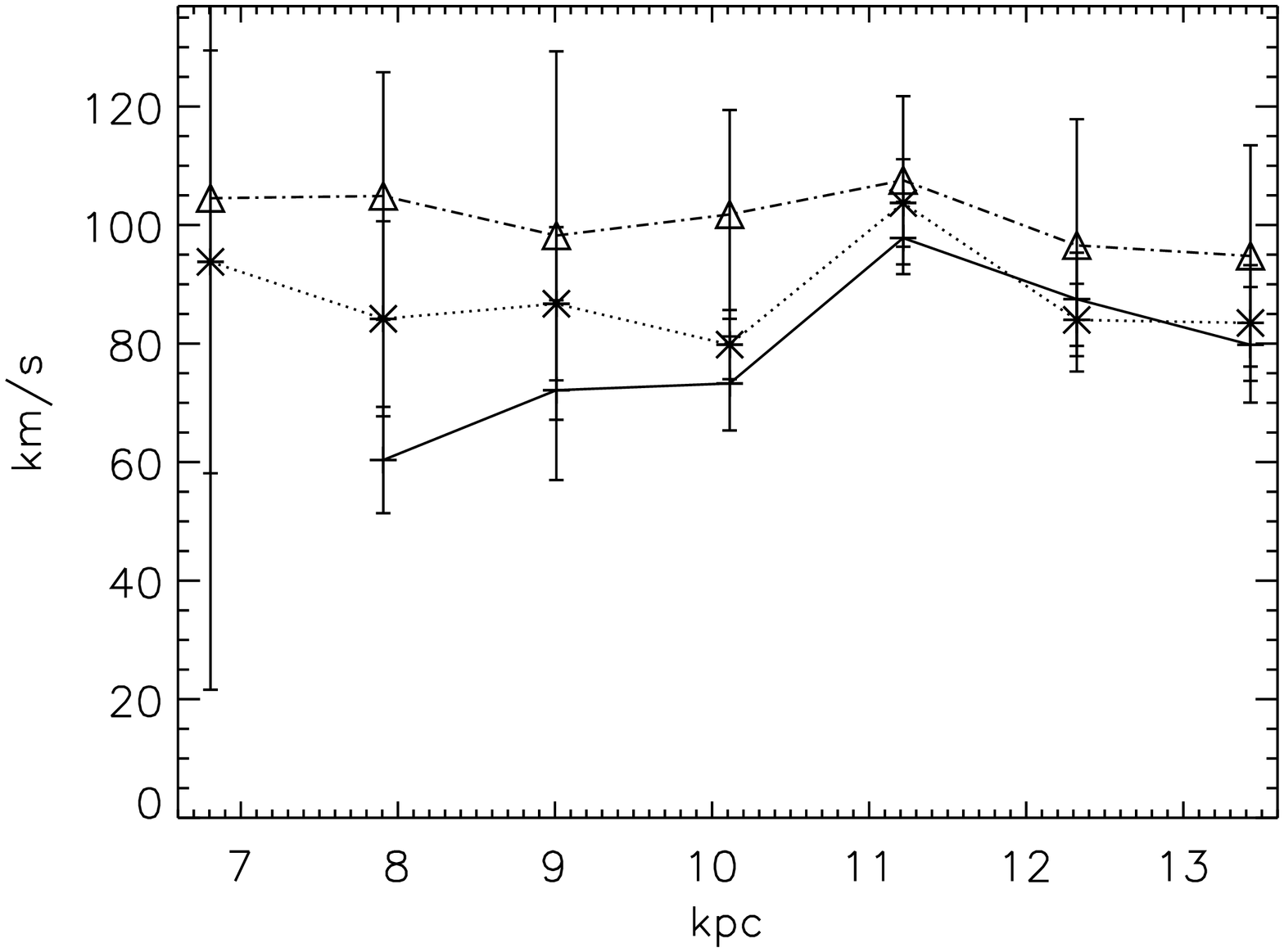}
  \vspace{0.5cm}
  \caption{SERSIC 159-03 Filaments. We show from top to bottom the flux, velocity and velocity dispersion along the filaments in S159. The left images show the northern filament and the right images the western filament. Each point shown represents a 1$\arcsec\times$1$\arcsec$ area along a filament. These areas are marked by the green squares in Fig. \ref{fig_h2_area_ss44}.  The distance of the points along the northern filament are given in $\mathrm{kpc}$ to the north-eastern tip of this filament. The distance of the points along the western filament are given in $\mathrm{kpc}$ from the nucleus. The plus, asterisk and triangle symbols indicate values obtained for the H$_{\mathrm{2}}$~1-0~S(1), 1-0~S(3) and the Pa~$\alpha$ lines respectively. The blue lines and symbols indicate values obtained for the X-ray emission as observed by \textit{CHANDRA}. These X-ray points have been normalised with respect to the maximum value of the H$_{\mathrm{2}}$~1-0~S(3) line.}\label{fig_film_sersic_ss22}
\end{figure*}

\begin{figure*}
    \includegraphics[width=0.45\textwidth, angle=90]{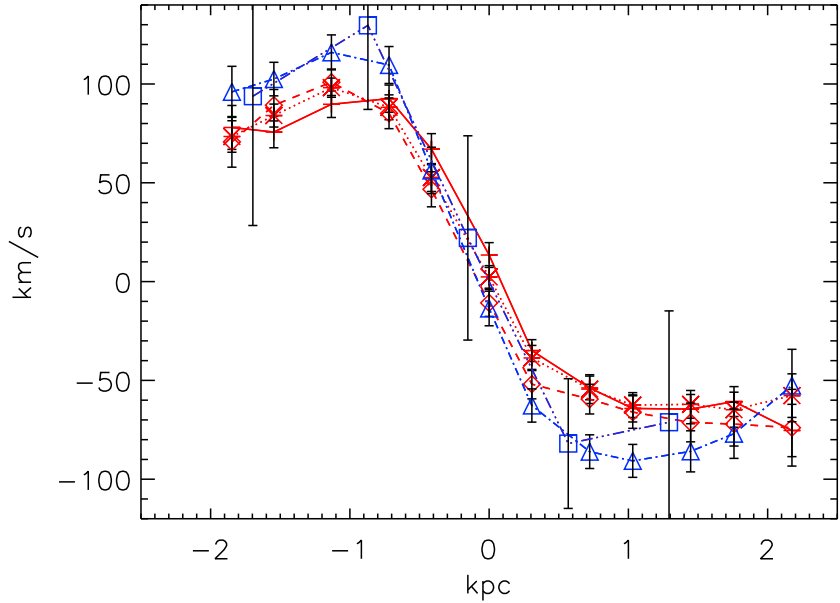}
  \vspace{0.5cm}
  \caption{ABELL 2597 Position-Velocity Diagram. Each point shown represents a 0.25$\arcsec\times$1.0$\arcsec$ area along a pseudo long slit with a PA of 105.5 degrees. The slit centre is placed 1~$\mathrm{kpc}$ south of the nucleus. This is done to avoid the velocity features just north-east and north-west of the nucleus. The plus, asterisk, diamond, triangle and square symbols indicate values obtained for the H$_{\mathrm{2}}$~1-0~S(1), 1-0~S(3), 1-0~S(5), Pa~$\alpha$ and Fe~II lines respectively. The molecular line velocities are shown in red and the ionised line velocities are shown in blue.}\label{fig_a2597_pv}
\end{figure*}

In the case of a gas in LTE, assuming a ortho:para abundance ratio of 3:1, there is a simple relation between the flux $\textit{F}$ and the temperature $T_{\mathrm{u}}$ corresponding to the energy of the upper state of a line, $F$~$\sim$~$h\nu~AN$~$\sim$~$h\nu~gAexp(-T_{\mathrm{u}}/T_{\mathrm{exc}})$, \citep[e.g., J05;][]{J01,W02,W05}. Normalising the flux of each H$_{\mathrm{2}}$ line flux by the flux of the corresponding H$_{\mathrm{2}}$~1-0~S(1) line, we find;
\\
\\
\begin{eqnarray}
ln(F) &=& ln\left(\frac{F_{i}\nu_{S1}A_{S1}g_{S1}}{F_{S1}\nu_{i}A_{i}g_{i}}\right)\\ 
 &=& \left(\frac{-1}{T_{\mathrm{exc}}}\right) \times (T_{\mathrm{u,i}}-T_{\mathrm{u,S1}})
\end{eqnarray}
\\
\\
Here $\textit{F}$ is the flux of the line, $\textit{A}$ its transition probability, $\nu$ its frequency and $\textit{g}$ the statistical weight of the transition. If the molecular gas is in LTE the H$_{\mathrm{2}}$ emission lines will lie on a straight line in a diagram of $\mathrm{ln(\textit{F})}$ vs. $\textit{T}_{\mathrm{u}}$. $\textit{T}_{\mathrm{exc}}$, the reciprocal of the slope, will then be the kinetic temperature of this gas. We have investigated relation (1) for the all regions selected in Abell 2597 in which we have detected at least 3 H$_{\mathrm{2}}$ lines. We show in Fig. \ref{fig_lte_mod_a2597_ss20} that the H$_{\mathrm{2}}$ lines detected in these regions are well fit by a thermal excitation model, with an average temperature of about 2300~$\mathrm{K}$. The derived excitation temperature for each region is given in Table \ref{tabl_line_area_a2597_ss20}.

Besides the best-fitting LTE model for the H$_{\mathrm{2}}$ line fluxes we plot three additional H$_{\mathrm{2}}$ models. These models are shown for qualitative comparison purposes only, and are not tuned exactly to our physical conditions, see Fig. \ref{fig_lte_mod_a2597_ss20}. The best-fitting LTE model is given by the black solid line. The red dotted line is a low-density UV fluorescence model by \citet{B87} that does not include collisions (their model 14; $\textit{n}$=3$\times$10$^{3}$ cm$^{-3}$, a temperature $\textit{T}$=100 $\mathrm{K}$ and a UV intensity $\textit{I}_{\mathrm{UV}}$=10$^{3}$ relative to $\textit{I}_{\mathrm{tot}}$). The blue dotted line is a high-density UV fluorescence model by \citet{S89} which does include collisions (their model 2D; $\textit{n}$=1$\times$10$^{6}$ $\mathrm{cm}^{-3}$, a temperature $\textit{T}$=1000 $\mathrm{K}$ and $\textit{I}_{\mathrm{UV}}$=10$^{2}$ relative to $\textit{I}_{\mathrm{tot}}$). Lastly the green dotted line is the cosmic ray model by \citet{F09}, which was developed for the gas filaments observed in the Perseus cluster.

The Black \& van Dishoeck low-density UV and the Ferland cosmic ray models have distinct features that make them deviate from a thermal model. Low-density UV fluorescence models tend to boost the higher S-transitions (2-1, 3-2,...) relative to the 1-0~S-transitions as compared to a LTE model. The Ferland model is observed to boost the even H$_{\mathrm{2}}$~1-0~S-transitions relative to uneven H$_{\mathrm{2}}$~1-0~S-transitions as compared to a LTE model. The high-density UV model by Sternberg \& Dalgarno simply shows that at high densities, collisions within the gas will cause it to become thermalised and thus the line ratios also produce a straight line in our excitation diagrams. We thus conclude, qualitatively, that out of the four models investigated here that a LTE model provides the best description of the data.

As the molecular line ratios in A2597 appear to be in LTE this implies that the density of this gas is near its critical density, $\textit{n}_{\mathrm{H_{2},crit}}$~$\approx$ 10$^{6}$~$\mathrm{cm}^{3}$ \citep{Sb82} and is dominated by collisional excitation. Information regarding the source of excitation is thus not obtainable from this data set.

There is a trend that on average we find higher LTE temperatures of the molecular gas in the filaments as compared to the nuclear region. We note though that within errors the temperatures agree for all regions, except for A6. Neither the H$_{\mathrm{2}}$~1-0 nor the H$_{\mathrm{2}}$~2-1~S-transitions lie exactly on a straight line. If we use only pairs of lines like the H$_{\mathrm{2}}$~1-0~S(1), 1-0~S(3) line pair or the H$_{\mathrm{2}}$~1-0~S(3), 1-0~S(5) line pair to determine an excitation temperature, assuming LTE, we find on average a temperature that is a few hundred degrees lower or higher than when all lines are used. Typically the first pair gives a lower temperature and the second pair a higher temperature.

If the H$_{\mathrm{2}}$ gas observed here is at its critical density then there is a pressure imbalance between the molecular gas $\textit{p}(H_{2})$~$\sim$~$\textit{nT}$~=~$\textit{n}_{\mathrm{crit}}\textit{T}_{\mathrm{exc}}$~$\approx$~10$^{9}$ and the ionised gas has $\textit{p}(HII)$~$\sim$~$\textit{nT}$~=~10$^{2}\times$10$^{4}$~=~10$^{6}$. That the molecular and ionised gas are not in pressure equilibrium has previously been suggested by J05 using similar arguments.

For Sersic 159-03 only two H$_{\mathrm{2}}$ lines were reliably detected in all regions. If we assume that the molecular gas here is in LTE, we can calculate an excitation temperature for this gas. The resulting excitation temperatures for the selected regions are given in Table \ref{tabl_line_area_sersic_ss22}. Again we find an average temperature of about 2300 K for the H$_{\mathrm{2}}$ gas in the filaments. In the nuclear region the strong increase of the H$_{\mathrm{2}}$~1-0~S(3)/H$_{\mathrm{2}}$~1-0~S(1) ratio leads to a very uncertain and physically unrealistic temperature above the dissociation temperature for H$_{\mathrm{2}}$. This either means that our line fits overestimate this ratio or that a different excitation mechanism is at work here.

In this work we have only considered the K-band H$_{\mathrm{2}}$ lines. These H$_{\mathrm{2}}$ lines do not show strong deviations from a single temperature thermal model with $\textit{T}$~$\approx$~2000-2500~$\mathrm{K}$. From recent Spitzer spectroscopy (Jaffe \& Bremer in prep.) we find that another H$_{\mathrm{2}}$ gas component exists in the nuclear regions of A2597 and S159 with a LTE temperature of about 300~$\mathrm{K}$. The situation in these clusters is thus similar to the situation in the Perseus cluster where multiple temperature LTE components have been invoked to explain the H$_{\mathrm{2}}$ line ratios \citep[e.g.,][]{W02,W05,J07}.

Whether the need for multiple LTE components to H$_{\mathrm{2}}$ line ratios hints at a multiphase gas, a difference in optical depth, or a different excitation mechanism is unclear currently. A more thorough modelling of the molecular gas, including all of the measured of the H$_{\mathrm{2}}$ lines in the infrared and mid-infrared, for A2597 and S159 will be presented by us in a future paper.  

\subsection{Luminosity of the Warm Molecular Gas}
If we assume that the H$_{\mathrm{2}}$ gas is in LTE and can be described by a single kinetic temperature of about 2000~$\mathrm{K}$, then the H$_{\mathrm{2}}$ 1-0 S(1) line luminosity represents about 10 percent of the total H$_{\mathrm{2}}$ luminosity. This is estimated using a list of 312 H$_{\mathrm{2}}$ lines, corresponding to all H$_{\mathrm{2}}$ emission lines with an intensity greater than 1 percent of H$_{\mathrm{2}}$~1-0~S(1) line flux for $\textit{T}$~$\leq$~4000K. From the total integrated H$_{\mathrm{2}}$~1-0~S(1) line fluxes in Tables \ref{tabl_line_summ_a2597_ss44} and \ref{tabl_line_summ_sersic_ss44} we thus find $\textit{L}$($\mathrm{H_{2}}$, A2597)~=~1.1$\times$10$^{42}$~$\mathrm{erg}$~$\mathrm{s}^{-1}$ and $\textit{L}$($\mathrm{H_{2}}$, S159)~=~1.2$\times$10$^{41}$~$\mathrm{erg}$~$\mathrm{s}^{-1}$ within the fields observed by us.

This is significantly below the total H$_{\mathrm{2}}$ luminosity  estimated by J05, i.e. $\textit{L}$($\mathrm{H_{2}}$)$\sim$10$^{43-44}$~$\mathrm{erg}$~$\mathrm{s}^{-1}$. J05 calculate the total H$_{\mathrm{2}}$ luminosity in the same manner as we do here. The difference follows from two simple arguments. Firstly J05 assumed that the H$_{\mathrm{2}}$ 1-0 S(1) line luminosity represents only about 1 percent of the total H$_{\mathrm{2}}$ luminosity. This is true for low-density UV excitation models such as the Black \& van Dishoeck models. However, when the gas is in LTE (see above) or in a XDR environment \citep{D92} a fraction of 10 percent is found. The J05 results thus need to be corrected down by a factor of 10 as consequence of this. The second argument that J05 made is that the H$_{\mathrm{2}}$ emission coexists with the H$\alpha$ emission. This argument follows from the HII/H$_{\mathrm{2}}$ ratio which is observed to remain rather constant over large areas (J05 and this work). J05 detect H$\alpha$ over a much larger area than the fields observed by us. Correcting our results upward for the area covered by the H$\alpha$ map in J05 increases the H$_{\mathrm{2}}$ luminosity given above by a factor of 10. With these corrections, we find that the H$_{\mathrm{2}}$ luminosity obtained in this work is in good agreement with the J05 results.

To conclude, upon correcting our results for the area covered by the total extent of the ionised gas nebulae, we find H$_{\mathrm{2}}$ luminosities $\textit{L}$($\mathrm{H_{2}}$, A2597)$\sim$10$^{43}$ $\mathrm{erg}$~$\mathrm{s}^{-1}$ and $\textit{L}$($\mathrm{H_{2}}$, S159)$\sim$10$^{42}$ $\mathrm{erg}$~$\mathrm{s}^{-1}$ for our clusters. The analysis performed above relies heavily on the assumption that the H$_{\mathrm{2}}$ gas is in a single temperature phase. If there is colder H$_{\mathrm{2}}$ gas present in these clusters, as expected from CO observations by \citet{E01} and mid-infrared spectroscopy (Jaffe \& Bremer in prep.) we underestimate the total H$_{\mathrm{2}}$ luminosity here.

\subsection{Mass of the Warm Molecular Gas}
In the previous section we found that a LTE model provides a good description of the observed H$_{\mathrm{2}}$ line ratios in A2597 and assumed that the same is true for a S159. Following \citet{S82} and \citet{S09} the mass of a H$_{\mathrm{2}}$ gas in LTE conditions with a single kinetic temperature, can be calculated using the following equation,
\\
\\
\begin{eqnarray}
M_\mathrm{{H_{2}}} &=& \frac{2m_{p}F_{\mathrm{H_{2} 1-0 S1}}4\pi D^{2}}{f_{\mathrm{\nu =1,J=3}}A_{\mathrm{H_{2} 1-0 S1}}h\nu}\\
 &=& (5.08 \times 10^{13}) \left(\frac{F_{\mathrm{H_{2} 1-0 S1}}}{\mathrm{erg} \mathrm{s}^{-1} \mathrm{cm}^{-2}}\right) \left(\frac{D}{\mathrm{Mpc}}\right)^{2}
\end{eqnarray}
\\
\\
Here $\textit{m}_{\mathrm{p}}$ is the proton mass, $\textit{h}$ is the Planck constant and $\textit{D}$ is the distance to the cluster core. $\textit{F}_{\mathrm{H_{2} 1-0 S1}}$ is the flux of the H$_{\mathrm{2}}$ 1-0 S(1) line and $\textit{f}_{\mathrm{\nu =1,J=3}}$ is its population fraction. $\textit{A}_{\mathrm{H_{2} 1-0 S1}}$ is the transition probability ($\textit{A}_{\mathrm{H_{2} 1-0 S1}}$~=~3.47$\times 10^{-7}$ $\mathrm{s}^{-1}$) and $\nu$ is the frequency of the H$_{\mathrm{2}}$~1-0~S(1) line. The right-hand side of equation (2) is obtained by assuming a vibration temperature $\textit{T}_{\mathrm{vib}}$~=~2000 $\mathrm{K}$ for the gas, in which case $\textit{f}_{\mathrm{\nu =1,J=3}}$~=~1.22$\times 10^{-2}$. In the above formula $\textit{M}_{\mathrm{H_{2}}}$ is given in units of solar masses.

The resulting H$_{\mathrm{2}}$ gas masses for the regions investigated here are given in Tables \ref{tabl_phys_area_a2597_ss20} and \ref{tabl_phys_area_sersic_ss22}. The total H$_{\mathrm{2}}$ gas mass integrated over all observed fields is 4.5$\times$10$^{4}$~$\pm$~4.2$\times$10$^{2}$~$\mathrm{M_{\odot}}$ for A2597 and 5.2$\times$10$^{3}$~$\pm$~2.9$\times$10$^{3}$~$\mathrm{M_{\odot}}$ for S159. Since our observations do not cover the entire cluster core these numbers underestimate the total amount of molecular gas present in these clusters. We furthermore note that there is evidence from mid-infrared spectroscopy from Jaffe \& Bremer (in prep.) that a colder H$_{\mathrm{2}}$ phase ($\textit{T}$~$\sim$~300~$\mathrm{K}$) exists in A2597 and S159. If this colder H$_{\mathrm{2}}$ gas coexists with the warmer H$_{\mathrm{2}}$ gas observed here this means that we strongly underestimate the H$_{\mathrm{2}}$ mass present in these clusters. CO observations by \citet{E01} indicate that even colder H$_{\mathrm{2}}$ gas is present in A2597 at a temperature of a few tens of kelvin. The total molecular gas mass inferred from CO for A2597 is 8$\times$10$^{9}$~$\mathrm{M_{\odot}}$, although this value relies heavily on the CO to H$_{\mathrm{2}}$ conversion factor. To our knowledge there is no published CO detection for S159.

\subsection{Mass of the Ionised Gas}
Following \citet{S09} the mass of the ionised hydrogen gas can be estimated as $\textit{M}_{\mathrm{HII}}$~=~$\textit{m}_{\mathrm{p}}\textit{n}_{\mathrm{e}}\textit{V}_{\mathrm{HII}}$. Here $\textit{m}_{\mathrm{p}}$ is the proton mass, $\textit{n}_{\mathrm{e}}$ is the electron density and $\textit{V}_{\mathrm{HII}}$ is the volume of the HII emitting region. Using $\textit{j}_{\mathrm{H\beta}}/\textit{n}_{\mathrm{e}}^{2}$~=~9.788$\times$10$^{-27}$~$\mathrm{erg}$~$\mathrm{s}^{-1}$~$\mathrm{cm}^{3}$ and $\textit{j}_{\mathrm{Pa\alpha}}/\textit{j}_{\mathrm{H\beta}}$~=~0.339 \citep{O06}, for $\textit{n}_{\mathrm{e}}$~=~10$^{2}$~$\mathrm{cm}^{-3}$ and $\textit{T}$~=~10$^{4}$~$\mathrm{K}$), we can write $\textit{F}_{\mathrm{Pa\alpha}}$~=~$\textit{j}_{\mathrm{Pa\alpha}}\textit{V}_{\mathrm{HII}}\textit{D}^{-2}$~=~3.32$\times$10$^{-27}\times$($\textit{n}_{\mathrm{e}}^{2}\textit{V}_{\mathrm{HII}}\textit{D}^{-2}$). Here $\textit{F}_{\mathrm{Pa\alpha}}$ is the flux of the Pa~$\alpha$ line, $\textit{j}_{\mathrm{Pa\alpha}}$ is the volume emissivity coefficient of Pa~$\alpha$ and $\textit{D}$ is the distance to the cluster core. The HII gas mass can be written as:
\\
\\
\begin{eqnarray}
M_{\mathrm{HII}} &=& m_{\mathrm{p}}n_{\mathrm{e}}V_{\mathrm{HII}}\\ 
 &=& (2.41 \times10^{18}) \left(\frac{F_{\mathrm{Pa\alpha}}}{\mathrm{erg} \mathrm{s}^{-1} \mathrm{cm}^{-2}}\right)\\ \nonumber
 &\times& \left(\frac{D}{\mathrm{Mpc}}\right)^{2} \left(\frac{n_{\mathrm{e}}}{\mathrm{cm}^{-3}}\right)^{-1}
\end{eqnarray}
\\
\\
Using the right-hand side of equation (3) the mass of the HII emitting gas, $\textit{M}_{\mathrm{HII}}$, is given in units of solar masses. \citet{H89} finds an electron density, $\textit{n}_{\mathrm{e}}$~=~200~$\mathrm{cm}^{-3}$, for the central regions of A2597. Thus assuming an electron density $\textit{n}_{\mathrm{e}}$~=~200~$\mathrm{cm}^{-3}$, the mass of the HII emitting gas in the selected regions of Abell 2597 and Sersic 159-03 can be calculated.

The resulting HII gas masses are given in Tables \ref{tabl_phys_area_a2597_ss20} and \ref{tabl_phys_area_sersic_ss22}. The total HII gas mass integrated over all observed fields is 9.7$\times$10$^{6}$~$\pm$~3.2$\times$10$^{5}$~$\mathrm{M_{\odot}}$ for A2597 and 7.6$\times$10$^{5}$~$\pm$ 8.4$\times$10$^{4}$~$\mathrm{M_{\odot}}$ for S159. Since our observations do not cover the entire cluster core these numbers underestimate the total amount of ionised gas present in these clusters. The total ionised gas mass to warm molecular gas mass is similar in both clusters; $\textit{M}_{\mathrm{HII}}/\textit{M}_{\mathrm{H_{2}}}$~$\sim$ 10$^{2}$. The HII gas mass derived here for A2597 agrees well with \citet{H89} who find $\textit{M}_{\mathrm{HII}}$~=~5.8$\times$10$^{6}$~$\mathrm{M_{\odot}}$ for a region with radius 2.5~$\mathrm{kpc}$ centered on the BCG nucleus.

\subsection{Stability of the Filaments}
From our observations we see that the ionised and molecular gas is locked up in large scale filamentary structures surrounding the BCG. If the filaments in our the clusters are similar to the ones observed in more nearby clusters, such as Perseus \citep{F08}, then we do not resolve their true physical thickness. Furthermore if we assume that the filaments observed here are connected to the global distribution of ionised gas in these clusters then its clear that these extent far beyond the distance observed here by us \citep[J05;][]{O04}.

This interpretation is strengthened by higher resolution images in for example H$\alpha$ and Ly~$\alpha$ emission \citep[D00;][]{O04}. These show the existence of narrow gas filaments coincidental with the larger scale structures observed here. Considering this, even if these large structures consist of many unresolved filaments it seems plausible that all of these have a similar origin. The currently favoured picture is that these structures are related to past and current AGN outflows. Another explanation may be that these structures arise in regions of these clusters where heating and cooling rates are not balanced and thus that we observe residual cooling here.

Not only is the origin of these structures unclear, there are many other puzzling aspects concerning them. On large scales these structures appear to connect all the way  down to the nucleus of the BCG. One may speculate on whether this gas falls in towards the black hole thereby feeding it and sustaining its activity. However, the velocities observed for the filaments studied here or in other clusters \citep{H05} does not agree with gas that is freely infalling from distances far away from the potential centre. We observe the gas to be moving at velocities of about 100~$\mathrm{km}$~$\mathrm{s}^{-1}$, whereas for freely infalling gas we would expect to observe velocities of about 600~$\mathrm{km}$~$\mathrm{s}^{-1}$.

We would expect tidal forces to rip these filaments apart as discussed in \citet{F08}. These authors suggest that magnetic fields may help stabilise these filaments against these gravitational forces. For a typical filament in Perseus, or one of our clusters, the required magnetic field is around a few hundred $\mu$G. From equipartition arguments the ICM typically has magnetic field strengths less than 10~$\mu$G. Larger magnetic field strengths, up to about 50~$\mu$G, have been found in the centres of clusters but these are usually related to current AGN outflows \citep{T07,G04}. Significantly higher magnetic field strengths in the filaments are necessary to stabilise them against gravitational forces.

Some of the gaseous filaments appear, in projection, to be related to the current AGN outflows, see Fig. \ref{fig_sign_maps_a2597_ss44} and \ref{fig_sign_maps_sersic_ss44}. These outflows maybe be responsible for pushing gas outwards, thereby in fact creating the observed filaments \citep{H05}. This may explain why some filaments do not show large infall velocities. However, there are also filaments that are not related to current radio outflows. If these filaments were deposited at their current location by previous AGN outflows, then this gas has had a long time to cool and should now be falling back at high velocities towards the potential centre. No signs of high velocity infalling gas is observed in either of our clusters. We will discuss the relation between the filaments and the radio emission in more detail below.

\section{X-ray and Radio Emission}
In the above sections we have shown that there is a strong spatial and dynamical relation between the HII and H$_{\mathrm{2}}$ gas phases in the cores of A2597 and S159. Below we investigate how these gas phases relate to X-ray and Radio emission.

\subsection{X-ray emission}
We find that globally the H$_{\mathrm{2}}$ and HII line emission in A2597 appears to match the X-ray emission as imaged by \textit{CHANDRA} rather well, see Fig. \ref{fig_area_a2597_ss44}. However, upon examining the detailed distribution, Fig. \ref{fig_film_a2597_ss20}, we find that the X-ray brightness profile is considerable shallower than that of the colder gas.

In S159, see Fig. \ref{fig_film_sersic_ss22}, we observe that the peak X-ray emission is significantly offset from the BCG nucleus and the peak H$_{\mathrm{2}}$, HII emission. One may perhaps argue for a relation between the X-ray and Pa~$\alpha$ emission along the northern filament, but Pa~$\alpha$ is difficult to detect in the nuclear region. There appears to be no relation between Pa~$\alpha$ and X-ray emission along the western filament in S159. This leads us to conclude that is no strong correlation between the X-ray and either the H$_{\mathrm{2}}$ or HII phase in A2597 and S159.

The total X-ray luminosity is $\textit{L}_{\mathrm{X}}$(A2597)~=~2.1$\times$10$^{44}$~$\mathrm{erg}$~$\mathrm{s}^{-1}$ \citep[][0.5-2.0~$\mathrm{keV}$]{dG99} and $\textit{L}_{\mathrm{X}}$(S159)~=~9.6$\times$10$^{43}$~$\mathrm{erg}$~$\mathrm{s}^{-1}$ \citep[][0.5-2.0~$\mathrm{keV}$]{dG99} in A2597 and S159 respectively. Taking into account the difference in areas between this work and that of J05, the total warm H$_{\mathrm{2}}$ luminosity for these clusters is a factor 10-100 times less than their X-ray luminosities and not of the same order of magnitude as previously claimed in J05. Adding the colder H$_{\mathrm{2}}$ gas to this analysis (Jaffe \& Bremer in prep.) to obtain a total warm plus cold H$_{\mathrm{2}}$ luminosity may change this conclusion. 

\subsection{Radio emission}
In Figs. \ref{fig_sign_maps_a2597_ss44}, \ref{fig_kinm_velc_maps_a2597_ss20}, \ref{fig_kinm_disp_maps_a2597_ss20}, \ref{fig_sign_maps_sersic_ss44}, \ref{fig_kinm_velc_maps_sersic_ss22} and \ref{fig_kinm_disp_maps_sersic_ss22}, we show VLA 8.4~$\mathrm{GHz}$ radio contours on top of the H$_{\mathrm{2}}$~1-0~S(3) and Pa~$\alpha$ surface brightness, velocity and velocity dispersion maps. The total radio power is 3.1$\times$10$^{42}$~$\mathrm{erg}$~$\mathrm{s}^{-1}$ for  A2597 and 2.1$\times$10$^{41}$~$\mathrm{erg}$~$\mathrm{s}^{-1}$ for S159 \citep[][integrating the radio spectrum between 10~$\mathrm{MHz}$ and 10~$\mathrm{GHz}$]{Bi08}.

Our observations of the molecular and ionised gas in A2597 and S159 do not have the spatial resolution to investigate the detailed sub-$\mathrm{kpc}$ scale correspondence between the AGN outflows and the gas. On $\mathrm{kpc}$ scales we note small enhancements in the HII and H$_{\mathrm{2}}$ intensity along the lower parts of the projected radio lobes in A2597. Most notably along the northern lobe. D00, using higher resolution HST imaging, have previously shown this in more detail for A2597. Similarly in S159 we observe that on $\mathrm{kpc}$-scales some of the gas appears to lie along the radio lobes. In neither S159 and A2597 is there a clear correlation between the radio emission and the cold gas.

It has been postulated that gas observed on the edges of radio lobes is gas that was uplifted and deposited here through AGN outflows \citep{H05}. Another possibility in which the AGN outflows may create the observed structures is via pressure driven compression of the in-situ thermal gas by the outward expanding non-thermal plasma. In this latter scenario the pressure driven compression increases the local gas density leading to an increase in the cooling rate here.

That the gas observed along the radio lobes is a purely projectional effect and has nothing to do with the radio outflow seems unlikely. Strong evidence that an AGN interacts with the gas in the nuclear region of a BCG comes, for example, from the flux enhancement and high velocity dispersion structure observed in A2597 here and in D00. These structures coincide with the current radio outflows. Many other observations, mostly done in X-rays, have shown that there is a global correlation between the radio outflows in BCGs and the hot X-ray gas surrounding it \citep[e.g.,][]{M01,F02,C05,Bi08}. We do caution the reader that even though this global correlation is found in many cool core clusters that there are also cases where one observes similar amounts of gas in structures that appear to have no relation at all to the current radio outflows. One example of this is the western filament in S159.  

\subsubsection{Alignment of the radio and gas rotation axes ?}
In A2597 we note that, in projection, there is a rough alignment between the radio axis and the axis of rotation for the HII and H$_{\mathrm{2}}$ gas. The same seems to be true for S159 although the evidence for gas rotation is less clear here. Three other clusters, Abell 1664, PKS 0745-19 and A2204, have recently been observed with SINFONI by \citet{W09}. They show that the HII and H$_{\mathrm{2}}$ gas in PKS 0745-19 rotates. High resolution radio images by \citet{Ba91} show us that again that the radio axis and gas rotation axis roughly agree. For the other two clusters the situation is unclear. The velocity fields of the gas in Abell 1664 and Abell 2204 do not show clear evidence for gas rotation, and radio images by \citet*{Go01} and \citet*{Sa09} do not allow for a reliable identification of the jet axis. Higher spatial resolution spectroscopy and radio imaging will be required to investigate the possible alignment of the radio axis and the gas rotation axis in more detail.

\subsubsection{Kinetic energy of the outflows}
It has been proposed in the literature that AGN feedback in the form of radio outflows, jets could deliver the required heat that keeps the gas in cluster cores from cooling \citep[e.g.,][]{Bi04,D06}. The kinetic power in the molecular and ionised phases may be a useful indicator for estimating the total energy input into the cluster medium by the current jet. Here we will estimate the kinetic power from the kinetic energy of the disturbed gas and its turbulent dissipation. The kinetic energy $\textit{E}_{\mathrm{K}}$ and kinetic (turbulent) power $\textit{P}_{\mathrm{K}}$ may be written as;
\\
\\
\begin{eqnarray}
E_{\mathrm{K}} &=& \frac{Mv^{2}}{2} = \frac{3 M\sigma^{2}}{2}\\
P_{\mathrm{K}} &=& \frac{Mv^{3}}{2r} = \frac{E_{\mathrm{K}}\sqrt{3}\sigma}{r}
\end{eqnarray}
\\
\\
Here $\textit{v}$~=~$\sqrt{3}\sigma$ with the $\sigma$ the velocity dispersion of the gas and $\textit{r}$ is the thickness of the high dispersion features. We will perform this estimate using region A1 in A2597. This region contains most of the gas whose dispersion is clearly related to the AGN as decribed in Section \ref{sect_a2597_kinm}. The mass of the HII gas in A1 is 4.1$\times$10$^{6}$~$\mathrm{M_{\odot}}$ and its velocity dispersion is 256~$\mathrm{km}$~$\mathrm{s}^{-1}$. We use only the HII gas here as this gas component dominates in mass over the warm H$_{\mathrm{2}}$ gas. We thus find that a total kinetic energy of $\textit{E}_{\mathrm{K}}$(A2597:A1)~=~8.1$\times$10$^{54}$~$\mathrm{erg}$ and a kinetic energy power $\textit{P}_{\mathrm{K}}$(A2597:A1)~=~5.8$\times$10$^{40}$~$\mathrm{erg}$~$\mathrm{s}^{-1}$, where we used $\textit{r}$~=~2~$\mathrm{kpc}$.

For S159 this calculation is more uncertain due to the lower significance of the Pa~$\alpha$ line in the nuclear region B1. If we use that the mass of the HII gas in B1 is 5.8$\times$10$^{4}$ $\mathrm{M_{\odot}}$ and its velocity dispersion is 127~$\mathrm{km}$~$\mathrm{s}^{-1}$. We then find that a total kinetic energy of $\textit{E}_{\mathrm{K}}$(S159:B1)~=~2.8$\times$10$^{52}$~$\mathrm{erg}$ and a kinetic power $\textit{P}_{\mathrm{K}}$(S159:B1)~=~1.0$\times$10$^{38}$~$\mathrm{erg}$~$\mathrm{s}^{-1}$ is injected into stirring up the gas, again using $\textit{r}$~=~2~$\mathrm{kpc}$. These numbers are rather low for S159. If we use the H$_{\mathrm{2}}$~1-0~S(3) velocity dispersion for the Pa~$\alpha$ line and increase the Pa~$\alpha$ flux such that the Pa~$\alpha$/H$_{\mathrm{2}}$~1-0~S(3) ratio is the same as in region A1 for A2597 we find that $\textit{E}_{\mathrm{K}}$ and $\textit{P}_{\mathrm{K}}$ both increase by about a factor of 10 for S159.

We compare this mechanical power injected into the cool gas to the current radio power. We assume that the stirring of the gas in the nuclei of A2597 and S159 is related to the most recent AGN outburst and thus we look only at the recent radio emission. This is done to avoid the older radio plasma, from previous outbursts, which will dominate the radio power due to its emissivity at low frequencies.

The 8.4~$\mathrm{GHz}$ radio emission is a good indicator of recent AGN activity. The synchrotron lifetime of electrons emitting at 8.4~$\mathrm{GHz}$ is about 10$^{6-7}$ years, for a magnetic field strength of 30~$\mu\mathrm{G}$ \citep{T07}. The 8.4~$\mathrm{GHz}$ specific luminosities $\textit{L}_{\nu}$ in regions A1 and B1 are 1.75$\times$10$^{31}$~$\mathrm{erg}$~$\mathrm{s}^{-1}$~$\mathrm{Hz}^{-1}$ and 3.88$\times$10$^{29}$~$\mathrm{erg}$~$\mathrm{s}^{-1}$~$\mathrm{Hz}^{-1}$. One can then estimate the total current radio power $\textit{P}_{\mathrm{rad}}$ in these regions by $\textit{P}_{\mathrm{rad}}$~=~$\nu\times\textit{L}_{\nu}$, with $\nu$~=~8.4~$\mathrm{GHz}$. We find $\textit{P}_{\mathrm{rad}}$(A2597:A1)~=~1.5$\times$10$^{41}$~$\mathrm{erg}$~$\mathrm{s}^{-1}$ and $\textit{P}_{\mathrm{rad}}$(S159:B1)~=~3.3$\times$10$^{39}$~$\mathrm{erg}$~$\mathrm{s}^{-1}$.

The kinetic power inserted into the cool gas is thus smaller than the radio power in A2597 by a factor 2-3. For S159 we find that $\textit{P}_{\mathrm{K}}$ is smaller than $\textit{P}_{\mathrm{rad}}$ by more than an order of magnitude. If we allow for the higher dispersion and flux as mentioned above then this ratio becomes similar to that observed in A2597. Adding the warm ($\textit{T}\sim$~2300~$\mathrm{K}$) H$_{\mathrm{2}}$ gas to the above analysis does not significantly change the derived kinetic powers due to its low mass. We thus find that within the nuclear regions the current radio jets in A2597 and S159 are radiating efficiently with respect to the warm HII and H$_{\mathrm{2}}$ gas.

The kinetic power deduced here should though be seen as a lower limit to the total kinetic power of the cool ($\textit{T}\leq$10$^{4}$~$\mathrm{K}$) gas in these clusters since the colder gas phases have not been taken into account. O94 find a kinetic energy $\textit{E}_{\mathrm{K}}\sim$10$^{56}$ $\mathrm{erg}$ and a kinetic power $\textit{P}_{\mathrm{K}}\sim$10$^{42}$ $\mathrm{erg} \mathrm{s}^{-1}$ for the HI gas in a region with radius 2.5 $\mathrm{kpc}$ centered on the BCG nucleus in A2597. Similarly the cold ($\textit{T}\leq$10$^{3}$~$\mathrm{K}$) molecular gas may contain a significant amount of kinetic energy. Studies of X-ray cavities find evidence for inefficiently radiating jets in that the mechanical power of the jet often can be orders of magnitude larger than the radio power. However, when looking at X-ray cavities one may be looking back in time towards an epoch in which the balance between the mechanical and radiative power of the radio source is different from the current one.

\subsubsection{Low velocity shocks}
The argument that AGN feedback is responsible for reheating the ICM gas in cool core clusters is made purely on the global energetics of the problem and does not consider the microphysics of the energy transfer between the non-thermal radio plasma and thermal ICM plasma, nor does it consider the various scales involved in this energy transfer. Fast shocks ($\textit{v}$~$\ge$~40~$\mathrm{km}$~$\mathrm{s}^{-1}$) can be ruled out based on the absence of typical shock tracers like the [O~III] 4363~$\mathrm{Angstrom}$ line in optical spectra (VD97) and the high H$_{\mathrm{2}}$/HII ratios \citep{D00}. However, another way of delivering the required heat for the cooler gas observed here is via low velocity shocks in a dense gas. Such a model was recently invoked to explain the strong H$_{\mathrm{2}}$ emission observed in 'Stephan's Quintet' \citep{Gu09}. From X-ray observations \citep[e.g.][]{M01,F06} and jet models \citep[e.g.]{Su07} it is found that AGN outflows may give rise to low velocity C-type shocks through a turbulent cascade of the expanding jet momentum in a inhomogeneous gas.

\begin{figure*}
    \includegraphics[width=0.45\textwidth, angle=90]{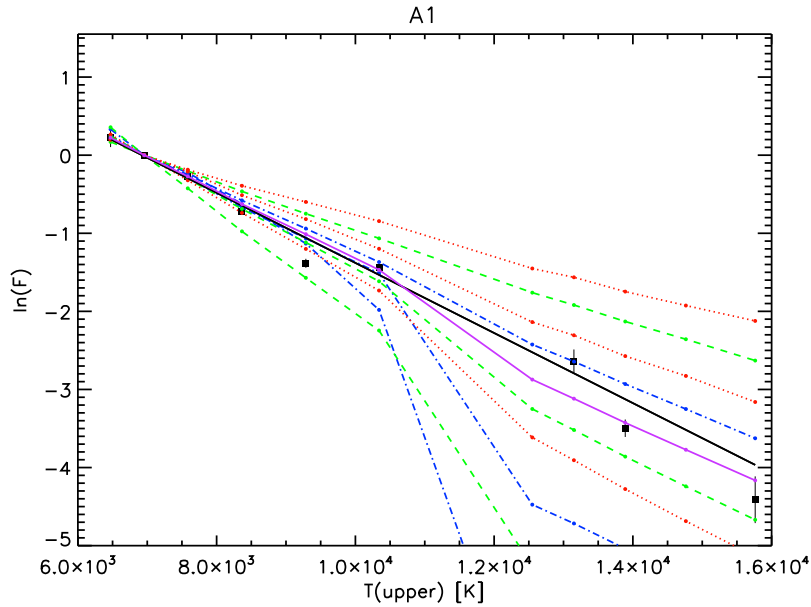}
  \vspace{0.5cm}
  \caption{ABELL 2597 Low velocity shock excitation for region A1. The figure shows the natural logarithm of the normalised line flux $\textit{ln(F)}$, normalisation with respect to the H$_{\mathrm{2}}$~1-0~S(1) line, versus the upper state temperature $\textit{T}_{\mathrm{u}}$. The best-fitting LTE model is given by the solid black line line. The best-fitting low velocity shock model (shock velocity $\textit{v}_{\mathrm{s}}$~=~22~$\mathrm{km}$~$\mathrm{s}^{-1}$, pre-shock density $\textit{n}_{\mathrm{H}}$~=~5$\times$10$^{6}$~$\mathrm{cm}^{-3}$ and magnetic field strength $\textit{B}$~=~5.6~$\mathrm{mG}$) is shown by the purple solid line. The blue dash-dot lines show what happens to the best-fitting model if we change the pre-shock density. Increasing the pre-shock density from 1$\times$10$^{5}$~$\mathrm{cm}^{-3}$, 1$\times$10$^{6}$~$\mathrm{cm}^{-3}$ to 1$\times$10$^{7}$~$\mathrm{cm}^{-3}$ the model becomes progressively flatter. Similarly increasing the shock velocity from 15~$\mathrm{km}$~$\mathrm{s}^{-1}$, 20~$\mathrm{km}$~$\mathrm{s}^{-1}$ to 30~$\mathrm{km}$~$\mathrm{s}^{-1}$, green dashed lines, flattens the model. However, increasing the magnetic field strength from 1.1~$\mathrm{mG}$, 3.4~$\mathrm{mG}$ to 7.8~$\mathrm{mG}$, red dotted lines, leads to a progressive steepening of the model.}\label{fig_a2597_lte_shockonly}
\end{figure*}

We have investigated low velocity C-type shocks by \citet{Kr07} and \citet{Fl03} to model the molecular ratios in region A1 for A2597. The following parameter range was allowed for, shock velocity $\textit{v}_{S}$~=~[15-30]~$\mathrm{km}$~$\mathrm{s}^{-1}$, pre-shock density $\textit{n}\mathrm{_{H}}$~=~[10$^{4}$-10$^{7}$]~$\mathrm{cm}^{-3}$ and magnetic field strength $\textit{B}$~=~[0.05-30]~$\mathrm{mG}$.  The best-fitting model to the molecular line ratios is $\textit{v}_{S}$~=~22~$\mathrm{km}$~$\mathrm{s}^{-1}$, $\textit{n}\mathrm{_{H}}$~=~5$\times$10$^{6}$~$\mathrm{cm}^{-3}$ and $\textit{B}$~=~5.6~$\mathrm{mG}$. This model is shown by the purple line in Fig. \ref{fig_a2597_lte_shockonly}. It implies a pre-shock density above the critical density and thus that the gas is already in LTE prior to the shock. This model furthermore implies a very high magnetic field in the molecular gas, something which has not yet been observed in our objects.

\section{Summary}
Warm molecular gas is observed in the central 20 $\mathrm{kpc}$ of the A2597 and S159 galaxy clusters. All of the H$_{\mathrm{2}}$ 1-0 S-transitions and H$_{\mathrm{2}}$ 2-1 S-transitions redshifted into the K-band and none of the higher ro-vibrational H$_{\mathrm{2}}$ transitions are detected in A2597. Ionised gas is observed in A2597 through the detection of the Pa~$\alpha$, Br~$\gamma$, Br~$\delta$ and Fe~II (1.81 $\mu \mathrm{m}$) lines. For S159 only two H$_{\mathrm{2}}$~1-0~S-transitions are detected. There is tentative evidence for the presence of the other H$_{\mathrm{2}}$~1-0~S-transitions, but these are strongly affected by atmospheric effects. None of the higher ro-vibrational H$_{\mathrm{2}}$ lines were detected. Ionised gas is observed in S159 through the detection of the Pa~$\alpha$ line. Higher ionisation lines, such as [Si IV], are not detected in either S159 or A2597.

Molecular and ionised gas is detected out to 22.5~$\mathrm{kpc}$ from the nucleus of PGC071390 in A2597 and 18.0~$\mathrm{kpc}$ from the nucleus of ESO291-G009 in S159. The fields observed by us do not cover the entire extent over which molecular and ionised emission has previously been detected in these clusters. Our observations thus miss a significant fraction of the total emission and gas mass. Based on the H$\alpha$ maps in J05 we estimate that we have observed about one tenth of the total HII and H$_{\mathrm{2}}$ gas present in A2597 and S159.

The gas morphology in A2597 is more diffuse than in S159. In both clusters the ionised and molecular gas is observed to be distributed in clumps and filamentary structures surrounding their BCGs. This observation brings these higher redshift cool core clusters closer to the situation observed in more nearby clusters, such as Perseus and Centaurus, that show a wealth of small scale filamentary structure \citep[e.g.][]{F08,Cr05}.

In both A2597 and S159 the ionised and molecular gas phases trace each other closely on $\mathrm{kpc}$-scales, in both extent and dynamical state. The H$_{\mathrm{2}}$ line ratios and the H$_{\mathrm{2}}$ to HII line ratios only vary slowly as a function of position throughout the investigated regions, with the exception of the nuclear regions. This implies that there is a tight coupling between these gas phases and that there may be a single excitation mechanism responsible for the observed emission.

In both clusters the Pa~$\alpha$ to H$_{\mathrm{2}}$ ratio is found to be lower within their nuclear regions than outside. In A2597 this decrease is small as the Pa~$\alpha$/H$_{\mathrm{2}}$ 1-0 S(3) ratio decreases from 0.94$\pm$0.04 in region A2 to 0.77$\pm$0.07 in the nuclear region A1. In S159 the decrease is much larger and we find that the Pa~$\alpha$/H$_{\mathrm{2}}$~1-0~S(3) ratio decreases from 0.66$\pm$0.12 in region B3 to 0.19$\pm$0.08 in the nuclear region B1. This may imply that the physical conditions differ within the nuclear regions as compared to further out.

The velocity structure of the ionised and molecular gas within the nuclear regions of A2597 and S159 is consistent with gas rotating around the nucleus. The data suggests a possible alignment between the axis of rotation and the radio axis. Further high spatial resolution, spectroscopic observations are required to disentangle this rotation from velocity gradients observed along the filamentary gas structures and to confirm the axis of rotation.

The high velocity dispersion of the gas within the nuclear regions clearly shows that the AGNs harboured by these BCGs interact with the molecular and ionised gas situated here. This is especially obvious in A2597 where the observed increase in the velocity dispersion is suggestive of a turbulent wake marking the passage of the radio lobes through the surrouding medium, or of direct acceleration of the dense material by the AGN jet.

We estimate that the mechanical power of the AGN flow is somewhat smaller than its radiative radio power within the nuclear regions of A2597 and S159 with respect to the ionised and warm molecular gas. If we were to include the mechanical energy contained within the colder atomic and molecular gas phases than this situation will reverse.

The velocities observed for the filamentary structures are less than 150~$\mathrm{km}$~$\mathrm{s}^{-1}$. This does not agree with freely infalling gas which should have much higher velocities, i.e. $\sim$600~$\mathrm{km}$~$\mathrm{s}^{-1}$. The filaments do show smooth velocity gradients. Possible explanations for this are; (i) we see a single filament that is being stretched and bended by its proper motion and gravitational forces, (ii) we see a blend of a number of unresolved filaments along the line of sight. From observations of nearby clusters such as Perseus the latter explanation appears most likely \citep[e.g.][]{F08,H05}.

The velocity dispersion of the ionised and molecular gas decreases with distance to the nucleus. It drops below 100~$\mathrm{km}$~$\mathrm{s}^{-1}$ at distances of a few $\mathrm{kpc}$ from the nucleus which implies that the gas is kinematically cold outside the nuclear region. The decrease in velocity dispersion is strongest within the first $\sim$2~$\mathrm{kpc}$ from the nucleus and slows down after that. This point may mark a transition in where the AGN is able to strongly interact with its surrounding medium.

The low velocity and velocity dispersion of the gas in the filamentary structures indicates that the gas here resides in well defined structures. The gas may still be clumpy but individual clumps must have a spread in velocity no larger than the derived dispersion. It also implies that these structures need some form of support, which can not be kinematical, to keep them from plummeting towards the potential centre. A possible source of support are magnetic fields as suggested by \citet{F08}. This requires magnetic field strengths of a few hundred $\mu$G. So far there is no observational evidence for the existence of these magnetic fields in galaxy clusters.

The molecular and ionised gas emission observed in A2597 and S159 does not match the observed X-ray emission very well. In S159 there is no match in the observed brightness profiles. In A2597 there is a rough global agreement in the peak emission but the detailed X-ray brightness profiles are significantly shallower than observed for the H$_{\mathrm{2}}$ and HII gas.

There is no evidence for a strong correlation between either H$_{\mathrm{2}}$ or HII and  the radio emission in A2597 and S159. If the AGN is main source of heating and excitation for these gas phases, then we would expect to find this gas near the current AGN outflows at all times. This is especially important for the H$_{\mathrm{2}}$ gas since it has a cooling time of order years. One prime example where we find H$_{\mathrm{2}}$ gas that appears to have no relation to the current outflows is the western filament in S159. The emission line spectra for all regions in S159 are similar. This then implies that the physical proces that is heating the gas in these regions is similar and likely not directly related to the current AGN outflows.

In A2597 the H$_{\mathrm{2}}$ line emitting gas is well described by an LTE model with a single excitation temperature $\textit{T}_{\mathrm{exc}}$. We find that $\textit{T}_{\mathrm{exc}}$ is about 2300 K for the different regions investigated. We find slightly higher excitation temperatures for the molecular gas in the filaments as compared to the nuclear region.

In S159 we have only been able to reliably measure two H$_{\mathrm{2}}$ lines. The temperature and excitation of the H$_{\mathrm{2}}$ gas in this cluster thus remains largely unconstrained. Assuming a single temperature LTE model, we find that $\textit{T}_{\mathrm{exc}}$ is about 2300 K for the different regions investigated. The temperature derived in this way for the nuclear region in S159 is above the dissociation temperature of H$_{\mathrm{2}}$ and considered unphysical. No clear trends in temperature with distance to the nucleus are observed for S159.

If the molecular gas in A2597 and S159 is in LTE then this implies that the H$_{\mathrm{2}}$ gas is not in pressure equilibrium with the HII gas, as has previously been pointed out by J05. We find that $\textit{M}_{\mathrm{HII}}/\textit{M}_{\mathrm{H_{2}}}$~$\sim$~10$^{2}$ in A2597 and S159.  We note that $\textit{M}_{\mathrm{H_{2}}}$ is likely to be seriously underestimated since only the K-band H$_{\mathrm{2}}$ lines tracing the warm molecular gas have been considered here.

For a region with radius 2.5~$\mathrm{kpc}$ centered on the nucleus of the A2597 BCG we can make an inventory of the cool gas. In agreement with \citet{H89} we find an ionised gas mass $\textit{M}_{\mathrm{HII}}\approx$6$\times$10$^{6}$~$\mathrm{M_{\odot}}$. We find a warm molecular gas mass $\textit{M}_{\mathrm{H_{2},warm}}\approx$3$\times$10$^{4}$~$\mathrm{M_{\odot}}$. \citet{O94} find an neutral hydrogen mass $\textit{M}_{\mathrm{HI}}\sim$10$^{8}$~$\mathrm{M_{\odot}}$. Using much lower spatial resolution \citet{E01} find a spatially unresolved cold molecular gas mass $\textit{M}_{\mathrm{H_{2},cold}}$=8$\times$10$^{9}$~$\mathrm{M_{\odot}}$ in the central 22$\arcsec$ of A2597.

We are able to fit the H$_{\mathrm{2}}$ line emitting gas by a low velocity shock model. Such a model requires high pre-shock densities and very high magnetic fields. Low velocity shocks may be a viable option for heating the H$_{\mathrm{2}}$ gas in cool core clusters, but this requires further modelling and observations.

Using recent Spitzer observations we find that another lower temperature H$_{\mathrm{2}}$ gas component exists at about 300 K in A2597 and S159 (Jaffe \& Bremer in prep.). This shows that a single temperature LTE model is not a good description for all of the H$_{\mathrm{2}}$ gas present in cool core clusters and again brings the situation in our clusters closer to that observed in more nearby cool core clusters \citep{J07}. This is also implies that we significantly underestimate the total H$_{\mathrm{2}}$ luminosity and gas mass by using only the H$_{\mathrm{2}}$ K-band spectrum.

Our K-band spectra of cool core clusters do not resemble typical K-band spectra of other line emitting, extragalactic objects such as Ultraluminous Infrared Galaxies (ULIRG) and AGNs that are not in BCGs. For comparisons we can look at the Br~$\gamma$ and H$_{\mathrm{2}}$~1-0~S(1) line. In ULIRGs theses line have a similar intensity and width and in AGNs that are not in BCGs the Br~$\delta$ line usually has a higher intensity and larger width than the H$_{\mathrm{2}}$~1-0~S(1) line \citep{D03,D05}. In our spectra for Abell 2597 we find that these lines have similar widths, but the H$_{\mathrm{2}}$~1-0~S(1) line is more than ten times brighter than the Br~$\gamma$ line.

\section{Conclusions}
Above we have presented the first K-band integral-field spectroscopic observations of the extended molecular and ionised gas distributions in the core regions of the A2597 and S159 galaxy clusters. These observations add one extra dimension to the analysis of the molecular and ionised gas in these clusters as compared to our previous investigation in J05. This allows us to study the distribution, kinematics and thermal structure of this gas over a much larger area than in our previous investigation. The spatial resolution of our observations limit us to studying this gas on $\mathrm{kpc}$-scales. Below we summarise our conclusions.
\begin{itemize}
\item Line emission from molecular and ionised gas is found in all observed fields. We confirm the conclusion in J05 that molecular and ionised gas are tightly coupled out to 20~$\mathrm{kpc}$ from the BCG nucleus in A2597 and S159. 
\item The molecular and ionised gas is distributed in filamentary structures surrounding the BCG. The gas morphology is more diffuse in A2597 than in S159. 
\item In all regions where we have sufficient signal to noise we find that the H$_{\mathrm{2}}$ gas can be described an by an LTE model with a temperature $\textit{T}$~$\sim$~2300~$\mathrm{K}$. LTE implies a high-density for this gas.
\item Kinematically there is a fairly clear separation between the central few $\mathrm{kpc}$, energetically dominated by the AGN, and an area outside where the support of the HII and H$_{\mathrm{2}}$ gas remains to be explained.
\item Within the nuclear region of A2597 and S159 the kinetic luminosity of the HII and warm H$_{\mathrm{2}}$ gas is somewhat smaller than the current radio luminosity.
\item The very high dispersion regions observed in H$_{\mathrm{2}}$ and HII may represent a turbulent wake of the relative motion of galaxy and surrounding medium, or the interaction of the current jet with this medium.
\item A2597 shows a better correlation between the morphology of the cool gas and X-ray emission than S159. The detailed radial brightness profiles of the X-ray and either the H$_{\mathrm{2}}$ or HII gas do not match in either cluster. 
\item The high frequency radio emission and cool gas have the same overall scale size. There is no strong correlation between either the H$_{\mathrm{2}}$ or HII gas and the radio structures, but there are weak enhancements in the surface brightness of this gas along the northern radio lobe in A2597.
\item The data suggests that the HII and H$_{\mathrm{2}}$ gas situated within a few $\mathrm{kpc}$ from the nucleus rotates around an axis parallel to the radio jet, this needs further confirmation.
\end{itemize}
A more detailed investigation into the heating and cooling of the intracluster medium will need to take into account the microphysics of the interaction between the various phases situated within the intracluster medium; the hot X-ray gas, the cool HII, H$_{\mathrm{2}}$ gas and the radio emitting plasma. Once such a model is in place, the next step will be to compare detailed maps of local heating versus local cooling in cool core clusters. We will follow this work up in the near future with optical integral field spectroscopy (Oonk et al. in prep.) and Spitzer IRS spectroscopy (Jaffe \& Bremer in prep.). A more detailed investigation of the excitation mechanisms for the molecular hydrogen gas will also be presented in a future paper.

\section*{Acknowledgments}
JBRO wishes to thank J. Reunanen and L. Vermaas for useful discussions regarding the reduction of the SINFONI data. We would like to thank Lars Kristensen for use of his large grid of 1D shock models The SINFONI observations were taken at the Very Large Telescope (VLT) facility of the European Southern Observatory (ESO) as part of project 075.A-0074. The National Radio Astronomy Observatory is a facility of the National Science Foundation operated under cooperative agreement by Associated Universities, Inc.

\begin{table*}
 \centering
  \begin{tabular}{|l|l|l|l|l|l|l|} \hline
  Cluster & $\textit{z}$ & $\textit{D}_{\mathrm{L}}$ [$\mathrm{Mpc}$]& $\Theta$ [$\mathrm{kpc}$ arcsec$^{-1}$] & $\textit{L}_{X}$ [10$^{44}$ $\mathrm{erg}$ $\mathrm{s}^{-1}$] & $\textit{L}_{1.4\mathrm{GHz}}$ [10$^{31}$ $\mathrm{erg}$ $\mathrm{s}^{-1}$ $\mathrm{Hz}^{-1}$] & $\textit{t}_{\mathrm{obs}}$ [$\mathrm{h}$] \\ \hline
  Sersic 159-03 & 0.0564 & 245 & 1.1 & 2.1 & 29.3 & 4.2\\
  Abell 2597    & 0.0821 & 363 & 1.5 & 1.0 & 1.6 & 8.5\\ \hline
  \end{tabular}
 \caption[]{Targets. Columns 1, 2 give the cluster name and the redshift of its BCG. In column 3 we give the luminosity distance to the BCG in units of $\mathrm{Mpc}$. In column 4 we give the angular size scale at the distance of the BCG. In column 5 we give the X-ray luminosity of the cluster in the 0.5-2.0~$\mathrm{keV}$ band from \citet{dG99} and in column 6 we give the radio power of the cluster at 1.4~$\mathrm{Ghz}$ from \citet{Bi08} In column 7 we give the on-source integration time in units of hours. An equal amount of time was spend off-source observing the sky.}\label{tabl_obs_summ}
\end{table*}

\begin{table*}
 \centering
  \begin{tabular}{|l|l|l|l|l|} \hline
  Resolution & no smoothing & ss20 & ss22 & ss44 \\ \hline
  Spatial [arcsec]  & 0.9 & 1.0 & 1.0 & 1.1 \\
  Spectral [$\mathrm{km}$ $\mathrm{s}^{-1}$] & 38 & 38 & 51 & 73 \\ \hline
  \end{tabular}
 \caption[]{Spatial and spectral Resolution. Dependence of the spatial and spectral resolution on smoothing. The spatial resolution is given in units of arcsec and the spectral resolution in terms of $\mathrm{km}$~$\mathrm{s}^{-1}$. Column 1 gives the type of resolution. Column 2 gives the resolution for no smoothing. Column 3 gives the resolution after smoothing with a two pixel FWHM Gaussian in the spatial domain. Column 4 gives the resolution after smoothing with two pixel FWHM Gaussians in both the spatial and spectral domain. Column 5 gives the resolution after smoothing with four pixel FWHM Gaussians in both the spatial and spectra domain.}\label{tabl_smooth_resolution}
\end{table*}

\begin{table*}
 \centering
  \begin{tabular}{|l|l|l|l|} \hline
  Field [10$^{-17}$ $\mathrm{erg}$ $\mathrm{s}^{-1}$ $\mathrm{cm}^{-2}$ $\mu \mathrm{m}^{-1}$] & no smoothing & ss20 & ss44 \\ \hline
  Central & 24.9 & 12.6 & 3.7 \\
  South-East & 40.0 & 19.8 & 5.4 \\
  South-West & 25.8 & 12.9 & 3.5 \\
  North & 30.1 & 15.0 & 4.0 \\ \hline
  \end{tabular}
 \caption[]{ABELL 2597 Sensitivity. The RMS noise is given in units of 10$^{-17}$~$\mathrm{erg}$~$\mathrm{s}^{-1}$ $\mathrm{cm}^{-2}$~$\mu \mathrm{m}^{-1}$ for an area of 0.125$\times$0.125 arcsec$^{2}$ which is equivalent to one spatial pixel. The RMS noise was calculated in the interval 2.07-2.30~$\mu \mathrm{m}$ with a sampling of 2.45$\times$10$^{-4}$~$\mu \mathrm{m}$ which is equivalent to one spectral pixel. Column 1 gives the observed field. Column 2 gives the RMS noise for no smoothing. Column 3 gives the RMS noise after smoothing with a two pixel FWHM Gaussian in the spatial domain. Column 4 gives the RMS noise after smoothing with a four pixel FWHM Gaussian in both the spatial and spectral domain. The RMS noise does not vary much across the observed field.}\label{tabl_smooth_sensitivity_a2597}
\end{table*}

\begin{table*}
 \centering
  \begin{tabular}{|l|l|l|l|} \hline
  Field [10$^{-17}$ $\mathrm{erg}$ $\mathrm{s}^{-1}$ $\mathrm{cm}^{-2}$ $\mu \mathrm{m}^{-1}$] & no smoothing & ss22 & ss44 \\ \hline
  South-East & 33.1 & 13.2 & 9.0 \\
  South-West & 30.5 & 10.2 & 4.3 \\
  North & 36.7 & 12.2 & 5.0 \\ \hline
  \end{tabular}
 \caption[]{SERSIC 159-03 Sensitivity. The RMS noise is given in units of 10$^{-17}$~$\mathrm{erg}$~$\mathrm{s}^{-1}$~$\mathrm{cm}^{-2}$~$\mu \mathrm{m}^{-1}$ for an area of 0.125$\times$0.125 arcsec$^{2}$ which is equivalent to one spatial pixel. The RMS noise was calculated in the interval 2.07-2.30~$\mu \mathrm{m}$ with a sampling of 2.45$\times$10$^{-4}$~$\mu \mathrm{m}$ which is equivalent to one spectral pixel. Column 1 gives the observed field. Column 2 gives the RMS noise for no smoothing. Column 3 gives the RMS noise after smoothing with two pixel FWHM Gaussians in both the spatial and spectral domain. Column 4 gives the RMS noise after smoothing with four pixel FWHM Gaussian in both the spatial and spectral domain. The RMS noise does not vary much across the observed field.}\label{tabl_smooth_sensitivity_sersic}
\end{table*}

\begin{table*}
 \centering
  \begin{tabular}{|l|l|l|} \hline
  Line Name & Flux [10$^{-17}$ $\mathrm{erg}$ $\mathrm{s}^{-1}$ $\mathrm{cm}^{-2}$] & Luminosity [10$^{39}$ $\mathrm{erg}$ $\mathrm{s}^{-1}$] \\ \hline
  H$_{\mathrm{2}}$ 1-0 S(0) (2.2235 $\mu \mathrm{m}$)$^{1}$ & 121.7$\pm$11.3 & 19.2$\pm$1.8 \\
  H$_{\mathrm{2}}$ 1-0 S(1) (2.1218 $\mu \mathrm{m}$)       & 669.0$\pm$14.6 & 105.5$\pm$2.3 \\
  H$_{\mathrm{2}}$ 1-0 S(2) (2.0338 $\mu \mathrm{m}$)       & 280.2$\pm$8.4  & 44.2$\pm$1.3 \\
  H$_{\mathrm{2}}$ 1-0 S(3) (1.9576 $\mu \mathrm{m}$)       & 732.0$\pm$19.3 & 115.4$\pm$3.0 \\
  H$_{\mathrm{2}}$ 1-0 S(4) (1.8920 $\mu \mathrm{m}$)       & 129.7$\pm$7.5  & 20.5$\pm$1.2 \\
  H$_{\mathrm{2}}$ 1-0 S(5) (1.8358 $\mu \mathrm{m}$)       & 462.2$\pm$11.7 & 72.9$\pm$1.8 \\
  H$_{\mathrm{2}}$ 2-1 S(2) (2.1542 $\mu \mathrm{m}$)$^{2}$ & 22.0$\pm$4.1   & 3.5$\pm$0.6 \\
  H$_{\mathrm{2}}$ 2-1 S(3) (2.0735 $\mu \mathrm{m}$)$^{1}$ & 38.0$\pm$2.2   & 6.0$\pm$0.3 \\
  H$_{\mathrm{2}}$ 2-1 S(5) (1.9449 $\mu \mathrm{m}$),Br $\delta$ (1.9451 $\mu m$)$^{1}$ & 39.3$\pm$4.2 & 6.2$\pm$0.7 \\ 
  Br $\gamma$ (2.1661 $\mu \mathrm{m}$)$^{2}$      & 30.1$\pm$4.6   & 4.7$\pm$0.7 \\
  Pa $\alpha$ (1.8756 $\mu \mathrm{m}$)           & 613.0$\pm$20.3 & 96.7$\pm$3.2 \\
  Fe II (1.8100 $\mu \mathrm{m}$)$^{1}$            & 86.5$\pm$5.6   & 13.6$\pm$0.9 \\ \hline
  \end{tabular}
 \caption[]{ABELL 2597 Integrated line fluxes and luminosities. The integrated line fluxes are obtained by collapsing the data cube into a single spectrum and fitting the lines by a single Gaussian. The data was smoothed by a four pixel FWHM in both the spatial and spectral domain. The results were inspected by eye and the errors were estimated using Monte-Carlo simulations. The spectrum also hints at the presence of the H$_{\mathrm{2}}$~2-1~S(4) line, but a reliable flux could not be derived for this line. Fluxes are given in units of 10$^{-17}$~$\mathrm{erg}$~$\mathrm{s}^{-1}$~$\mathrm{cm}^{-2}$ and luminosities are given in units of 10$^{39}$~$\mathrm{erg}$~$\mathrm{s}^{-1}$.\\
$^{1}$ The H$_{\mathrm{2}}$~1-0~S(0), H$_{\mathrm{2}}$~2-1~S(3), H$_{\mathrm{2}}$~2-1~S(5), Br~$\delta$ and Fe~II lines have been integrated over the central field only.\\
$^{2}$ The H$_{\mathrm{2}}$~2-1~S(2) and Br~$\gamma$ lines have been integrated over the combined areas A1 and A2 only, see Fig. \ref{fig_h2_area_ss44}.}\label{tabl_line_summ_a2597_ss44}
\end{table*}

\begin{table*}
 \centering
  \begin{tabular}{|l|l|l|} \hline
  Line Name & Flux [10$^{-17}$ $\mathrm{erg}$ $\mathrm{s}^{-1}$ $\mathrm{cm}^{-2}$] & Luminosity [10$^{39}$ $\mathrm{erg}$ $\mathrm{s}^{-1}$] \\ \hline
  H$_{\mathrm{2}}$ 1-0 S(1) (2.1218 $\mu \mathrm{m}$) & 171.0$\pm$11.9 & 12.3$\pm$0.9 \\
  H$_{\mathrm{2}}$ 1-0 S(3) (1.9576 $\mu \mathrm{m}$) & 217.9$\pm$18.1 & 15.7$\pm$1.3 \\
  Pa $\alpha$ (1.8756 $\mu m$)     & 104.4$\pm$11.6 & 7.5$\pm$0.8 \\ \hline
  \end{tabular}
 \caption[]{SERSIC 159-03 Integrated line fluxes and luminosities. The integrated line fluxes are obtained by collapsing the data cube into a single spectrum and fitting the lines by a single Gaussian. The data was smoothed by a four pixel FWHM in both the spatial and spectral domain. The results were inspected by eye and the errors were estimated using Monte-Carlo simulations. The spectrum also hints at the presence of the H$_{\mathrm{2}}$~1-0~S(0), 1-0~S(2) and 1-0~S(4) lines, but reliable fluxes could not be derived for these lines. Fluxes are given in units of 10$^{-17}$~$\mathrm{erg}$~$\mathrm{s}^{-1}$~$\mathrm{cm}^{-2}$ and luminosities are given in units of 10$^{39}$ $\mathrm{erg}$ $\mathrm{s}^{-1}$.}\label{tabl_line_summ_sersic_ss44}
\end{table*}

\begin{table*}
 \centering
  \begin{tabular}{|l|l|l|l|l|l|l|l|} \hline
  Line/Area [10$^{-17}$ $\mathrm{erg}$ $\mathrm{s}^{-1}$ $\mathrm{cm}^{-2}$] & A1 & A2 & A3 & A4 & A5 & A6 & A7 \\ \hline
  H$_{\mathrm{2}}$ 1-0 S(0) &  69.0$\pm$8.5 & 18.6$\pm$3.0 & -           & -          & -           & -            & - \\
  H$_{\mathrm{2}}$ 1-0 S(1) & 332.0$\pm$6.6 & 84.1$\pm$3.0 & 4.9$\pm$1.2 & 21.8$\pm$6.1 & 4.6$\pm$1.1 & 12.0$\pm$0.9 & 3.2$\pm$1.1 \\
  H$_{\mathrm{2}}$ 1-0 S(2) & 129.0$\pm$4.1 & 31.0$\pm$2.1 & (2.1)       & 14.6$\pm$1.7 & -           & -            & - \\
  H$_{\mathrm{2}}$ 1-0 S(3) & 335.3$\pm$9.2 & 85.2$\pm$3.0 & 6.0$\pm$1.0 & 32.6$\pm$4.3 & 4.8$\pm$0.7 & 10.2$\pm$1.3 & 2.6$\pm$0.9 \\
  H$_{\mathrm{2}}$ 1-0 S(4) &  69.7$\pm$4.0 & 21.5$\pm$2.0 & 1.9$\pm$0.9 & 10.9$\pm$2.1 & -           & -            & - \\
  H$_{\mathrm{2}}$ 1-0 S(5) & 223.1$\pm$5.4 & 53.0$\pm$2.5 & 3.7$\pm$1.1 & 25.4$\pm$4.5 & 3.5$\pm$0.7 & -            & 1.5$\pm$0.7 \\
  H$_{\mathrm{2}}$ 2-1 S(2) &  16.2$\pm$2.4 &  5.5$\pm$2.2 & -           & -            & -           & -            & - \\
  H$_{\mathrm{2}}$ 2-1 S(3) &  26.9$\pm$3.0 &  7.5$\pm$1.0 & -           & -            & -           & -            & - \\
  H$_{\mathrm{2}}$ 2-1 S(5) &  13.7$\pm$4.1 &  2.8$\pm$1.0 & -           & -            & -           & -            & - \\
  Br $\delta$     &  14.1$\pm$4.1 &  4.3$\pm$1.0 & -           & -            & -           & -            & - \\
  Br $\gamma$     &  21.1$\pm$2.5 &  6.5$\pm$2.1 & -           & -            & -           & -            & - \\
  Pa $\alpha$     & 257.9$\pm$8.8 & 79.7$\pm$3.6 & 5.3$\pm$1.2 & 29.0$\pm$7.0 & 3.6$\pm$0.7 & -            & 2.4$\pm$0.8 \\
  Fe II           &  50.2$\pm$7.2 &  5.0$\pm$2.1 & -           & -            & -           & -            & - \\ \hline
  \end{tabular}
 \caption[]{ABELL 2597 Integrated line fluxes for regions A1-A7. The integrated line fluxes are obtained by collapsing the data cube, within the selected region, into a single spectrum and fitting the lines by a single Gaussian. The smoothing performed for a given line is specified in the caption of the line spectra shown in Appendix \ref{app_tb}. The results were inspected by eye and the errors were estimated using Monte-Carlo simulations. If Br~$\gamma$ is detected the H$_{\mathrm{2}}$~2-1~S(5), Br~$\delta$ complex is disentangled by assuming that the Br~$\gamma$/Br~$\delta$ ratio is 1.5 (Osterbrock \& Ferland 2006). The uncertain H$_{\mathrm{2}}$~1-0~S(2) flux in region A3 could not be fitted well. The H$_{\mathrm{2}}$~1-0~S(2) flux given for A3 is based on fixing the width of this line to that of the H$_{\mathrm{2}}$~1-0~S(4) line. Fluxes are given in units of 10$^{-17}$~$\mathrm{erg}$~$\mathrm{s}^{-1}$~$\mathrm{cm}^{-2}$.}\label{tabl_line_area_a2597_ss20}
\end{table*}

\begin{table*}
 \centering
  \begin{tabular}{|l|l|l|l|l|l|l|l|} \hline
  Line/Area [$\mathrm{km}$ $\mathrm{s}^{-1}$] & A1 & A2 & A3 & A4 & A5 & A6 & A7 \\ \hline
  v(H$_{\mathrm{2}}$ 1-0 S(3)) & +2.9$\pm$6.1 & -29.3$\pm$5.7 & +36.0$\pm$19.1 & -10.2$\pm$12.9 & +19.8$\pm$14.2 & +174.0$\pm$16.7 & +156.1$\pm$24.5 \\
  $\sigma$(H$_{\mathrm{2}}$ 1-0 S(3)) & 214.3$\pm$6.0 & 145.8$\pm$6.1 & 97.6$\pm$19.9 & 88.8$\pm$13.4 & 103.4$\pm$14.9 & 119.5$\pm$17.4 & 60.1$\pm$14.9 \\
  v(Pa $\alpha$) & -21.2$\pm$7.5 & -59.2$\pm$8.3 & +58.9$\pm$26.7 & -3.8$\pm$29.4 & +21.6$\pm$21.8 & -            & +152.4$\pm$51.8 \\
  $\sigma$(Pa $\alpha$) & 255.9$\pm$8.7 & 176.9$\pm$8.3 & 108.1$\pm$27.5 & 118.2$\pm$29.1 & 115.9$\pm$23.1 & -            & 109.0$\pm$56.0 \\ \hline
  \end{tabular}
 \caption[]{ABELL 2597 Kinematics for regions A1-A7. For each line the top row gives the velocity, with respect to the systemic velocity, and the bottom row gives the velocity dispersion. These are derived from a single Gaussian line fit to the collapsed spectrum for a selected region. Here we present results only for the H$_{\mathrm{2}}$~1-0~S(3) and Pa~$\alpha$ lines. All other lines follow the behaviour observed in these two lines within errors. The smoothing performed for a given line is specified in the caption of the line spectra shown in Appendix \ref{app_tb}. The results were inspected by eye and the errors were estimated using Monte-Carlo simulations. Velocity and velocity dispersion are both given in units of $\mathrm{km}$~$\mathrm{s}^{-1}$.}\label{tabl_kinm_area_a2597_ss20}
\end{table*}

\begin{table*}
 \centering
  \begin{tabular}{|l|l|l|l|l|l|l|l|} \hline
  Line/Area & A1 & A2 & A3 & A4 & A5 & A6 & A7 \\ \hline
  T$_{\mathrm{exc,H_{2}}}$ [K] & 2229$\pm$38 & 2141$\pm$59  & 2792$\pm$808 & 2660$\pm$481 & 2678$\pm$640 & 1586$\pm$264 & 1863$\pm$592\\
  M$_{\mathrm{H_{2}}}$ [10$^{2}$ M$_{\odot}$] & 222.2$\pm$4.4  & 56.3$\pm$2.0  & 3.3$\pm$0.8 & 14.6$\pm$4.1 & 3.1$\pm$0.7 & 8.0$\pm$0.6  & 2.1$\pm$0.6 \\
  M$_{\mathrm{HII}}$ [10$^{4}$ M$_{\odot}$] & 410.0$\pm$14.0  & 126.7$\pm$5.7 & 8.4$\pm$1.9  & 46.1$\pm$11.1  & 5.7$\pm$1.1 & -            & 3.8$\pm$1.3 \\ \hline
  \end{tabular}
 \caption[]{ABELL 2597 Gas temperatures and masses for regions A1-A7. For each region an excitation temperature, T$_{\mathrm{exc,H_{2}}}$ in units of Kelvin, is calculated for the H$_{\mathrm{2}}$ gas. Molecular and ionised gas masses are calculated using the equations in the text. The molecular gas mass M$_{\mathrm{H_{2}}}$ is given in units of 10$^{2}$~M$_{\odot}$. The ionised gas mass M$_{\mathrm{HII}}$ is given in units of 10$^{4}$~M$_{\odot}$, using n$_{e}$~=~200~$\mathrm{cm}^{-3}$ and $\textit{T}_{\mathrm{HII}}$~=~10$^{4}$~$\mathrm{K}$. The temperature and mass calculations for the H$_{\mathrm{2}}$ gas assume LTE conditions.}\label{tabl_phys_area_a2597_ss20}
\end{table*}

\begin{table*}
 \centering
  \begin{tabular}{|l|l|l|l|l|l|l|l|} \hline
  Line/Area [10$^{-17}$ $\mathrm{erg}$ $\mathrm{s}^{-1}$ $\mathrm{cm}^{-2}$] & B1 & B2 & B3 & B4 & B5 & B6 & B7 \\ \hline
  H$_{\mathrm{2}}$ 1-0 S(1) & 26.6$\pm$2.6 & 20.4$\pm$1.8 & 16.4$\pm$1.7 & 3.1$\pm$0.9 & 35.7$\pm$2.6 & 12.2$\pm$5.6 & 6.0$\pm$1.2 \\
  H$_{\mathrm{2}}$ 1-0 S(3) & 42.2$\pm$5.3 & 24.1$\pm$2.0 & 18.8$\pm$1.4 & 3.5$\pm$0.7 & 40.2$\pm$2.8 & 12.6$\pm$3.3 & 7.1$\pm$0.9 \\
  Pa $\alpha$     &  8.0$\pm$2.4 & 12.9$\pm$1.5 & 12.4$\pm$1.3 & 3.1$\pm$0.8 & 25.2$\pm$3.2 & 11.4$\pm$5.4 & 5.6$\pm$1.2 \\ \hline
  \end{tabular}
 \caption[]{SERSIC 159-03 Integrated line fluxes for regions B1-B7. The integrated line fluxes are obtained by collapsing the data cube, within the selected region, into a single spectrum and fitting the lines by a single Gaussian. The smoothing performed for a given line is specified in the caption of the line spectra shown in Appendix \ref{app_tb}. The results were inspected by eye and the errors were estimated using Monte-Carlo simulations. Fluxes are given in units of 10$^{-17}$~$\mathrm{erg}$~$\mathrm{s}^{-1}$~$\mathrm{cm}^{-2}$.}\label{tabl_line_area_sersic_ss22}
\end{table*}

\begin{table*}
 \centering
  \begin{tabular}{|l|l|l|l|l|l|l|l|} \hline
  Line/Area [$\mathrm{km}$ $\mathrm{s}^{-1}$] & B1 & B2 & B3 & B4 & B5 & B6 & B7 \\ \hline
  v(H$_{\mathrm{2}}$ 1-0 S(3)) & -26.3$\pm$28.6 & +109.1$\pm$12.2 & +101.8$\pm$8.3 & -48.7$\pm$17.1 & -23.4$\pm$7.8 & -62.9$\pm$20.5 & -19.1$\pm$9.0 \\
  $\sigma$(H$_{\mathrm{2}}$ 1-0 S(3)) & 249.9$\pm$31.1 & 132.2$\pm$12.6 & 103.1$\pm$8.3 & 77.8$\pm$17.6 & 100.0$\pm$7.8 & 65.1$\pm$14.3 & 61.6$\pm$4.4 \\
  v(Pa $\alpha$) & -6.7$\pm$39.6 & +120.9$\pm$15.0 & +105.6$\pm$9.6 & -41.8$\pm$34.7 & -27.6$\pm$13.1 & -58.1$\pm$48.8 & -19.8$\pm$20.6 \\
  $\sigma$(Pa $\alpha$) & 126.5$\pm$39.8 & 120.3$\pm$15.5 & 95.1$\pm$9.6 & 121.6$\pm$36.1 & 103.5$\pm$15.1 & 97.9$\pm$53.3 & 85.7$\pm$20.3\\ \hline
  \end{tabular}
 \caption[]{SERSIC 159-03 Kinematics for regions B1-B7. For each line the top row gives the velocity, with respect to the systemic velocity, and the bottom row gives the velocity dispersion. These are derived from a single Gaussian line fit to the collapsed spectrum for a selected region. Here we present results only for the H$_{\mathrm{2}}$~1-0~S(3) and Pa~$\alpha$ lines. All other lines follow the behaviour observed in these two lines within errors. The smoothing performed for a given line is specified in the caption of the line spectra shown in Appendix \ref{app_tb}. The results were inspected by eye and the errors were estimated using Monte-Carlo simulations. Velocity and velocity dispersion are both given in units of $\mathrm{km}$~$\mathrm{s}^{-1}$.}\label{tabl_kinm_area_sersic_ss22}
\end{table*}

\begin{table*}
 \centering
  \begin{tabular}{|l|l|l|l|l|l|l|l|} \hline
  Line/Area & B1 & B2 & B3 & B4 & B5 & B6 & B7 \\ \hline
  T$_{\mathrm{exc,H_{2}}}$ [K] & 5655$\pm$3632 & 2520$\pm$546 & 2391$\pm$518 & 2331$\pm$1359 & 2321$\pm$385 & 2032$\pm$1550 & 2527$\pm$1073 \\
  M$_{\mathrm{H_{2}}}$ [10$^{2}$ M$_{\odot}$] & 8.0$\pm$0.8  & 6.2$\pm$0.5 & 5.0$\pm$0.5 & 0.9$\pm$0.3 & 10.9$\pm$0.8 & 3.7$\pm$1.7  & 1.8$\pm$0.4 \\
  M$_{\mathrm{HII}}$ [10$^{4}$ M$_{\odot}$] & 5.8$\pm$1.7  & 9.3$\pm$1.1 & 9.0$\pm$0.9 & 2.2$\pm$0.6 & 18.2$\pm$2.3 & 8.3$\pm$3.9  & 4.1$\pm$0.9 \\ \hline
  \end{tabular}
 \caption[]{SERSIC 159-03 H$_{\mathrm{2}}$ Gas temperatures and masses for regions B1-B7. For each region an excitation temperature, $\textit{T}_{\mathrm{exc,H_{2}}}$ in units of Kelvin, is calculated for the H$_{\mathrm{2}}$ gas. Molecular and ionised gas masses are calculated using the equations in the text. The molecular gas mass $\textit{M}_{\mathrm{H_{2}}}$ is given in units of 10$^{2}$~$\mathrm{M_{\odot}}$. The ionised gas mass $\textit{M}_{\mathrm{HII}}$ is given in units of 10$^{4}$~$\mathrm{M_{\odot}}$, using $\textit{n}_{\mathrm{e}}$~=~200~$\mathrm{cm}^{-3}$ and $\textit{T}_{\mathrm{HII}}$~=~10$^{4}$~$\mathrm{K}$. The temperature and mass calculations for the H$_{\mathrm{2}}$ gas assume LTE conditions.}\label{tabl_phys_area_sersic_ss22}
\end{table*}

\clearpage

\clearpage

\clearpage

\clearpage

\clearpage
\appendix

\section[]{Abell 2597 Surface Brightness Maps}\label{app_lmp_a2597}
\begin{figure*}
    \includegraphics[width=0.33\textwidth, angle=90]{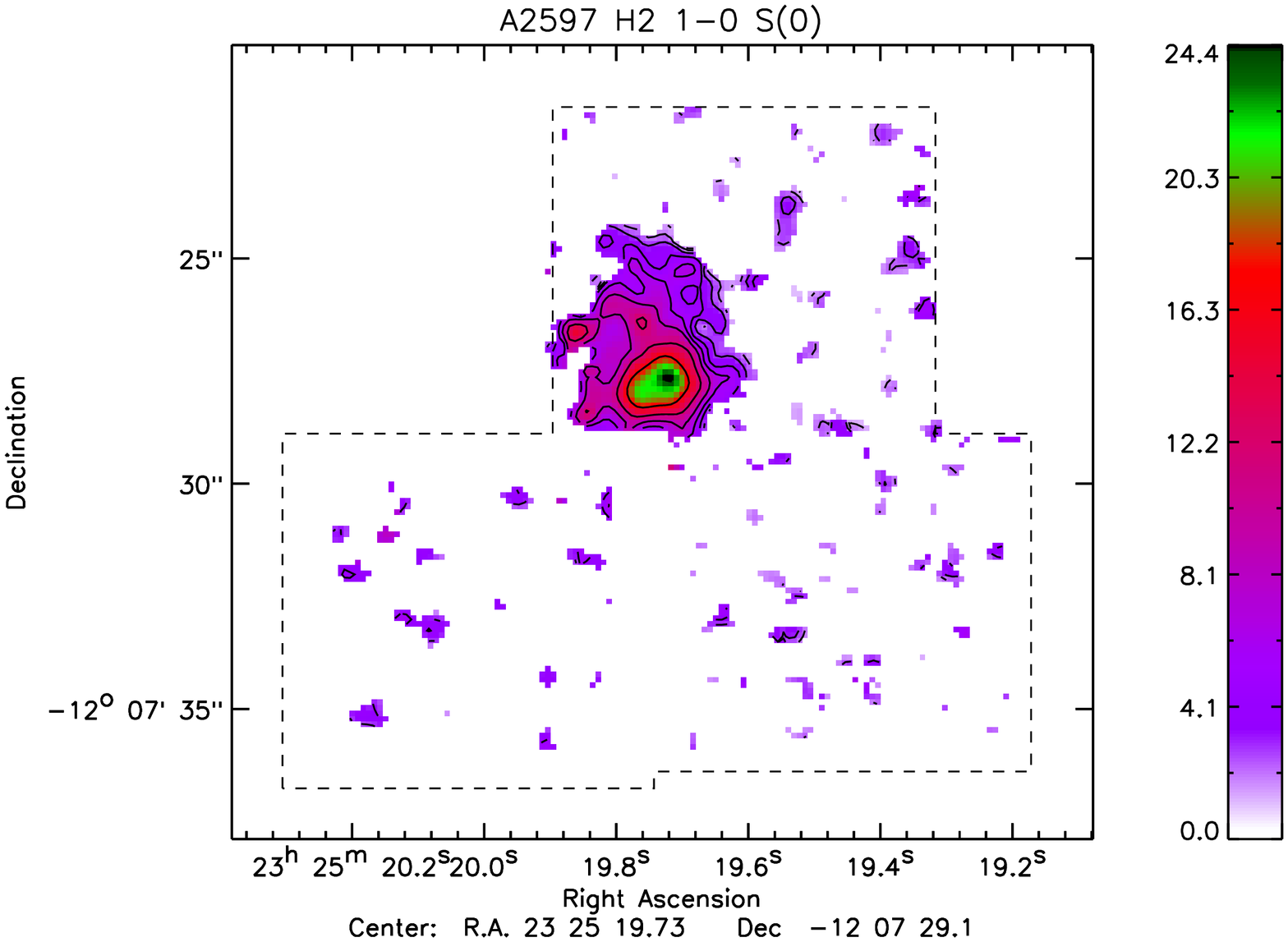}
    \includegraphics[width=0.33\textwidth, angle=90]{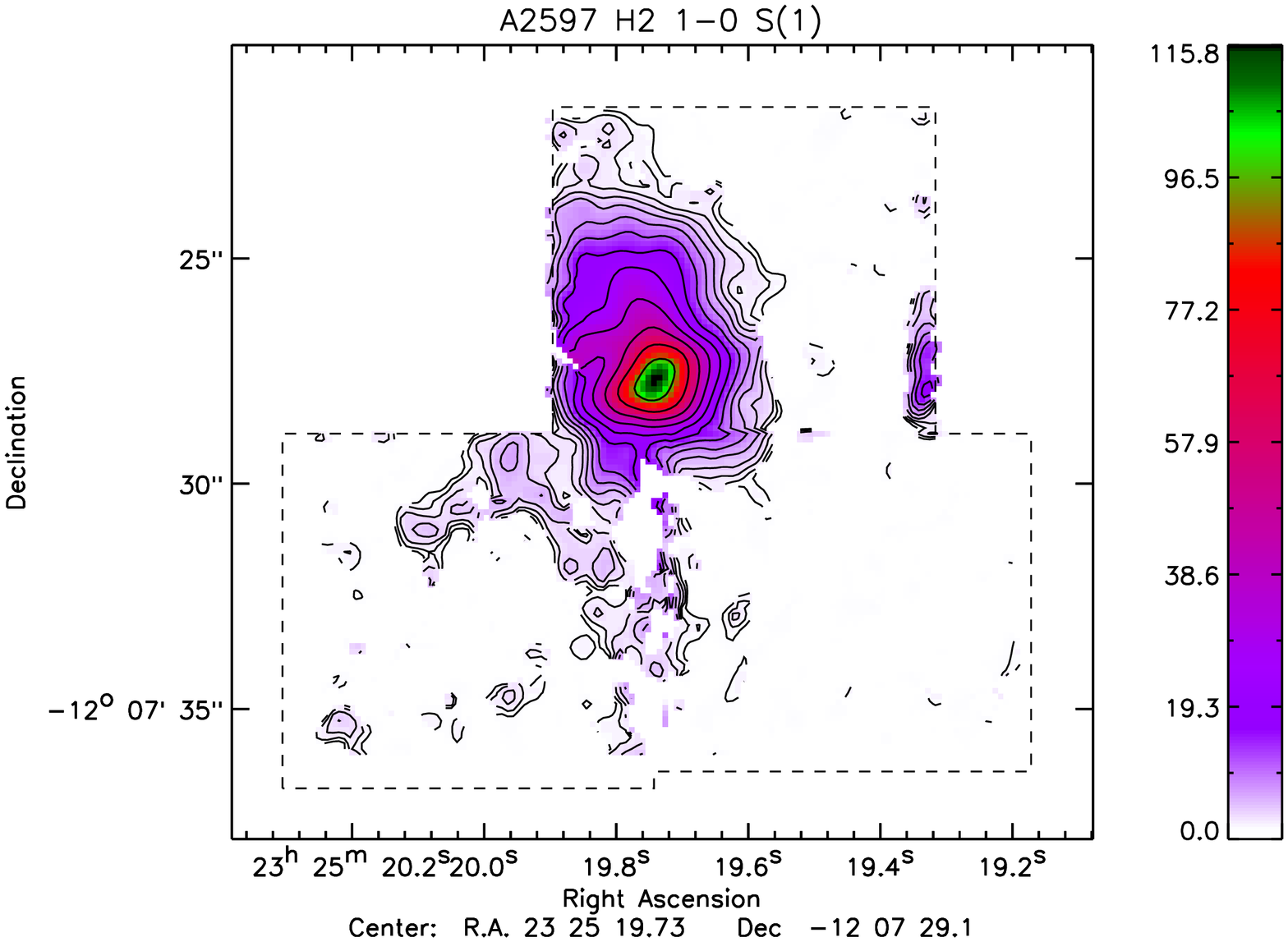}
    \includegraphics[width=0.33\textwidth, angle=90]{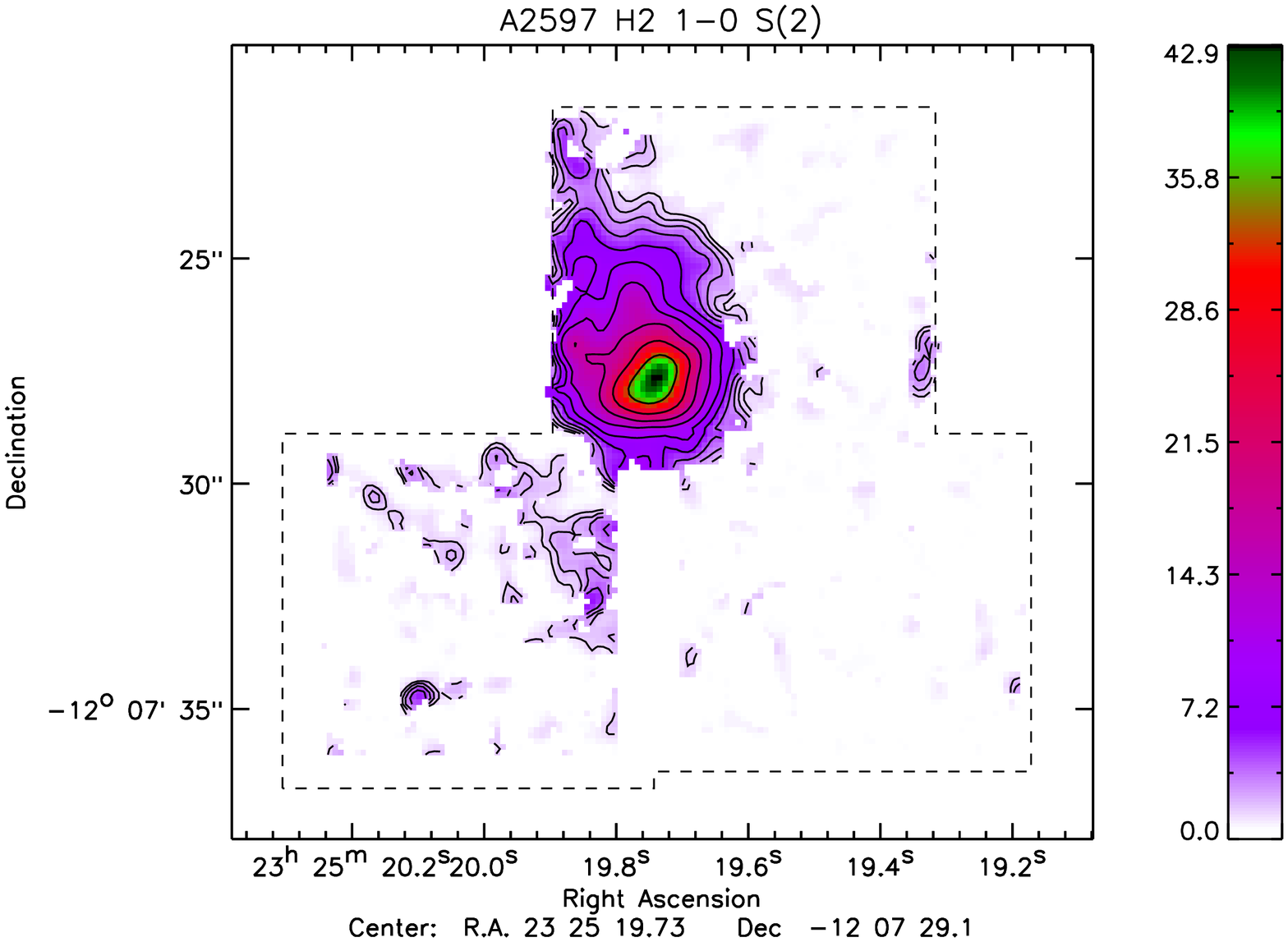}
    \includegraphics[width=0.33\textwidth, angle=90]{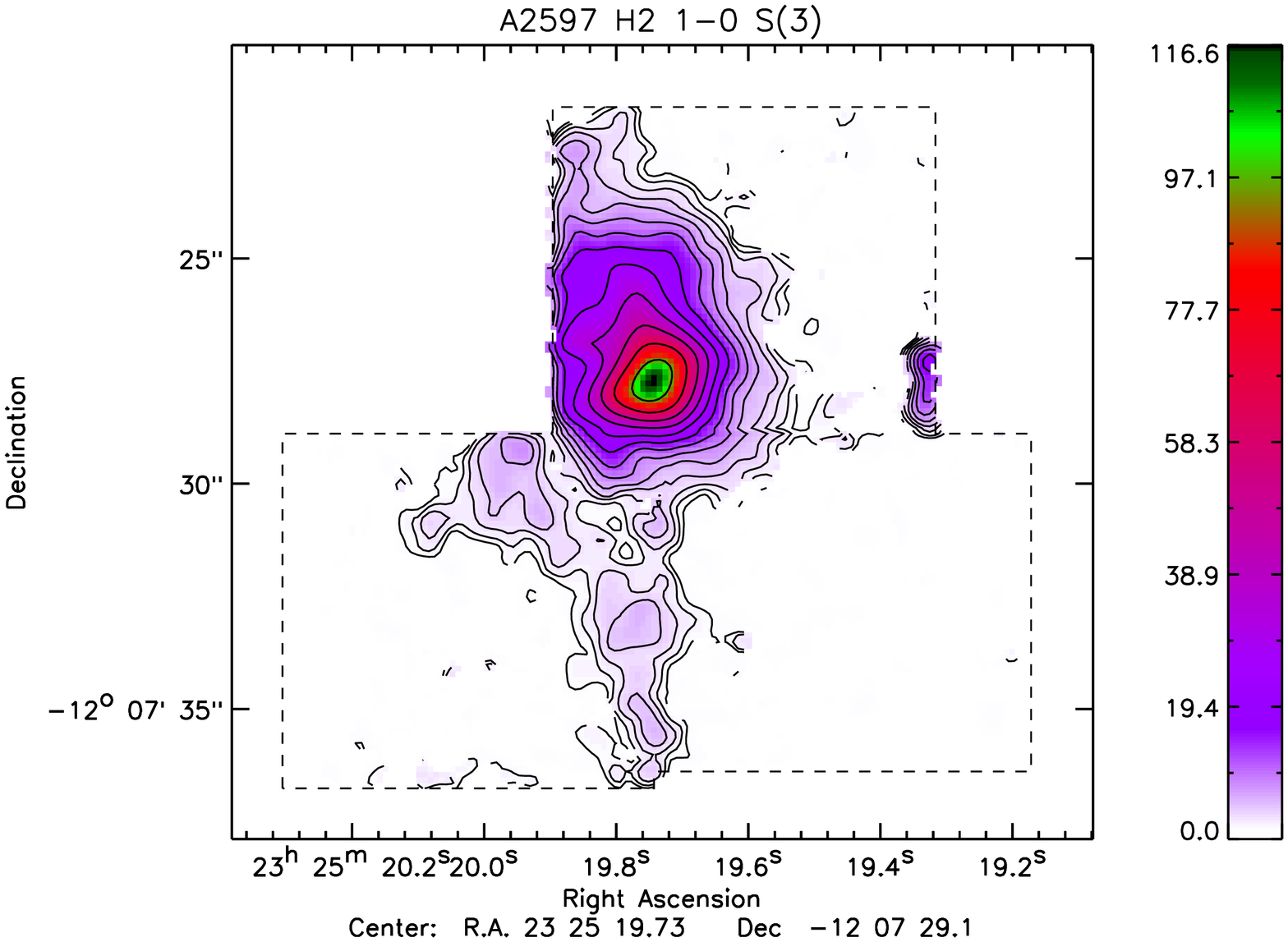}
    \includegraphics[width=0.33\textwidth, angle=90]{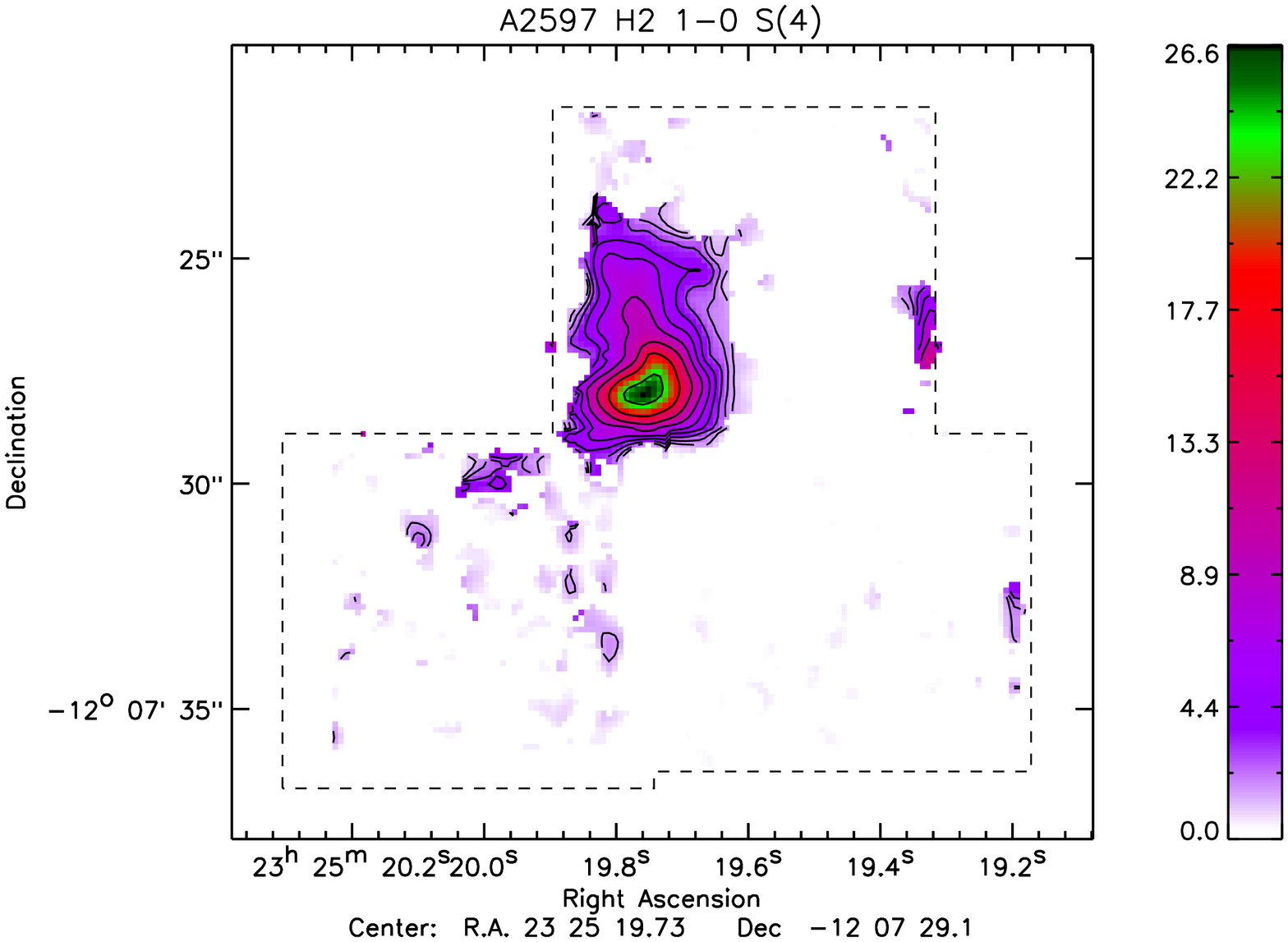}
    \includegraphics[width=0.33\textwidth, angle=90]{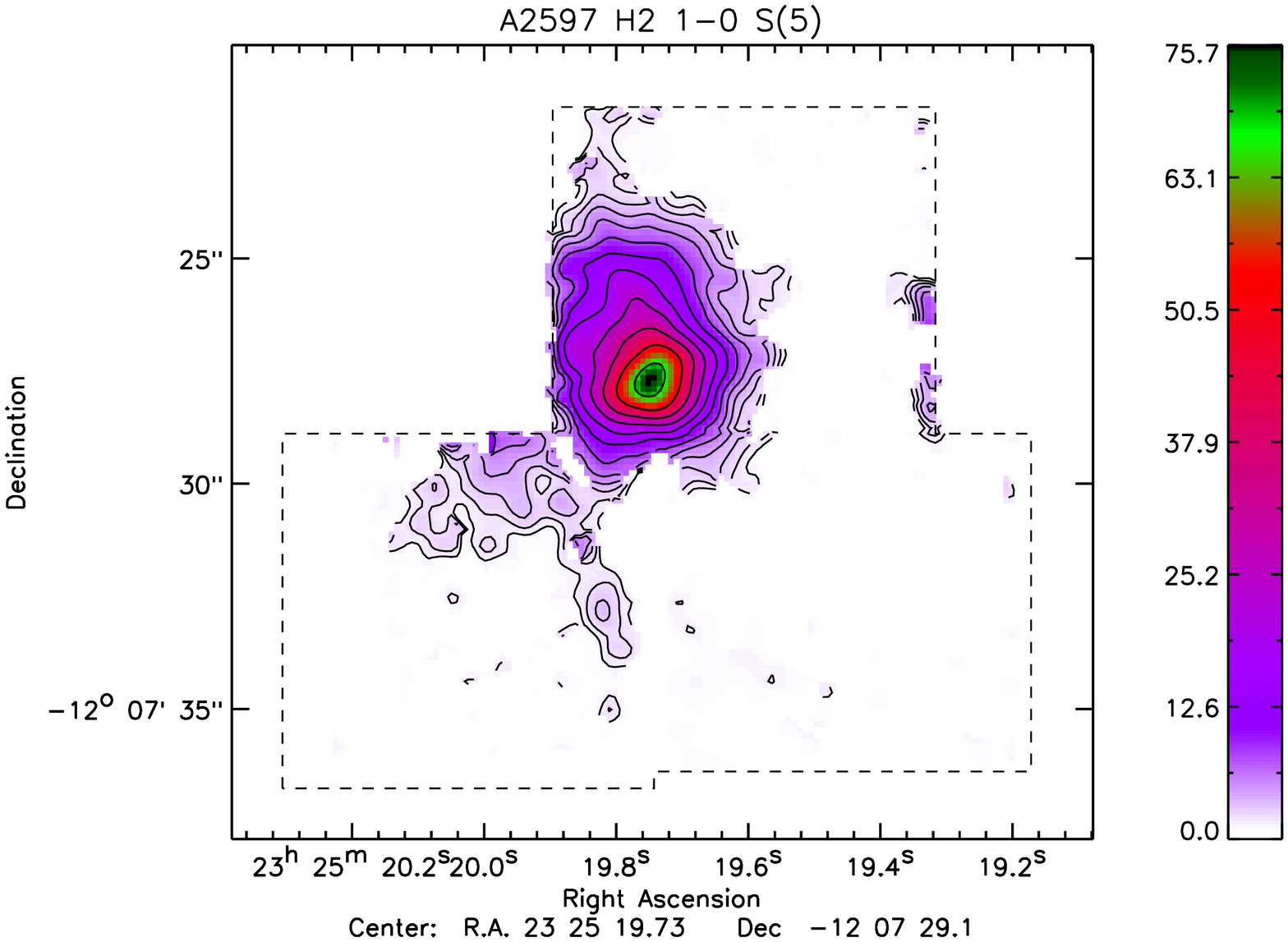}
  \vspace{0.5cm}
  \caption{ABELL 2597 Surface Brightness Maps. We show surface brightness maps for all detected lines that could be mapped on a pixel to pixel basis. The surface brightness is given in units of 10$^{-17}$~$\mathrm{erg}$~$\mathrm{s}^{-1}$~$\mathrm{cm}^{-2}$~$\mathrm{arcsec}^{-2}$ and contours are drawn starting at 1.5$\times$10$^{-17}$~$\mathrm{erg}$~$\mathrm{s}^{-1}$~$\mathrm{cm}^{-2}$~arcsec$^{-2}$ in steps of 2$^{\mathrm{n/2}}$ with n=0,1,2,... The maps are obtained by fitting the spectrum for each spatial pixel in the data cube by a single Gaussian. The data was smoothed by four pixels in both the spatial and spectral planes. Maps are shown for H$_{\mathrm{2}}$~1-0~S(0), 1-0~S(1), 1-0~S(2), 1-0~S(3), 1-0~S(4), 1-0~S(5), 2-1~S(3), Pa~$\alpha$, Br~$\delta$ and Fe~II (1.81~$\mu m$). The Br~$\delta$ map includes the flux due to H$_{\mathrm{2}}$~2-1~S(5). The Br~$\gamma$, H$_{\mathrm{2}}$~2-1~S(2) and H$_{\mathrm{2}}$~2-1~S(4) lines are too weak and are thus not shown here. The northern-most observed field is not shown here.}\label{app_fig_sign_maps_a2597_ss44}
\end{figure*}

\addtocounter{figure}{-1}

\begin{figure*}
    \includegraphics[width=0.33\textwidth, angle=90]{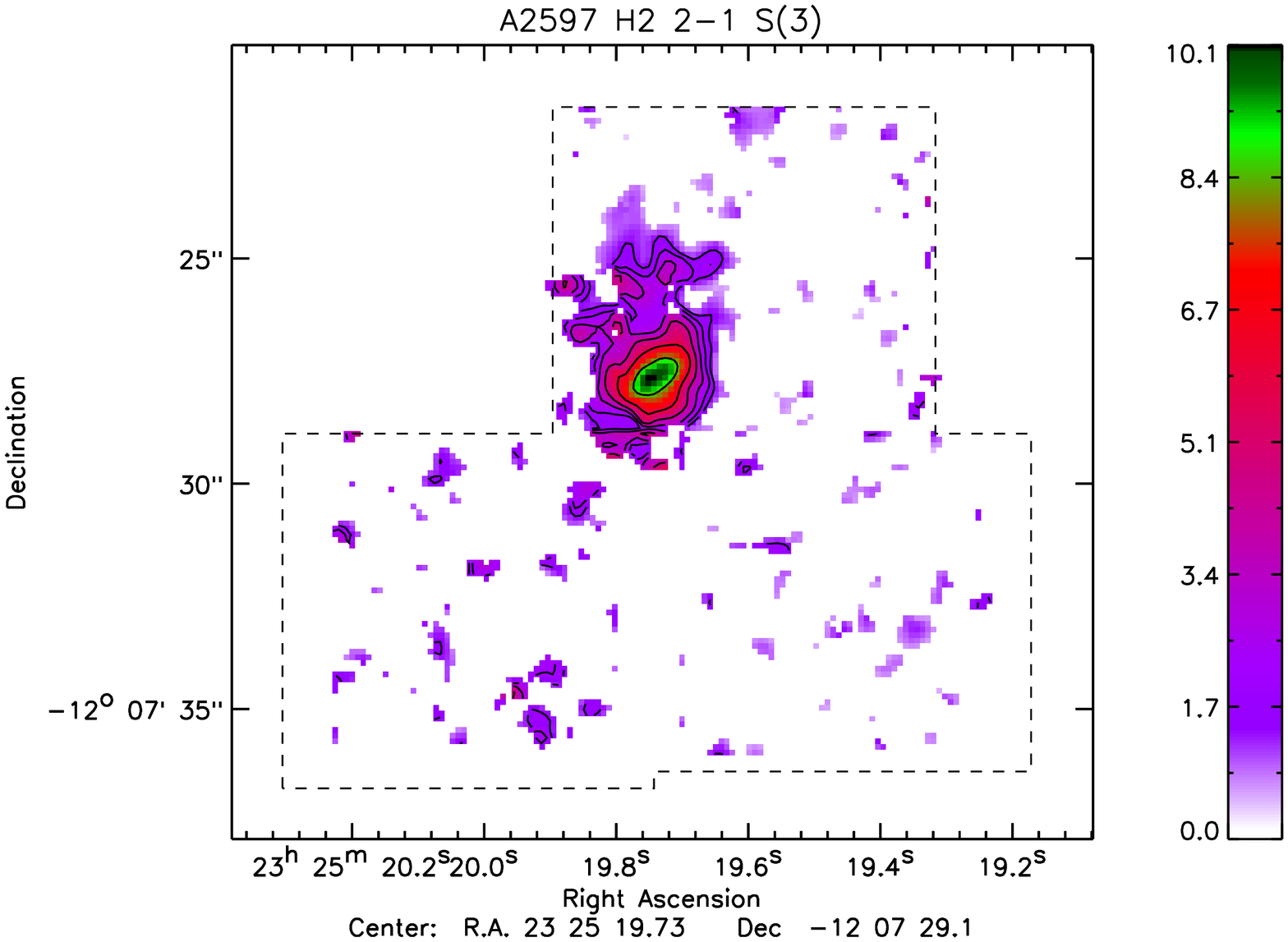}
    \includegraphics[width=0.33\textwidth, angle=90]{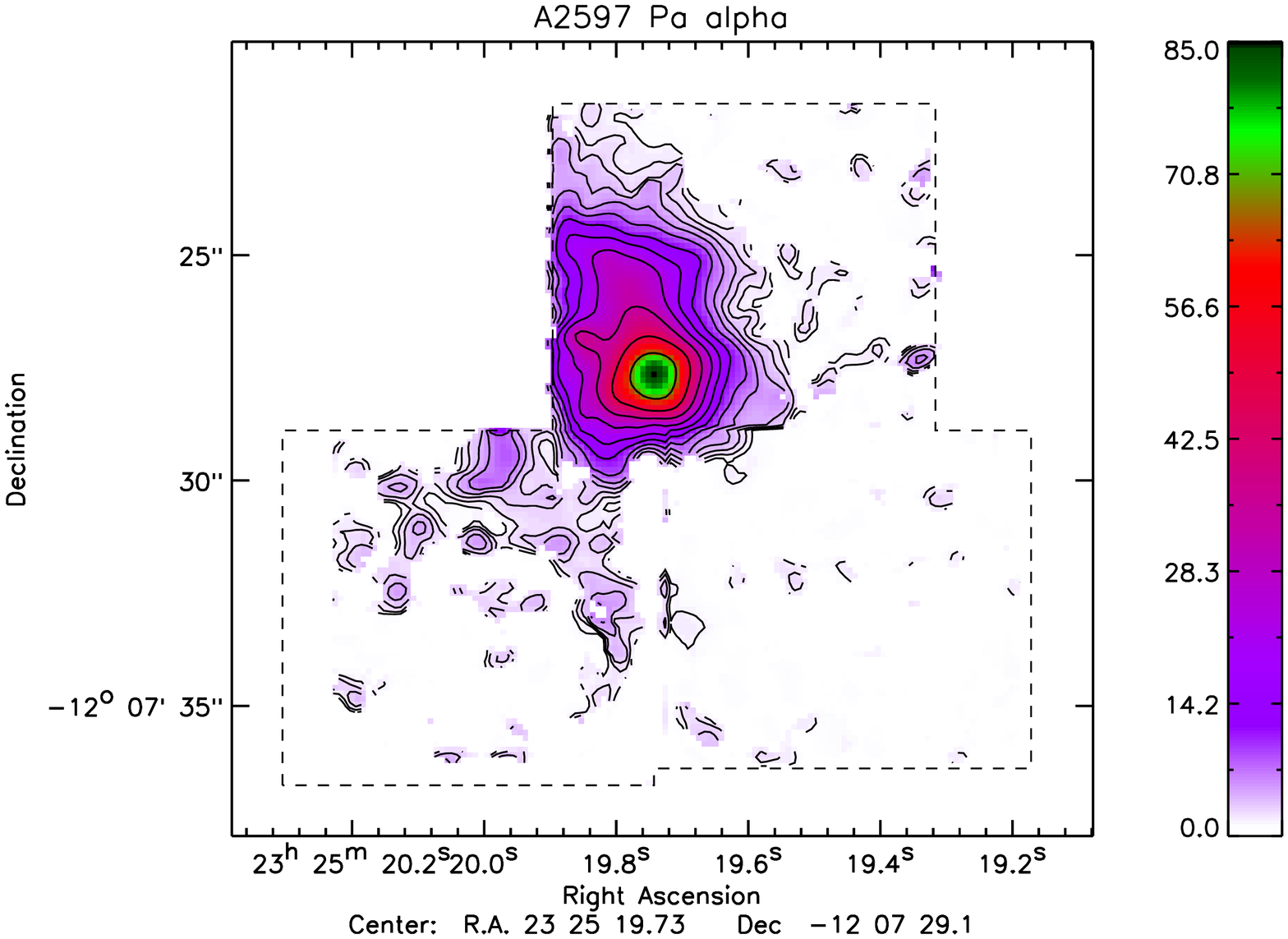}
    \includegraphics[width=0.33\textwidth, angle=90]{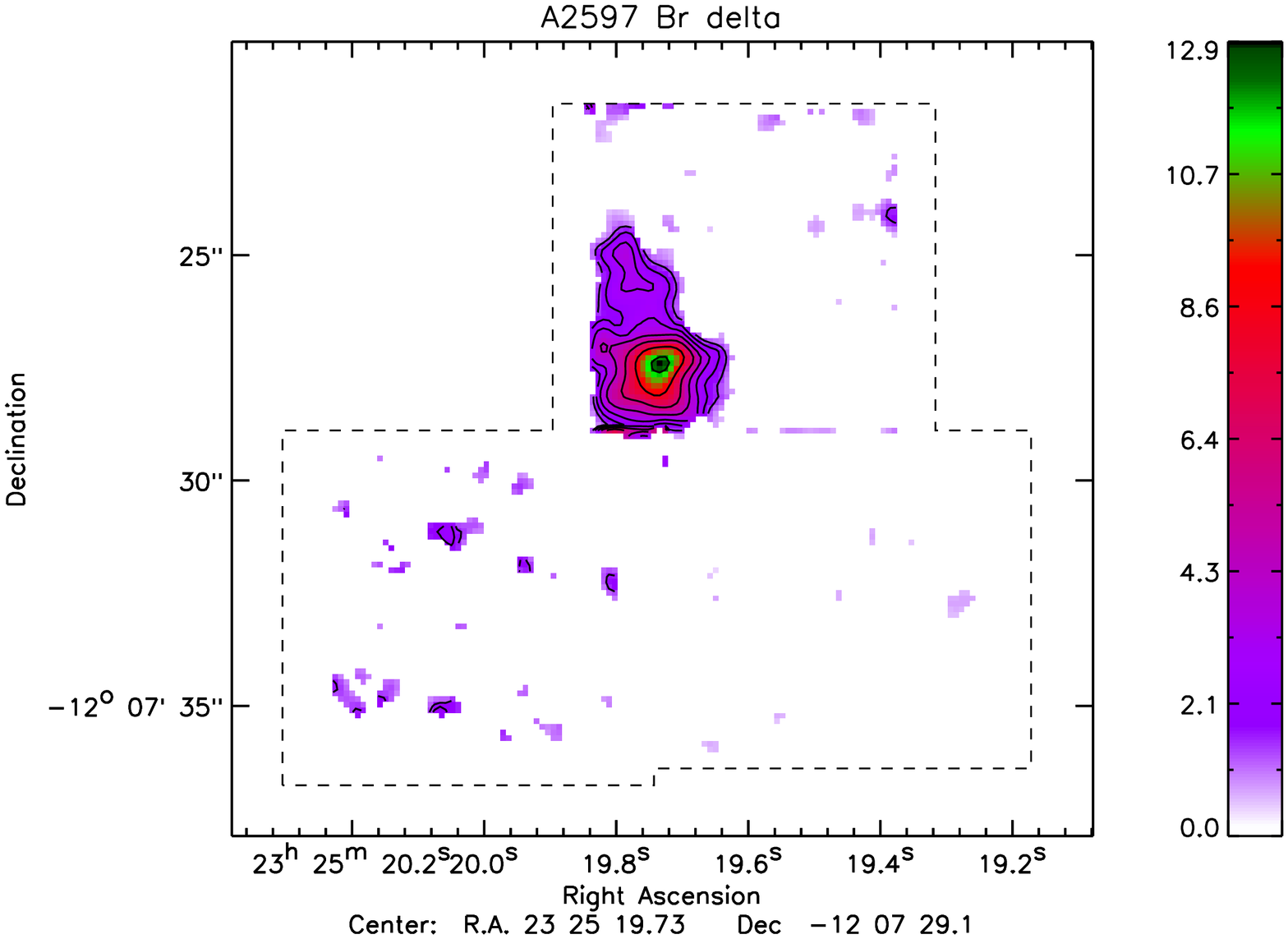}
    \includegraphics[width=0.33\textwidth, angle=90]{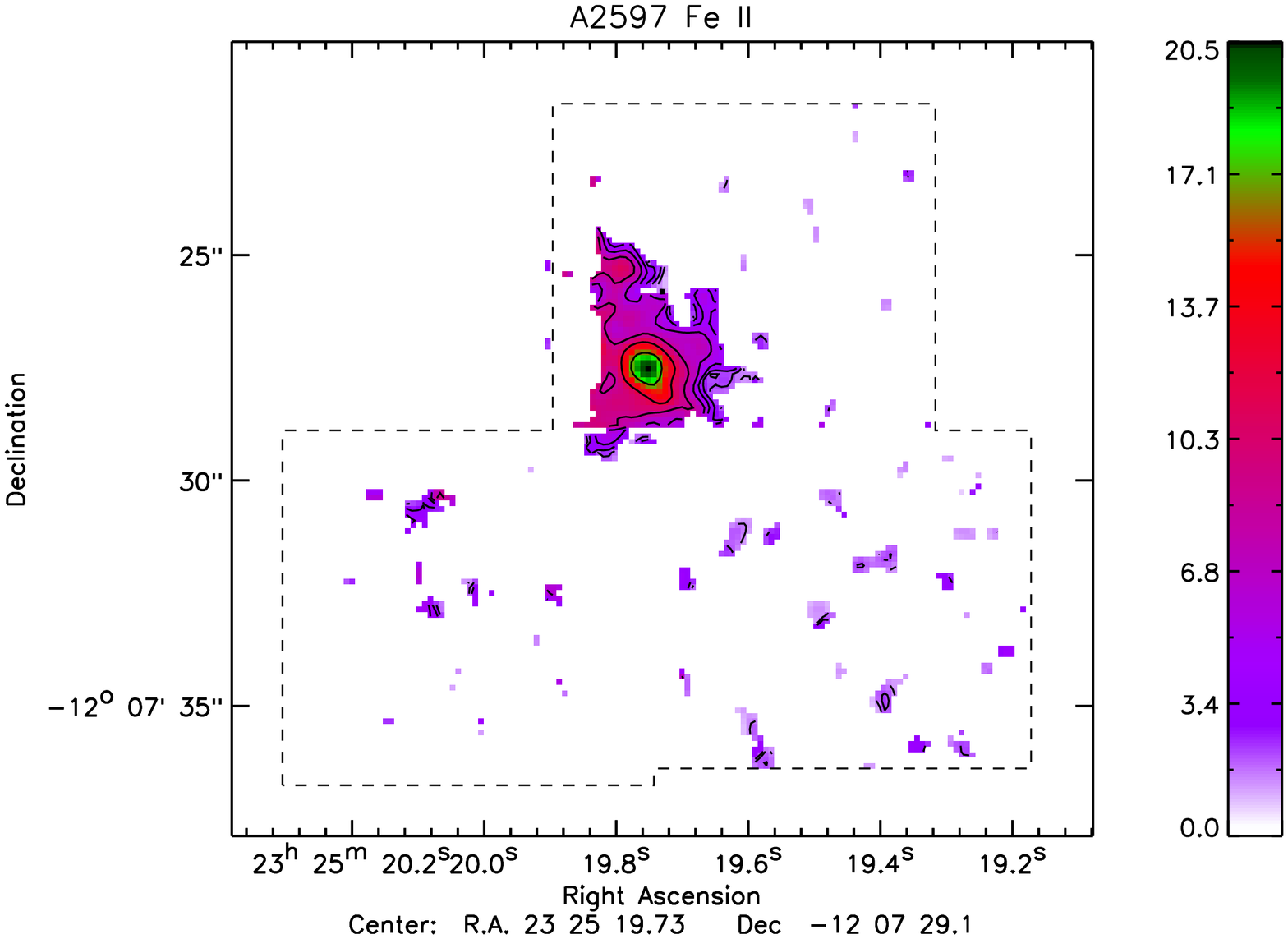}
  \vspace{0.5cm}
  \caption{\it continued}
\end{figure*}

\section[]{Sersic 159-03 Surface Brightness Maps}\label{app_lmp_sersic}
\begin{figure*}
    \includegraphics[width=0.33\textwidth, angle=90]{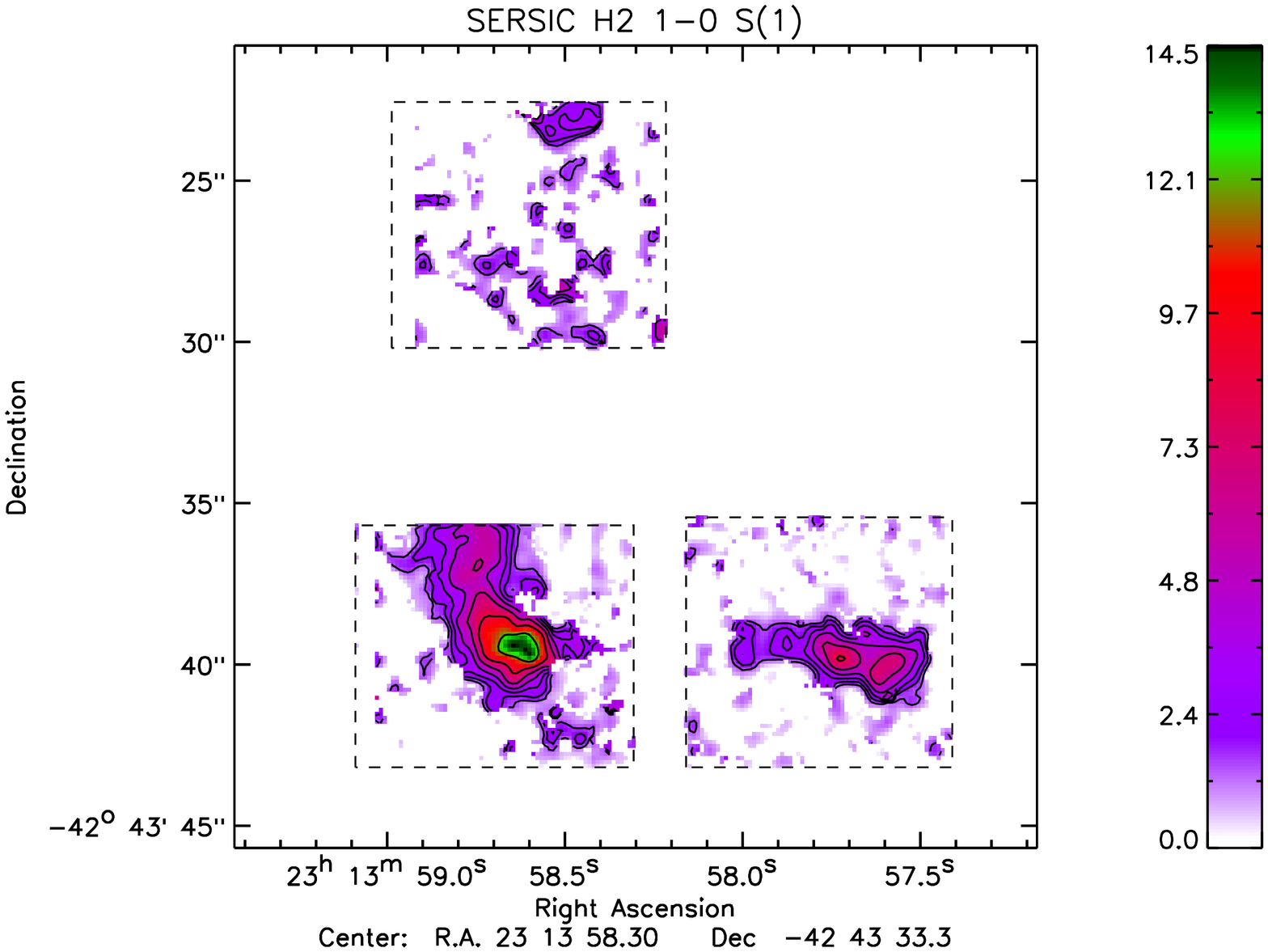}
    \includegraphics[width=0.33\textwidth, angle=90]{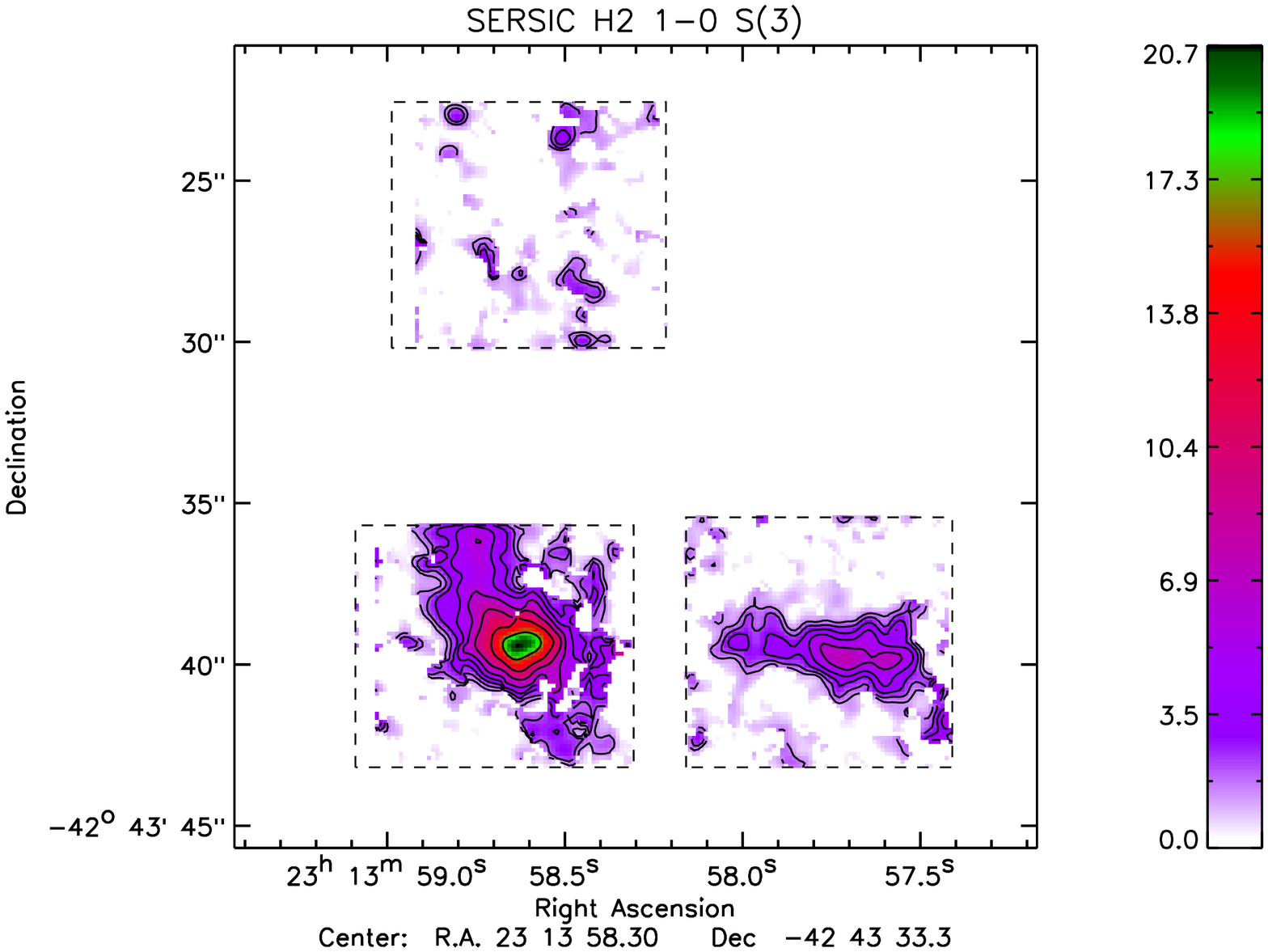}
    \includegraphics[width=0.33\textwidth, angle=90]{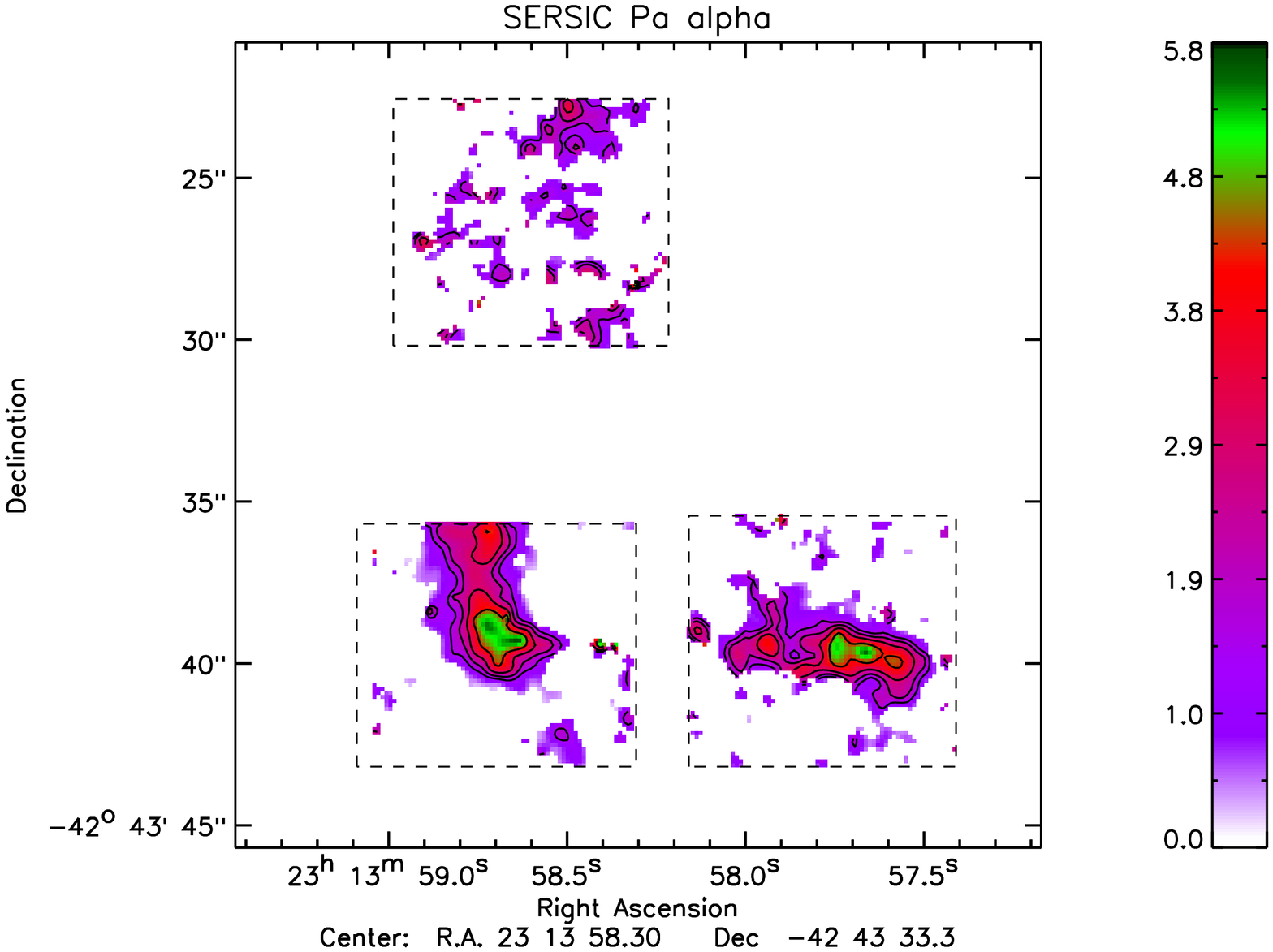}
  \vspace{0.5cm}
  \caption{SERSIC 159-03 Surface Brightness Maps. We show surface brightness maps for all detected lines that could be mapped on a pixel to pixel basis. The surface brightness is given in units of 10$^{-17}$~$\mathrm{erg}$ $\mathrm{s}^{-1}$ $\mathrm{cm}^{-2}$ $\mathrm{arcsec}^{-2}$ and contours are drawn starting at 1.5$\times$10$^{-17}$~$\mathrm{erg}$~$\mathrm{s}^{-1}$~$\mathrm{cm}^{-2}$~arcsec$^{-2}$ in steps of 2$^{\mathrm{n/2}}$ with $n$=0,1,2,... The maps are obtained by fitting the spectrum for each spatial pixel in the data cube by a single Gaussian. The data was smoothed by four pixels in both the spatial and spectral planes. Maps are shown for H$_{\mathrm{2}}$~1-0~S(1) 1-0~S(3) and Pa~$\alpha$.}\label{app_fig_sign_maps_sersic_ss44}
\end{figure*}

\section[]{Selected Regions: Line Profiles and Gaussian Fits}\label{app_tb}
\begin{figure*}
    \includegraphics[width=0.17\textwidth, angle=90]{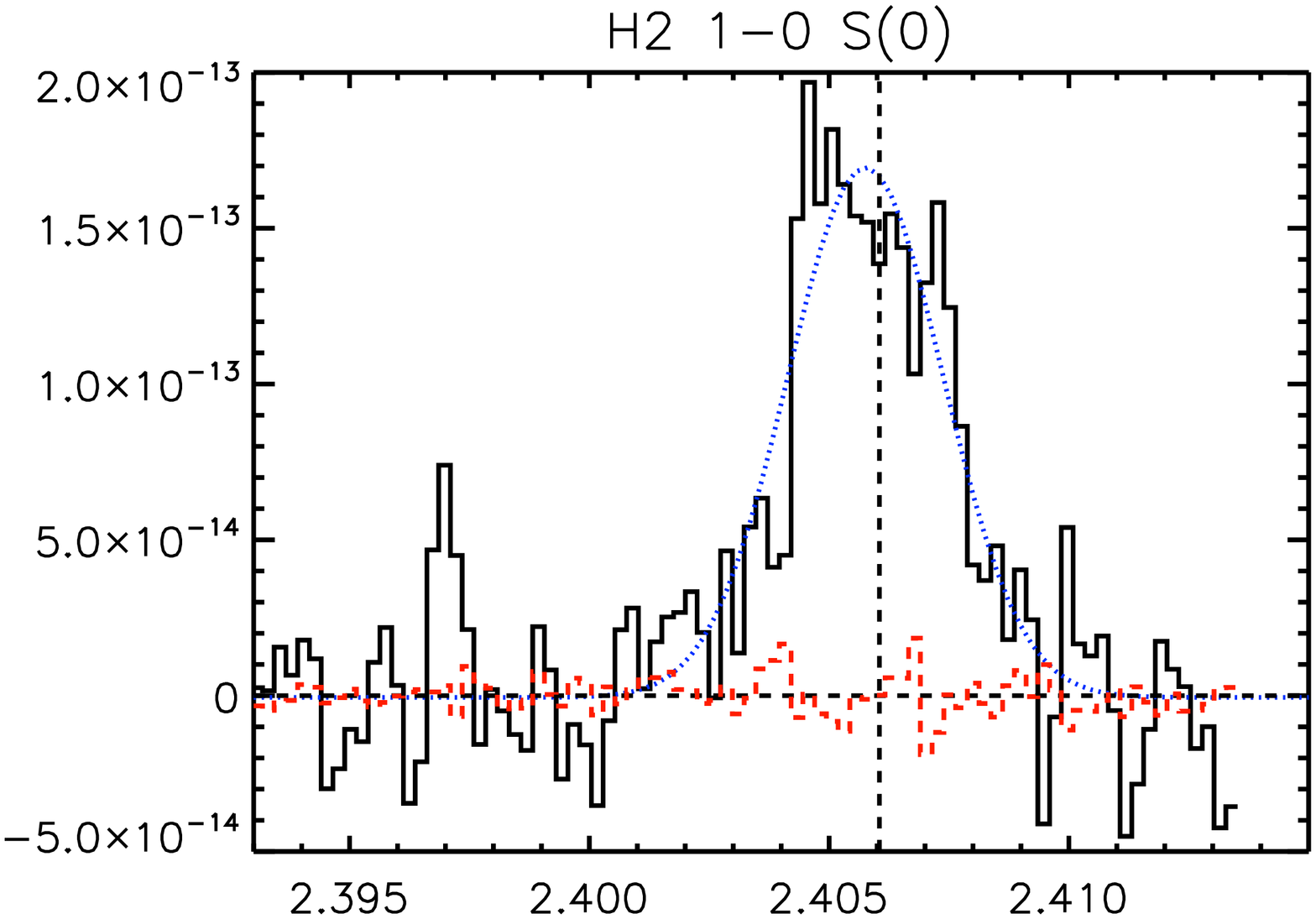}
    \includegraphics[width=0.17\textwidth, angle=90]{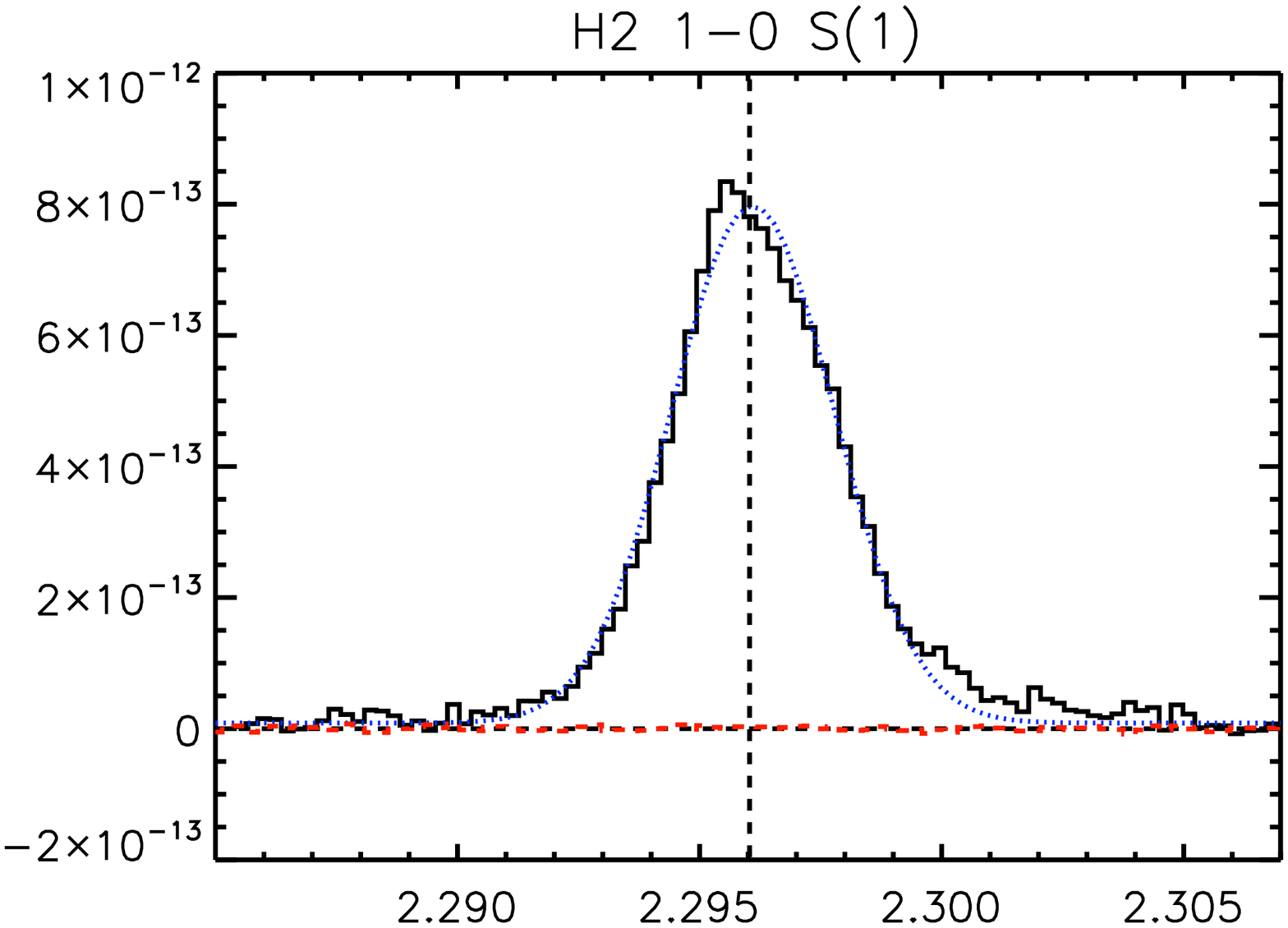}
    \includegraphics[width=0.17\textwidth, angle=90]{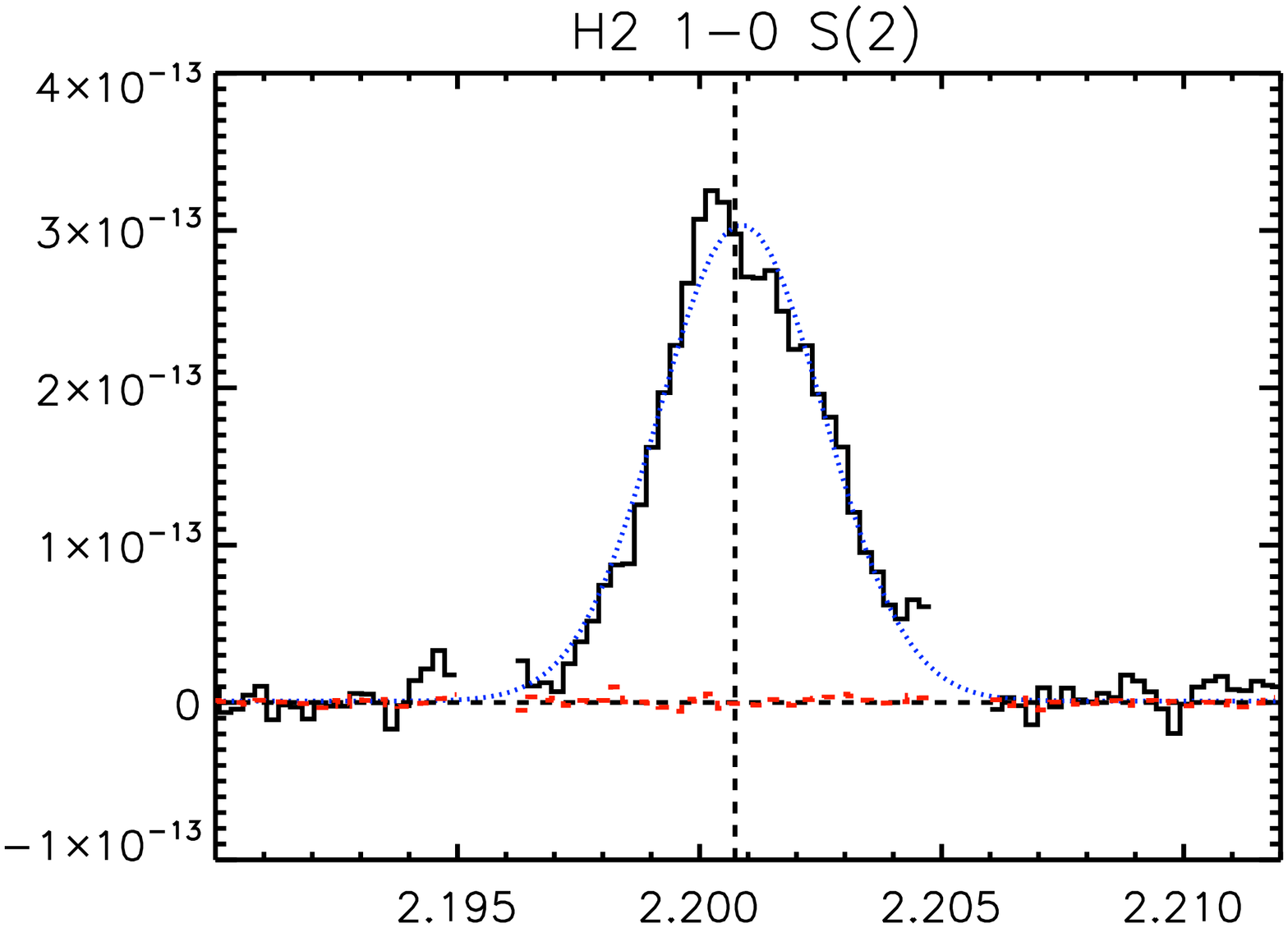}
    \includegraphics[width=0.17\textwidth, angle=90]{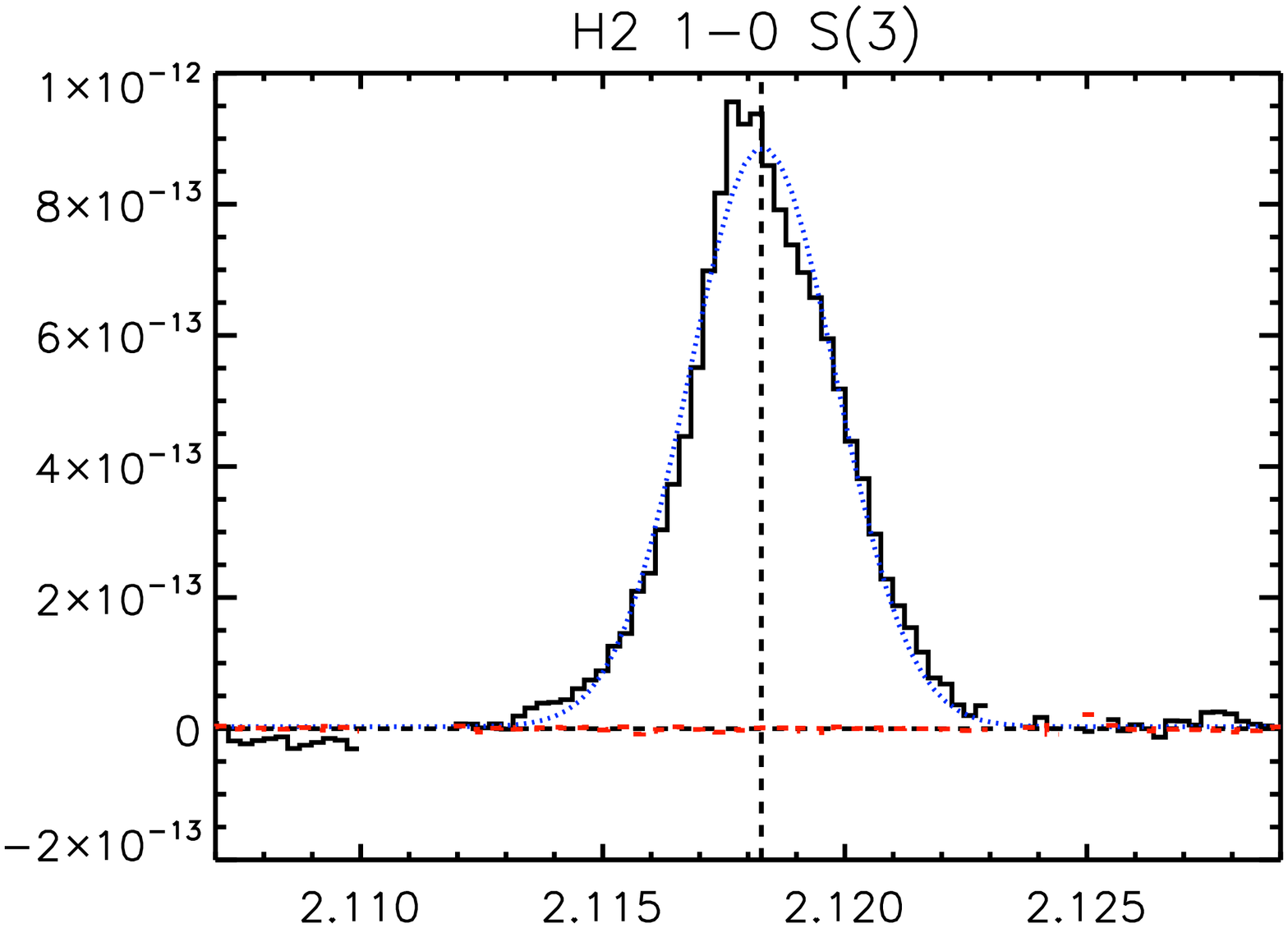}
    \includegraphics[width=0.17\textwidth, angle=90]{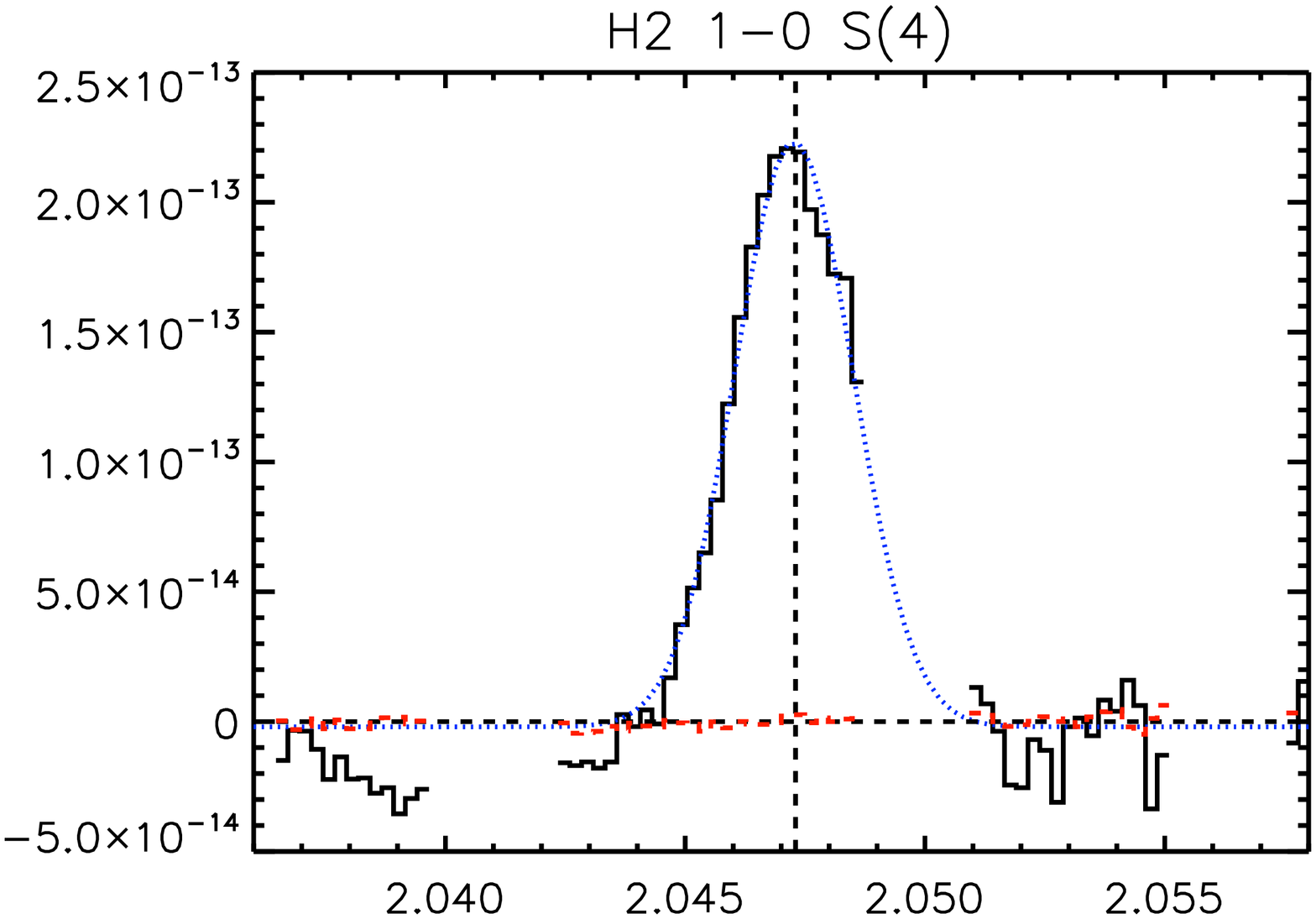}
    \includegraphics[width=0.17\textwidth, angle=90]{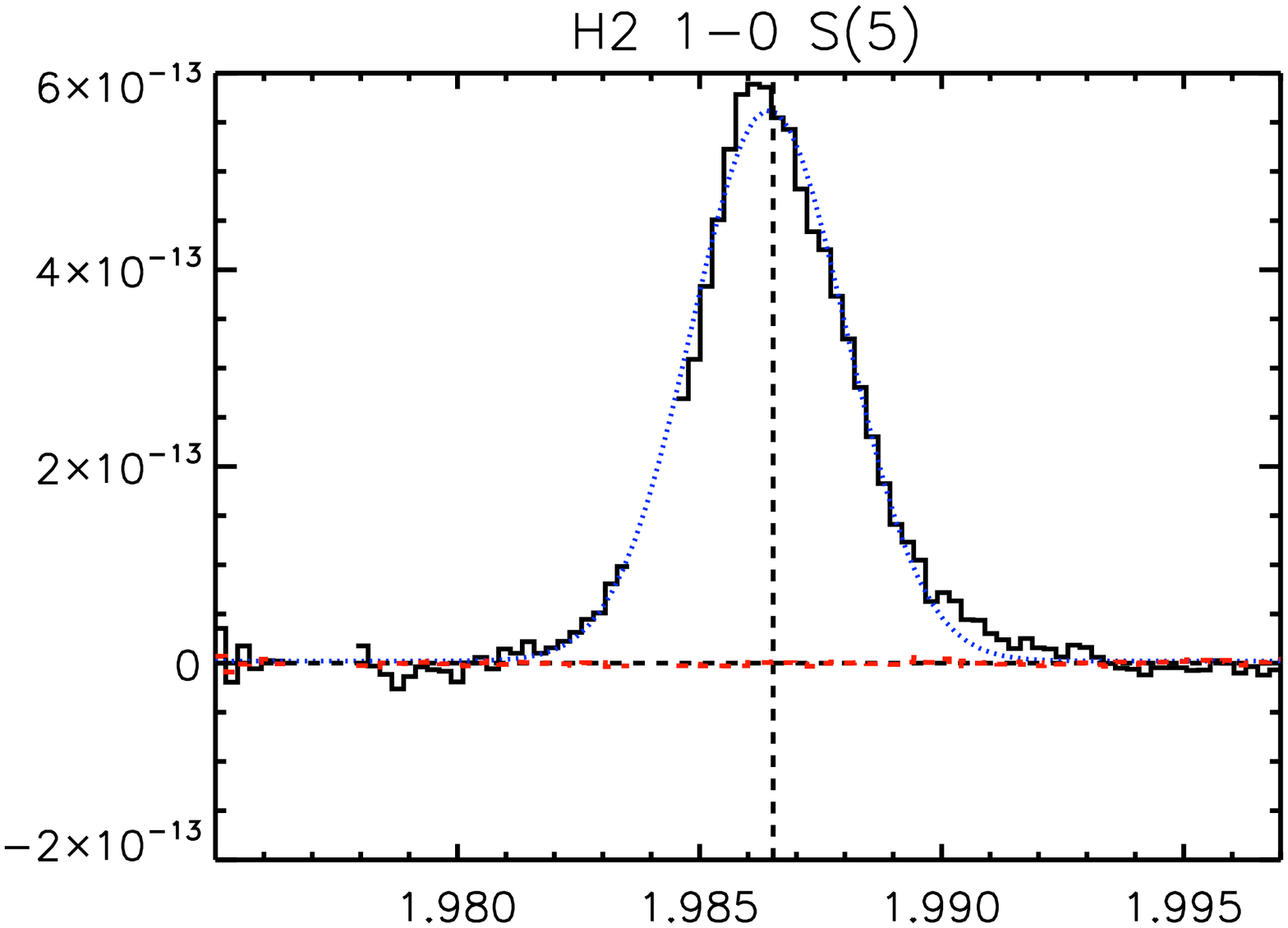}
    \includegraphics[width=0.17\textwidth, angle=90]{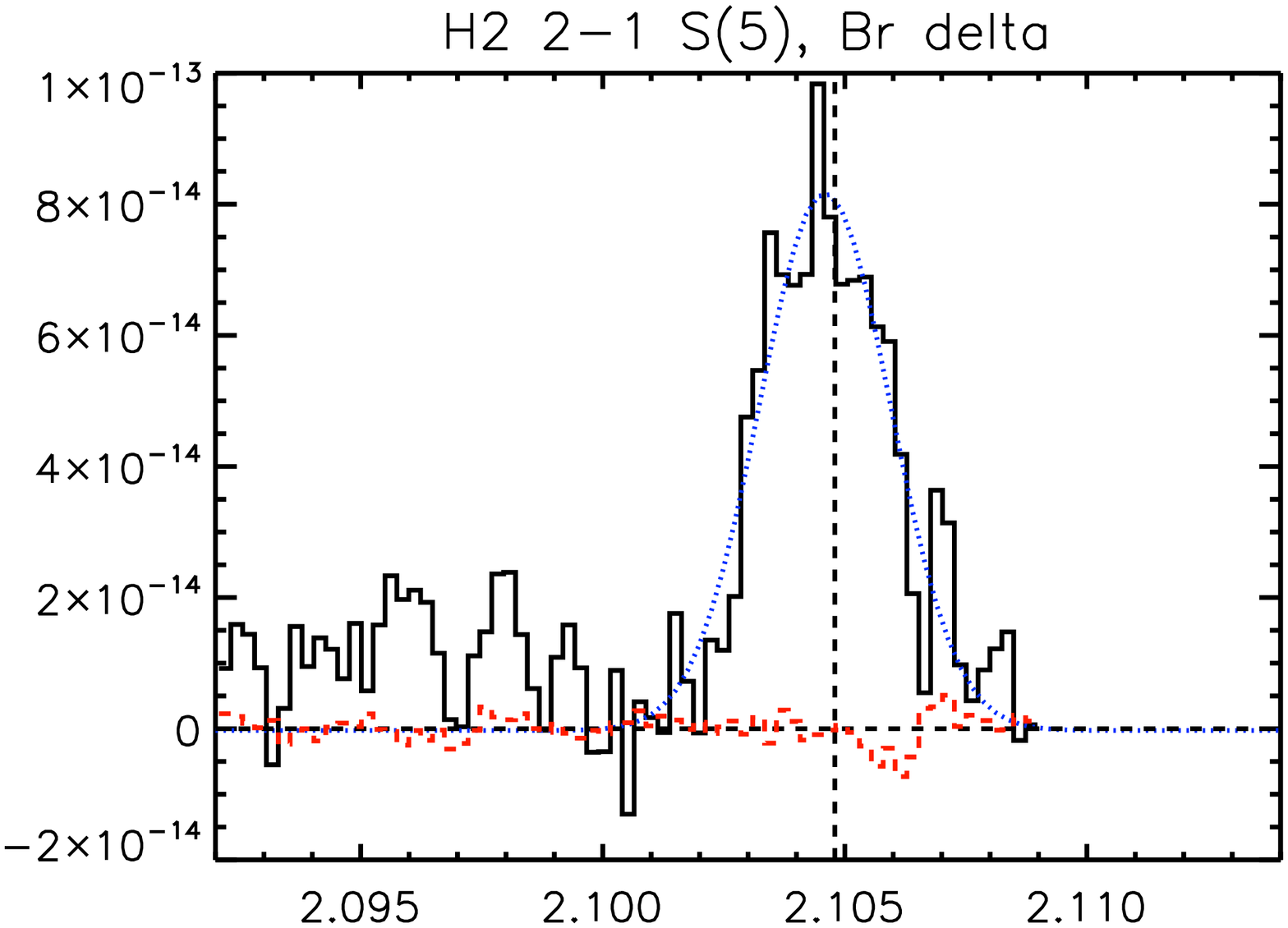}
    \includegraphics[width=0.17\textwidth, angle=90]{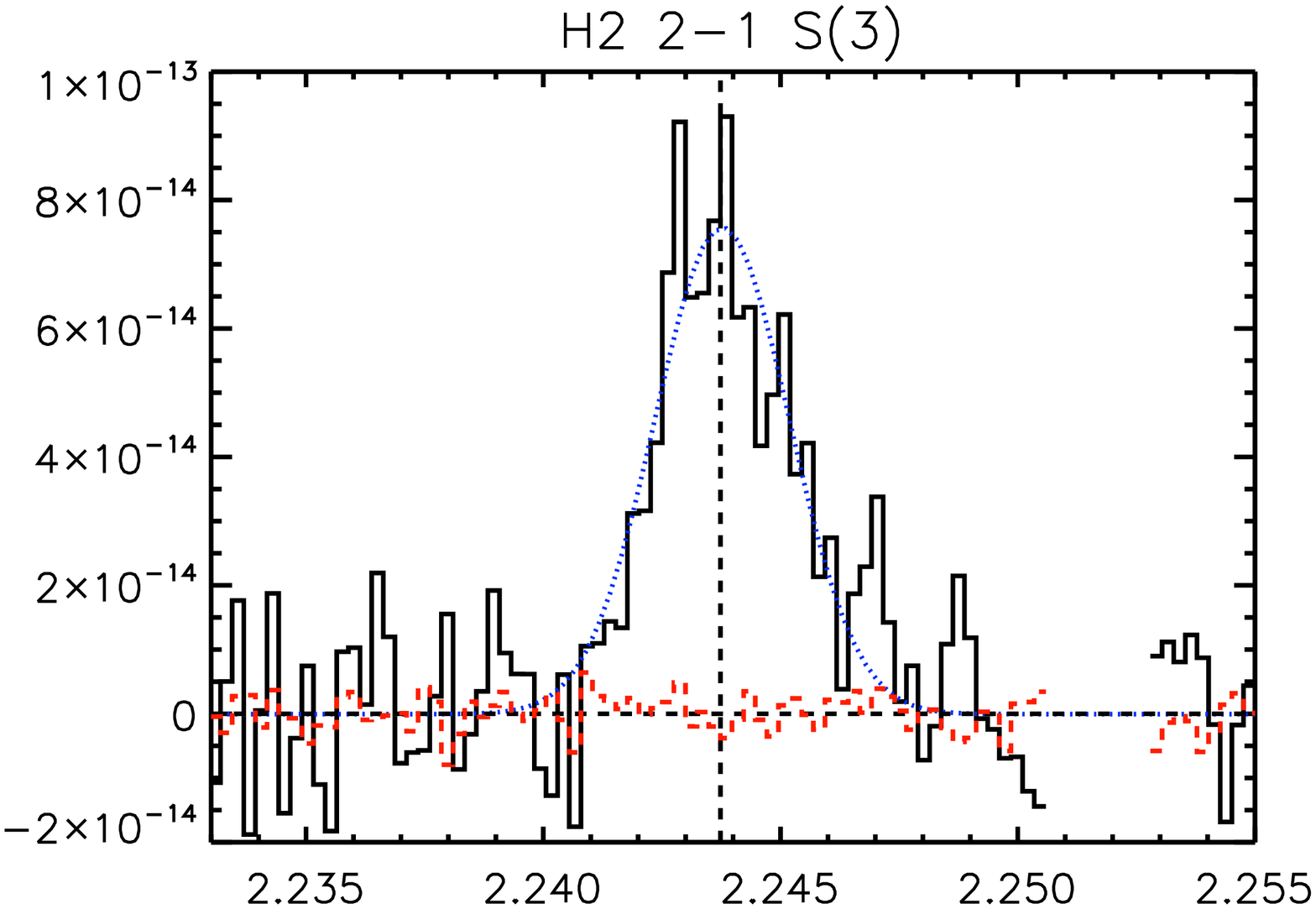}
    \includegraphics[width=0.17\textwidth, angle=90]{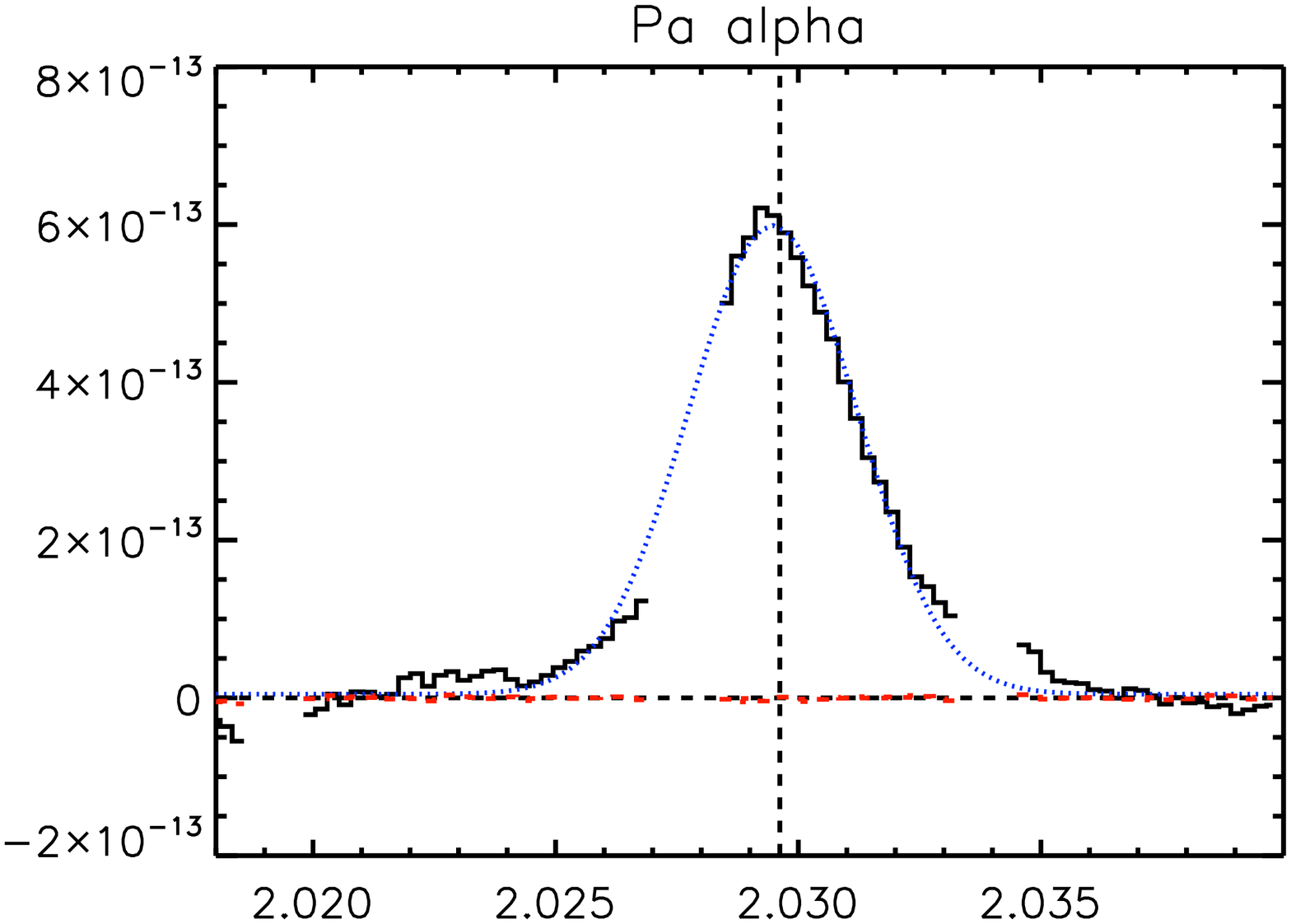}
    \includegraphics[width=0.17\textwidth, angle=90]{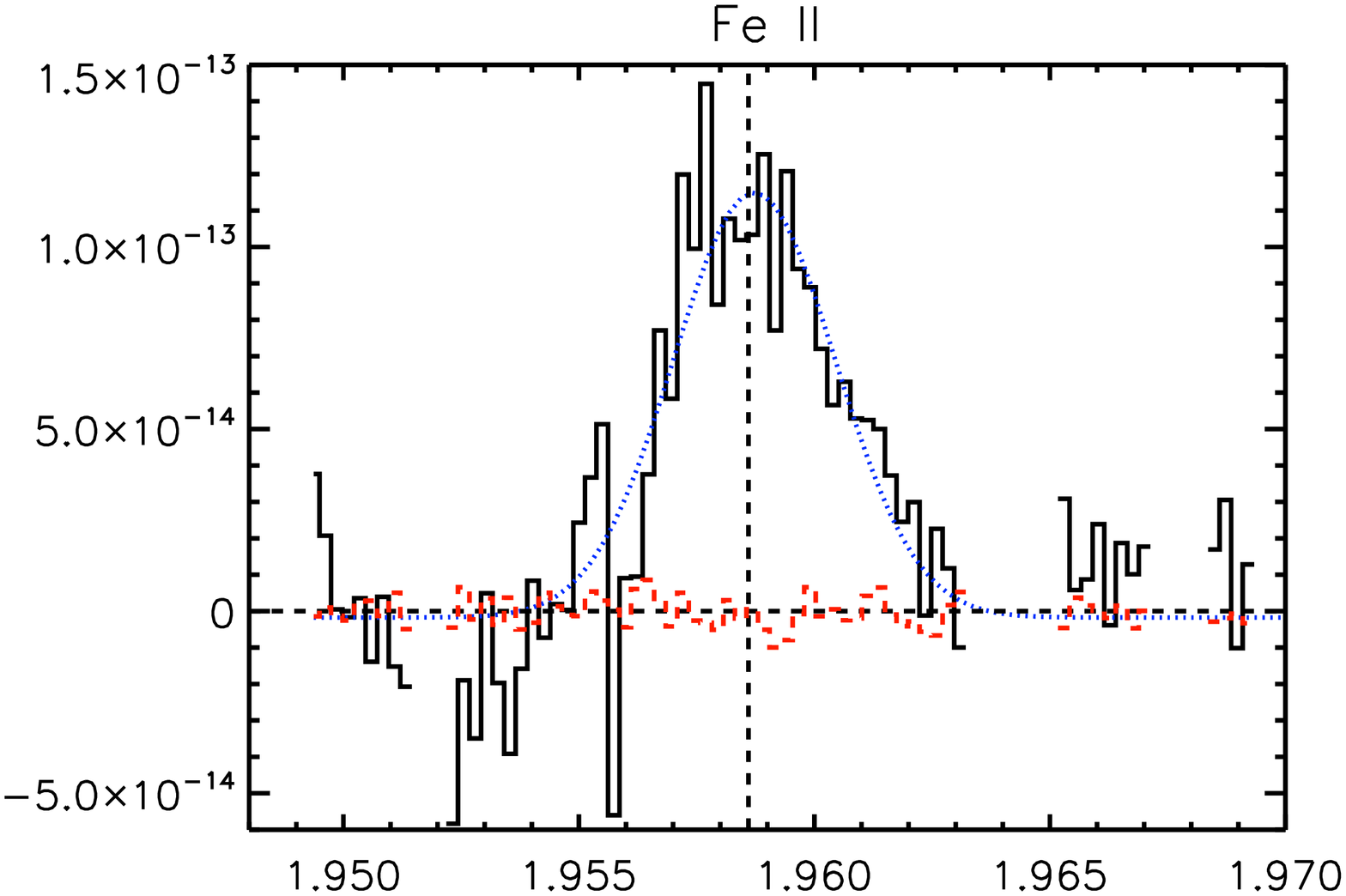}
    \includegraphics[width=0.17\textwidth, angle=90]{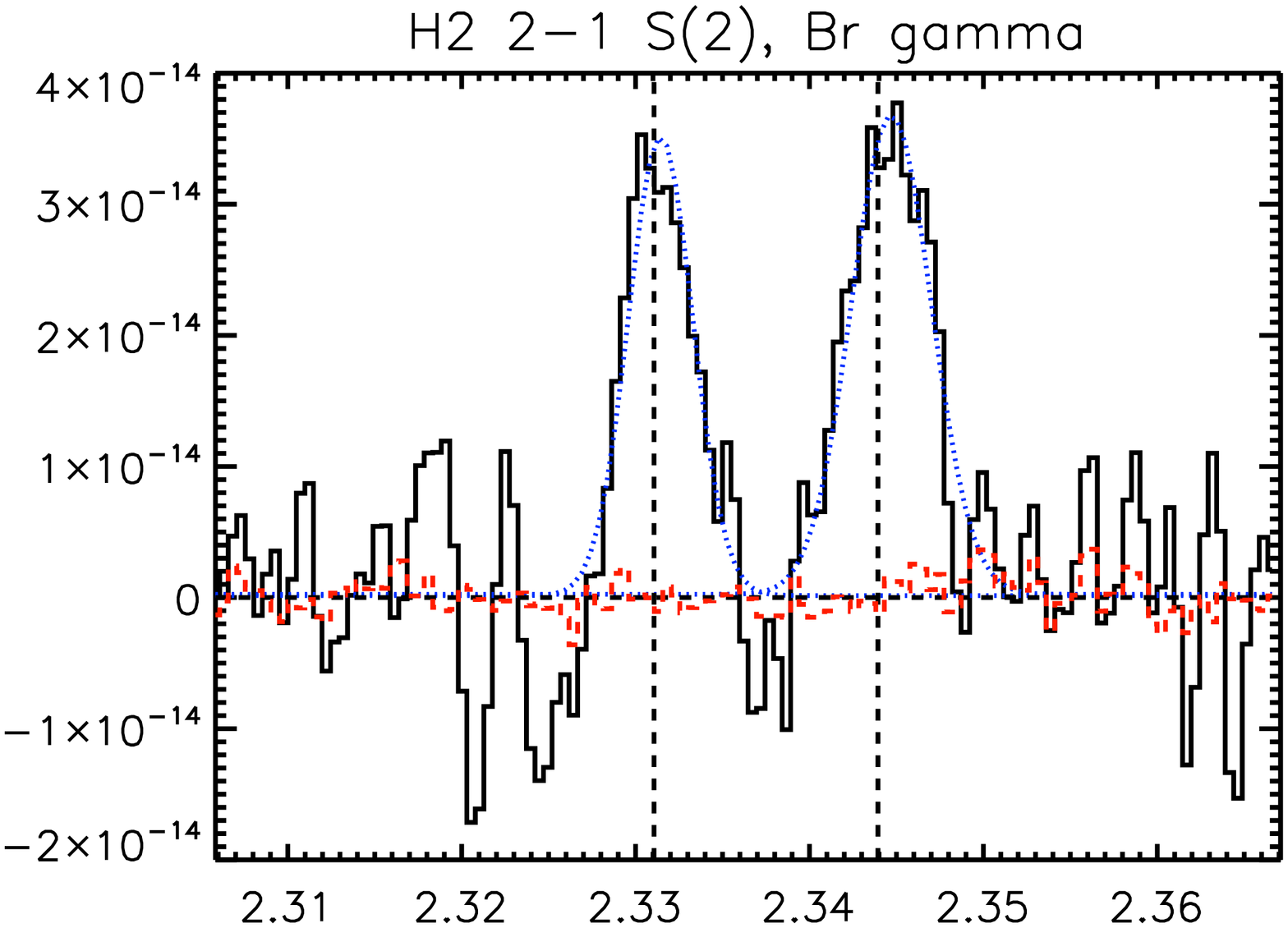}
  \caption{ABELL 2597 Line Spectra (Region A1). For the Br~$\gamma$ and H$_{\mathrm{2}}$~2-1~S(2) lines we used a four pixel spatial and spectral smoothing. For all other lines a two pixel spatial smoothing was used. The solid black line shows the measured line spectrum. The dotted blue line shows the Gaussian fit to the line profile. The dashed red line shows the spectrum from the corresponding spatial median region (this spectrum has been scaled to match the area of the selected region).}\label{fig_a2597_line_tb_a1}
\end{figure*}

\begin{figure*}
    \includegraphics[width=0.17\textwidth, angle=90]{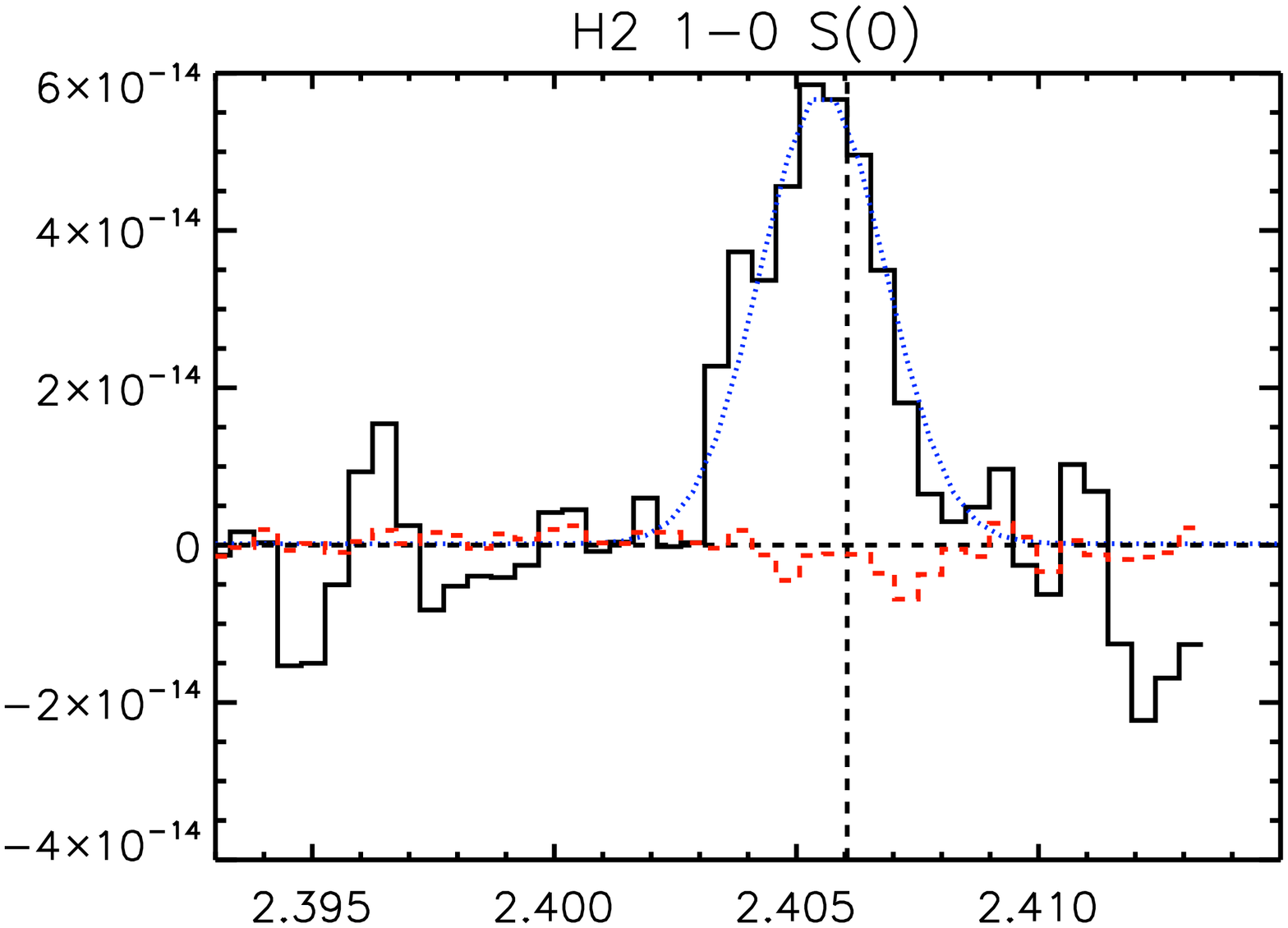}
    \includegraphics[width=0.17\textwidth, angle=90]{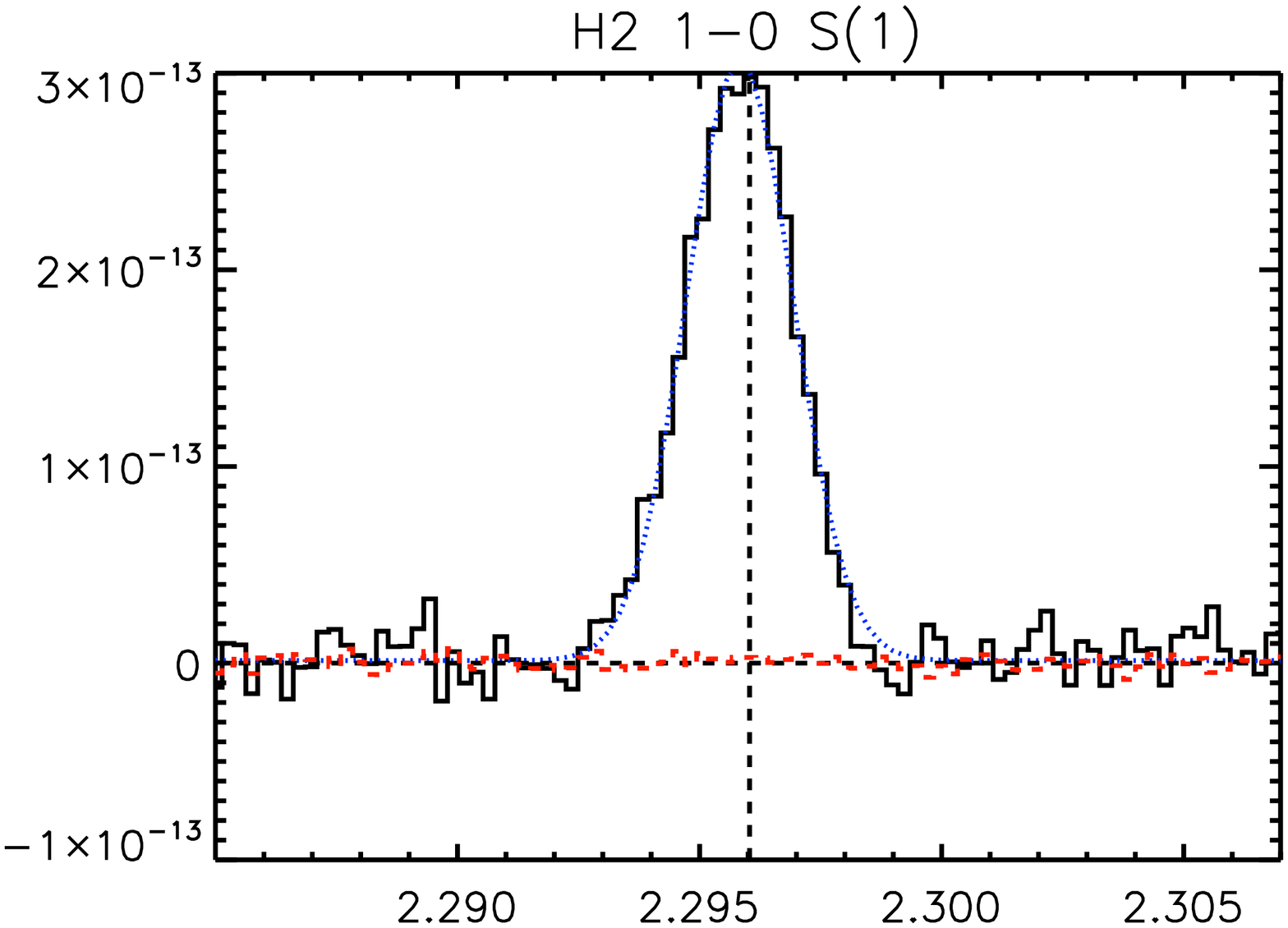}
    \includegraphics[width=0.17\textwidth, angle=90]{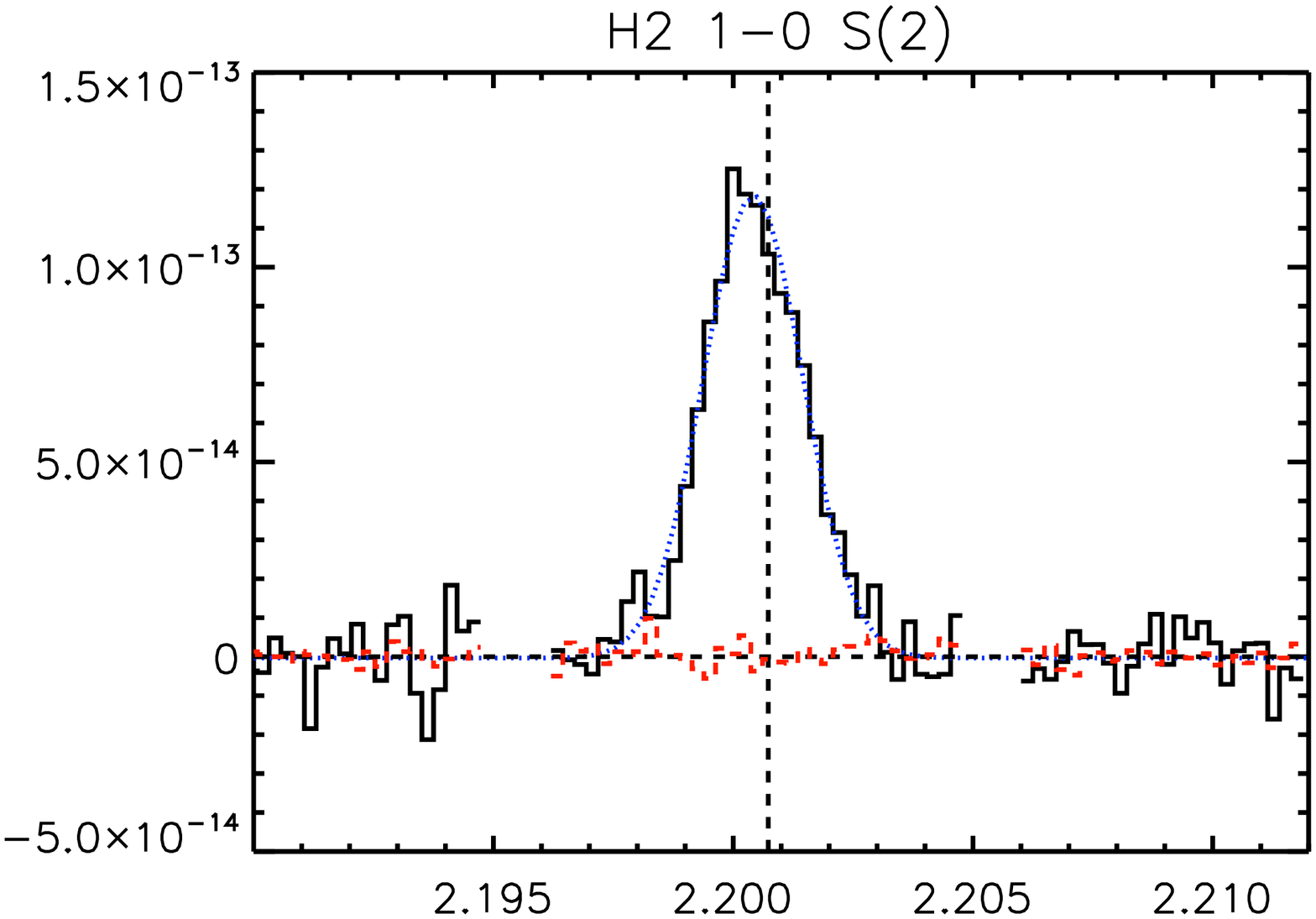}
    \includegraphics[width=0.17\textwidth, angle=90]{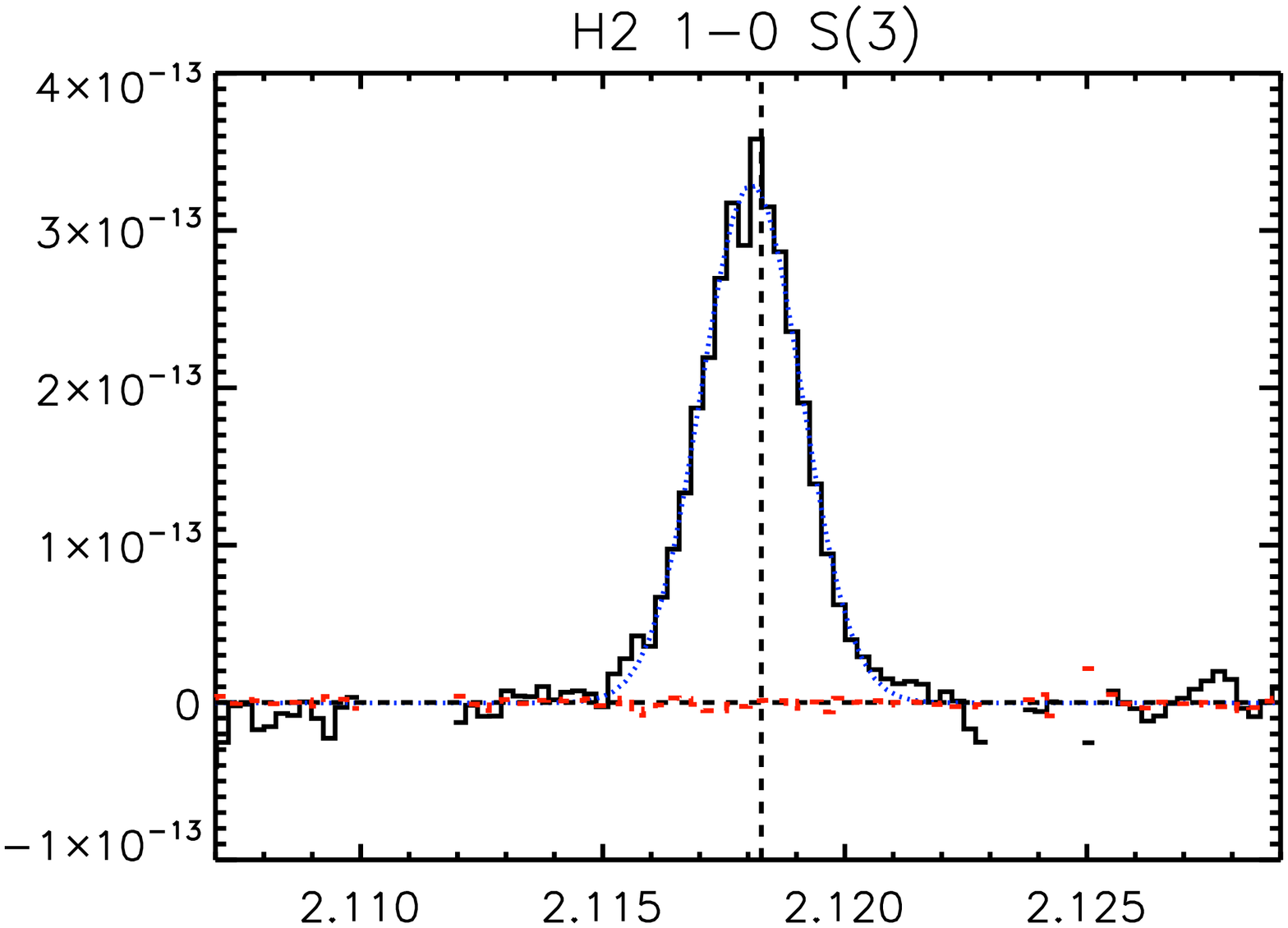}
    \includegraphics[width=0.17\textwidth, angle=90]{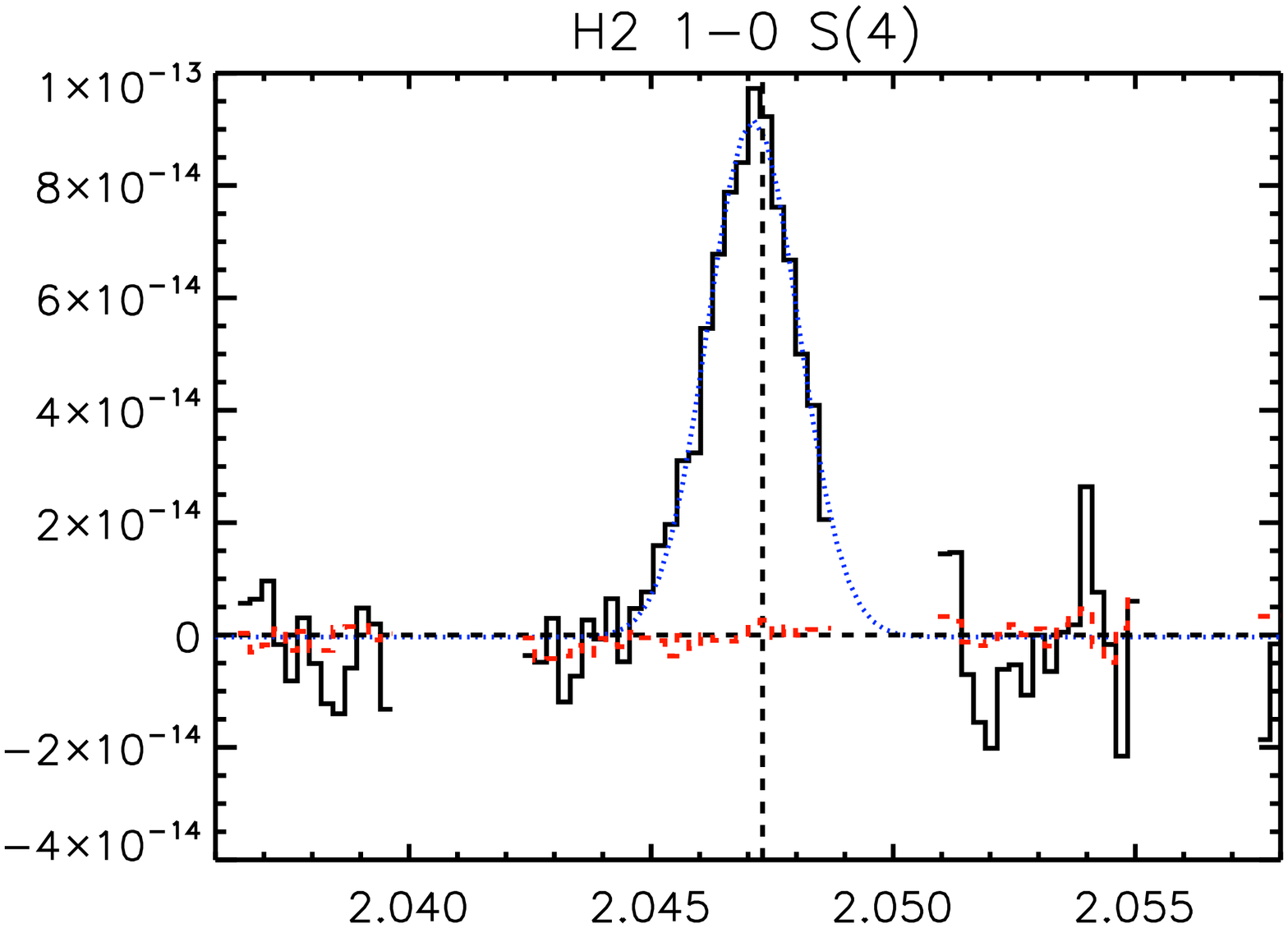}
    \includegraphics[width=0.17\textwidth, angle=90]{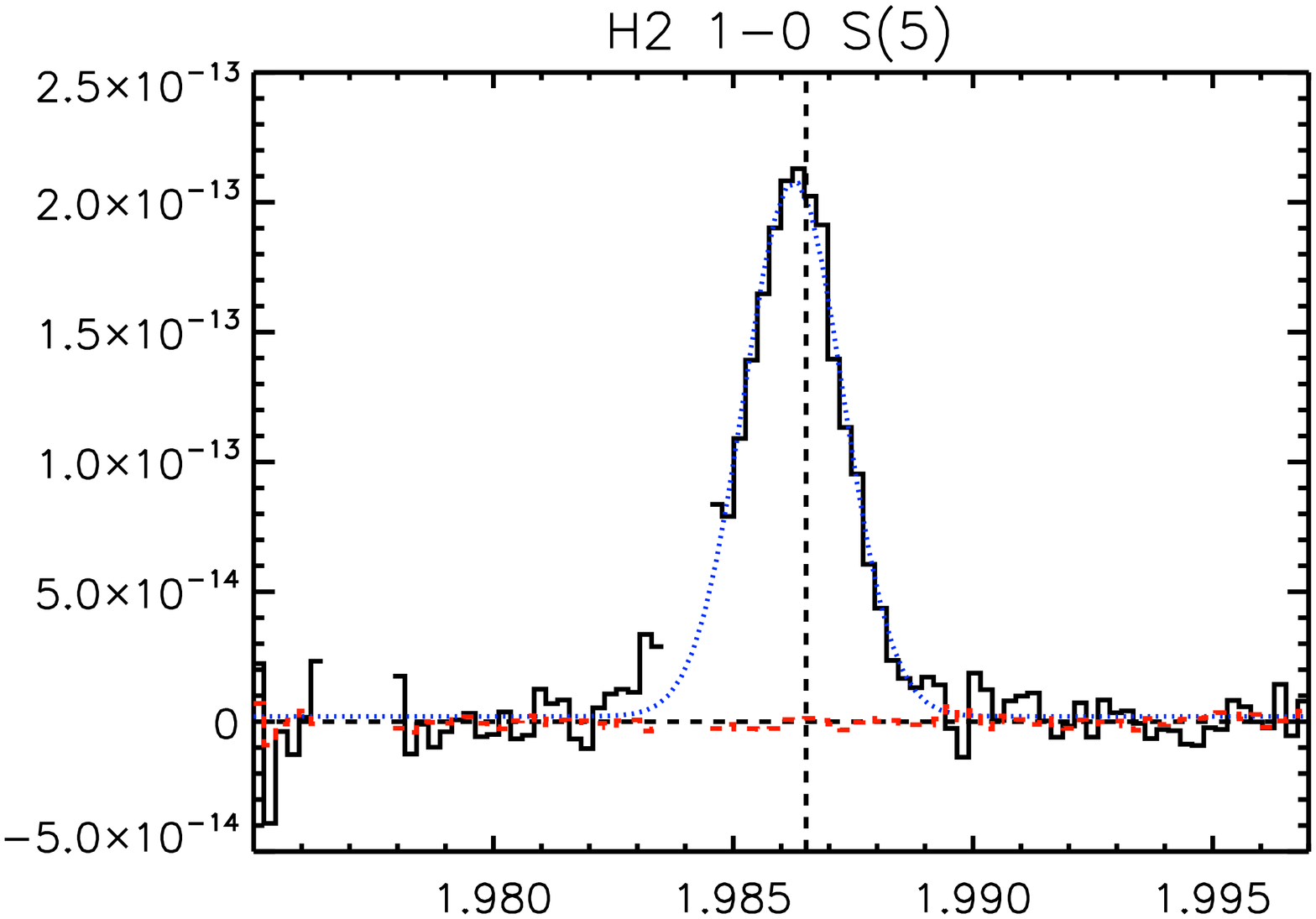}
    \includegraphics[width=0.17\textwidth, angle=90]{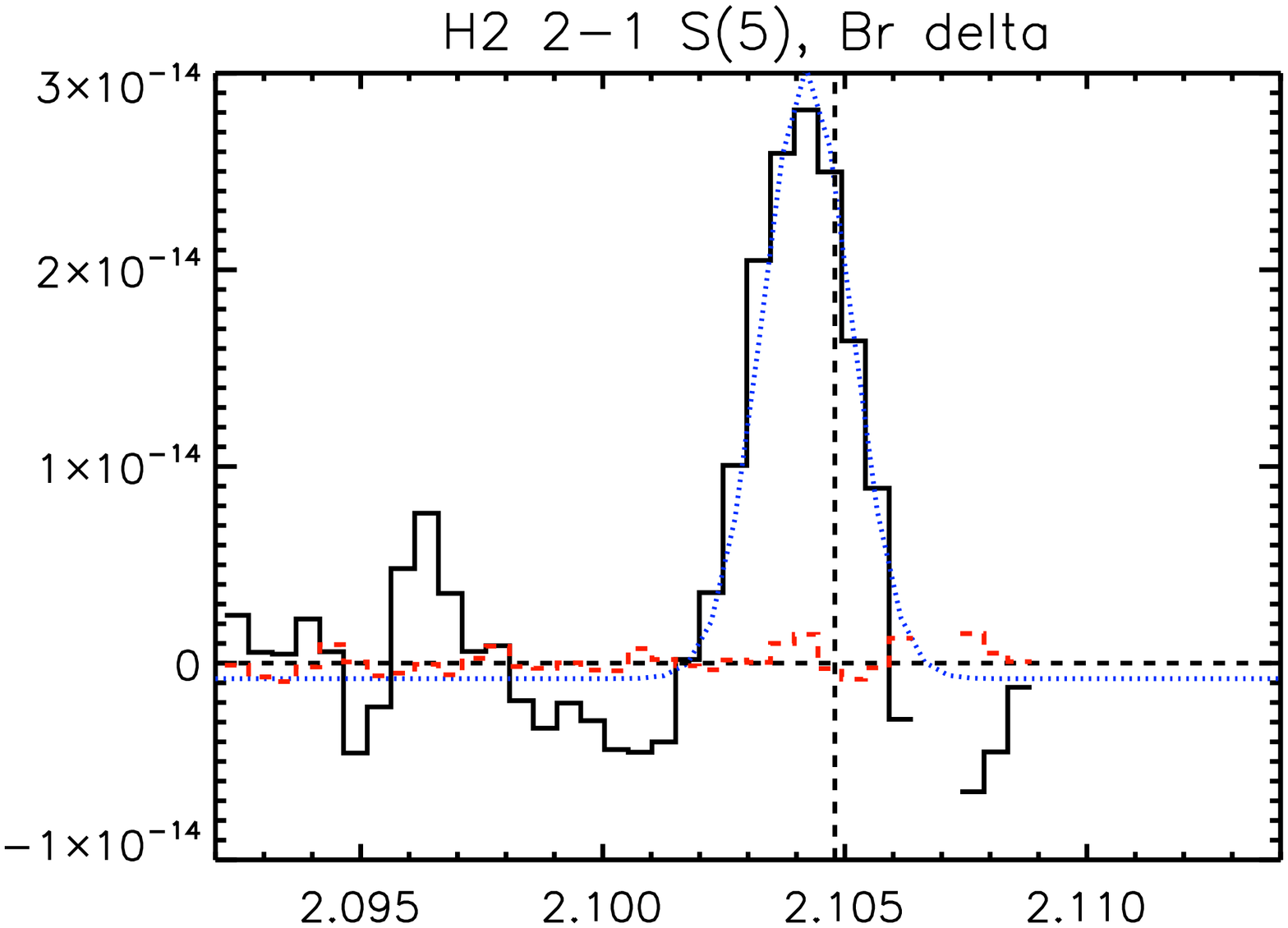}
    \includegraphics[width=0.17\textwidth, angle=90]{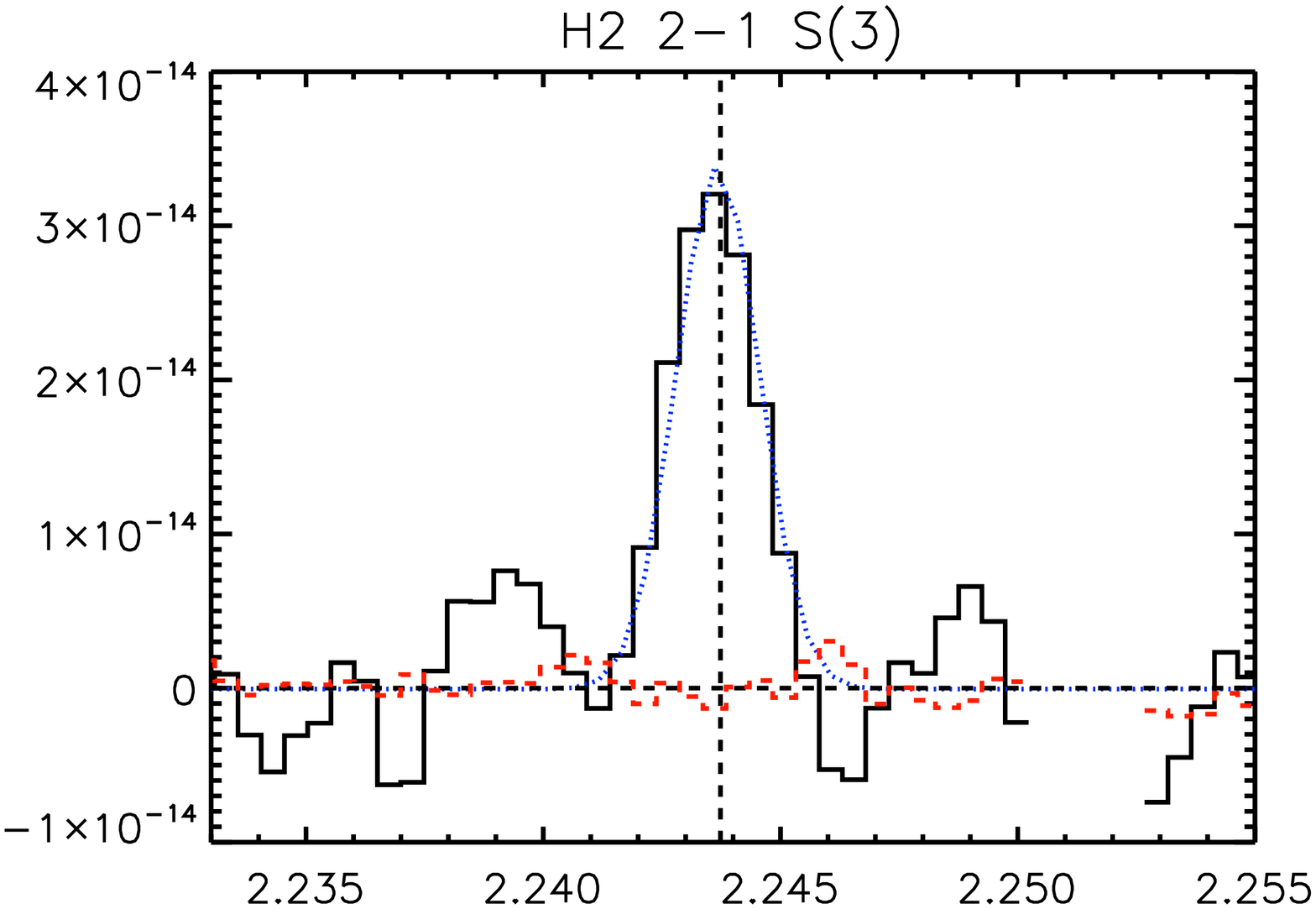}
    \includegraphics[width=0.17\textwidth, angle=90]{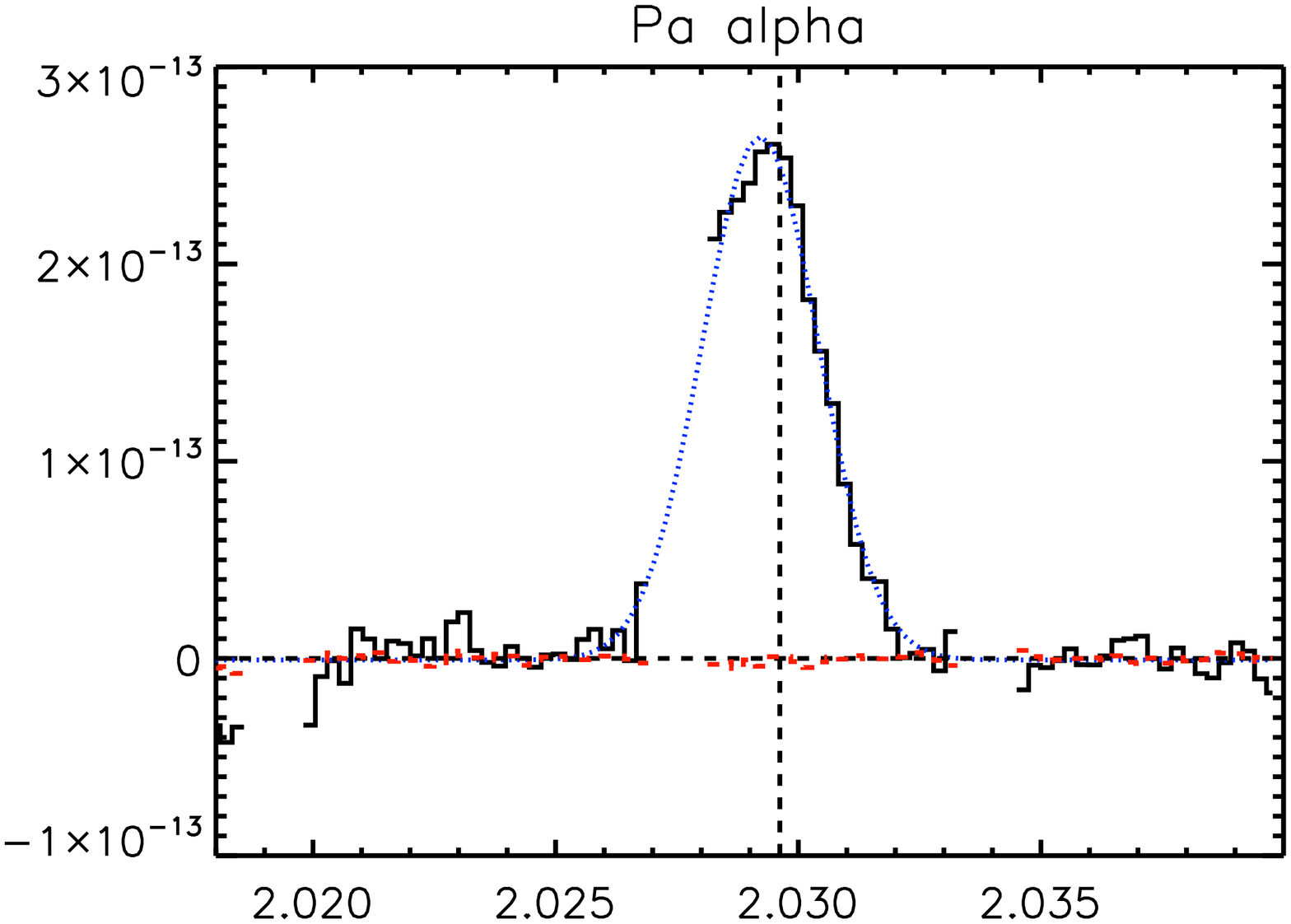}
    \includegraphics[width=0.17\textwidth, angle=90]{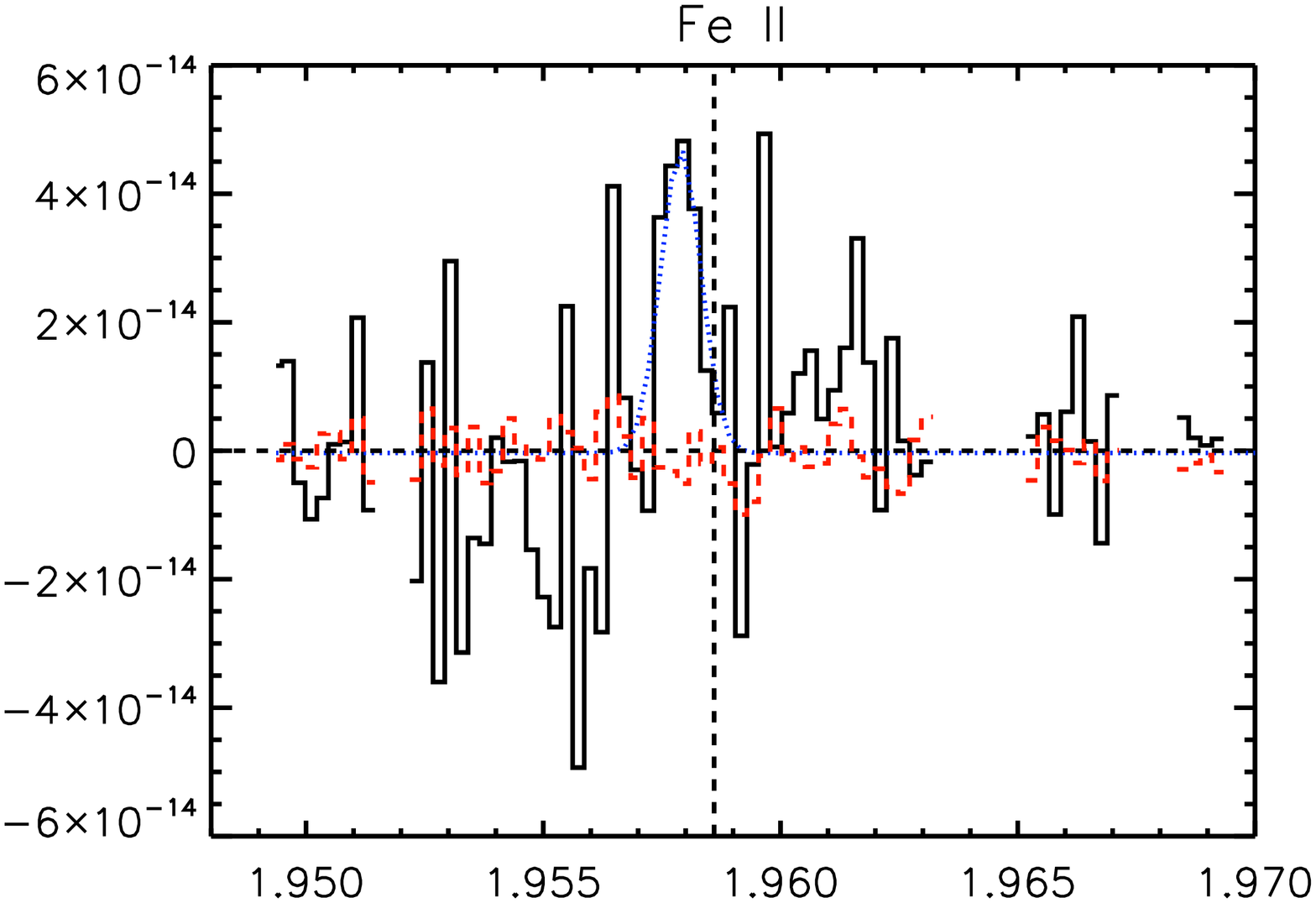}
    \includegraphics[width=0.17\textwidth, angle=90]{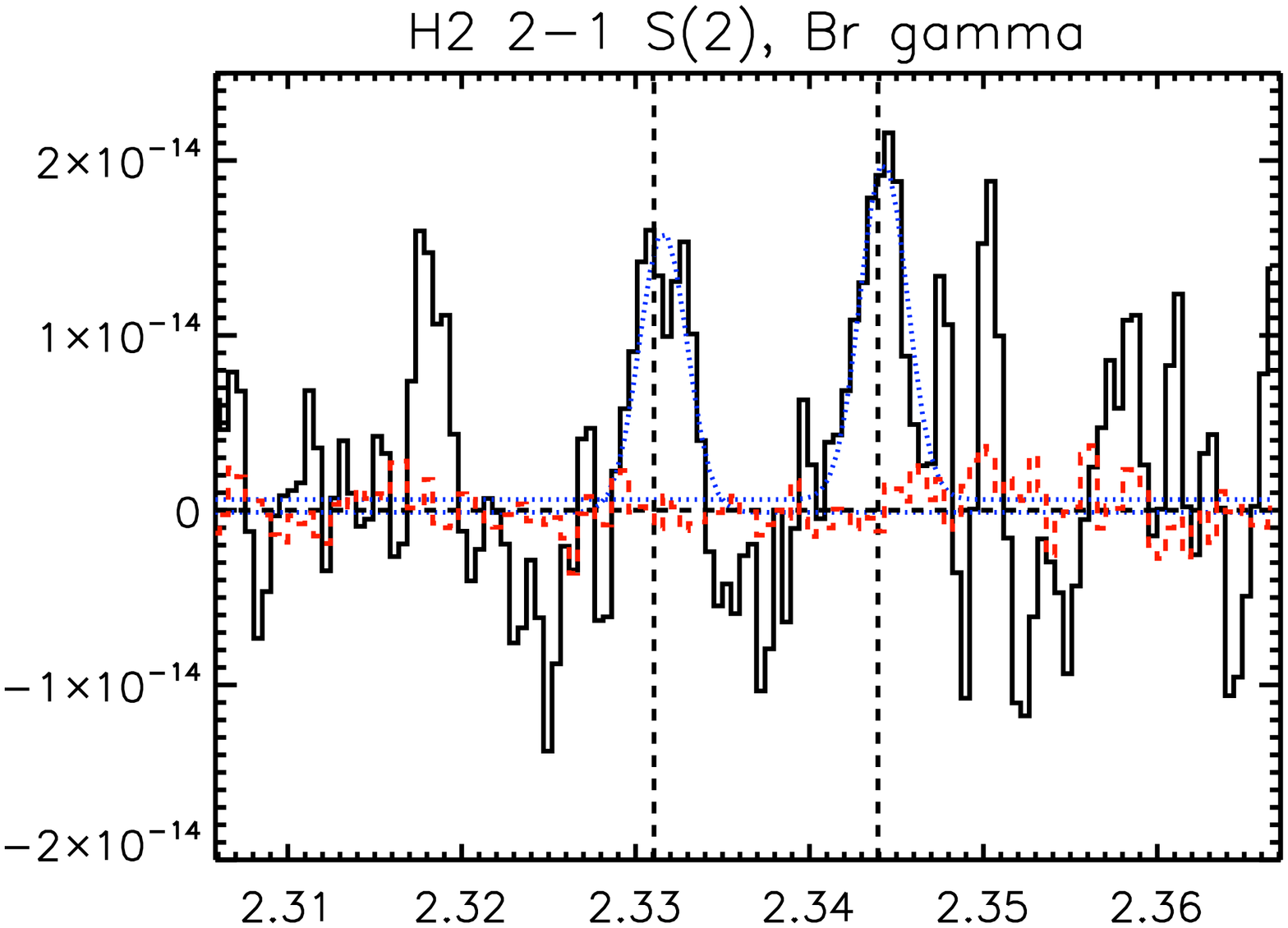}
  \caption{ABELL 2597 Line Spectra (Region A2). For H$_{\mathrm{2}}$~1-0~S(0), 2-1~S(3), 2-1~S(2), Br~$\gamma$, Br~$\delta$ we used a four pixel spatial and spectral smoothing. For all other lines a two pixel spatial smoothing was used. The lines and symbols used are the same as in Fig. \ref{fig_a2597_line_tb_a1}.}
\end{figure*}

\begin{figure*}
    \includegraphics[width=0.17\textwidth, angle=90]{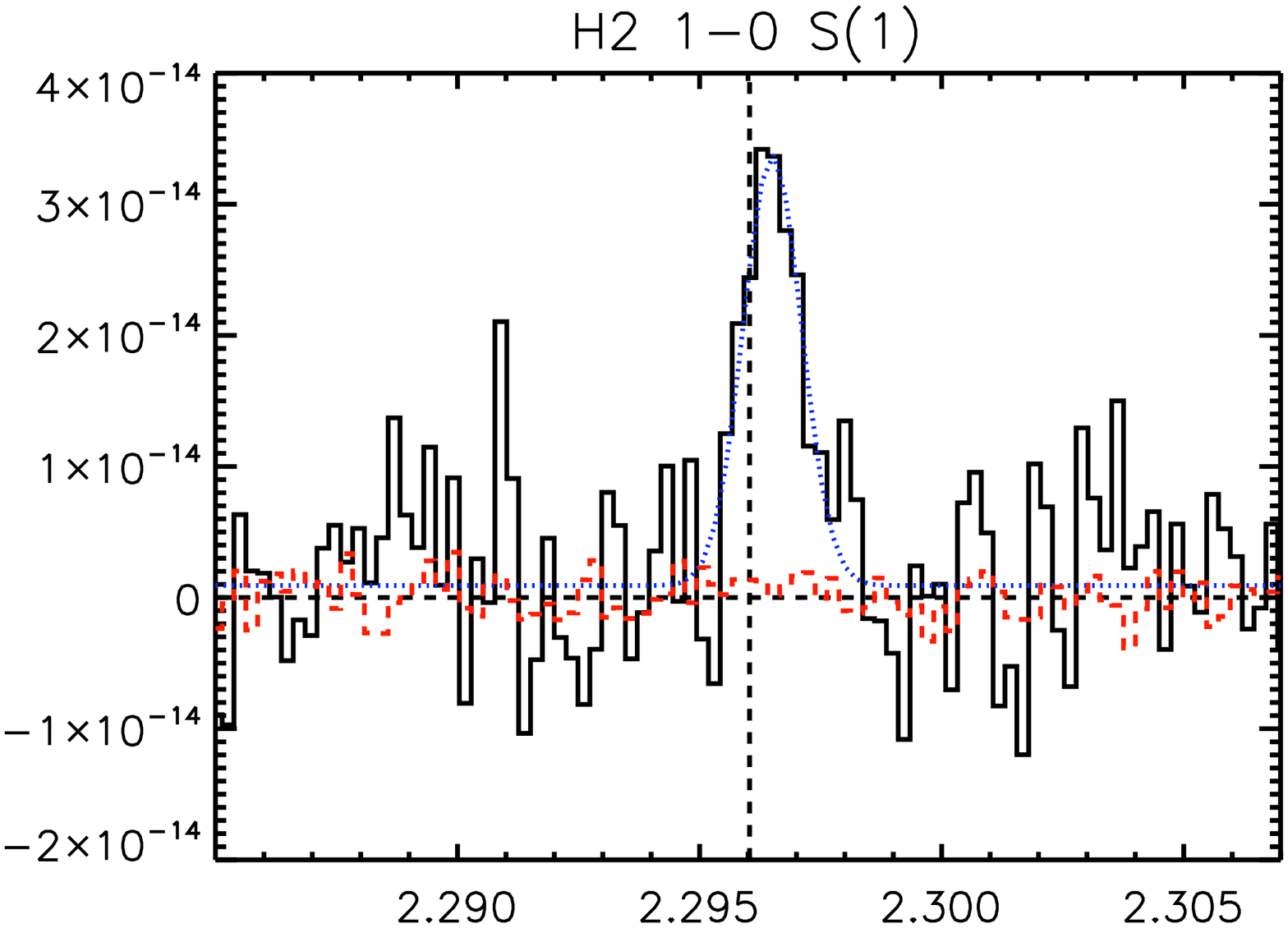}
    \includegraphics[width=0.17\textwidth, angle=90]{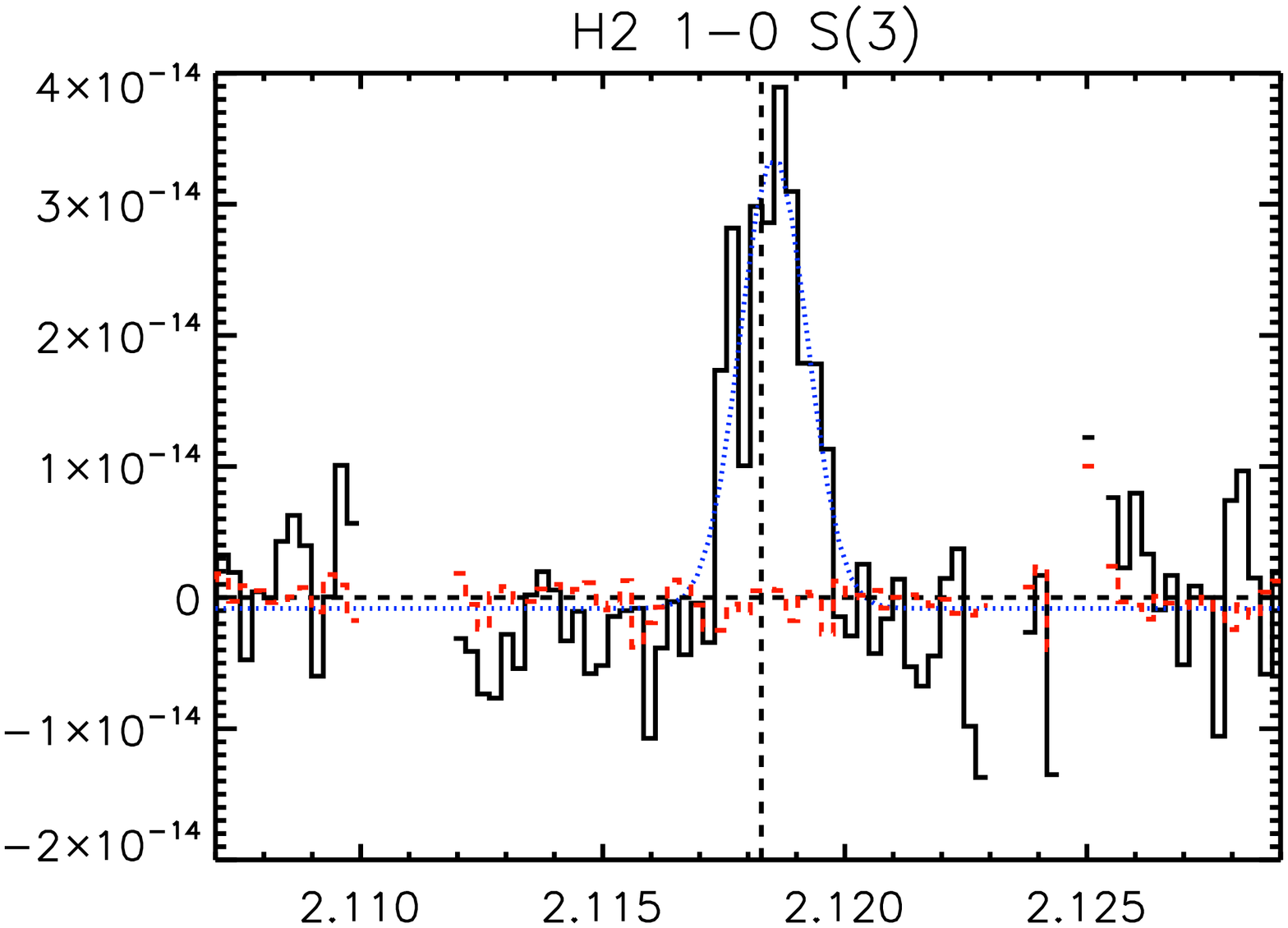}
    \includegraphics[width=0.17\textwidth, angle=90]{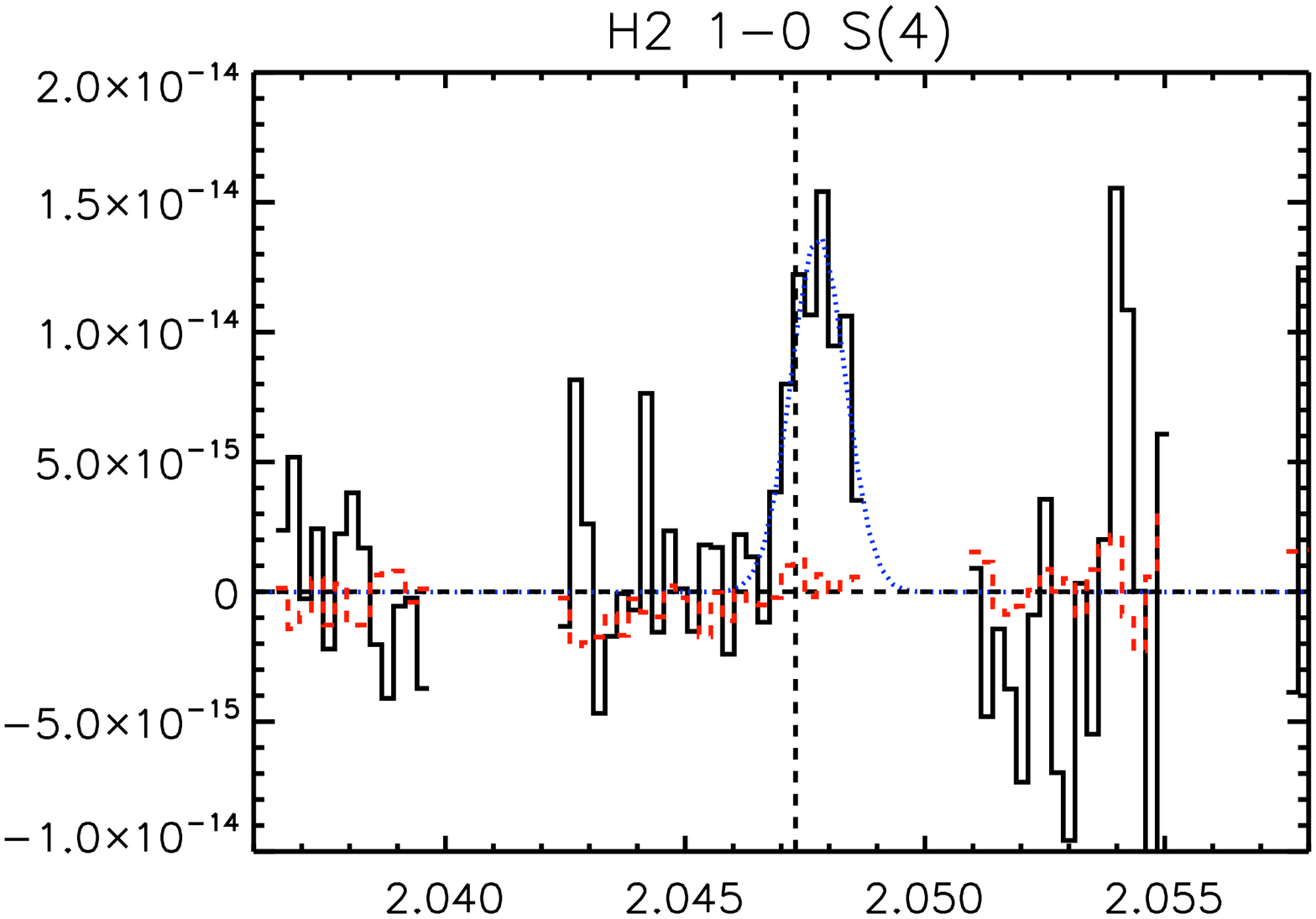}
    \includegraphics[width=0.17\textwidth, angle=90]{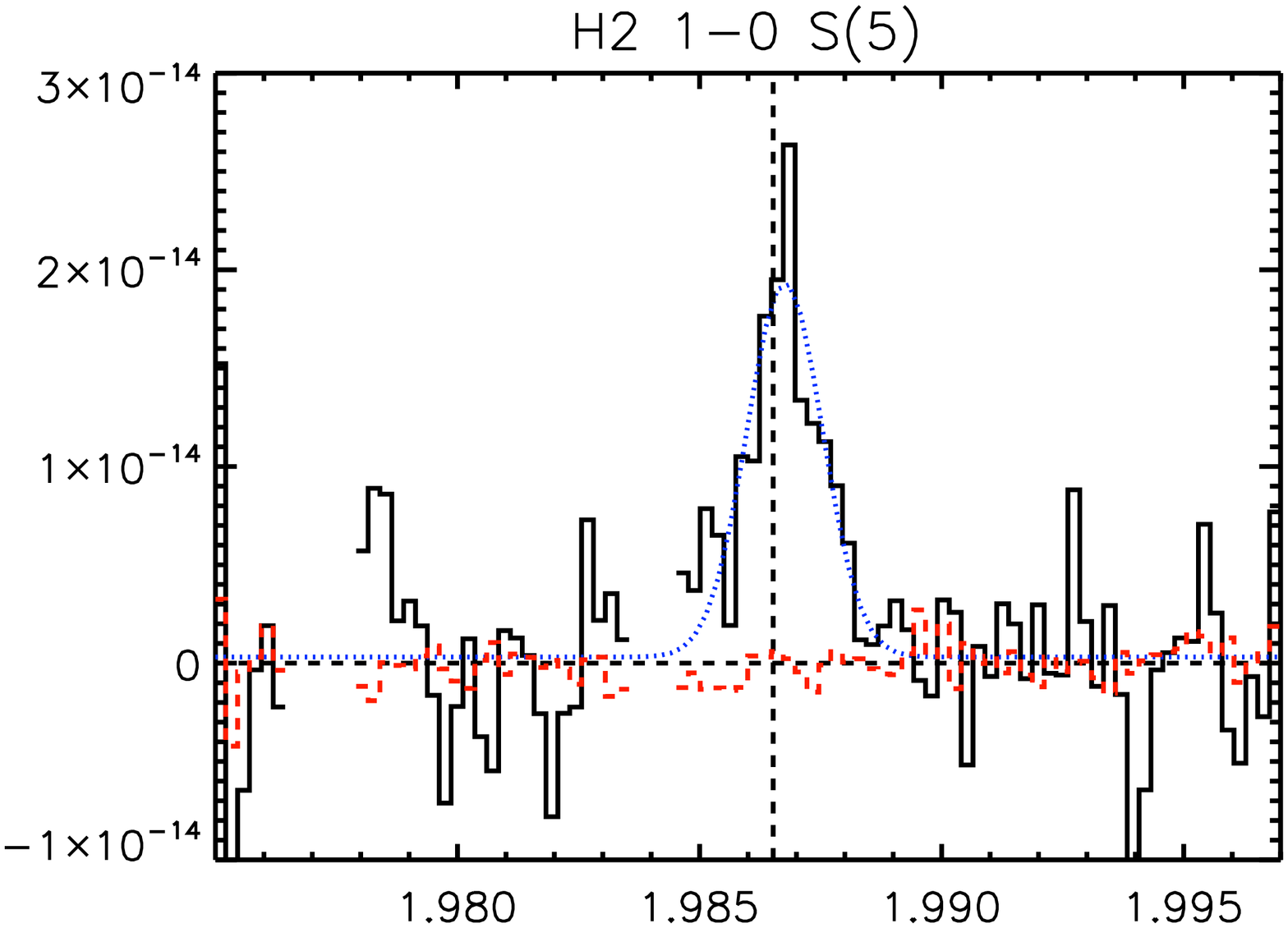}
    \includegraphics[width=0.17\textwidth, angle=90]{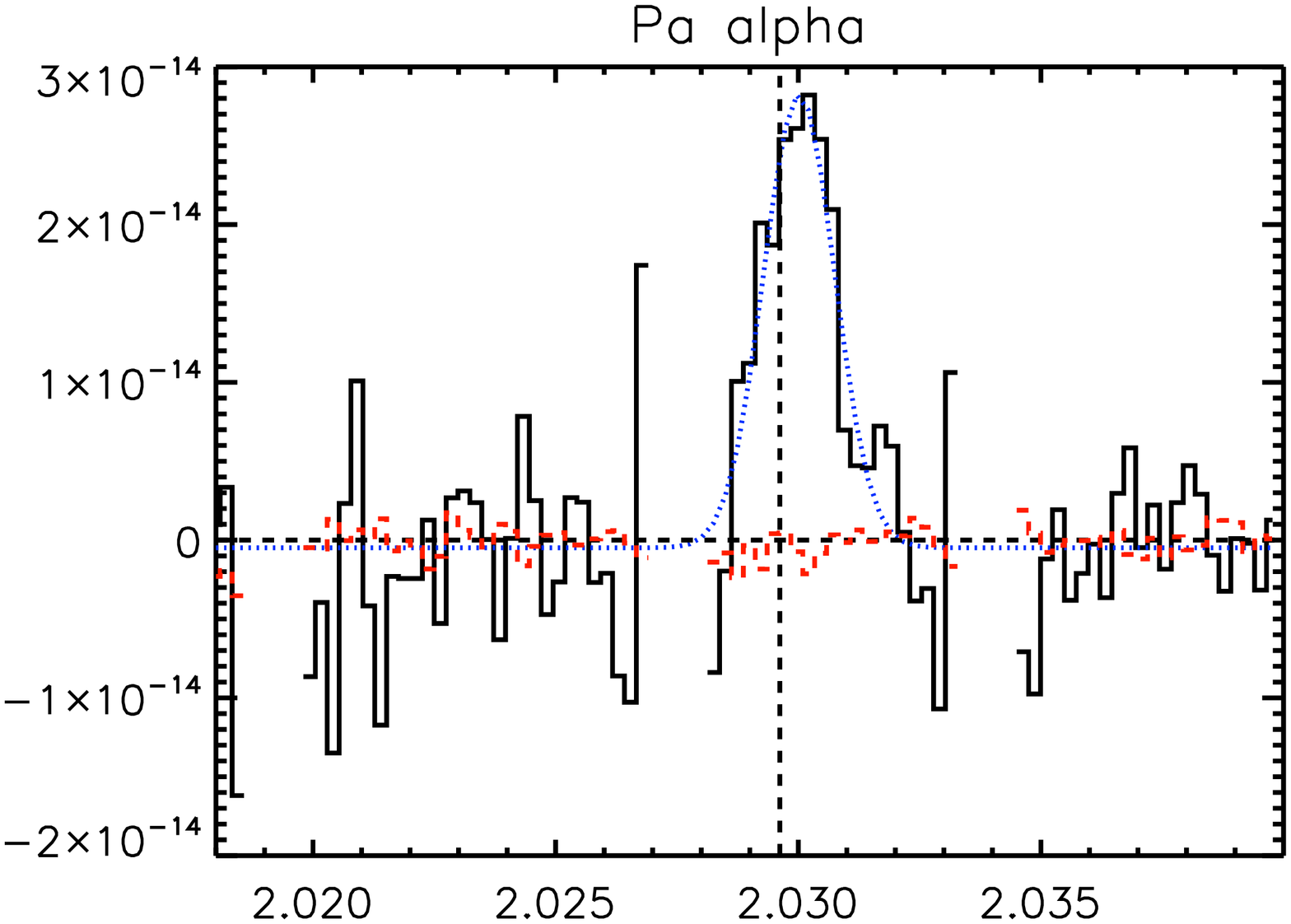}
  \caption{ABELL 2597 Line Spectra (Region A3). For all lines we used a two pixel spatial smoothing. The lines and symbols used are the same as in Fig. \ref{fig_a2597_line_tb_a1}.}
\end{figure*}

\begin{figure*}
    \includegraphics[width=0.17\textwidth, angle=90]{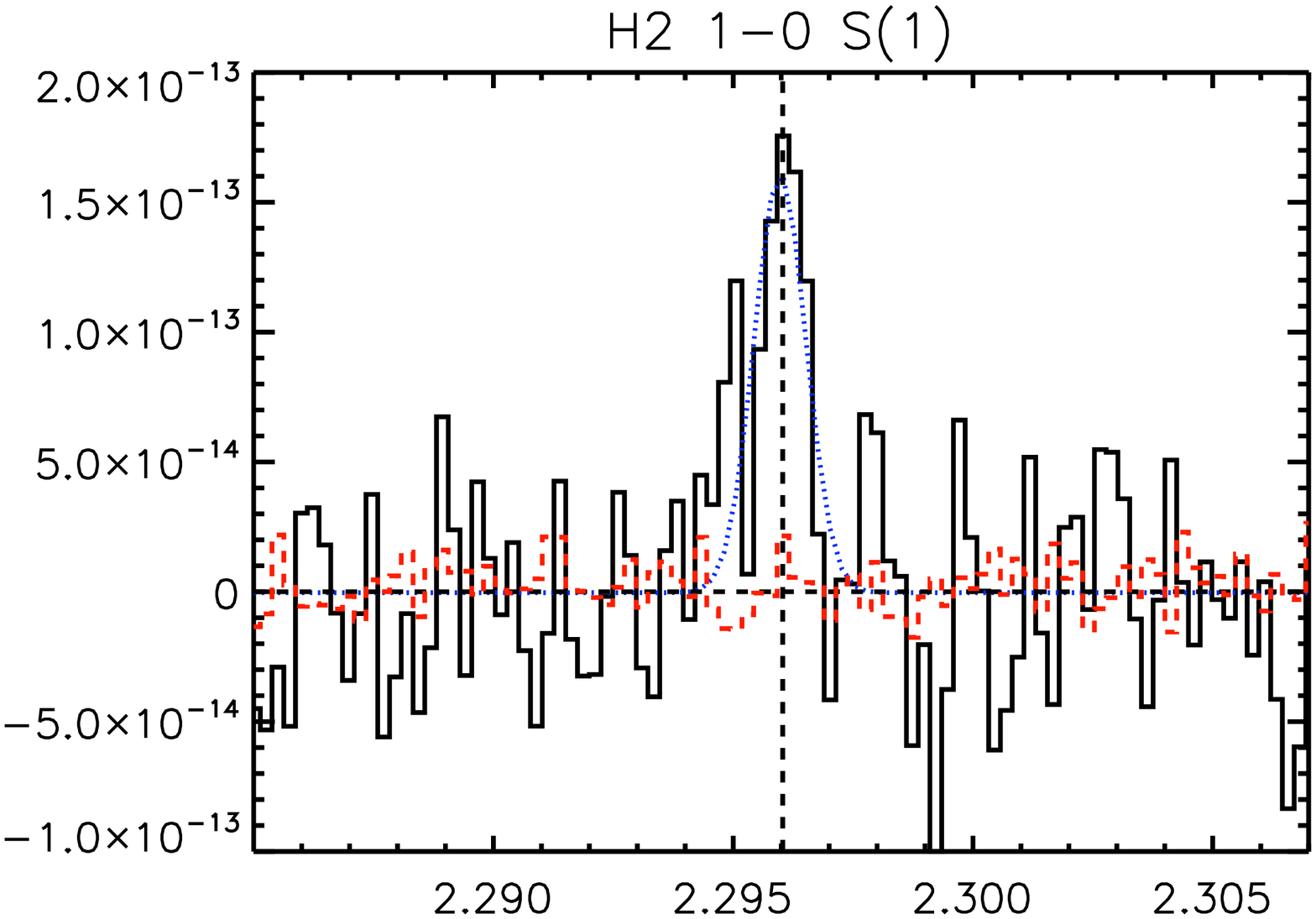}
    \includegraphics[width=0.17\textwidth, angle=90]{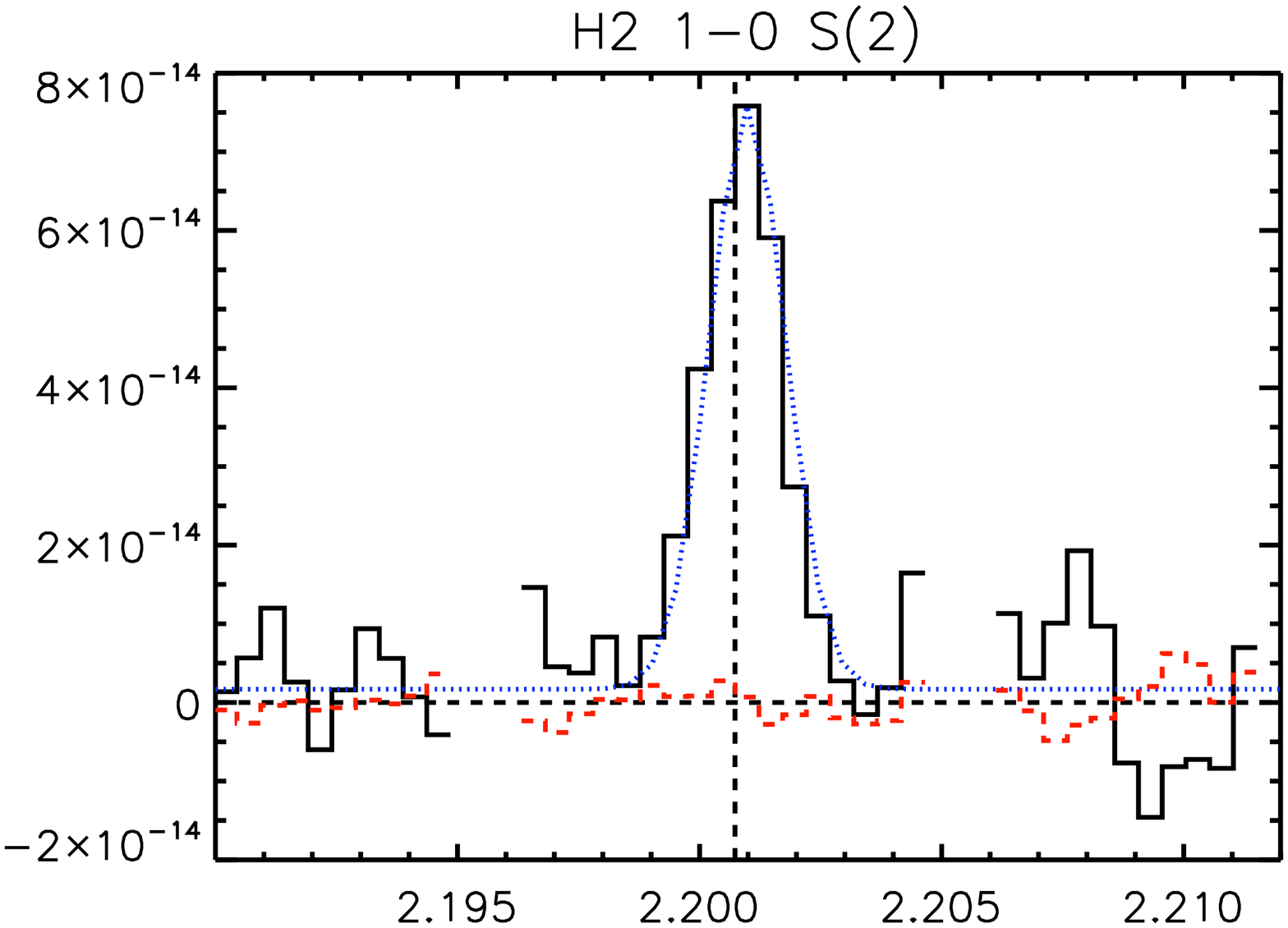}
    \includegraphics[width=0.17\textwidth, angle=90]{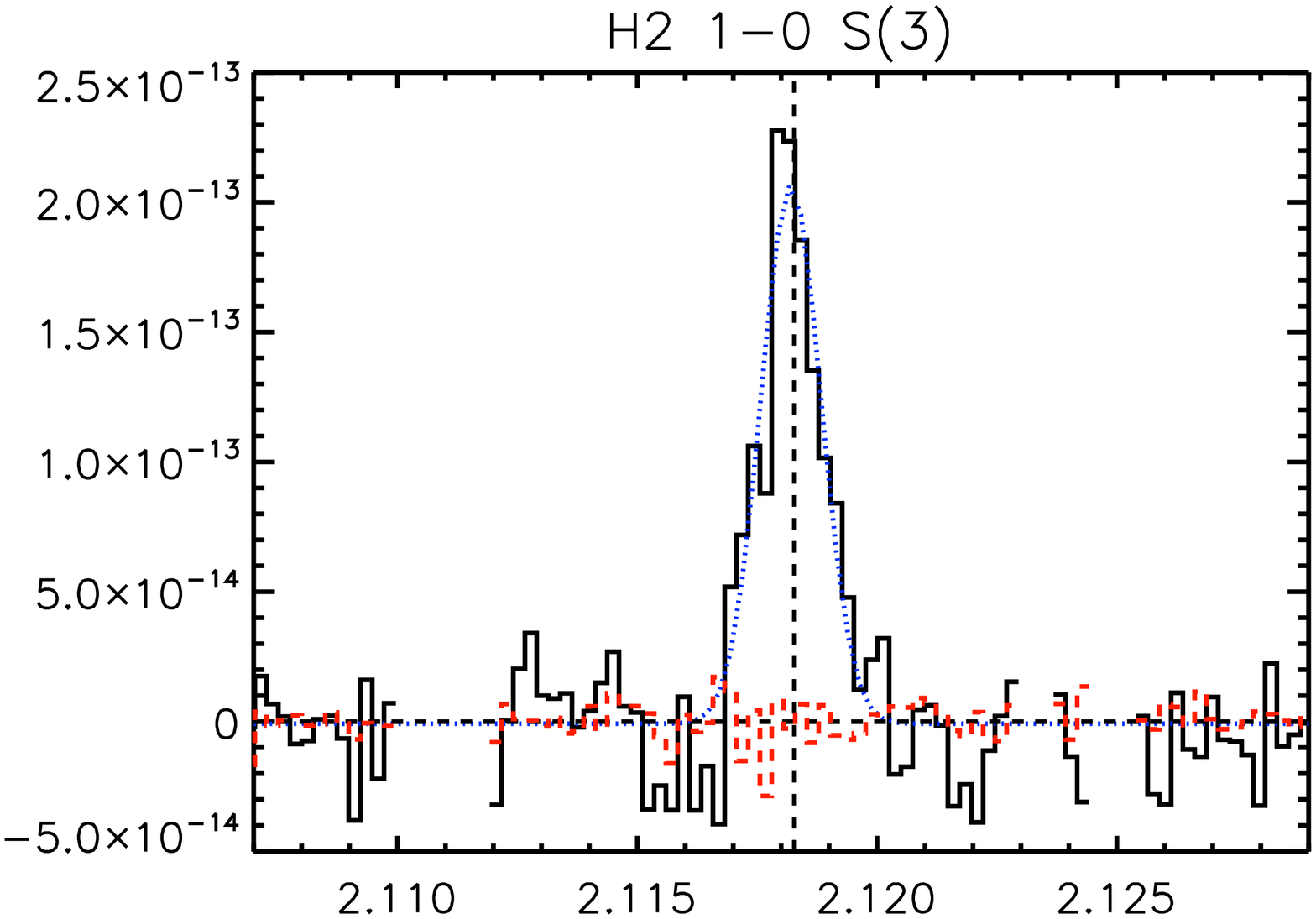}
    \includegraphics[width=0.17\textwidth, angle=90]{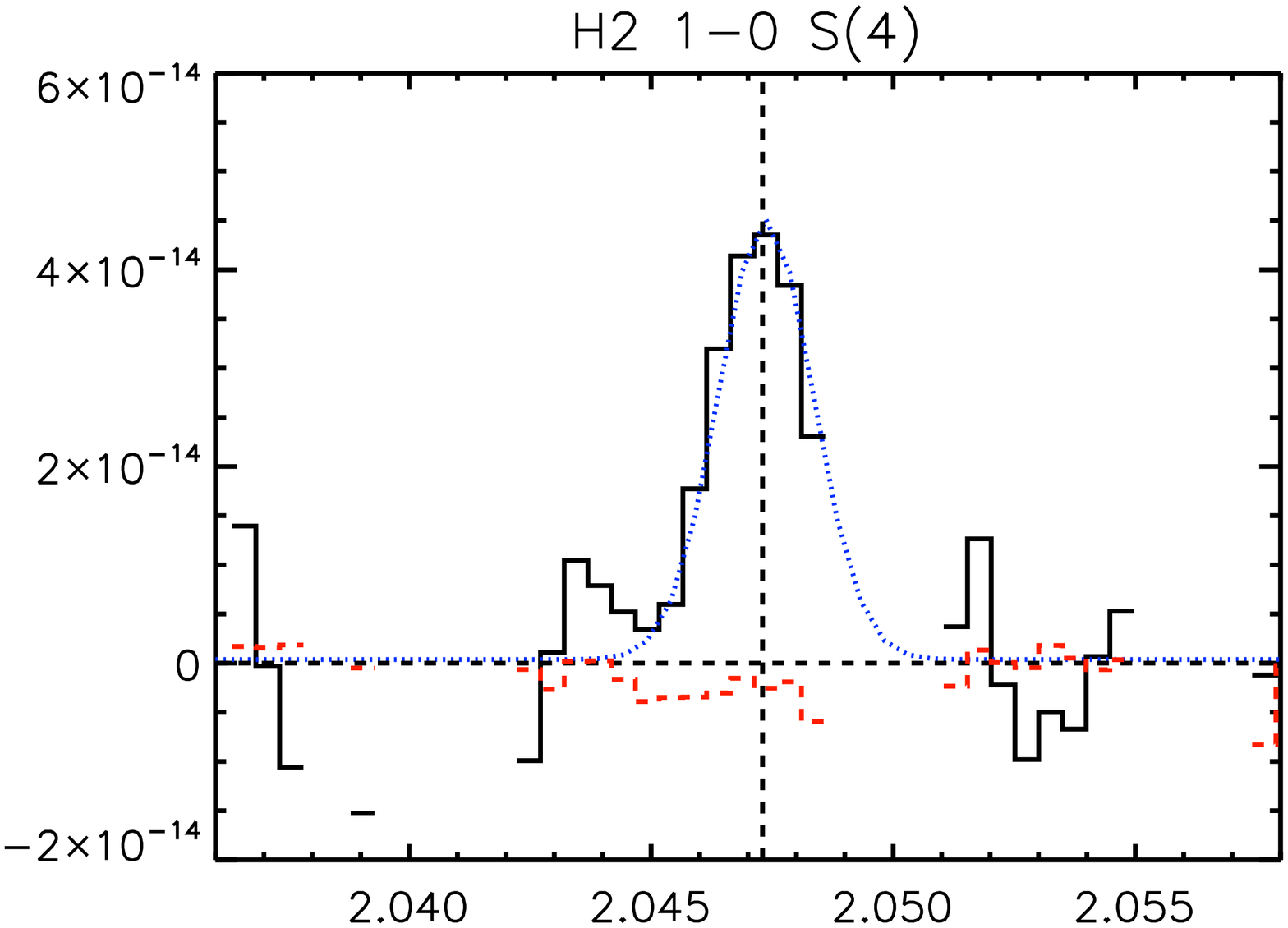}
    \includegraphics[width=0.17\textwidth, angle=90]{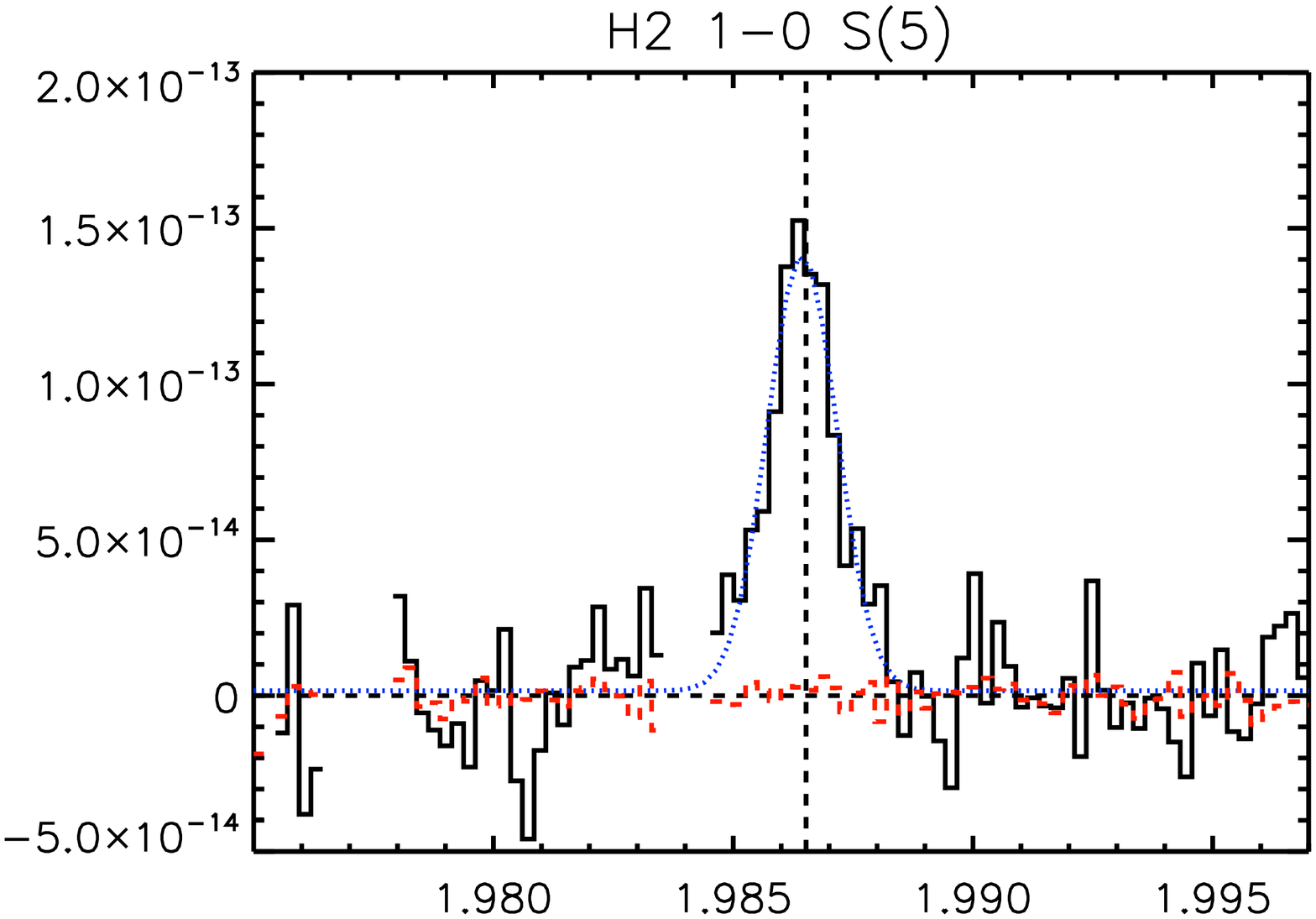}
    \includegraphics[width=0.17\textwidth, angle=90]{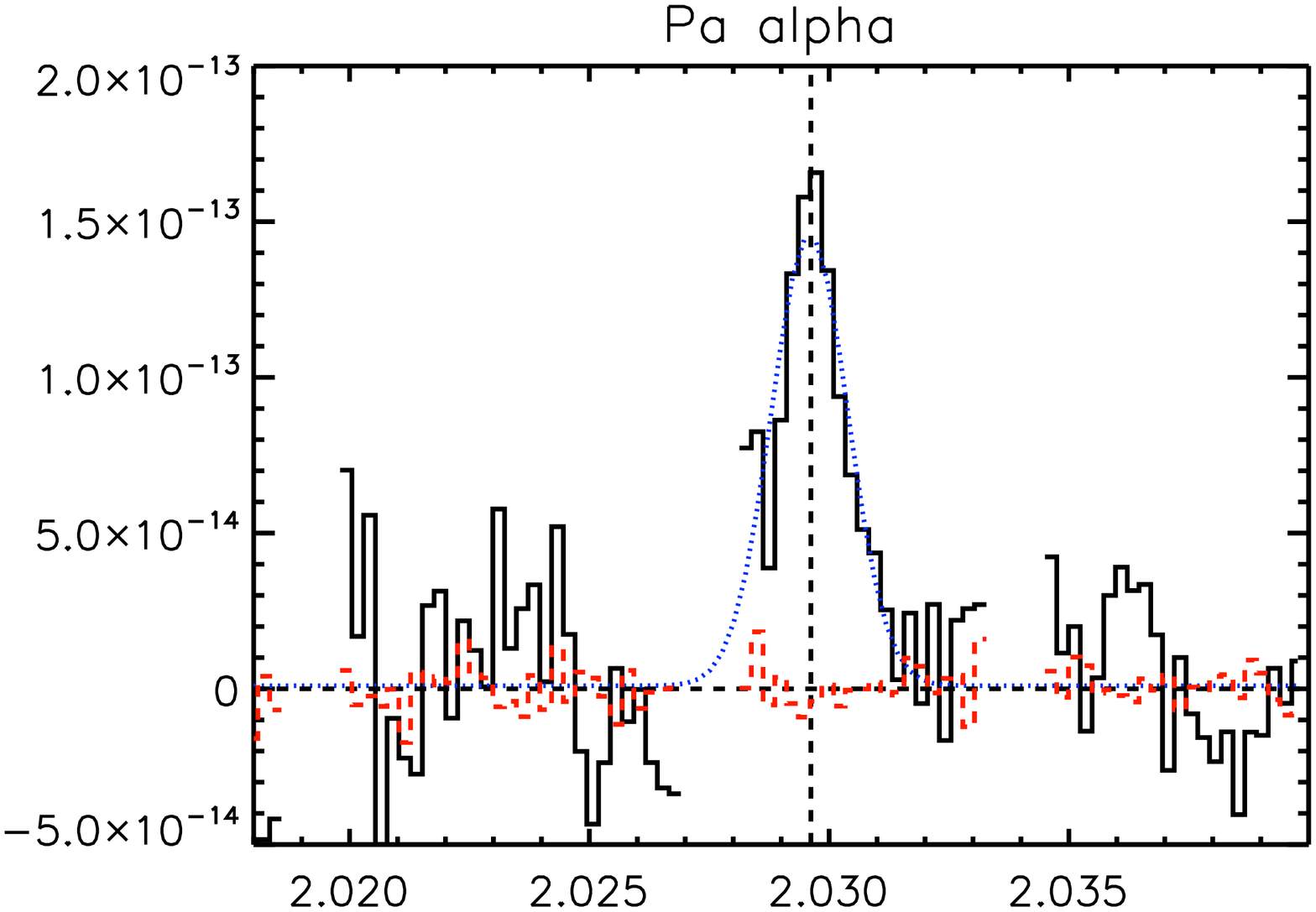}
  \caption{ABELL 2597 Line Spectra (Region A4). For H$_{\mathrm{2}}$~1-0~S(2) and 1-0~S(4) we used a four pixel spatial and spectral smoothing. For all other lines a two pixel spatial smoothing is used. The lines and symbols used are the same as in Fig. \ref{fig_a2597_line_tb_a1}.}
\end{figure*}

\begin{figure*}
    \includegraphics[width=0.17\textwidth, angle=90]{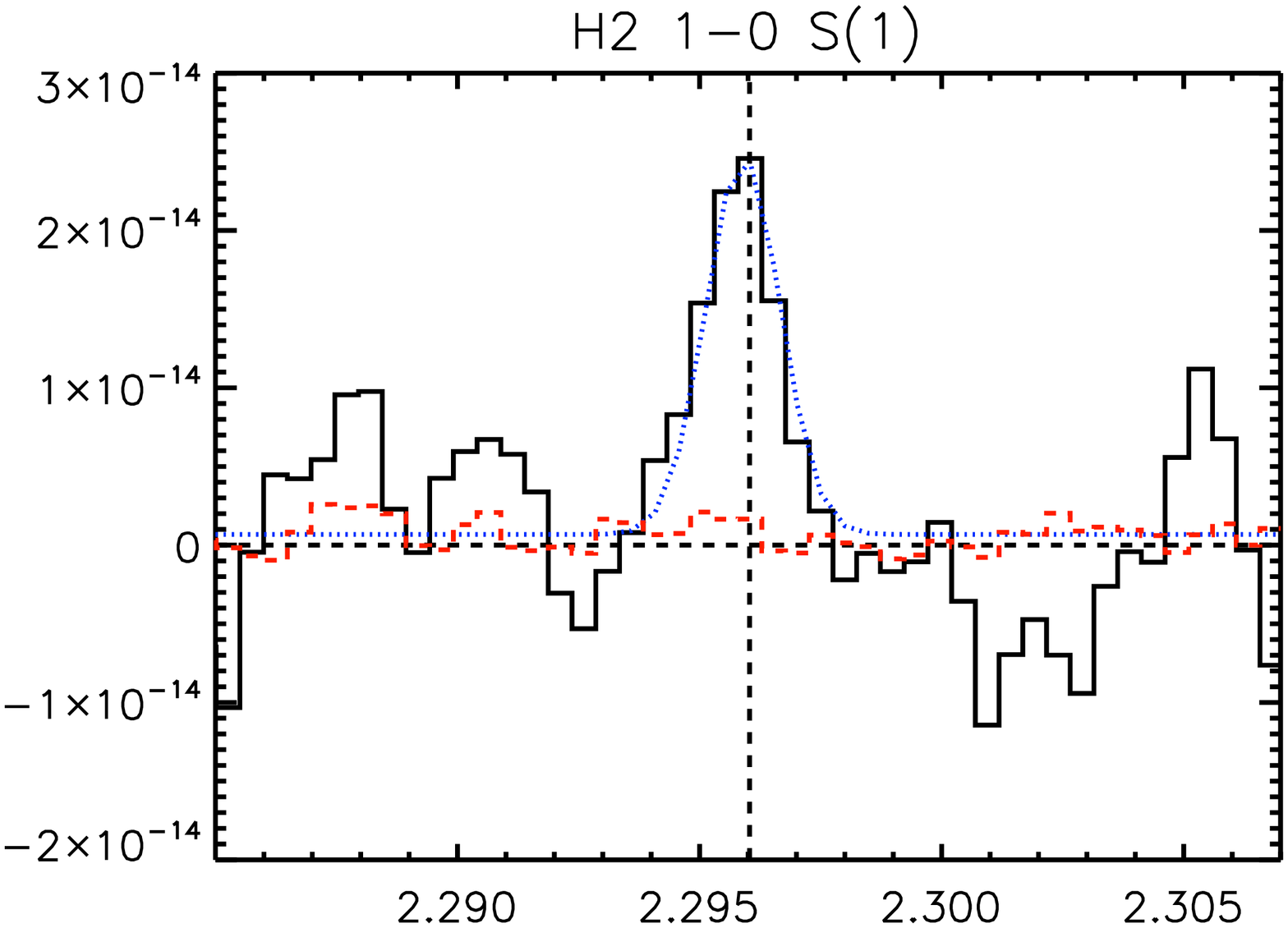}
    \includegraphics[width=0.17\textwidth, angle=90]{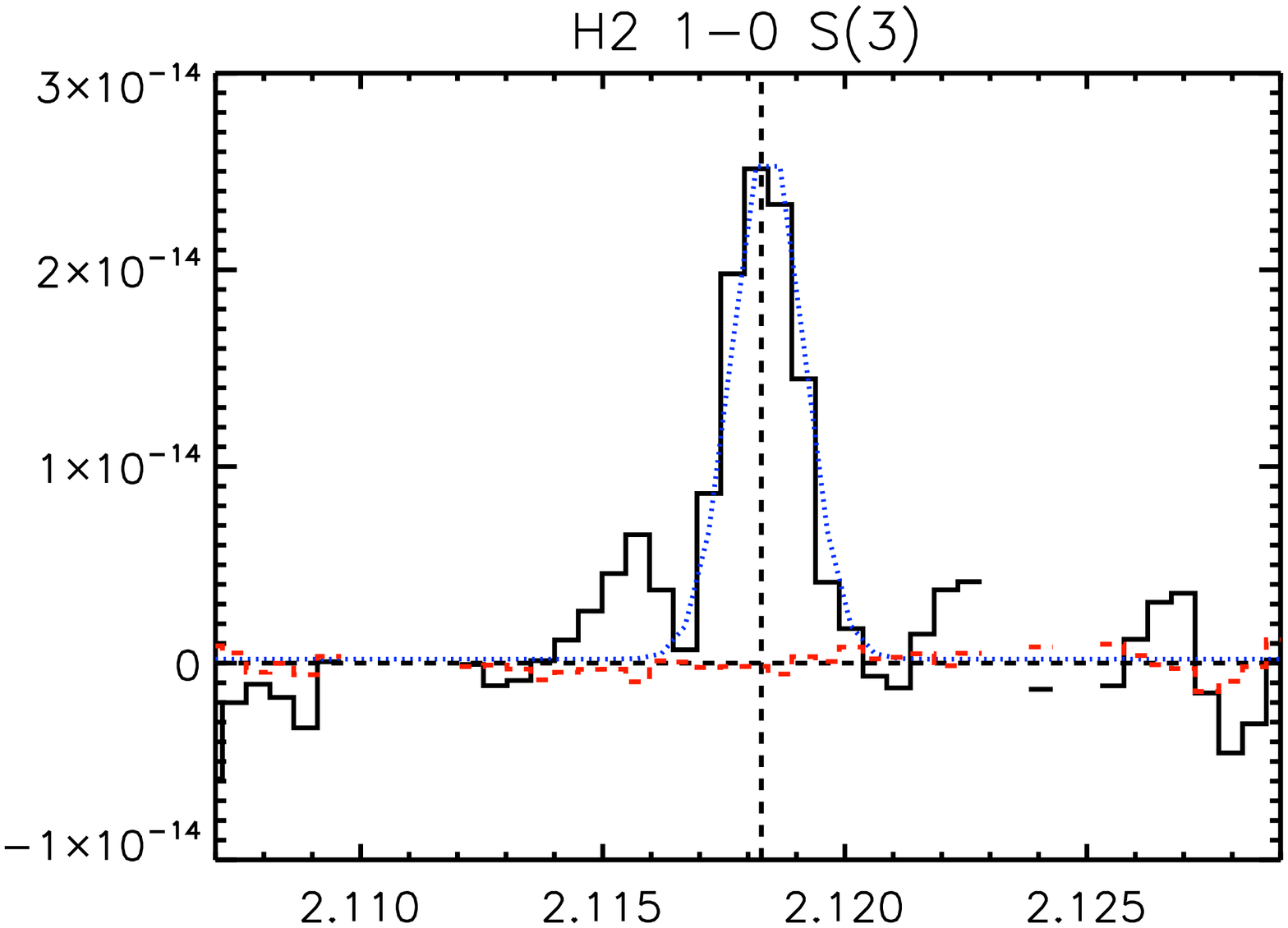}
    \includegraphics[width=0.17\textwidth, angle=90]{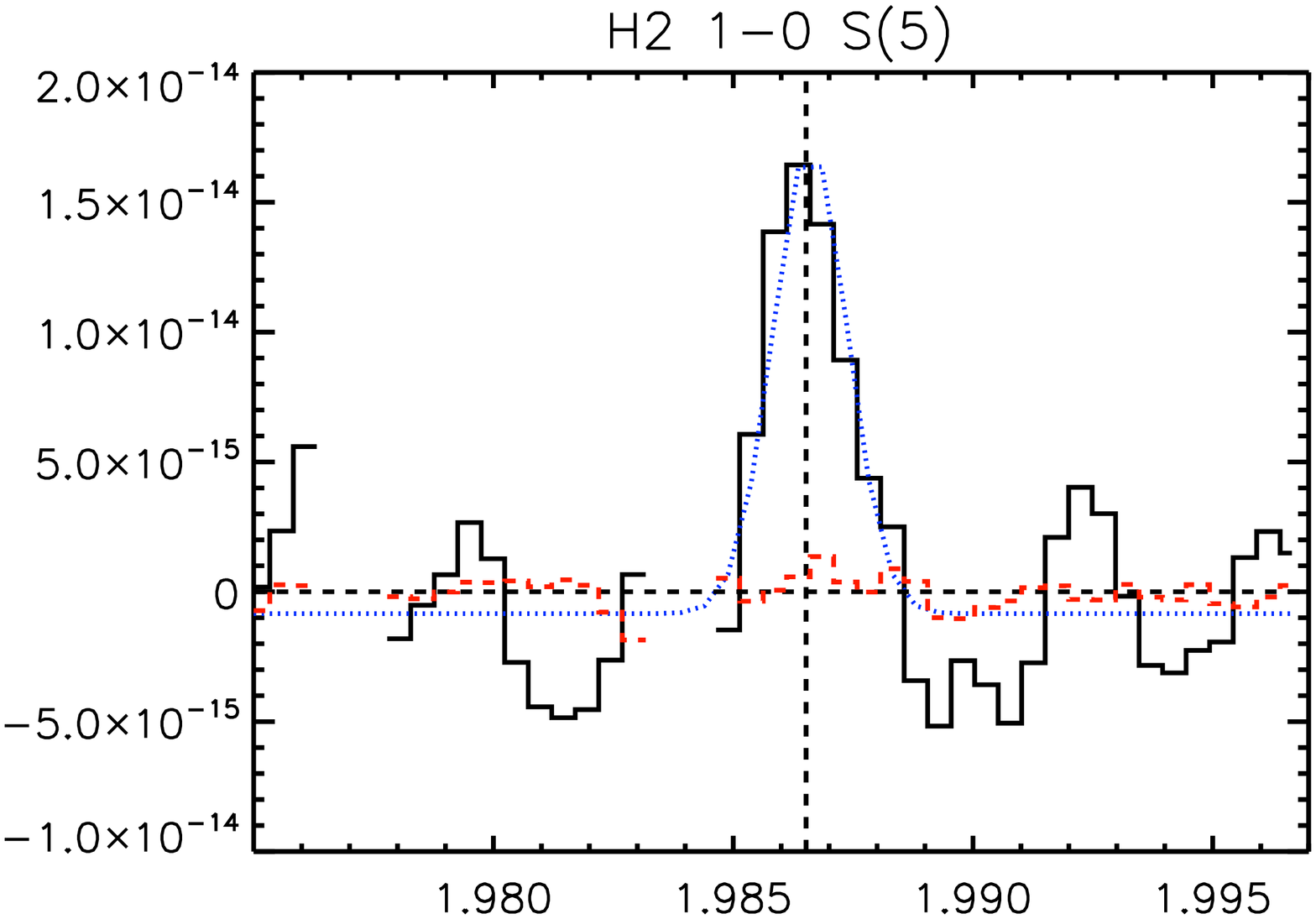}
    \includegraphics[width=0.17\textwidth, angle=90]{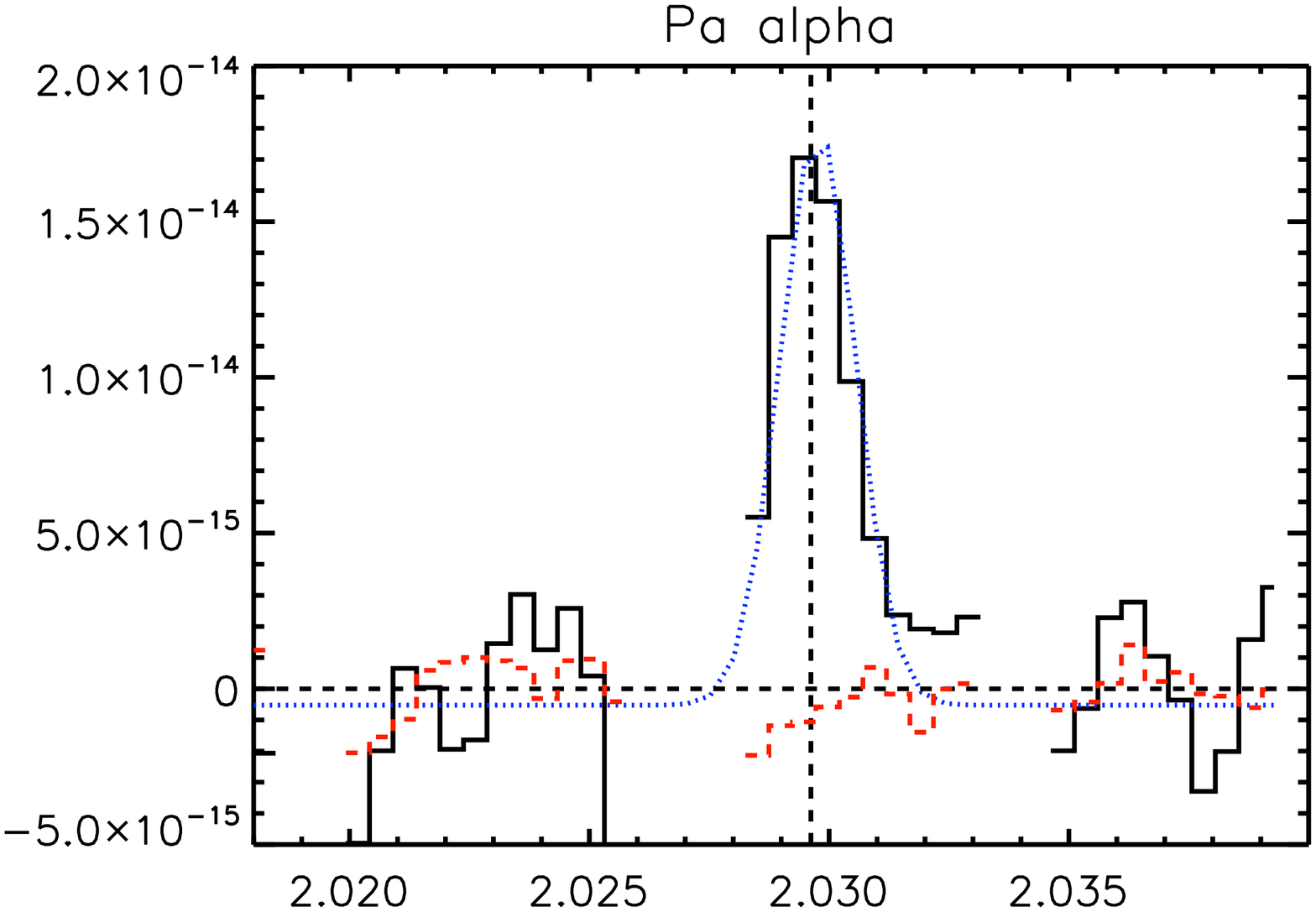}
  \caption{ABELL 2597 Line Spectra (Region A5). For all lines we used a four pixel spatial and spectral smoothing. The lines and symbols used are the same as in Fig. \ref{fig_a2597_line_tb_a1}.}
\end{figure*}

\clearpage

\begin{figure*}
    \includegraphics[width=0.17\textwidth, angle=90]{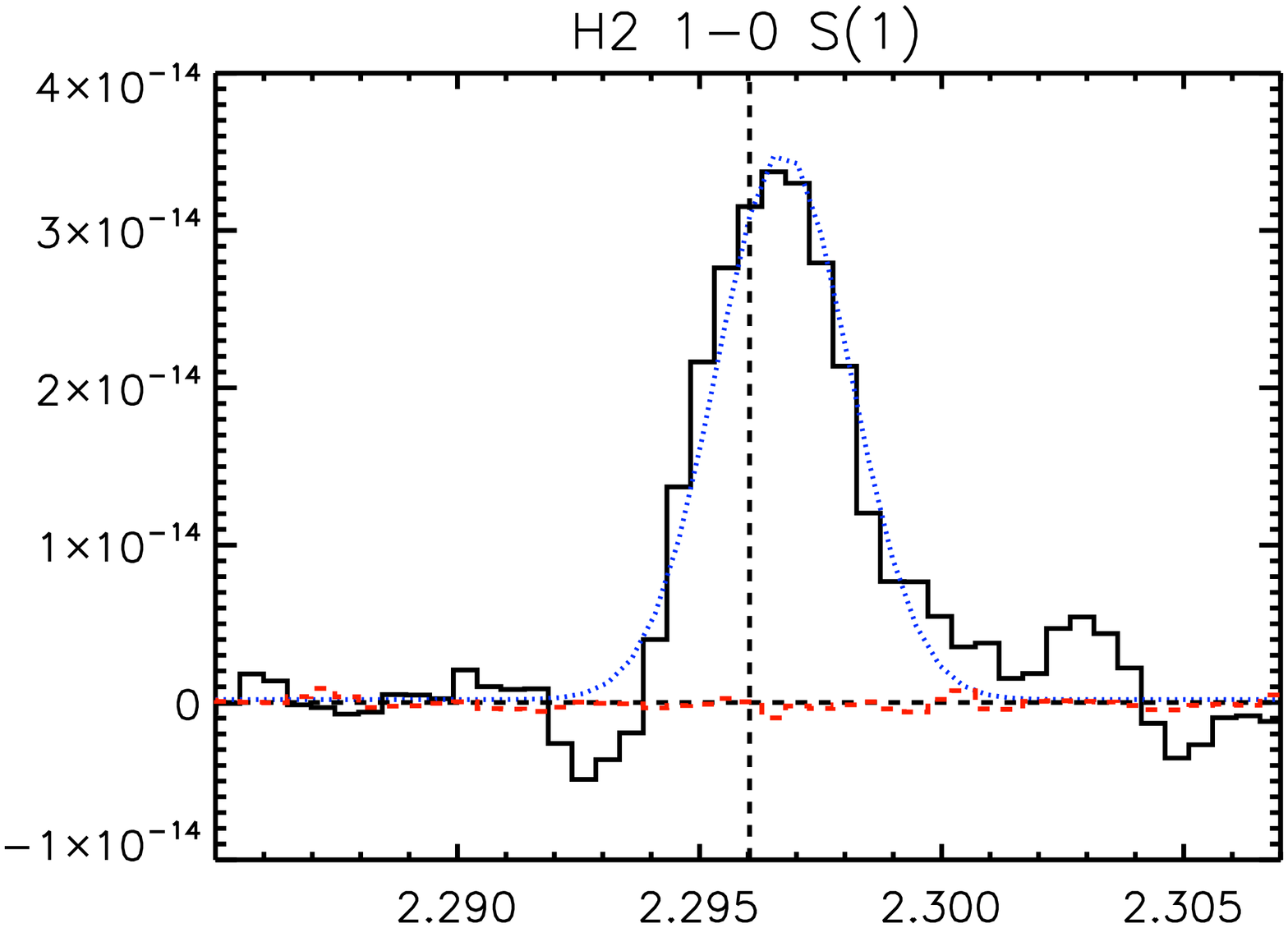}
    \includegraphics[width=0.17\textwidth, angle=90]{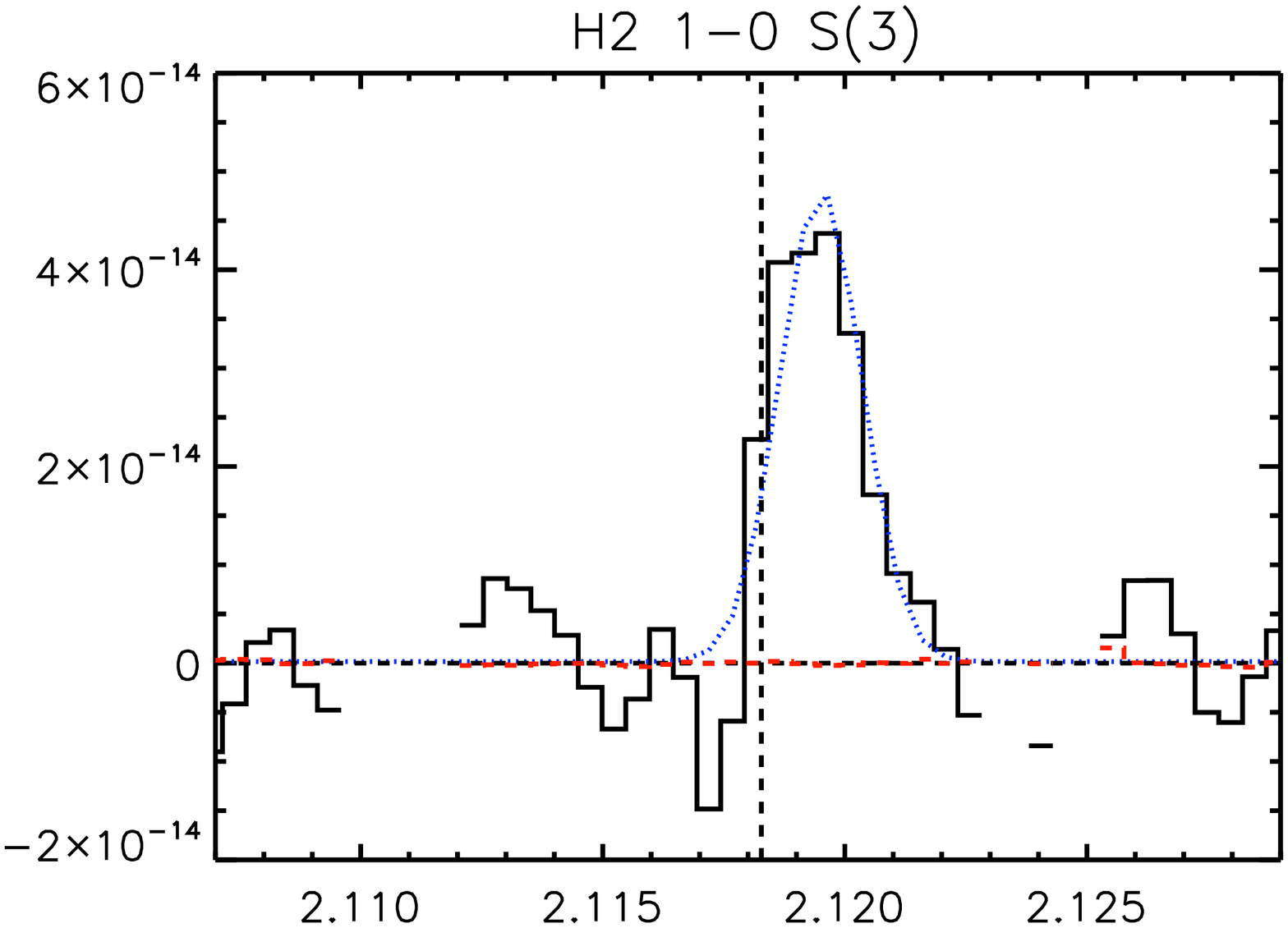}
  \caption{ABELL 2597 Line Spectra (Region A6). For all lines we used a four pixel spatial and spectral smoothing. The lines and symbols used are the same as in Fig. \ref{fig_a2597_line_tb_a1}.}
\end{figure*}

\begin{figure*}
    \includegraphics[width=0.17\textwidth, angle=90]{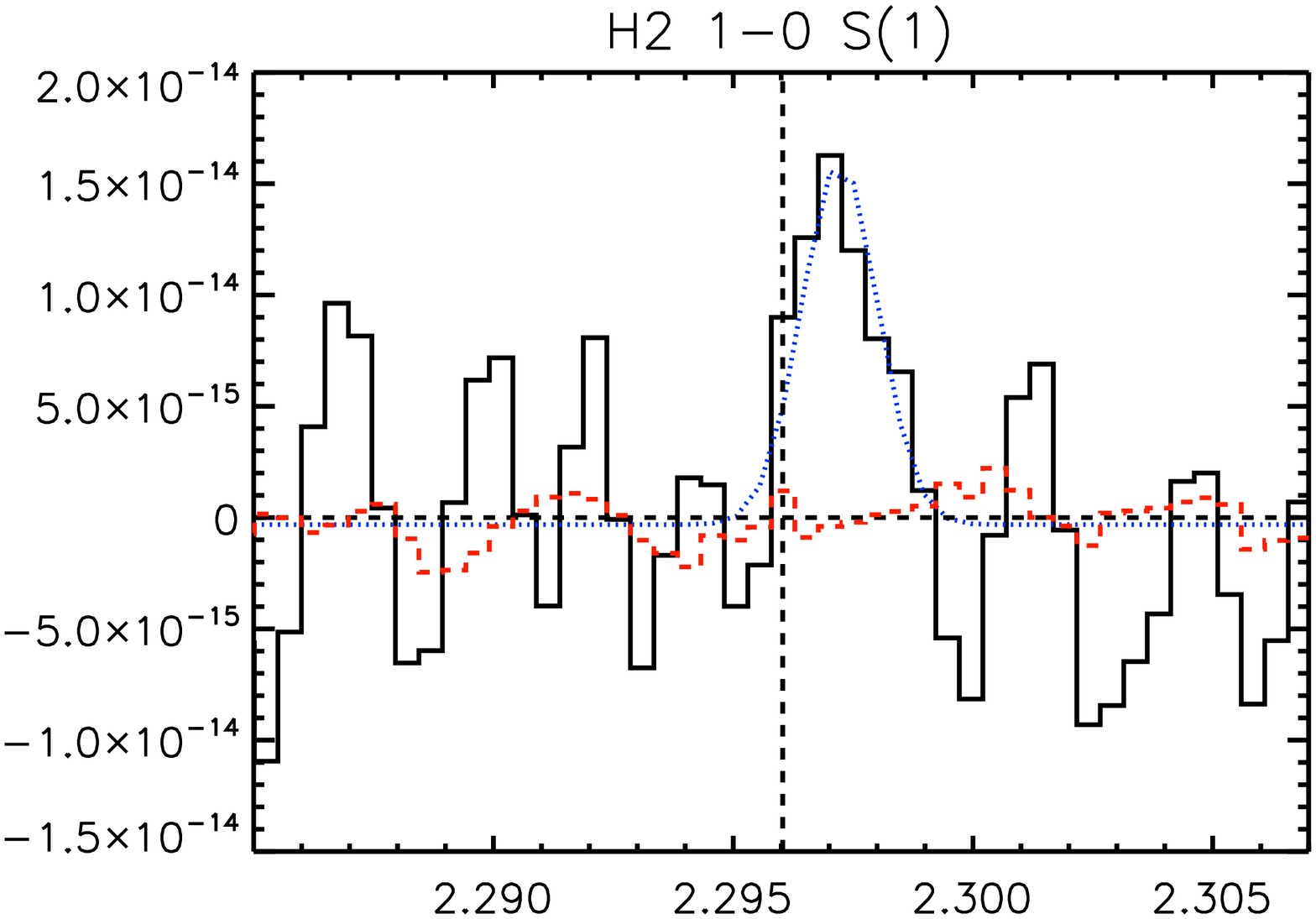}
    \includegraphics[width=0.17\textwidth, angle=90]{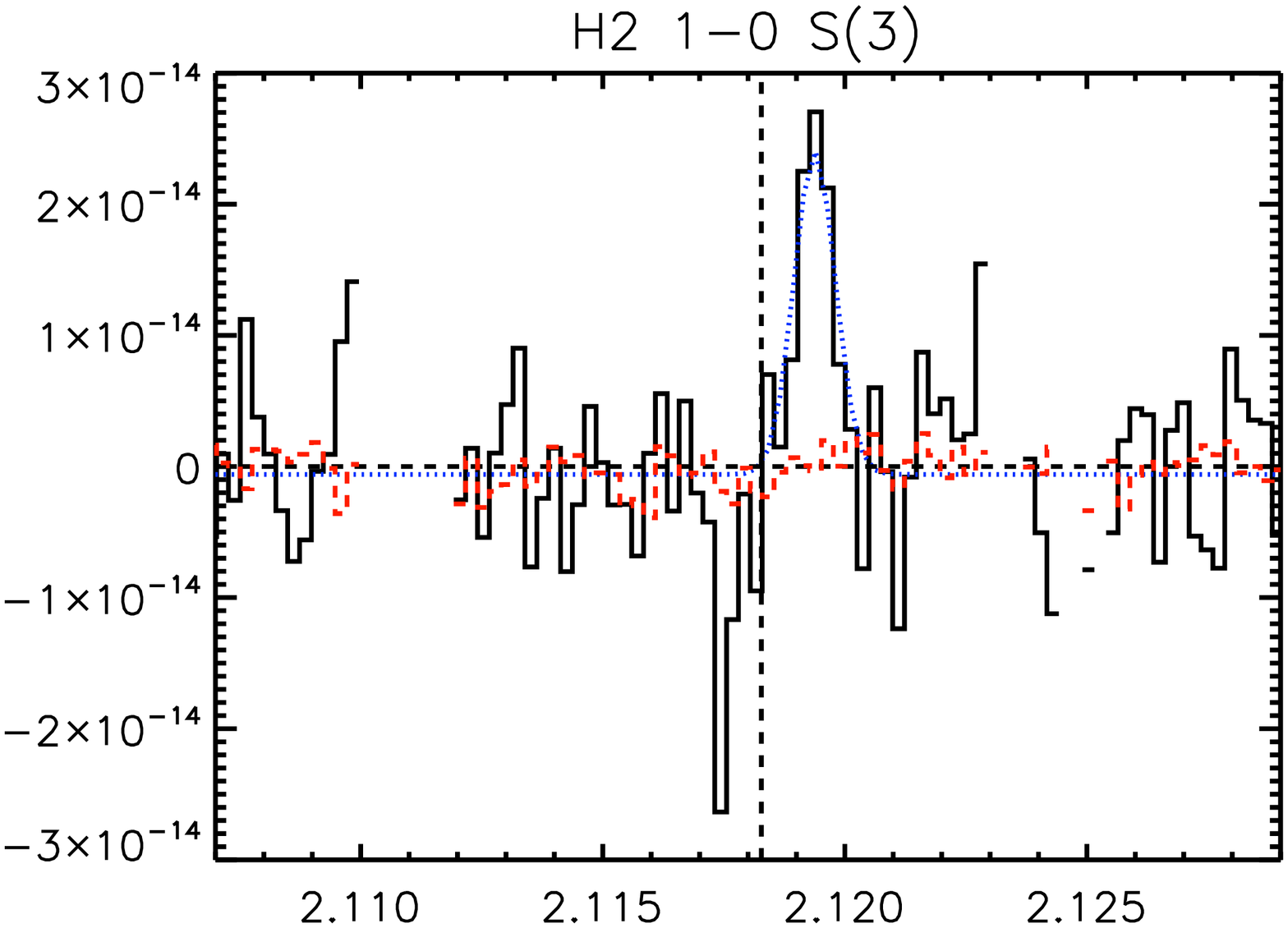}
    \includegraphics[width=0.17\textwidth, angle=90]{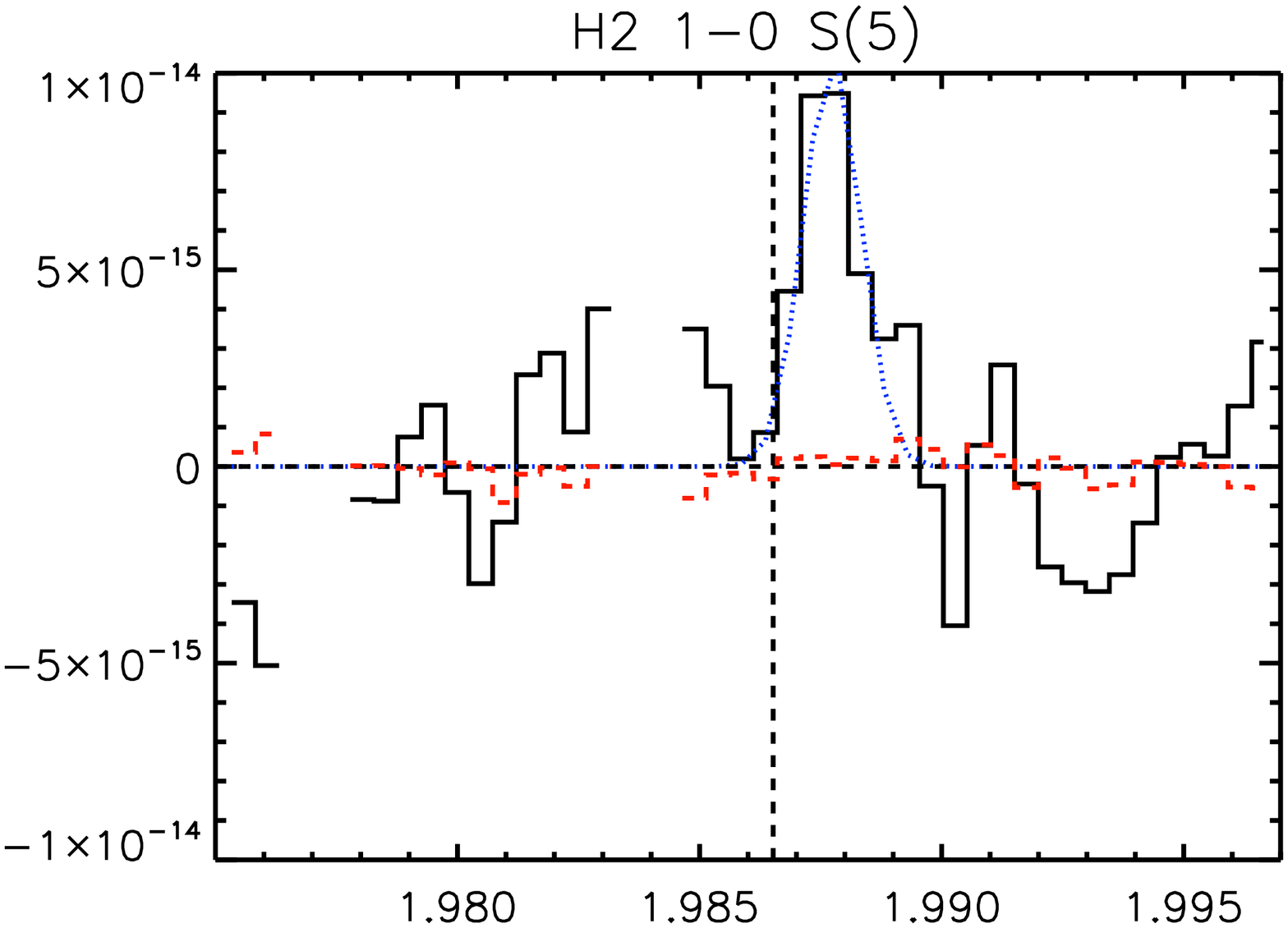}
    \includegraphics[width=0.17\textwidth, angle=90]{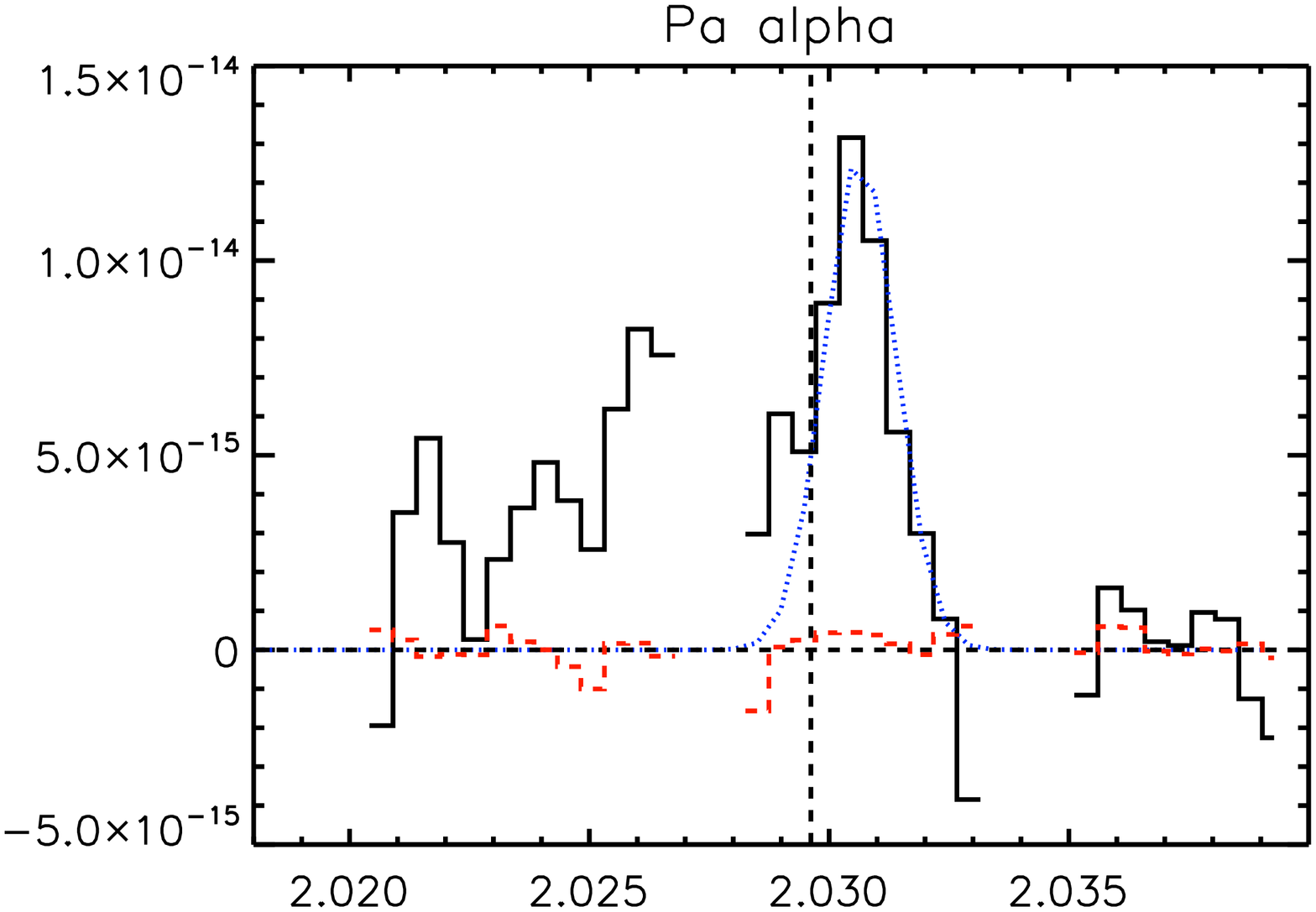}
  \caption{ABELL 2597 Line Spectra (Region A7). For the H$_{\mathrm{2}}$~1-0~S(3) we used a two pixel spatial smoothing. For all other lines a four pixel spatial and spectral smoothing was used. The lines and symbols used are the same as in Fig. \ref{fig_a2597_line_tb_a1}.}
\end{figure*}

\begin{figure*}
    \includegraphics[width=0.17\textwidth, angle=90]{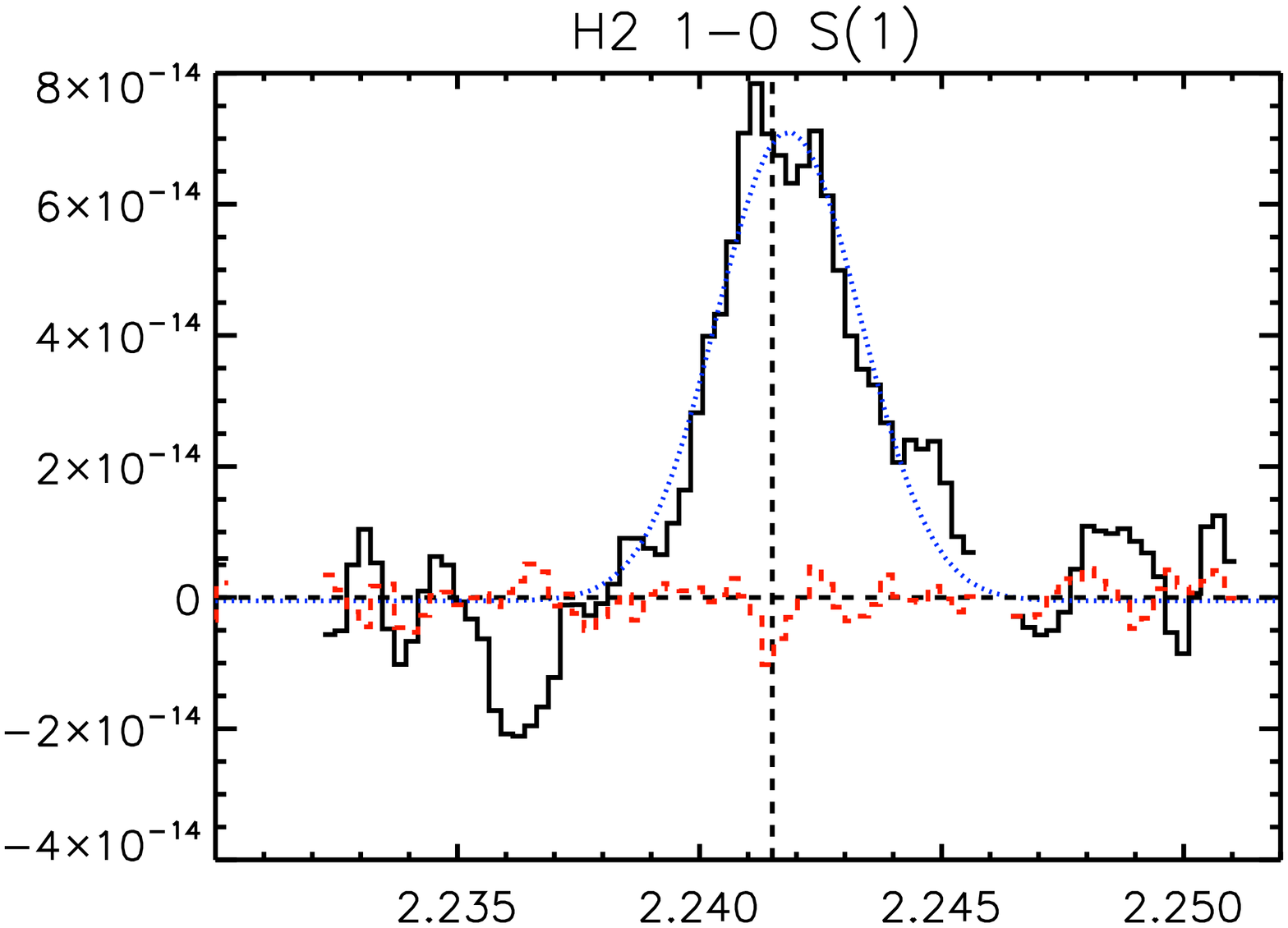}
    \includegraphics[width=0.17\textwidth, angle=90]{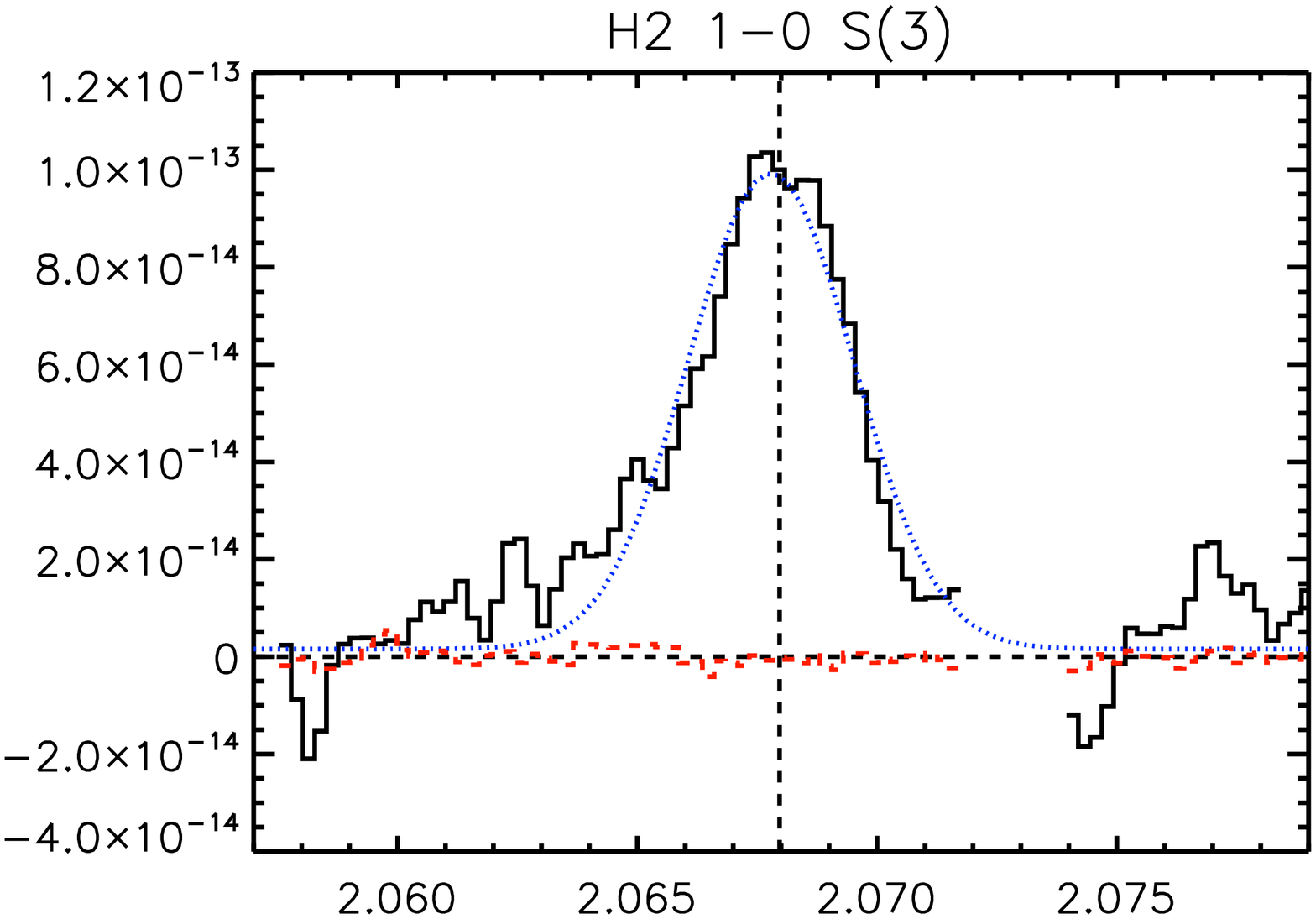}
    \includegraphics[width=0.17\textwidth, angle=90]{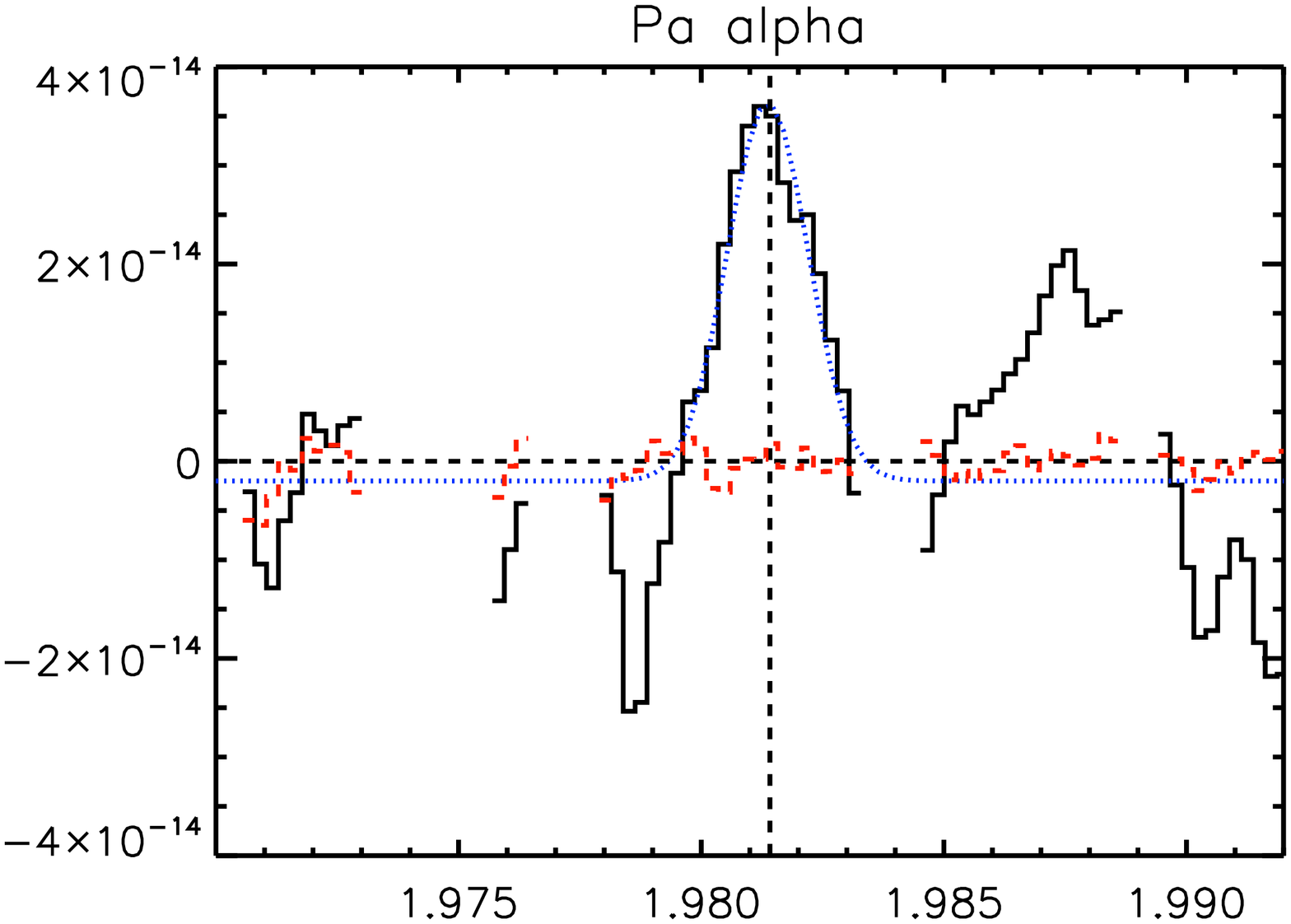}
  \caption{SERSIC 159-03 Line Spectra (Region B1). For all lines a two pixel spatial and spectral smoothing was used. The solid black line shows the measured line spectrum. The dotted blue line shows the Gaussian fit to the line profile. The dashed red line shows the spectrum from the corresponding spatial median region (this spectrum has been scaled to match the area of the selected region).}\label{fig_sersic_line_tb_s1}
\end{figure*}

\begin{figure*}
    \includegraphics[width=0.17\textwidth, angle=90]{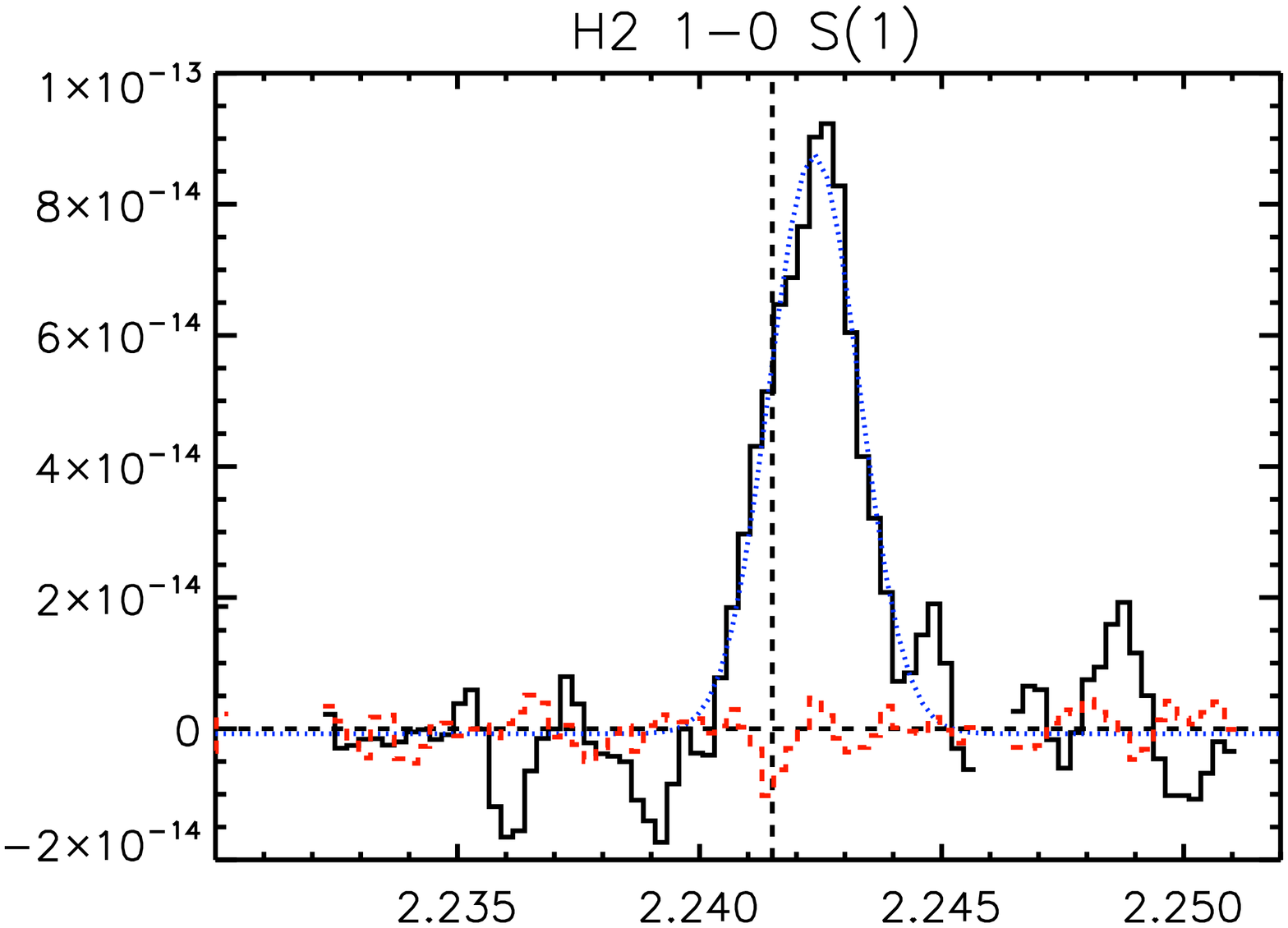}
    \includegraphics[width=0.17\textwidth, angle=90]{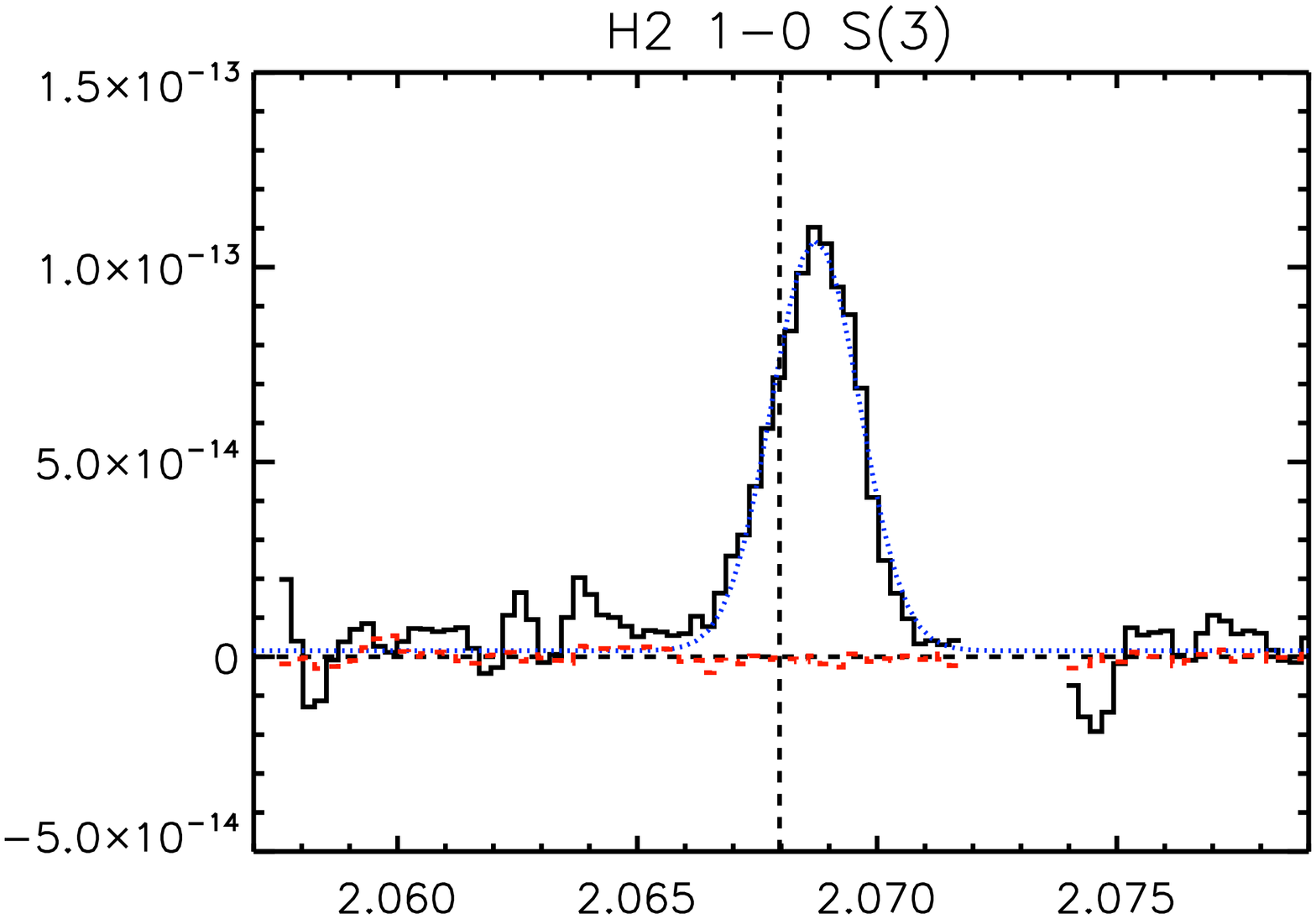}
    \includegraphics[width=0.17\textwidth, angle=90]{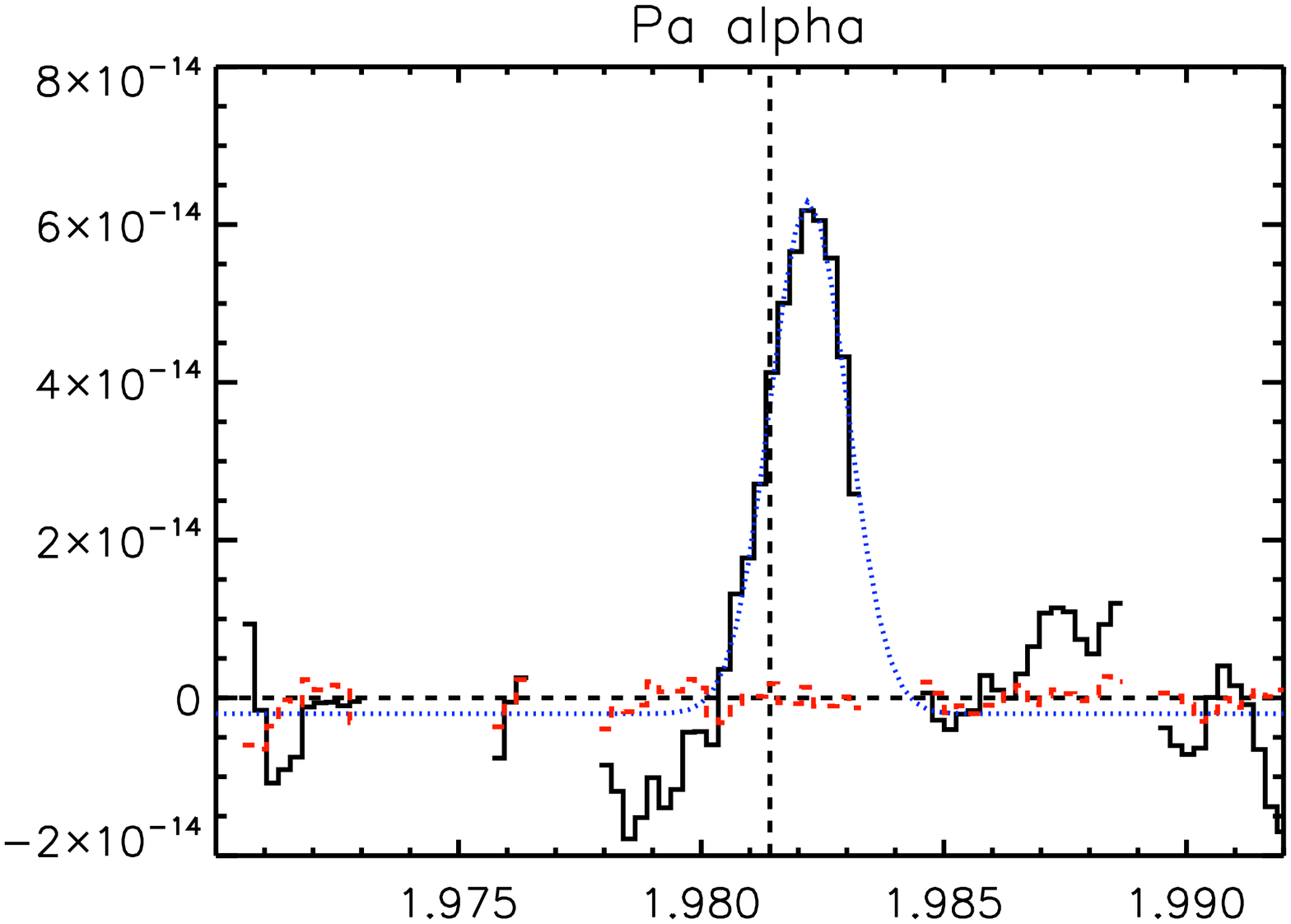}
  \caption{SERSIC 159-03 Line Spectra (Region B2). For all lines a two pixel spatial and spectral smoothing was used. The lines and symbols used are the same as in Fig. \ref{fig_sersic_line_tb_s1}.}
\end{figure*}

\begin{figure*}
    \includegraphics[width=0.17\textwidth, angle=90]{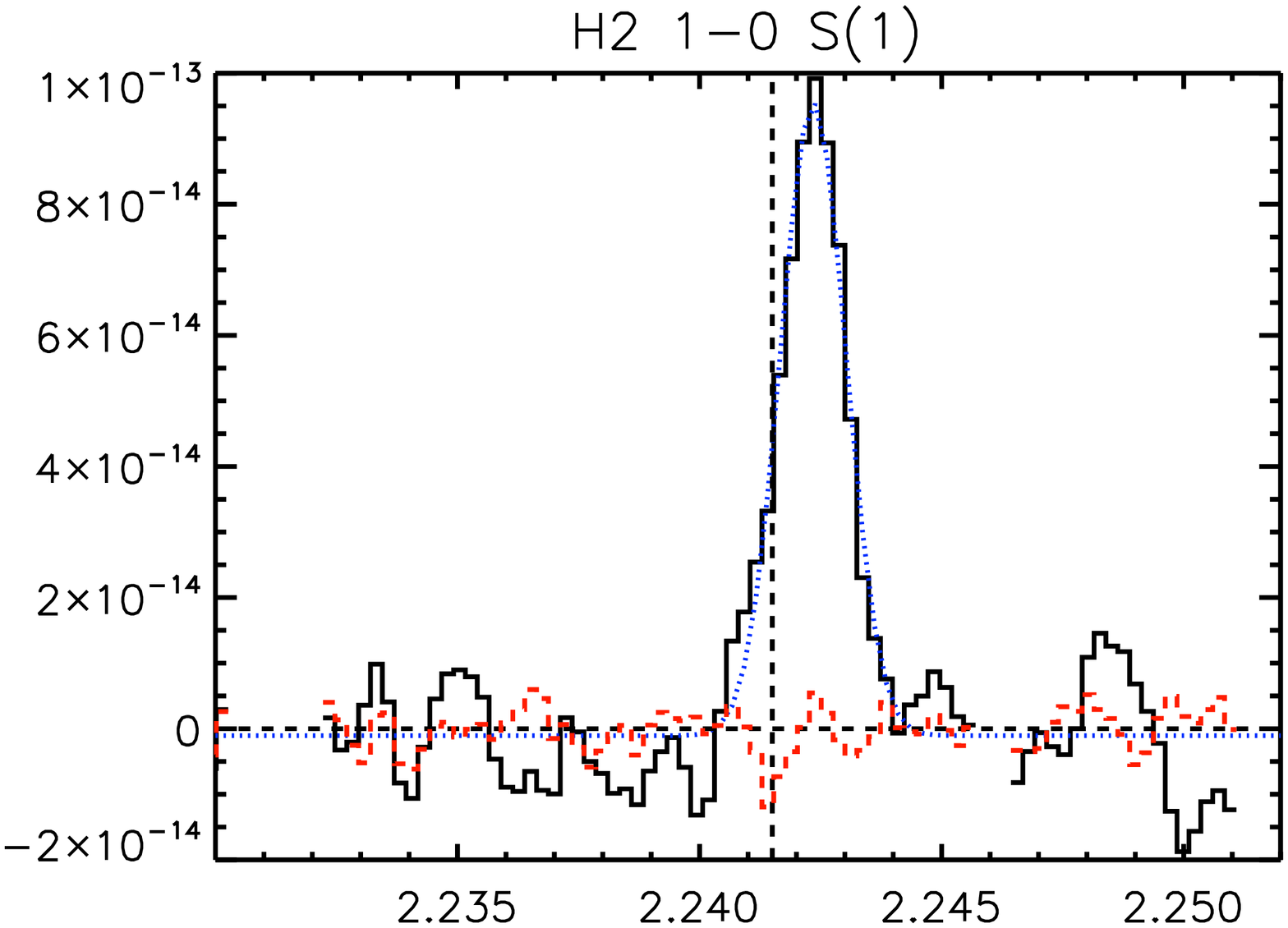}
    \includegraphics[width=0.17\textwidth, angle=90]{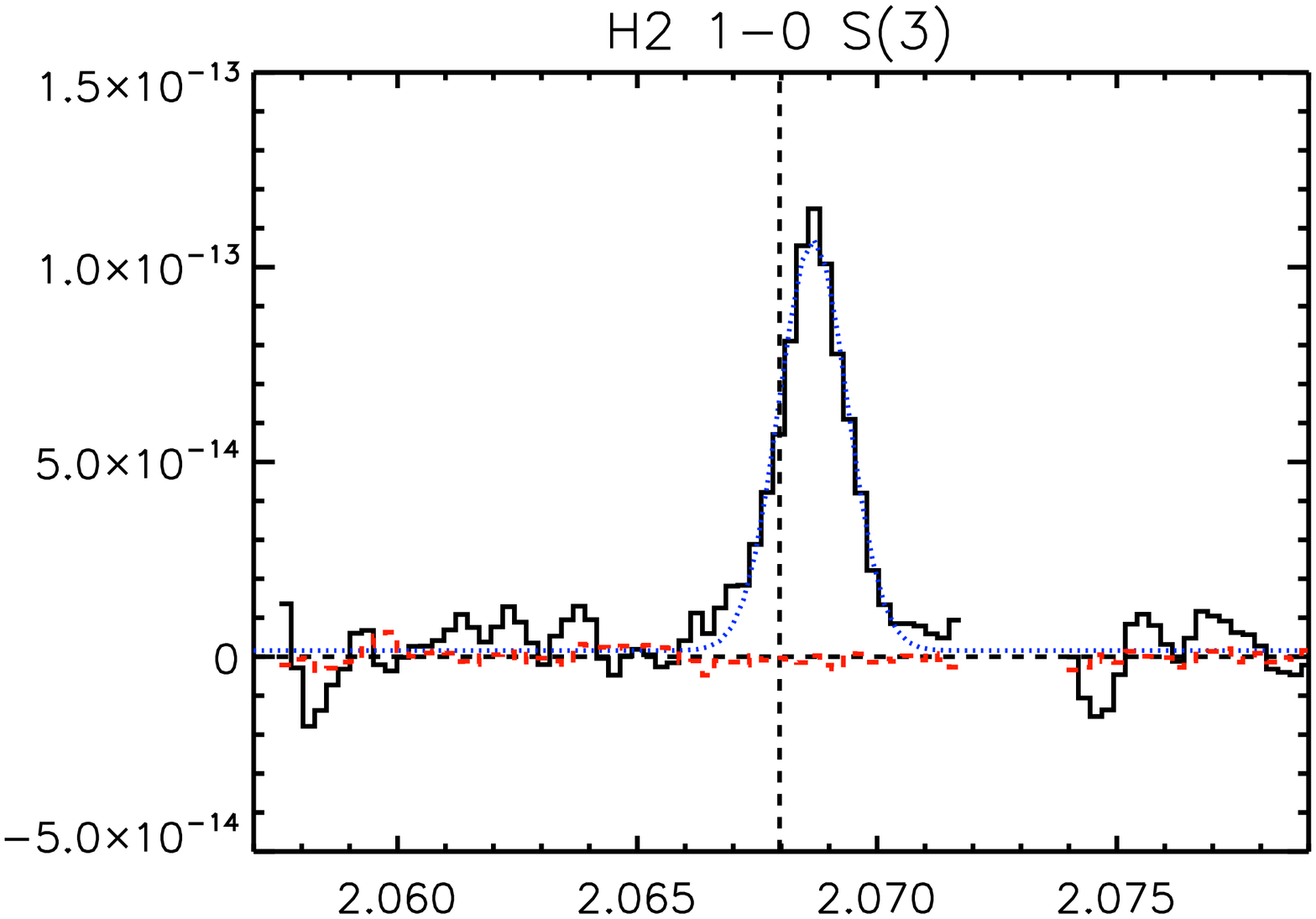}
    \includegraphics[width=0.17\textwidth, angle=90]{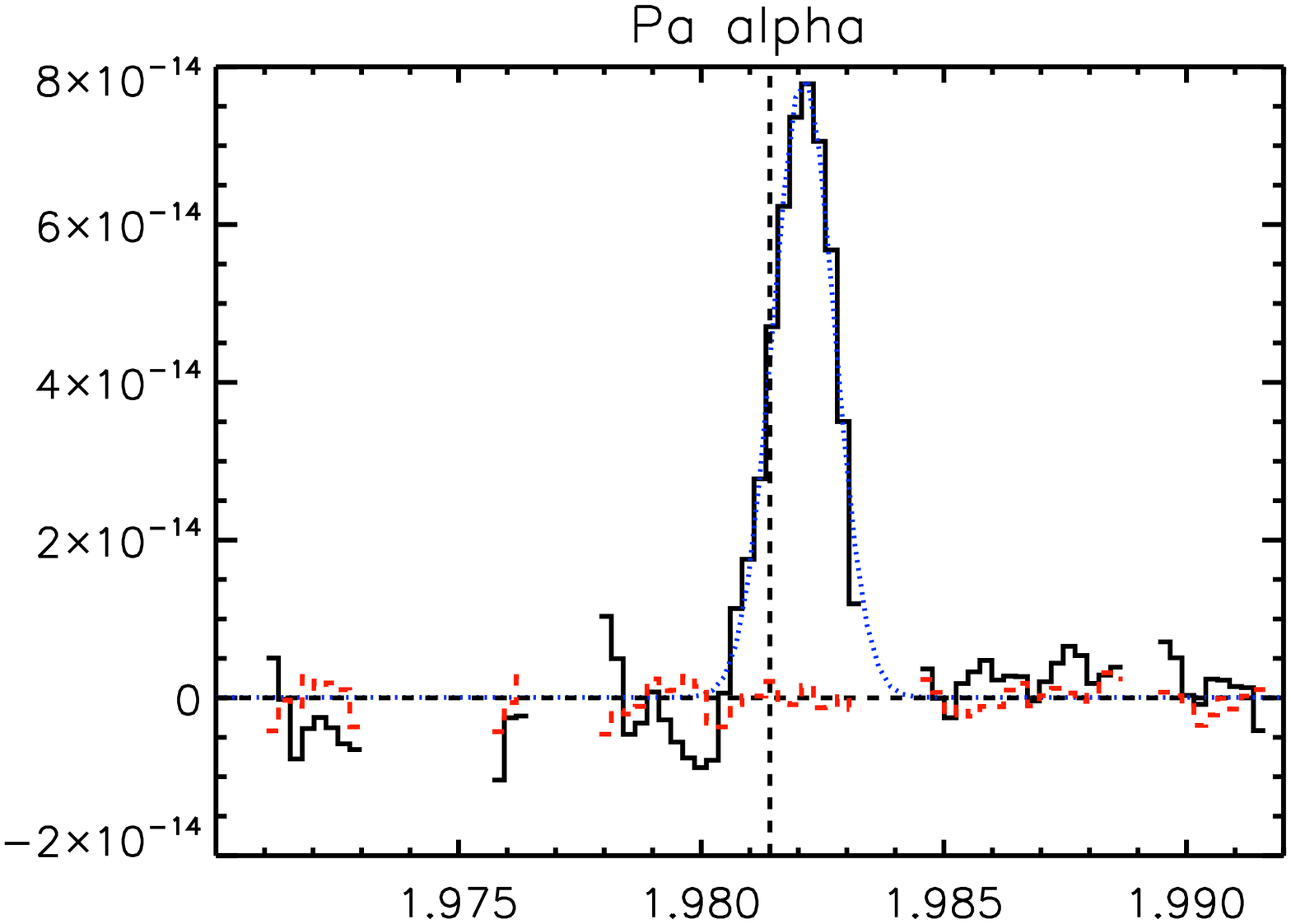}
  \caption{SERSIC 159-03 Line Spectra (Region B3). For all lines a two pixel spatial and spectral smoothing was used. The lines and symbols used are the same as in Fig. \ref{fig_sersic_line_tb_s1}.}
\end{figure*}

\begin{figure*}
    \includegraphics[width=0.17\textwidth, angle=90]{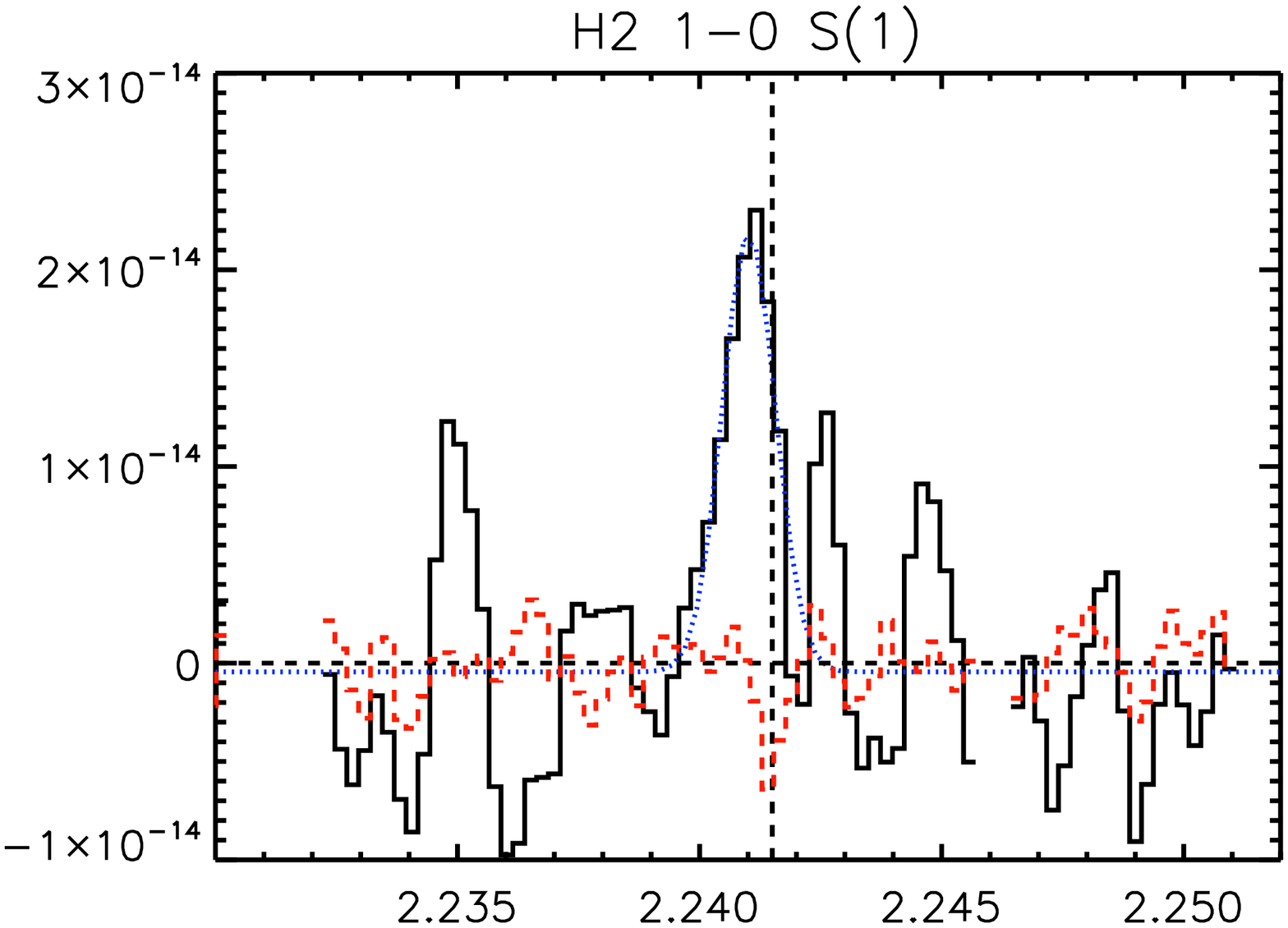}
    \includegraphics[width=0.17\textwidth, angle=90]{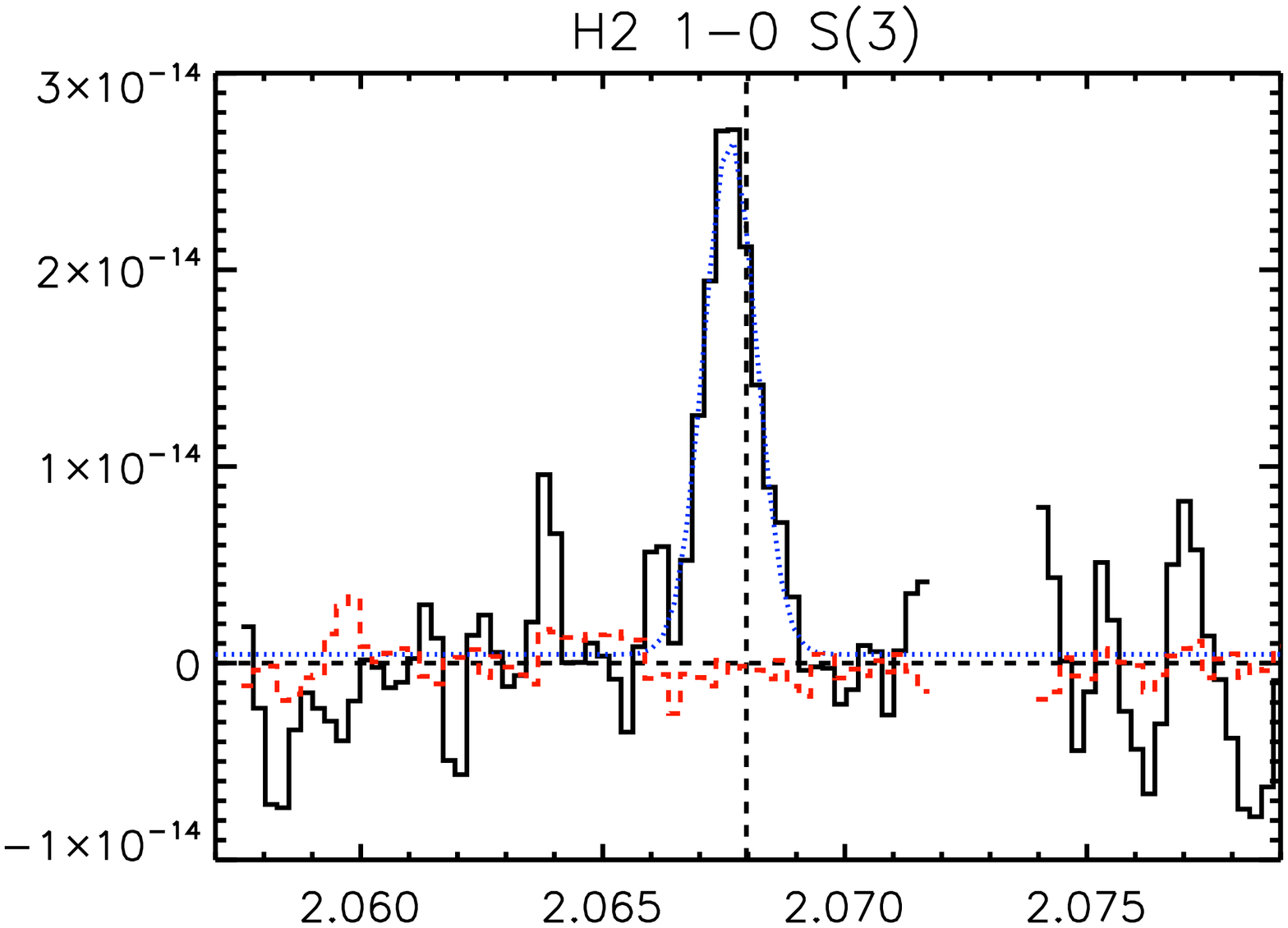}
    \includegraphics[width=0.17\textwidth, angle=90]{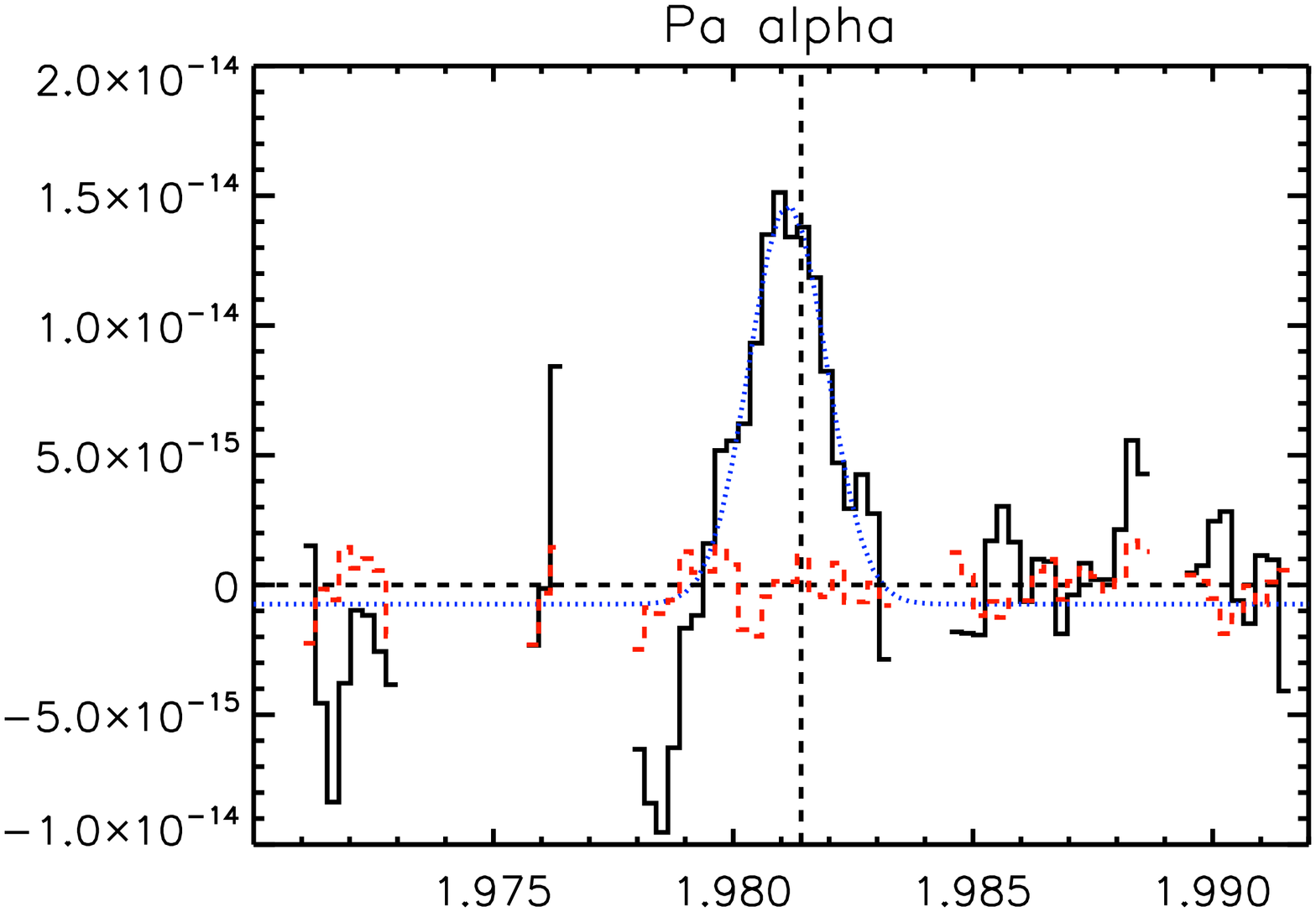}
  \caption{SERSIC 159-03 Line Spectra (Region B4). For all lines a two pixel spatial and spectral smoothing was used. The lines and symbols used are the same as in Fig. \ref{fig_sersic_line_tb_s1}.}
\end{figure*}

\begin{figure*}
    \includegraphics[width=0.17\textwidth, angle=90]{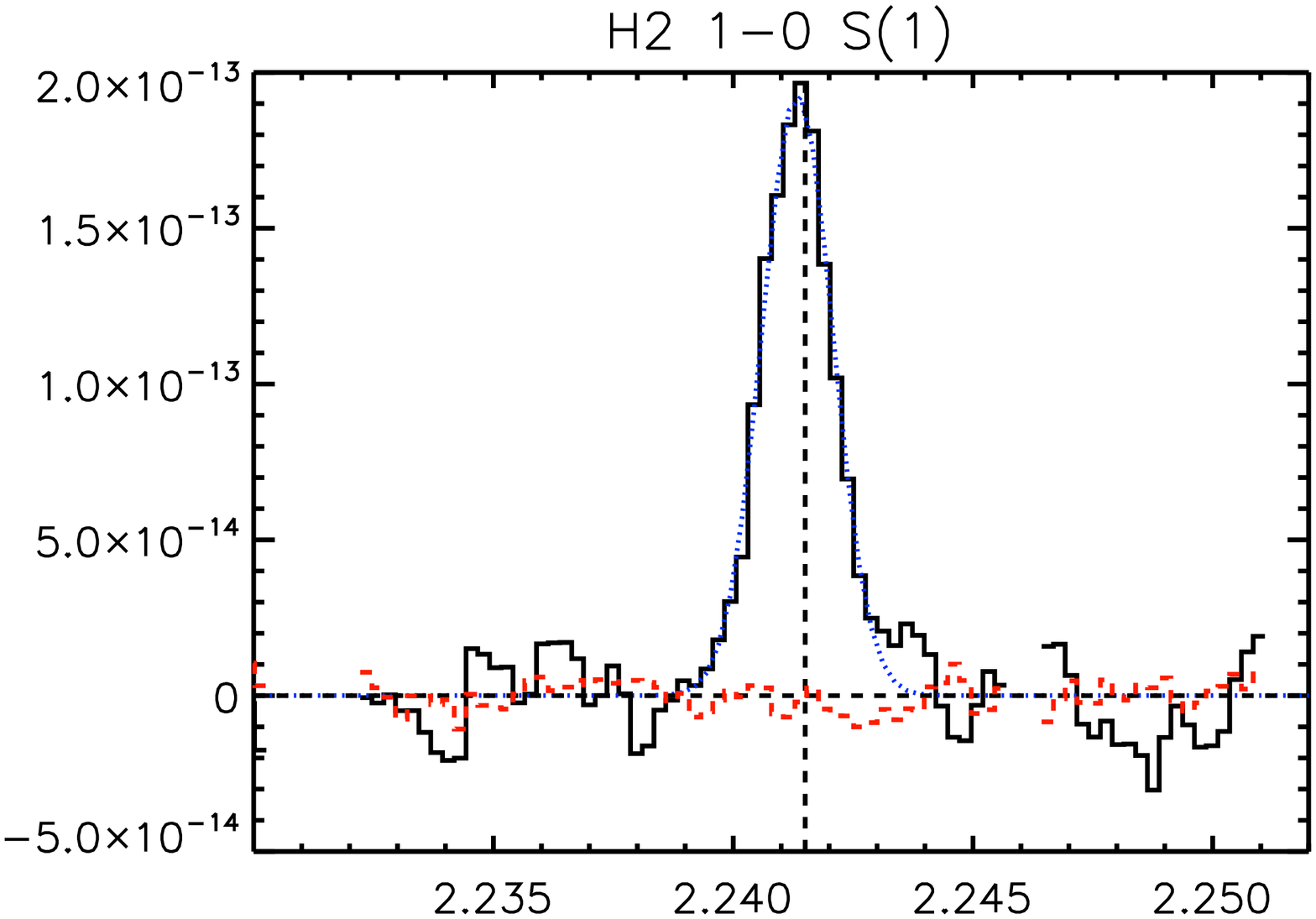}
    \includegraphics[width=0.17\textwidth, angle=90]{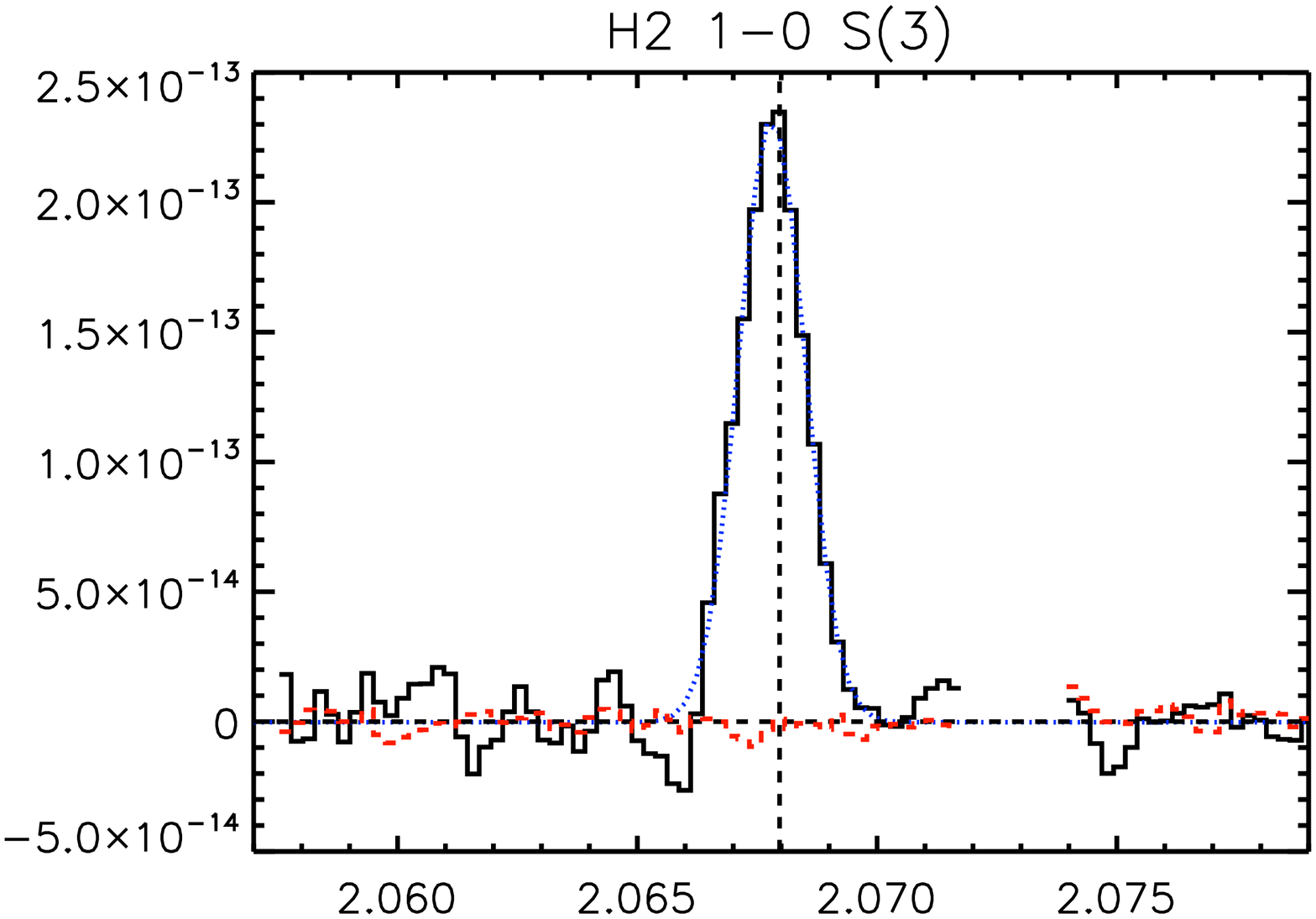}
    \includegraphics[width=0.17\textwidth, angle=90]{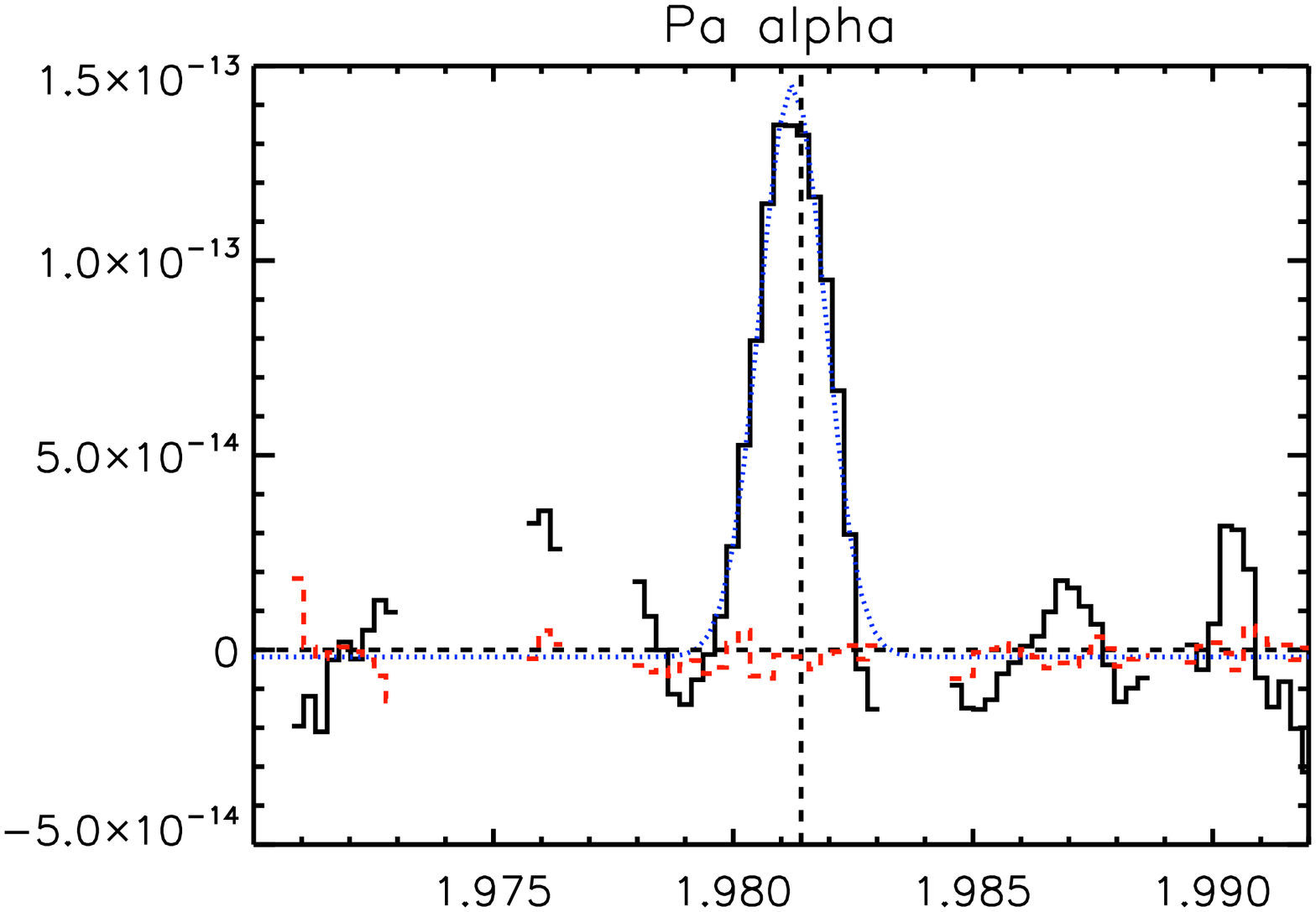}
  \caption{SERSIC 159-03 Line Spectra (Region B5). For all lines a two pixel spatial and spectral smoothing was used. The lines and symbols used are the same as in Fig. \ref{fig_sersic_line_tb_s1}.}
\end{figure*}

\begin{figure*}
    \includegraphics[width=0.17\textwidth, angle=90]{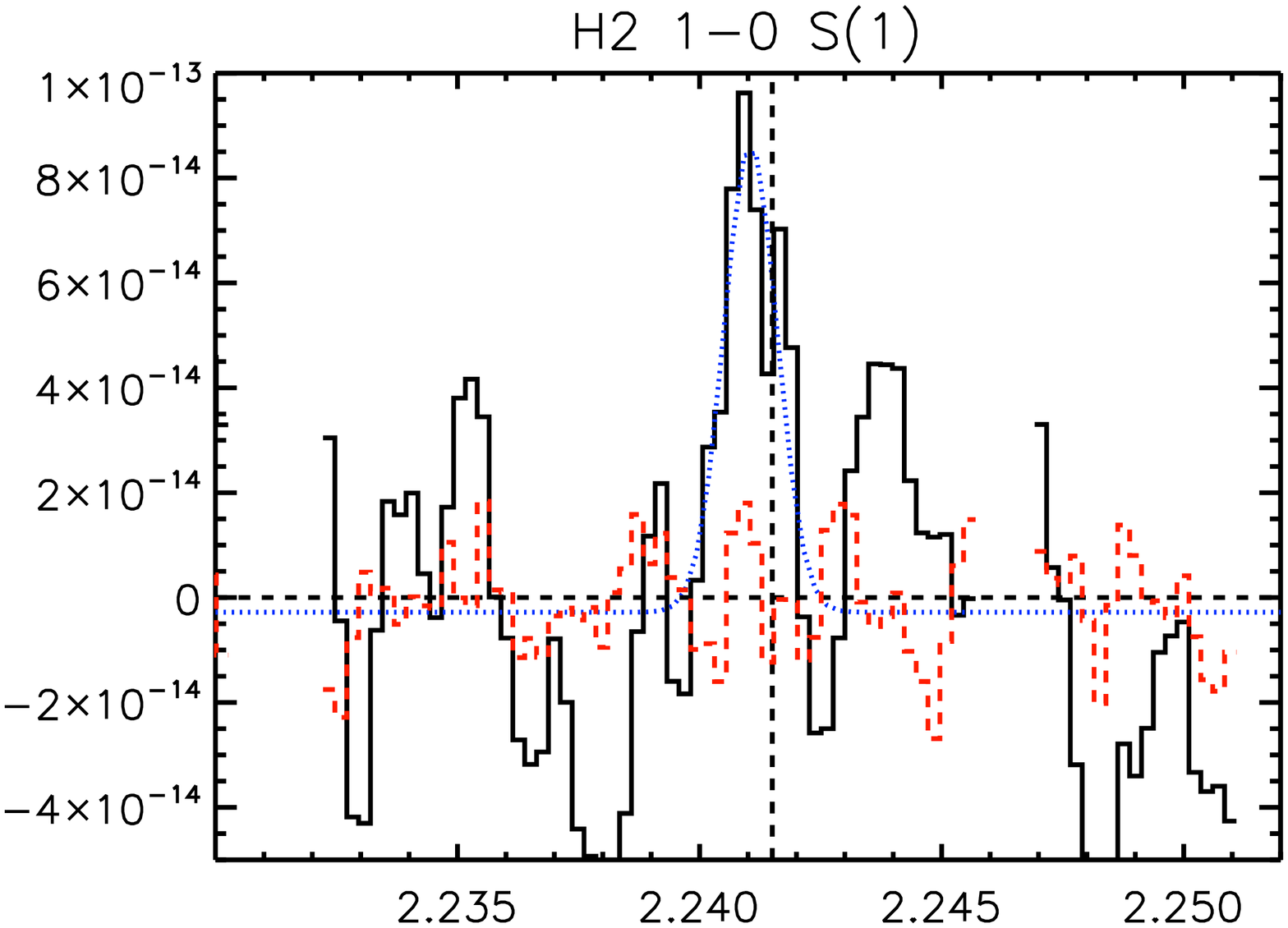}
    \includegraphics[width=0.17\textwidth, angle=90]{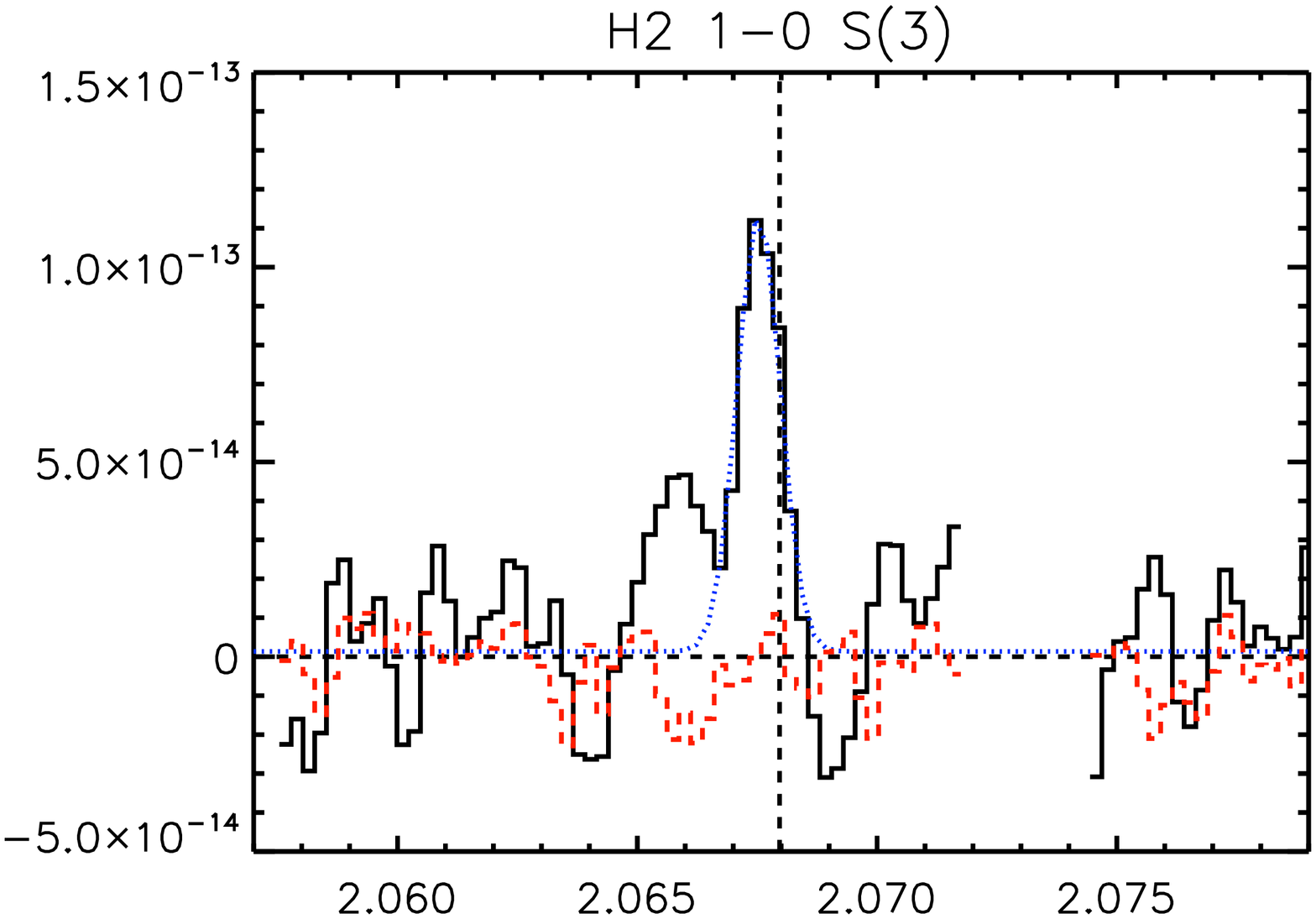}
    \includegraphics[width=0.17\textwidth, angle=90]{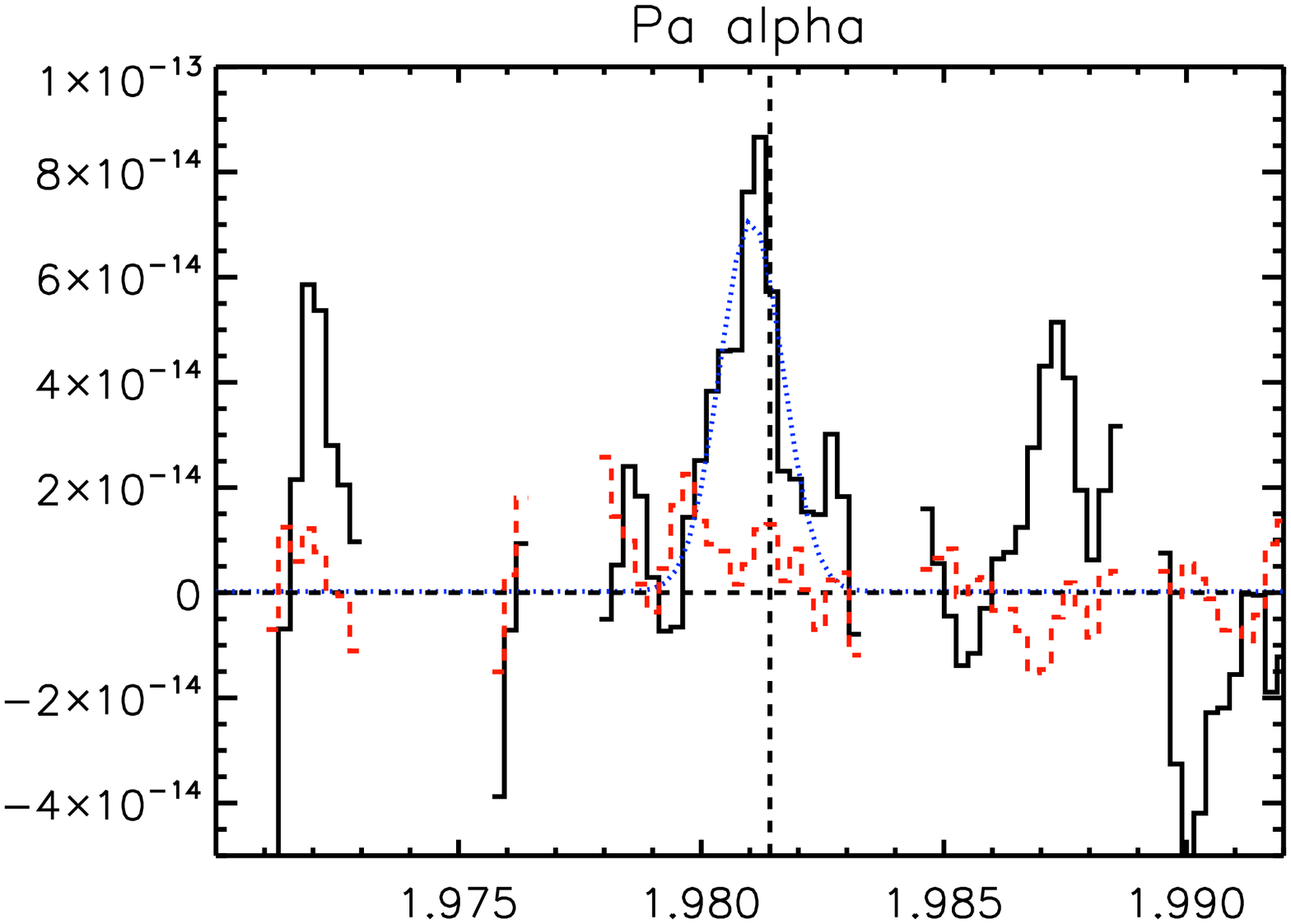}
  \caption{SERSIC 159-03 Line Spectra (Region B6). For all lines a two pixel spatial and spectral smoothing was used. The lines and symbols used are the same as in Fig. \ref{fig_sersic_line_tb_s1}.}
\end{figure*}

\begin{figure*}
    \includegraphics[width=0.17\textwidth, angle=90]{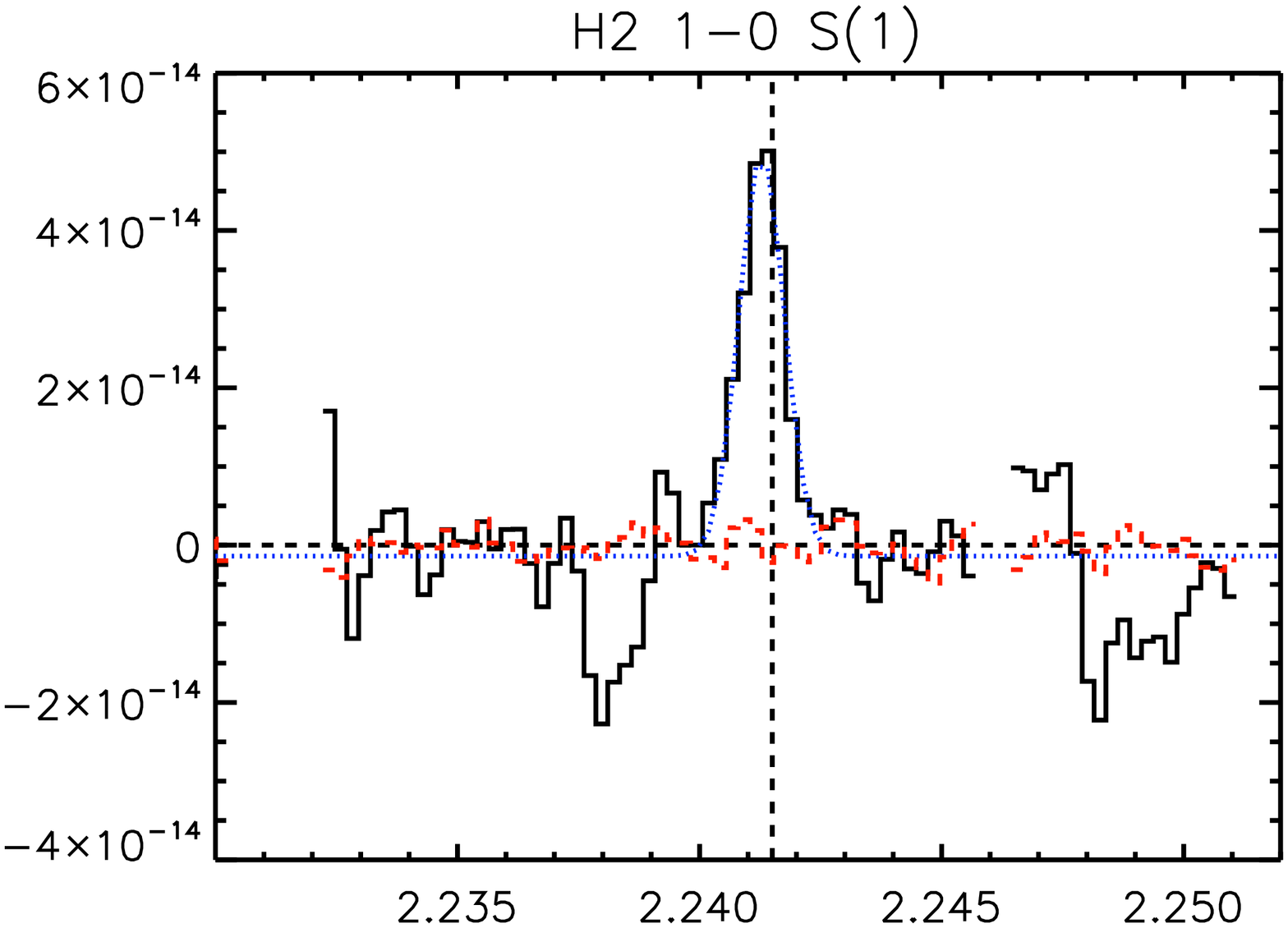}
    \includegraphics[width=0.17\textwidth, angle=90]{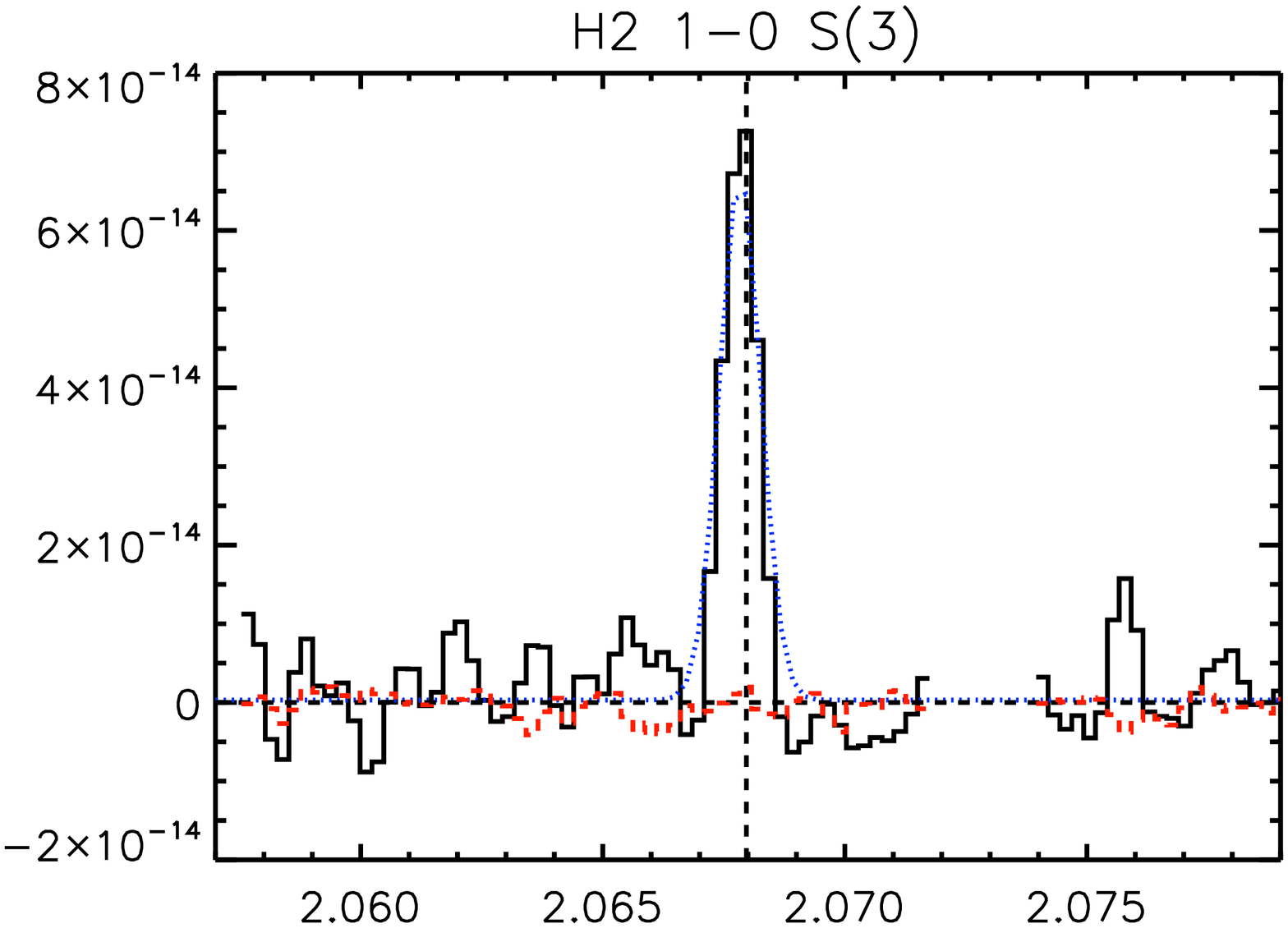}
    \includegraphics[width=0.17\textwidth, angle=90]{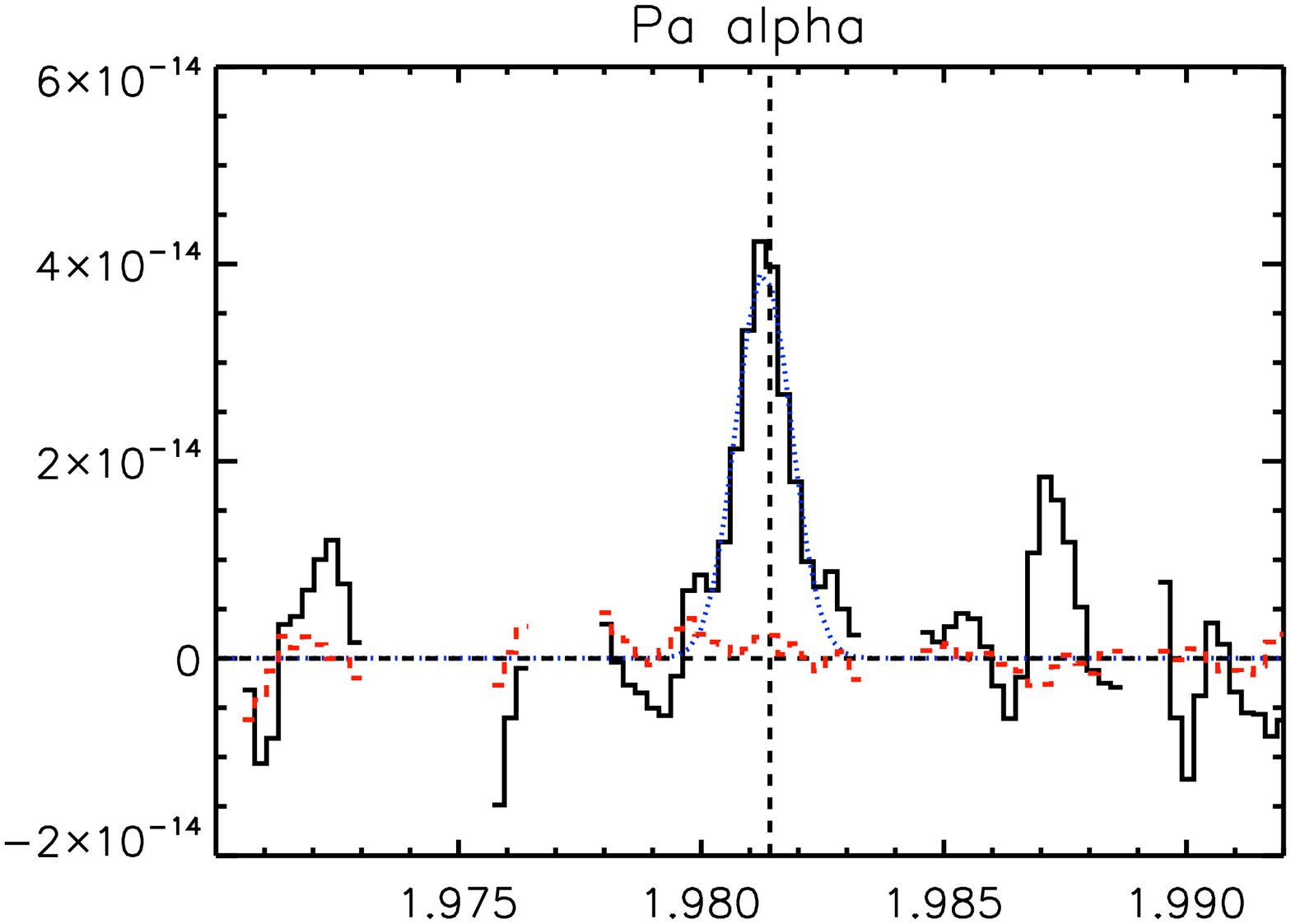}
  \caption{SERSIC 159-03 Line Spectra (Region B7). For all lines a two pixel spatial and spectral smoothing was used. The lines and symbols used are the same as in Fig. \ref{fig_sersic_line_tb_s1}.}
\end{figure*}

\bsp

\label{lastpage}

\end{document}